\documentclass{elsart}
\usepackage{rotating}

\usepackage{epsfig}

\usepackage{amssymb}

\def\be{\begin{equation}}
\def\ee{\end{equation}}
\def\bea{\begin{eqnarray}}
\def\eea{\end{eqnarray}}
\def\deu{{\delta}^{(1)}}
\def\ded{{\delta}^{(2)}}
\def\H{\,{\mathcal  H}}
\def\cal H{\,{\mathcal  H}}

\newcommand\bfx{{\bf x}}

\def\Aa{\frac{a^{\prime}}{a}}
\def\Ab{\Big(\frac{a^{\prime}}{a}\Big)^2}
\def\Ac{\frac{a^{\prime\prime}}{a}}

\def\deu{{\delta}^{(1)}}
\def\ded{{\delta}^{(2)}}

\def\wpsi{\widetilde{\Psi}}
\def\simlt{\stackrel{<}{{}_\sim}}
\def\simgt{\stackrel{>}{{}_\sim}}
\def\R{{\mathcal R}}
\def\S{{\mathcal S}}
\def\C{{\mathcal C}}
\def\P{{\mathcal P}}
\def\La{\nabla^2}
\def\LA{\nabla^2}

\def\mycaption#1{
\vskip 0.1truecm 
\rightskip=3truepc
\leftskip=3truepc
\baselineskip=13pt
\noindent{\small#1}
}

\begin{document}

\begin{frontmatter}


\title{Non--Gaussianity from Inflation:\\ Theory and Observations}


\author{N. Bartolo$^{(1)}$, E. Komatsu$^{(2)}$, S. Matarrese$^{(3,4)}$
and A. Riotto$^{(4)}$}

\address{$^{(1)}$Astronomy Centre, University of Sussex
Falmer, Brighton, BN1 9QH, U.K.}
\address{$^{(2)}$ Department of Astronomy, The University of Texas
at Austin,\\ Austin, TX 78712, USA}
\address{$^{(3)}$ Dipartimento di Fisica ``G. Galilei'', Universit\`a
di Padova, \\
via Marzolo 8, I--35131 Padova, Italy}
\address{$^{(4)}$ INFN,
Sezione di Padova, via Marzolo 8, I--35131 Padova, Italy}

\tableofcontents

\begin{abstract}
This is a review of models of inflation and of their predictions for the 
primordial non--Gaussianity in the density perturbations which are thought 
to be at the origin of structures in the Universe. Non--Gaussianity emerges 
as a key observable to discriminate among competing scenarios for the 
generation of cosmological perturbations and is one of the primary targets of 
present and future Cosmic Microwave Background satellite missions. 
We give a detailed presentation of 
the state--of--the--art of the subject of non--Gaussianity, both from the 
theoretical and the observational point of view, and provide all the tools 
necessary to compute at second order in perturbation theory the level of 
non--Gaussianity in any model of cosmological perturbations. We discuss the 
new wave of models of inflation, which are firmly rooted in modern particle 
physics theory and predict a significant amount of non--Gaussianity.
The review is addressed to both astrophysicists and particle physicists and 
contains useful tables which summarize the theoretical and observational 
results regarding non--Gaussianity.
\end{abstract}

\end{frontmatter}

\newpage


$$
$$
$$
$$
$$
$$
$$
$$
\begin{flushright}
{\large ``... {\it the linear perturbations are so surprisingly
simple that a perturbation analysis to second order
may be feasible} ...'' }
\end{flushright}

\vskip 1cm
\begin{flushright}
{\large (Sachs and Wolfe 1967)}
\end{flushright}

\newpage

\section{\bf Introduction}

One of the relevant ideas in modern cosmology is represented by the 
inflationary paradigm. It is widely belevied that there was an early epoch 
in the history of the Universe -- 
before the epoch of primordial nucleosynthesis -- when the 
Universe expansion was accelerated. Such a period of 
\emph{cosmological inflation} can be attained if the energy density of 
the Universe is dominated by the vacuum energy density associated with the 
potential of a scalar field $\varphi$, called the inflaton field. Through 
its kinematic properties, namely the acceleration of the Universe, the 
inflationary paradigm can elegantly solve the flatness, the horizon and the 
monopole problems 
of the standard Big--Bang cosmology, and in fact the first model of 
inflation by Guth in 1981~\cite{guth81} 
was introduced to address such problems. 
However all over the years 
inflation has become so popular also because 
of another compelling feature. It 
can explain the production of the first density perturbations in the early 
Universe which are the seeds for the Large--Scale Structure (LSS) 
in the distribution of galaxies 
and the underlying dark matter and for the Cosmic Microwave Background (CMB) 
temperature anisotropies that we observe today. 
In fact inflation has become the dominant paradigm to understand the 
initial conditions for structure formation and CMB anisotropies. 
In the inflationary picture, 
primordial density and gravity--wave fluctuations are
created from quantum fluctuations ``redshifted'' out of the horizon during an
early period of superluminal expansion of the Universe, where they
are ``frozen'' 
\cite{muk81,guth82,hawking82,Linde1982,starobinsky82,BardeenSteinTur}. 
Perturbations at the surface of last scattering are observable as temperature 
anisotropy in the CMB, which was first detected by the Cosmic Background 
Explorer ({\sl COBE}) satellite~\cite{smoot92,bennett/etal:1996,gorski96}.
The last and most impressive confirmation of the inflationary paradigm has 
been recently provided by the data 
of the Wilkinson Microwave Anisotropy Probe 
({\sl WMAP}) mission~\cite{bennett/etal:2003b}.
The {\sl WMAP} collaboration has produced a full--sky map of the angular 
variations 
of the CMB, with unprecedented accuracy.
{\sl WMAP} data confirm the inflationary mechanism as responsible for the
generation of curvature (adiabatic) superhorizon fluctuations \cite{ex}. 

Since the primordial cosmological perturbations are tiny, the generation
and evolution of fluctuations during inflation has been studied 
within linear perturbation theory. Within this approach, 
the primordial density perturbation is 
Gaussian; in other words, its Fourier components
are uncorrelated and have random phases.
Despite the simplicity of the inflationary paradigm, the mechanism
by which  cosmological adiabatic perturbations are generated  is not
yet established. In the standard slow--roll scenario associated
to one--single field models of inflation, the observed density 
perturbations are due to fluctuations of the inflaton field itself when it
slowly rolls down along its potential. 
When inflation ends, the inflaton $\varphi$ oscillates about the minimum of its
potential $V(\varphi)$  and decays, thereby reheating the Universe. 
As a result of the fluctuations
each region of the Universe goes through the same history but at slightly
different times. The final temperature anisotropies are caused by 
inflation lasting for different amounts of time in 
different regions of the Universe
leading to adiabatic perturbations. Under this hypothesis, 
the {\sl WMAP} dataset already allows
to extract the parameters relevant 
for distinguishing among single--field inflation models~\cite{ex,ex2}.

An alternative to the standard scenario is represented by the curvaton 
mechanism~\cite{Mollerach,curvaton1,LW2,curvaton3,LUW}
 where the final curvature perturbations
are produced from an initial isocurvature perturbation associated with the
quantum fluctuations of a light scalar field (other than the inflaton), 
the curvaton, whose energy density is negligible during inflation. The 
curvaton isocurvature perturbations are transformed into adiabatic
ones when the curvaton decays into radiation much after the end 
of inflation. 

Recently, other  mechanisms for the generation of cosmological
perturbations have been proposed, the inhomogeneous reheating scenario 
\cite{gamma1,gamma2,gamma3,MatRio,allahverdi}, the ghost inflationary scenario
\cite{ghost} and the D--cceleration scenario~\cite{tong2}, just to mention 
a few. For instance, the inhomogeneous reheating scenario acts during the 
reheating stage after inflation if superhorizon spatial fluctuations in the 
decay rate of the inflaton field are induced during inflation, causing 
adiabatic perturbations in the final reheating temperature
in different regions of the Universe.

The generation of gravity--wave fluctuations 
is a generic prediction of an accelerated  de Sitter expansion of the Universe
whatever mechanism for the generation of cosmological
perturbations is operative. 
Gravitational waves, whose possible observation might come from the  
detection of  the $B$-mode of polarization in the
CMB anisotropy~\cite{polreview1,polreview2},   
may be viewed as  ripples of space--time around the  background metric.

Since curvature  fluctuations are (nearly)
frozen on superhorizon scales,
a way of characterizing them is to compute
their spectrum on scales larger than the horizon. 
In the standard slow--roll inflationary models where the fluctuations 
of the inflaton field
$\varphi$ are responsible for the curvature perturbations, 
the power--spectrum $\P_\R$ of the comoving curvature
perturbation $\R$ (which is a measure of the spatial curvature
as seen by comoving observers)  is given by 

\begin{equation}
\P_{\R}(k)=
\frac{1}{2 M_{\rm P}^2\epsilon}\left(\frac{H_*}{2\pi}\right)^2
\left(\frac{k}{a H_*}\right)^{n_\R-1}\, ,
\end{equation}
where 
$n_\R=1-6\epsilon+2\eta\simeq 1$ is the spectral index, 
$M_{\rm P} \equiv (8\pi G_{\rm N})^{-1/2}\simeq 
2.4\times 10^{18}$ GeV is the reduced Planck scale. Here
\begin{eqnarray}
\label{potter1}
\epsilon &=& \frac{M_{\rm P}^2}2 
\left(\frac{V^\prime}{V}\right)^2 \, ,\nonumber \\
\eta     &=& M_{\rm P}^2 
\left(\frac{V^{\prime\prime}}{V}\right)
\end{eqnarray}
are the so--called slow--roll parameters ($\epsilon,\eta\ll 1$ during 
inflation), $H_*=\dot a/a$ indicates the Hubble rate during inflation
and primes here denote derivatives with respect to $\varphi$. 
The WMAP has determined the amplitude of the power--spectrum
as $\P_{\R}(k)\simeq 2.95\times 10^{-9}A$ 
where $A=0.6-1$ depending on the model under consideration 
\cite{ex,spergel/etal:2003}, which implies that
\be
\frac{1}{2 M_{\rm P}^2\epsilon}\left(\frac{H_*}{2\pi}\right)^2
 \simeq (2-3)\times 10^{-9},
\ee
or
\be
H_* \simeq (0.9-1.2)\times 10^{15}~\epsilon^{1/2}~{\rm GeV}.
\ee
The Friedmann equation in the slow--roll limit, $H^2=V/(3M_{\rm P}^2)$,
then gives ``the energy scale of inflation'',
\be
\label{potter}
V^{1/4} \simeq (6.3-7.1)\times 10^{16}~\epsilon^{1/4}~{\rm GeV}.
\ee

On the other hand, the power--spectrum of gravity--wave modes $h_{ij}$ 
is given by
\be
\nonumber
\P_{T}(k)= \frac{k^3}{2\pi^2}\langle h^*_{ij}h^{ij} \rangle
= \frac{8}{M_{\rm P}^2}\left(\frac{H_*}{2\pi}\right)^2
\left(\frac{k}{a H_*}\right)^{n_T},
\ee
where $n_T = -2\epsilon$ is the tensor spectral index.
Since the fractional change of the power--spectra with scale is much 
smaller than unity, one can safely 
consider the power--spectra as being roughly constant on the scales relevant 
for the CMB anisotropy and define a tensor--to--scalar amplitude ratio 

\begin{equation}
\label{q}
r = \frac{\P_{T}}{\P_\R}=16 \epsilon\, .
\end{equation}
The spectra $\P_\R(k)$ and $\P_T(k)$
provide the contact between theory and observation. 
The present {\sl WMAP} dataset  allows
 to extract an upper bound, $r<1.28$ (95\%) \cite{ex,ex2},
or $\epsilon<0.08$. 
This limit together with Eq.~(\ref{potter}) 
 provides an upper bound on the energy scale of inflation,

\begin{equation}
\label{cobe}
 V^{1/4}< 3.8\times 10^{16}\,{\rm GeV}\, .
\end{equation}
The corresponding upper bound on the Hubble rate
during inflation is $H_*<3.4\times 10^{14}$ GeV.
A positive detection of the $B$--mode in CMB polarization, and therefore
an indirect evidence of gravitational waves from inflation, 
once foregrounds due to gravitational lensing from local sources have been 
properly treated, requires $\epsilon > 10^{-5}$ corresponding to 
$V^{1/4}> 3.5\times 10^{15}$ GeV and $H_*>3\times 10^{12}$ 
GeV \cite{r1,r2,r3}. \footnote{If ``cleaning'' of the gravitational lensing 
effect can be achieved down to the level envisaged in Ref.~\cite{r3},   
then another source of $B$--mode polarization will   
limit our ability to detect the signature of primordial gravitational waves. 
This comes from vector and tensor modes arising from the second--order 
evolution of scalar perturbations~\cite{MHarM} and represents the 
ultimate barrier to gravitational--wave detection if $\epsilon<10^{-7}$.} 

However, {\it what if} the curvature perturbation is generated through
the quantum fluctuations of a scalar field other than the inflaton? 
Then, what is the expected amplitude of gravity--wave fluctuations in 
such scenarios? Consider,
for instance, the curvaton scenario and the inhomogeneous reheating 
scenario. They liberate the inflaton from the responsibility
of generating the cosmological curvature perturbation
and therefore  avoid slow--roll conditions. Their
basic assumption  is that the initial
curvature perturbation due to the inflaton field is negligible. 
The common lore to achieve such a 
condition is to assume that the energy scale of the inflaton potential
is too small to match the observed amplitude of CMB anisotropy, 
that is $V^{1/4}\ll 10^{16}\,{\rm GeV}$. 
Therefore --  while certainly useful to construct low--scale models
of inflation -- it is usually thought that these mechanisms predict
an amplitude  of gravitational waves which is far too small to be
detectable by future satellite experiments aimed at
observing the $B$-mode of the CMB polarization (see however Ref.~\cite{prz}). 
This implies that a future detection of the $B$-mode of the CMB polarization
would favour the slow--roll models of inflation as generators of the 
cosmological perturbations. On the othe hand, the lack of
signal of gravity waves in the CMB anisotropies will not
give us any information about the mechanism by which cosmological perturbations
are created. 

A precise measurement of the spectral index
of comoving curvature perturbations will be a powerful tool to constrain
inflationary models.
Slow--roll inflation predicts $|n_\R-1|$ significantly below 1. Deviations
of $n_\R$ from unity are generically (but not always) 
proportional to $1/N$, where $N$ is the number of $e$-folds
till the end of inflation.
The predictions of different models for the spectral index $n_\R$,
and for its scale--dependence,  are well summarised in the review
\cite{lrreview} within slow--roll inflationary models.
Remarkably, the eventual accuracy $\Delta n_\R
\sim 0.01$ offered by the {\sl Planck}
satellite \footnote{See, for instance, 
http://www.rssd.esa.int/index.php?project=PLANCK} 
is just what one might have specified in order
to distinguish between 
various slow--roll models of inflation. 
Observation will discriminate strongly between slow--roll models of 
inflation in next ten or fifteen years. If cosmological
perturbations are due to the inflaton field, then in ten or fifteen years 
there may be a consensus about the form of the inflationary
potential, and at a deeper level we may have learned something valuable
about the nature of the fundamental interactions beyond the Standard 
Model. 
However, {\it what if} Nature has chosen the other mechanisms for the 
creation of the cosmological perturbations, which generically predict 
a value of $n_\R$ very close to unity with a negligible scale dependence? 
Then, it implies that a precise measurement of the spectral index will not
allow us to efficiently discriminate among different scenarios.

These ``{\it what if}'' options would be discouraging if they turn out to 
be true.  
They would imply that all  future efforts for
measuring tensor modes in the CMB anisotropy and the spectral index
of adiabatic perturbations are of no use
to disentangle the various scenarios for the creation
of the cosmological perturbations. 

\begin{figure}
 \begin{center}
  \leavevmode\epsfxsize=9cm \epsfbox{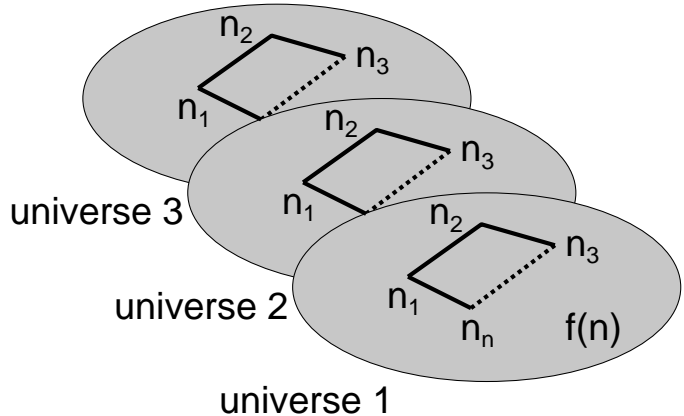}
 \end{center}
 \caption
 {\bf Ensemble Average of Angular Correlation Function} 
 \mycaption
 {A schematic view of the ensemble average of the $n$--point angular 
 correlation function,  
 $f(\hat{\mathbf{n}}_1)f(\hat{\mathbf{n}}_2)f(\hat{\mathbf{n}}_3)
 \dots f(\hat{\mathbf{n}}_n)$. 
 We measure it on each Universe, and then average it over many
 Universes.}
\label{fig:ensemble}
\end{figure}

There is, however, a third observable which will prove 
fundamental in providing information about the
mechanism chosen by Nature to produce the structures we see today. It
is the deviation from a pure Gaussian statistics, {\it i.e.}, the
presence of higher--order connected correlation functions of CMB anisotropies. 
The angular $n$--point correlation function
\begin{equation}
 \label{eq:n_corr}
  \left<f(\hat{\mathbf{n}}_1)f(\hat{\mathbf{n}}_2)
   \dots f(\hat{\mathbf{n}}_n)\right>,
\end{equation}
is a simple statistic characterizing a clustering pattern of
fluctuations on the sky, $f(\hat{\mathbf{n}})$.
The bracket denotes the ensemble average, and
Figure~\ref{fig:ensemble} sketches its meaning.
If the fluctuation is Gaussian, then the two--point correlation function
specifies all the statistical properties of $f(\hat{\mathbf{n}})$, for
the two--point correlation function is the only parameter in a Gaussian 
distribution. 
If it is not Gaussian, then we need higher--order correlation 
functions to determine the statistical properties.

For instance, a non--vanishing
three--point function of scalar perturbations, or its Fourier transform,
the bispectrum, is an indicator of a non--Gaussian feature in the
cosmological perturbations. 
The importance of the bispectrum comes from the fact that it represents
the lowest order statistics able to distinguish non--Gaussian from Gaussian
perturbations.
An accurate calculation of the primordial bispectrum of cosmological
perturbations has become an extremely important issue, as a number of
present and future experiments, such as {\sl WMAP} and {\sl Planck}, 
will allow to constrain or detect non--Gaussianity of CMB anisotropy 
with high precision.
A phenomenological way of parametrizing the level of non--Gaussianity in the
cosmological perturbations is to introduce a non--linearity parameter
$f_{\rm NL}$ through Bardeen's gravitational potential\footnote{Non--Gaussian 
models containing quadratic non--linearities, as in Eq.~(\ref{fund}), 
were introduced in the study of inflationary perturbations in 
Refs.~\cite{hodges,KBHP,FRS}, and have become a sort of  
``standard lore'' for the comparison of theoretical predictions on primordial 
non--Gaussianity to CMB and LSS observational
data~\cite{colbar,mosca,weic,scherrer,mosca2,wk,verde/etal:2000,majive,verde/etal:2001,ks,avila,sefusa}.}

\begin{equation}
\label{fund}
\Phi=\Phi_{\rm L}+f_{\rm NL}\star\left(\Phi_{\rm L}\right)^2\, ,
\end{equation}

where $\Phi_{\rm L}$ represents the gravitational potential
at linear order and the $\star$-products reminds the fact that
the non--linearity parameter might have a non--trivial scale dependence.
As Eq.~(\ref{fund}) shows, in order to compute and keep track of the 
non--Gaussianity of the cosmological perturbations throughout the different 
stages of the evolution of the Universe, one has to perform a perturbation 
around the homogeneous background \emph {up to second order}. 

The non--Gaussianity in the primordial cosmological perturbations 
and, in particular, theoretical and observational 
determinations of the non--linearity
parameter, $f_{\rm NL}$, are the subject of this
review. Our goals are to present a thorough, detailed and updated review
of the state--of--the--art on the subject of non--Gaussianity, 
both from the theoretical and  observational
point of view, and to provide  the reader with all the tools necessary
to compute the level of non--Gaussianity in any model of cosmological
perturbations. 

Surprisingly, despite the importance of the subject of non--Gaussianity
in the  cosmological perturbations and despite the fact that
its detection is one of the primary goals of the present and future
satellite missions such as {\sl WMAP} and {\sl Planck}, not much
attention has been devoted to this issue on the theoretical side. For
instance, no
firm theoretical predictions were available till about 
two years ago for the case of slow--roll models of inflation. Spurred by
the large amount of data available in the next future, a 
new wave of models, which are firmly 
rooted in modern particle theory, have been recently proposed to generate
a large and detectable amount of non--Gaussianity from inflation.  

Probably, one of the reasons why the theoretical investigations of the 
primordial non--Gaussianity were so limited was because no direct 
observational constraints on $f_{\rm NL}$ were available until the year 2001.
Many authors have shown that CMB temperature anisotropy is consistent with 
Gaussianity since the very first detection of anisotropy in the {\sl COBE} DMR 
data (see Ref.~\cite{komatsu:phd} and references therein); 
however, very little 
attention has been paid to put {\it quantitative} constraints on degrees to 
which the data are consistent with Gaussianity.
Since non--Gaussian fluctuations have infinite degrees of freedom as opposed 
to Gaussian fluctuations, testing a Gaussian hypothesis is a very difficult 
task; one statistical method showing CMB {\it consistent} with Gaussianity 
does not mean that CMB is {\it really} Gaussian.
If one does not have any specific, physically motivated non--Gaussian models 
to constrain (such as those described above), then one cannot learn anything 
about the nature of temperature fluctuations from the statement that just says,
``the CMB is consistent with Gaussianity". Rather, 
``{\it How Gaussian is it?~}" 
is a more relevant question, when we try to constrain (and exclude) certain 
non--Gaussian models.

The first direct comparison between the inflationary non--Gaussianity and 
observational data was attempted for the {\sl COBE} DMR data in 2001, 
using the angular bispectrum, the harmonic counterpart of the 
three--point correlation 
function~\cite{komatsu/etal:2002}. A very weak constraint,
$|f_{\rm NL}|<1500$ ($68\%$) was found. Although this constraint is 
still too weak to be
useful, it explicitely demonstrated that measurements of
non--Gaussianity can put quantitative constraints on inflationary models.
The angular bispectrum of the CMB is particularly useful in finding a limit 
on $f_{\rm NL}$, as the exact analytical calculation of the bispectrum from 
inflationary non--Gaussianity is possible~\cite{ks}.
While the {\sl COBE} DMR data constrain non--Gaussianity on large scales 
($\sim 7^\circ$), in Ref.~\cite{santos/etal:2003} a constraint is obtained 
on small scales 
($\sim 10'$), $|f_{\rm NL}|<950$ ($68\%$), using the {\sl MAXIMA} data.
A recent analysis of the bispectrum of the {\sl VSA} data~\cite{smith04}
gives an upper bound of 5400 on the value of $|f_{\rm NL}|$ ($95\%$). 
The {\sl WMAP} team has measured the bispectrum to obtain the 
tightest limit to date, $-58<f_{\rm NL}<134$ ($95\%$) \cite{k}.

\begin{figure}
\begin{center}
\leavevmode\epsfxsize=10cm \epsfbox{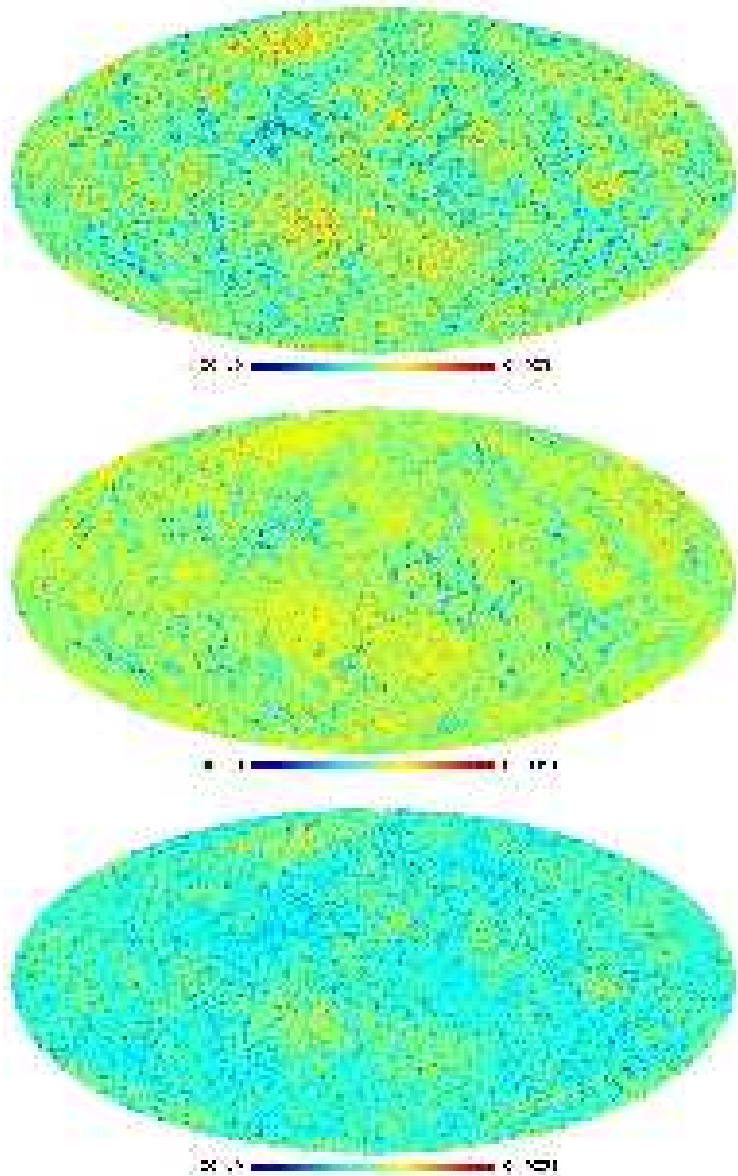}
 \end{center}
 \caption
 {\bf {\sl Planck}--resolution Simulations of Non--Gaussian CMB Maps} 
 \mycaption
  {From top to bottom, the three panels show a Gaussian and two
  non--Gaussian simulations of CMB maps, at the {\sl Planck} resolution 
  ($FWHM = 5^\prime$). The non--Gaussian maps are 
  obtained from the model of Eq.~(\ref{fund}), using the 
  spherical--coordinates algorithm of Ref.~\cite{liguori1}, with   
  initial power--spectrum and radiation transfer function of 
  a standard ``concordance'' ($\Lambda$CDM) model. The values of the 
  non--linearity parameter are  $f_{\rm NL}=0$ (top), $f_{\rm NL}=3000$ 
  (middle) and $f_{\rm NL}=-3000$ (bottom); the high values of $|f_{\rm NL}|$ 
  are chosen to make the non--Gaussian effects visible by eye 
  (note that color--scales are calibrated to the temperature interval of 
  each map).} 
  \label{fig:maps_2048}
 \end{figure}

What about other statistical tools? Currently, analytical predictions exist 
only for the bispectrum~\cite{ks} and the trispectrum (the harmonic 
counterpart of the four--point function) \cite{okamoto/hu:2002}.
Predictions for other tools, e.g., Minkowski functionals, are usually
much more difficult; however, one 
can still use other statistical tools to constrain $f_{\rm NL}$ by 
using a Monte Carlo method: direct comparison between 
measurements on the observed sky maps and those on simulated 
non--Gaussian sky maps. The authors of Ref.~\cite{cayon/etal:2003} have 
measured the spherical Mexican--hat wavelets on the {\sl COBE} DMR data, 
and compared them to simulated measurements on non--Gaussian maps 
(which include only the Sachs--Wolfe~\cite{SWolfe} effect), 
finding $|f_{\rm NL}|<1100$ ($68\%$). 
This methodology can be applied to any other statistics, if we have accurate 
simulations taking into account not only the Sachs--Wolfe effect but also the 
full effect of the Integrated Sachs--Wolfe effect and baryon--photon fluid 
dynamics. The {\sl WMAP} 
team has simulated such full--sky non--Gaussian maps that 
include all the relevant effects. By comparing the Minkowski functionals 
measured on the {\sl WMAP} maps and those on the simulated maps, they obtain 
$f_{\rm NL}<139$ ($95\%$) \cite{k}.
Using the same simulations and {\sl WMAP} map, the authors of 
Ref.~\cite{mukherjee/wang:2004} find 
$f_{\rm NL}<220$ ($95\%$) with the spherical Mexican--hat wavelets.
Also, in Ref.~\cite{cabella/etal:2004} the local curvature of the CMB 
on the {\sl WMAP} map has been measured and compared to 
non--Gaussian CMB map simulations~\cite{liguori1}, finding
$-180<f_{\rm NL} <240$ ($95\%$).
Despite these statistical tools being very different and complementary to 
some extent, they give similar constraints on $f_{\rm NL}$.
(Although \cite{gazta} find a much tigher limit on $f_{\rm NL}$,
a direct comparison is not straightforward as their definition of $f_{\rm NL}$ 
differs from ours.)
A theoretical study suggests that the inflationary non--Gaussianity can be 
detected with the bispectrum, if $f_{\rm NL}>20$ and 5 for the {\sl WMAP} and 
Planck data, respectively~\cite{ks}. The current limits from the {\sl WMAP} 
data are weaker than the theoretical expectation, probably because of 
the current 
measurements treating the effects of inhomogeneous noise and Galaxy cut 
sub--optimally. An optimal method for measuring the bispectrum is still very 
time consuming~\cite{santos/etal:2003}, while other statistics may have a 
better chance to overcome this issue. For this study, having accurate 
non--Gaussian simulations is crucial. In Figure~\ref{fig:maps_2048}, 
we show some examples of non--Gaussian sky maps at the {\sl Planck} 
resolution, simulated by the spherical--coordinates method of 
Ref.~\cite{liguori1}.

So far we have talked only about measuring non--Gaussianity from temperature 
maps. On the other hand, adding polarization information will help to improve 
our sensitivity to $f_{\rm NL}$, as polarization probes a part of the 
spectrum of primordial fluctuations that cannot be measured by temperature 
alone. More specifically, the polarization radiation transfer function is 
non--zero at wavenumbers $k$ for which the temperature transfer function
is zero; thus, polarization contains information which is maximally 
complementary to temperature~\cite{KSW}. Therefore, one could measure 
$f_{\rm NL}$ as small as $\sim 3$ by combining the temperature and 
polarization bispectra. In addition, if we combine the bispectrum with other 
statistics, then sensitivity would further improve, depending on the extent 
to which those statistics are complementary~\cite{aghanim/etal:2003}.
It is important to keep improving our sensitivity until we reach a critical 
sensitivity, $f_{\rm NL}\sim 1$, which is set by non--Gaussian contributions 
from ubiquitous second--order perturbations.\\

Before concluding this Introduction, let us mention another important source 
of primordial non--Gaussianity which will not been covered in this review. 
The topological defects, cosmic strings in particular, are potential sources 
of strong non--Gaussianity. Although the current observations have ruled out
the topological defects as being the primary source of cosmological 
perturbations, it is still quite possible that topological defects 
do exist and contribute to a part of the perturbations (see, e.g., 
Ref.~\cite{kun,pogosian/etal:2003} for the latest results from the {\sl WMAP} 
data). Since the topological defects are intrinsically very non--Gaussian,
even a modest energy density of defects may give rise to a detectable 
level of non--Gaussianity. In particular, the small--scale
CMB experiments at an angular scale of $\sim 1'$ have good chance to 
test (or detect) non--Gaussianity from the cosmic strings, via the so--called 
Kaiser--Stebbins effect (see Refs.~\cite{kaiser/stebbins:1984,gott:1985} for 
the temperature and Ref.~\cite{benabed/bernardeau:2000} for the 
polarization). Accurate numerical simulations are needed to search for
signatures of topological defects through non--Gaussianity. 
Only recently, improved simulations of CMB sky maps from cosmic strings have 
become available by solving the full Boltzmann equations 
\cite{landriau/shellard:2003a,landriau/shellard:2004}.
It is very important to improve the dynamical range of the cosmic strings 
simulations and make accurate predictions for the CMB sky maps (both 
in terms of temperature and polarization),
which can be compared with future small--scale 
CMB experiments (recent progress on making a map on small scales has been 
reported in Ref.~\cite{landriau/shellard:2003b}).\\

We end this Introduction with an overview of the present article.
The article is addressed to a wide audience, including both cosmologists and 
particle physicists. To cope with this problem, we have tried to make each 
section reasonably homogeneous regarding the background knowledge that is 
taken for granted, while at the same time allowing considerable variation 
from one section to another.

Section 2 contains a brief review of the inflationary paradigm
and an introduction to the theory of quantum fluctuations 
for a generic scalar field evolving in a fixed de Sitter background.
Correlation functions up to order three are evaluated for an interacting
scalar field. 

Section 3 and 4  are devoted to the theory of cosmological perturbations
at first and second order including gravity. This treatment is done
in a gauge--invariant way and the equations up to second order necessary 
to follow the evolution of non--linearities are provided.

Section 5 deals with the standard slow--roll scenario where cosmological
fluctuations are due to the inflaton field. The goal of this section
is to show that the main contribution to the non--Gaussian
signal comes from the post--inflationary evolution.

Section 6 and 7 are devoted to the non--Gaussianity predicted in the 
curvaton and the inhomogeneous reheating scenarios, respectively.

Section 8 contains all the necessary tools to relate the
level of non--Gaussianity parametrized by $f_{\rm NL}$ 
deduced from the
true measurements  to the one predicted theoretically within a given model.

Section 9 contains a mini-review of alternative models of inflation
with the respective predictions of non--Gaussianity.

All the results of the previous sections are summarized in 
Table~\ref{Table1}.

Section 10 contains a mini-review of the present observational
constraints on non--linearities in the cosmological perturbations
and a thorough discussion of the future prospects on the detectability
of non--Gaussianity. Our conclusions are drawn in Section 11. Finally
we provide the reader with three Appendices where she/he could find
the full derivation of second--order geometric quantities
and Einstein and Klein--Gordon equations as well as the Wigner--$3j$ symbols.

\section{\bf The inflationary paradigm}

As we have mentioned in the Introduction, one 
of the relevant ideas of modern cosmology is represented by the 
inflationary paradigm. Here we just summarize some of the basics of inflation. 
For more details the reader is referred to some reviews on the 
subject~\cite{Lindebook,LythLiddle,Lidseyetal,lrreview,LythLiddlebook,MFB,Tonilectures}. 

As far as the dynamics of Inflation is concerned one can consider a 
homogeneous and isotropic Universe described by the 
Friedmann--Robertson--Walker (FRW) metric 
\be \label{RW}
ds^2=-dt^2+a^2(t)\left[\frac{dr^2}{1-K r^2}+r^2\left(d\theta^2
+\sin^2\theta d\phi^2\right)\right] \,,
\ee 
where $t$ is the cosmic time, $r$, $\theta$, $\phi$ are the comoving (polar) 
coordinates, $a(t)$ is the scale--factor of the Universe, and 
$K$ is the curvature constant of 3--dimensional hypersurfaces. 
If the Universe is filled with matter described by the energy--momentum 
tensor $T_{\mu \nu}$ of a perfect fluid with energy density $\rho$ 
and pressure $P$, the Einstein equations 
\be
\label{Einstein}
G_{\mu\nu}=8\pi G_{\rm N}~T_{\mu\nu} \, ,
\ee
with $G_{\mu\nu}$ the Einstein tensor and $G_{\rm N}$ the Newtonian 
gravitational constant give the Friedmann equations 
\be \label{E2}
H^2=\frac{8 \pi G_{\rm N}}{3}\, \rho - \frac{K}{a^2}\, ,
\ee
\be  \label{E1}   
\frac{\ddot{a}}{a}=-\frac{4 \pi G_{\rm N}}{3} (\rho+3 P)\, , 
\ee
where $H=\dot{a}/{a}$ is the Hubble expansion parameter and dots denote
differentiation with respect to cosmic time $t$. Eq.~(\ref{E1}) shows that a 
period of inflation is possible if the pressure $P$ is negative with 
\be
\label{condition:Inflation}
P < - \frac{\rho}{3}\, .
\ee
In particular a period of the history of Universe during which $P=-\rho$ is 
called a \emph{de Sitter stage}. From the energy 
continuity equation $\dot{\rho}+3 H (\rho +P)=0$ and Eq.~(\ref{E2}) 
(neglecting the curvature $K$ which is soon redshifted away 
as $a^{-2}$) we see that 
in a de Sitter phase $\rho=\textrm{constant}$ and 
\be
H=H_I=\textrm{constant}\, . 
\ee
Solving Eq.~(\ref{E1}) 
we also see the scale--factor grows exponentially 
\be
a(t)=a_{i}\,  e^{H_I (t-t_i)}\, ,
\ee
where $t_i$ is the time inflation starts. 
In fact the condition~(\ref{condition:Inflation}) 
can be satisfied by a scalar field, the inflaton $\varphi$. 

The action for a minimally--coupled scalar field $\varphi$ is given by 
\be
\label{inflatonaction}
S= \int d^4x \sqrt{-g} \mathcal{L} =  
\int d^4x \sqrt{-g} \left[-\frac{1}{2} g^{\mu \nu} \partial_\mu \varphi 
\partial_\nu \varphi -V(\varphi) \right]\, ,
\ee
where $g$ is the determinant of the metric tensor $g_{\mu \nu}$, 
$g_{\mu \nu}$ is the contravariant metric tensor, such that  
$g_{\mu\nu}g^{\nu\lambda}=\delta_\mu^\lambda$; finally   
$V(\varphi)$ specifies the scalar field potential. By varying 
the action with respect to $\varphi$ one obtains the Klein--Gordon equation 
\be
\label{KleinG0}
\square \varphi = \frac{\partial V}{\partial \varphi} \, ,
\ee
where $\square$ is the covariant D'Alembert operator 
\be
\label{def:Dalembert}
\square \varphi \,=\, \frac{1}{\sqrt {- g}}\,\partial_\nu \left( \sqrt{-
g}\,g^{\mu\nu}\,\partial_\mu \varphi \right)\, .
\ee
In a FRW Universe described by the metric~(\ref{RW}), 
the evolution equation for $\varphi$ becomes
\begin{equation}
\ddot{\varphi}+ 3H\dot{\varphi}-\frac{\nabla^2\varphi}{a^2}+V'(\varphi)=0\, ,
\label{nabla}
\end{equation}
where $V'(\varphi)=\left(dV(\varphi)/d\varphi\right)$. Note, in particular, 
the appearance of the friction term $3H\dot{\varphi}$: a scalar field
rolling down its potential suffers a friction due to the
expansion of the Universe. 
The energy--momentum tensor for a minimally--coupled 
scalar field $\varphi$ is given by
\be
T_{\mu \nu}=-2\frac{\partial \mathcal{L}}{\partial g^{\mu \nu}}+g_{\mu \nu}
\mathcal{L}=
\partial_{\mu}\varphi \partial_{\nu}\varphi+
g_{\mu \nu} \left[-\frac{1}{2} g^{\alpha \beta} \partial_\alpha \varphi 
\partial_\beta \varphi -V(\varphi) \right] \, .
\ee

We can now split the inflaton field as 
$$
\varphi(t,{\bf x})=\varphi_{0}(t)+\delta\varphi(t,{\bf x}),
$$
where $\varphi_{0}$ is the `classical' (infinite wavelength) field, that is 
the expectation value of the inflaton field 
on the initial isotropic and
homogeneous state, while $\delta\varphi(t,{\bf x})$ represents the quantum
fluctuations around $\varphi_{0}$.
In this section, we will be only concerned with the evolution of the
classical field $\varphi_0$. The next section will be devoted to the
crucial issue of the evolution of quantum perturbations during inflation.
This separation is justified by the fact that quantum fluctuations are much
smaller than the classical value and therefore negligible when looking at the 
classical evolution. To not be overwhelmed by the notation, we will
keep indicating
from now on the classical value of the inflaton field by $\varphi$.
A homogeneous scalar field $\varphi(t)$ behaves like a perfect fluid with 
background energy density and pressure given by
\begin{eqnarray}
\rho_{\varphi}=\frac{\dot{\varphi}^2}{2} + V(\varphi)\\
P_{\varphi}=\frac{\dot{\varphi}^2}{2} - V(\varphi).
\end{eqnarray}
Therefore if 
\be
V(\varphi) \gg \dot{\varphi}^2
\ee
we obtain the following condition
\be
P_\varphi\simeq -\rho_\varphi\, .
\ee
From this simple calculation, 
we realize that a scalar field whose energy is dominant
in the Universe and whose potential energy  
dominates over the kinetic term gives inflation. Inflation
is driven by the vacuum energy of the inflaton field.
Notice that the ordinary matter fields, in the form of a radiation fluid, 
and the spatial curvature $K$ are 
usually neglected during inflation because their contribution to the 
energy density is redshifted away during the accelerated expansion.\footnote{
For the very same reason also any small inhomogeneities are wiped out 
as soon as inflation sets in, thus justifying the use of the background 
FRW metric.} Moreover 
the basic picture we have discussed here refers only to the simplest models 
of inflation, where only a single scalar field is present. We will consider 
later on also some non--standard models of inflation involving more than one 
scalar field (multiple--field inflation). 
\subsection{The slow--rolling inflaton field}
Let us now better quantify under which circumstances a scalar field
may give rise to a period of inflation. 
The equation of motion of the homogeneous scalar field is
 \begin{equation}
 \ddot{\varphi}+3H\dot{\varphi}+V'(\varphi)=0\, .
\label{poi}
 \end{equation}
If we require that $\dot{\varphi}^2\ll V(\varphi)$, the scalar field 
slowly rolls down
its potential. Such a {\it slow-roll}  
period can be achieved if the inflaton field $\varphi$ 
is in a region where the potential is sufficiently flat.
We may also expect that -- being the potential flat -- 
$\ddot{\varphi}$ is negligible as well. We
will assume that this is true and we will quantify this condition soon.
The Friedmann equation (\ref{E2}) becomes
\be
H^2\simeq \frac{8\pi G_{\rm N}}{3}\,V(\varphi),
\ee
where we have assumed that the inflaton field dominates the
energy density of the Universe.
The new equation of motion becomes
\be
\label{srapprox}
 3H\dot{\varphi}=-V'(\varphi)\, ,
\label{friction}
\ee
which gives $\dot{\varphi}$ as a function of $V'(\varphi)$.
Using Eq.~(\ref{friction}) the slow--roll conditions then require
\be
\label{V'}
\dot\varphi^2 \ll  V(\varphi)   \\  \Longrightarrow  \\  \frac{(V')^2}{V} \ll
H^2 \label{slowroll1}
\ee
and
\be
\label{V''}
\ddot{\varphi} \ll 3H\dot{\varphi} \\  \Longrightarrow  \\  V'' \ll H^2.  
\label{slowroll2}
\ee
Eqs.~(\ref{V'}) and~(\ref{V''}) represent the flatness conditions 
on the potential which are conveniently parametrized in terms of the 
the so--called \emph{slow--roll parameters},  
which are built from $V$ and its derivatives $V^\prime$, $V^{\prime\prime}$, 
$V^{\prime\prime\prime}$, $V^{(n)}$,   
with respect to $\varphi$~\cite{Lidseyetal,LiddleParsons,lrreview}. 
In particular, one can define the two 
slow--roll parameters~\cite{Lidseyetal,LiddleParsons,lrreview} 
in Eq.~(\ref{potter1}).

Achieving a successful period of inflation requires the 
slow--roll parameters to be $ \epsilon, |\eta| \ll 1$. Indeed there exists 
a hierarchy of slow--roll 
parameters~\cite{LiddleParsons}. For example one can define the slow--roll 
parameter related to the third-derivative of the potential 
$\xi^2=1/(8\pi G_{\rm N})\, (V^{(1)}V^{(3)}/V^{2})$ which is a 
second--order slow--roll parameter (note that $\xi^2$ can be negative).
The parameter $\epsilon$ can also be written as $\epsilon=-\dot{H}/H^2$, 
thus it quantifies how much the
Hubble rate $H$ changes with time during inflation. In particular 
notice that, since
$$
\frac{\ddot a}{a}=\dot H+H^2=\left(1-\epsilon\right)H^2,
$$
inflation can be attained only if $\epsilon<1$. 
As soon as this condition fails, inflation ends. At first--order in the 
slow--roll parameters $\epsilon$ and $\eta$ can be considered  
constant, since the potential is very flat. In fact it is easy to see that 
that $\dot\epsilon,\dot\eta={\mathcal{O}}
\left(\epsilon^2,\eta^2\right)$. \footnote{With $\mathcal{O}(\epsilon,\eta)$ 
and $\mathcal{O}(\epsilon^2,\eta^2)$ we indicate general combinations of 
the slow--roll parameters of lowest order and next order respectively.}

Despite the simplicity of the inflationary paradigm, the number of 
inflationary models that have been proposed so far is 
enormous, differing for the kind of potential and for the underlying 
particle physics theory. In that respect the reader is referred to 
the review~\cite{lrreview}. We just want to mention here that a useful 
classification in connection with the observations   
may be the one in which the single--field inflationary models are divided into 
three broad groups as ``small field'', ``large field'' (or chaotic) 
and ``hybrid'' type, according to the region occupied in the 
$(\epsilon-\eta)$ space by a given inflationary potential~\cite{DKK}.
Typical examples of the large--field models ($0 < \eta < 2 \epsilon$) 
are polynomial potentials $V\left(\varphi\right) = \Lambda^4
\left({\varphi / \mu}\right)^p$, and exponential potentials, 
$V\left(\varphi\right) = \Lambda^4 \exp\left({\varphi / 
\mu}\right)$. The small--field potentials  ( $\eta < - \epsilon$ 
) are typically of the form $V\left(\varphi\right) = 
\Lambda^4 \left[1 - \left({\varphi / \mu}\right)^p\right]$, while generic 
hybrid potentials ($ 0 < 2 \epsilon < \eta $) are of the form 
$V\left(\varphi\right) 
= \Lambda^4 \left[1 + \left({\varphi / \mu}\right)^p\right]$.      
In fact according to such a scheme, 
the {\it WMAP} dataset already allows
to extract the parameters relevant 
for distinguishing among single--field inflation models~\cite{ex,ex1,ex2,ex3}.

Here we want to make an important comment. A crucial quantity for the 
inflationary dynamics and for understanding the generation of the primordial 
perturbations during inflation    
is the Hubble radius (also called the Hubble horizon size)  
$R_H = H^{-1}$. The Hubble radius represents a characteristic
length scale beyond which causal processes cannot operate. 
A key point is that during 
inflation the comoving Hubble horizon, $(a H)^{-1}$, decreases 
in time as the scale--factor, $a$, grows
quasi--exponentially, and the Hubble radius remains almost constant 
(indeed the decrease of $(a H)^{-1}$ is a consequence of the
accelerated expansion, $\ddot a > 0$, characterizing inflation).   
Therefore, a given comoving length scale, $L$, will become larger than
the Hubble radius and {\it leaves the Hubble horizon}.
On the other hand, the comoving Hubble radius increases as
$(a H)^{-1}\propto a^{1/2}$ and $a$ during
radiation and matter dominated era, respectively.

Previously, we have defined inflation as a period of 
accelerated expansion of the Universe; however, 
this is actually not sufficient.
A successful inflation must last for a long enough period 
 in order to solve the horizon and flatness problems.
By ``a long enough period'' we mean a period of accelerated expansion 
of the Universe long enough that 
a small, smooth patch of size that is smaller than the Hubble radius
can grow to encompass {\it at least} the entire observable Universe.
Typically the amount of inflation is measured in terms of the number of 
e--foldings, defined as 
\be
N_{\rm TOT} = \int_{t_i}^{t_f} H dt\, ,
\ee
where $t_i$ and $t_f$ are the time inflation starts and ends respectively.
To explain smoothness of the observable Universe, we impose that the largest 
scale we observe today, the present horizon $H_0^{-1}$ ($\sim 4200$ Mpc),
was reduced  during inflation to a value $\lambda_{H_0}$ at $t_i$,
which is smaller than $H_I^{-1}$ during inflation.
Then, it follows that we must have 
$N_{\rm TOT} > N_{\rm min}$, where $ N_{\rm min} \approx 60$ is 
the number of e--foldings before the end of inflation 
when the present Hubble radius leaves the horizon.
A very useful quantity is the number of e--foldings from the time 
when a given wavelength $\lambda$ leaves the horizon during inflation 
to the end of inflation,  
\be
N_{\lambda}=\int_{t(\lambda)}^{t_f} H dt=
\ln\left( \frac{a_{f}}{a_{\lambda}} \right)\, ,
\ee
where $t(\lambda)$ is the time when $\lambda$ leaves the horizon during 
inflation 
and  $a_\lambda=a(t(\lambda))$. The cosmologically interesting scales 
probed by the CMB anisotropies correspond to $N_{\lambda} \simeq 40$ -- $60$. 

Inflation ends when the inflaton field starts to roll fast along its 
potential. During this regime $V'' > H^2$ (or $\eta > 1$). 
The scalar field will reach 
the minimum of its potential and will start to oscillate 
around it. By this time any other contribution to the energy density 
and entropy of the Universe has been redshited away by the inflationary
expansion. However we know that the Universe must be repopulated by a 
hot radiation fluid in order for the standard Big-Bang cosmology to set in. 
This is achieved through a process, called \emph{reheating}, by 
which the energy of the inflaton field is transferred to radiation during 
the oscillating phase.
In the ordinary scenario of reheating~\cite{Albrecht,Lindereh,abbot,Dolgov}
such a transfer corresponds to the decay of the inflaton field into other 
lighter particles to which it couples through a decay rate $\Gamma_\varphi$. 
Such a decay damps the inflaton oscillations and when the decay products 
thermalize and form a thermal background the Universe is 
finally reheated. Alternatively, 
reheating may occur through preheating~\cite{preheating}).

\subsection{Inflation and  cosmological perturbations}

Besides the background inflationary dynamics, it is of crucial importance 
to discuss the issue of the evolution of the quantum fluctuations of the 
inflaton field 
$\delta \varphi(t,{\bf x})$. In the inflationary paradigm associated with 
these vacuum fluctuations there are primordial energy density 
perturbations, which 
survive after inflation and are the origin of all the structures 
in the Universe. Our current understanding of the origin of structure in the 
Universe is that once the Universe became matter dominated ($z\sim 3200$) 
primeval density inhomogeneities 
($\delta\rho /\rho \sim 10^{-5}$) were amplified by gravity and
grew into the structure we see today~\cite{peebles:1980,ColesLucchin}. 
The existence of these inhomogeneities was in fact confirmed
by the {\sl COBE} discovery of CMB anisotropies. 
In this section we just want to summarize in a qualitative way the process 
by which such ``seed'' perturbations are generated during inflation. This 
would also help the reader to better appreciate the alternative mechanisms 
that have been proposed recently to the inflationary scenario in order 
to explain the primordial density perturbations. 

First of all, in order for 
structure formation to occur via gravitational
instability, there must have been small preexisting fluctuations on
relevant physical length scales (say, a galaxy scale $\sim 1$~Mpc)
which left the Hubble radius in the 
radiation--dominated and matter--dominated 
eras. However in the standard Big--Bang model these small perturbations
have to be put in by hand, because it is impossible to produce
fluctuations on any length scales larger than the horizon size.
Inflation is able to provide a mechanism to generate both density 
perturbations and gravitational waves. As we mentioned in the previous 
section, a key ingredient of this mechanism is the fact that during inflation 
the comoving Hubble horizon $(a H)^{-1}$ decreases with time.
Consequently, the wavelength of a quantum fluctuation in the
scalar field whose potential energy drives inflation 
soon exceeds the Hubble radius. The quantum fluctuations 
arise on scales which are much smaller than the 
comoving Hubble radius $(aH)^{-1}$, which is the scale beyond which causal 
processes cannot operate. On such small scales one can use the 
usual flat space--time quantum field theory to describe 
the scalar field vacuum 
fluctuations. The inflationary expansion then stretches the wavelength
of quantum fluctuations to outside
the horizon; thus, gravitational effects become more and more important and 
amplify the quantum fluctuations, 
the result being that a net
number of scalar field particles are created by the changing cosmological 
background~\cite{muk81,guth82,hawking82,Linde1982,starobinsky82}. 
On large scales the perturbations just follow a classical evolution.
Since microscopic physics does not affect the evolution of fluctuations
when its wavelength is outside the horizon, the amplitude of fluctuations
is ``frozen-in'' and fixed at some nonzero value $\delta\varphi$ at the
horizon crossing, because of a large
friction term $3H\dot\varphi$ in the equation of motion of the
field $\varphi$.
The amplitude of the fluctuations on super-horizon scales then remains 
almost unchanged for a very long time, whereas its wavelength grows 
exponentially. Therefore, the appearance of such frozen fluctuations
is equivalent to the appearance of a classical field
$\delta\varphi$ that does not vanish after having averaged over some 
macroscopic interval of time.
Moreover, the same mechanism also generates
stochastic gravitational waves \cite{Starogw,AbbotWise}.

The fluctuations of the scalar field produce primordial 
perturbations in the energy 
density, $\rho_\varphi$, which are then inherited by the radiation and matter
to which the inflaton field decays during reheating after inflation.
Once inflation has ended,
however, the Hubble radius increases faster than the scale--factor, so
the fluctuations eventually reenter the Hubble radius during the
radiation or matter--dominated eras. The fluctuations that exit around
60 $e$-foldings or so before reheating reenter with physical
wavelengths in the range accessible to cosmological observations.
These spectra, therefore, preserve signature of inflation, giving
us a direct observational connection to physics of inflation.
We can measure inflationary fluctuations by a variety different ways,
 including the analysis of CMB anisotropies. The {\sl WMAP} collaboration has  
produced a full--sky map of the angular variations 
of the CMB, with unprecedented accuracy.
The {\sl WMAP} data confirm the detection of adiabatic 
\emph{super-horizon fluctuations} 
which are a distinctive signature of an early 
epoch of acceleration~\cite{ex}.

\begin{figure}
 \begin{center}
  \leavevmode\epsfxsize=9cm \epsfbox{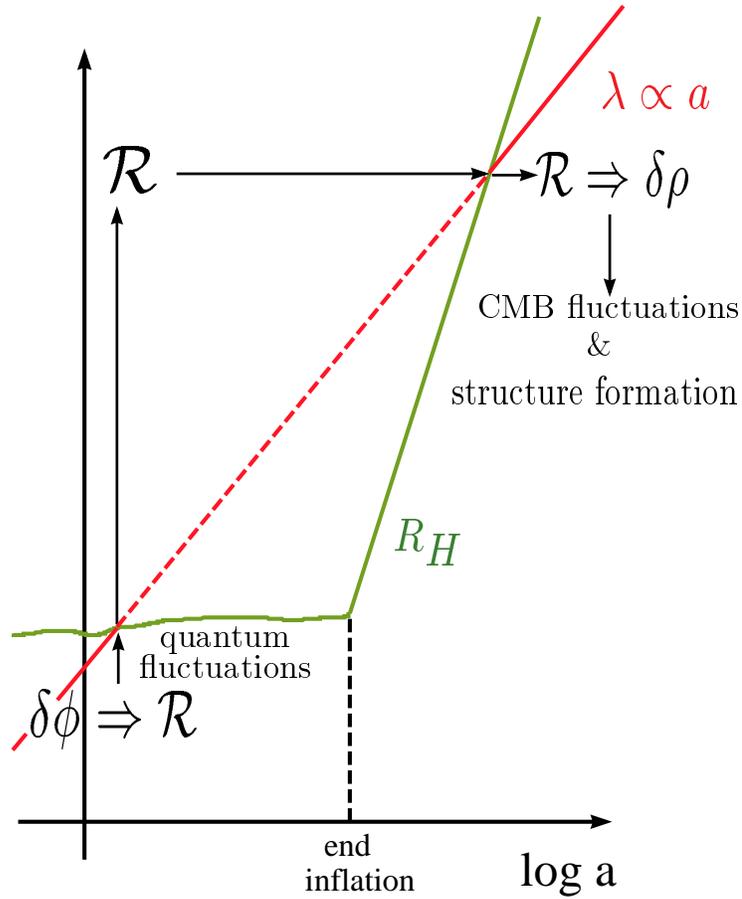}
 \end{center}
 \caption
 {\bf Stretching of cosmological perturbations during inflation} 
 \mycaption
 {Quantum perturbations in the curvature, ${\mathcal R}$, are created 
 during inflation and 
 their wavelengths, $\lambda$, are stretched from microscopic scales to 
 astronomical scales during inflation.}
\label{figpert}
\end{figure}

The physical inflationary processes which give rise to the structures we 
observe today are illustrated in Fig.~\ref{figpert}.

Quantum fluctuations of the inflaton field are generated during inflation.
Since gravity acts on any component of the Universe, small fluctuations
of the inflaton field are intimately related to fluctuations of the
space--time metric, giving rise to perturbations of the curvature
${\mathcal {R}}$ (or $\zeta$, which will be defined in the following; 
the reader may loosely think of them as a gravitational potential). 
The physical wavelengths $\lambda$ of these perturbations grow exponentially 
and leave the horizon when $\lambda> H^{-1}$. 
On superhorizon scales, curvature 
fluctuations are frozen in and considered as classical. Finally, 
when the
wavelength of these fluctuations reenters the horizon, at some
radiation or matter--dominated epoch, the curvature (gravitational
potential) perturbations
of the space--time give rise to matter (and temperature) 
perturbations $\delta\rho$
via the Poisson equation. These fluctuations will then start growing, thus 
giving rise to the structures we observe today.

In summary, two are the key ingredients for understanding 
the observed structures in the Universe within the
inflationary scenario:

\begin{itemize}

\item Quantum fluctuations of the inflaton field 
are excited during inflation and stretched to cosmological scales. At the
same time, as the inflaton fluctuations couple to the metric perturbations 
through  Einstein's equations, 
ripples on the metric are also excited and stretched to cosmological 
scales.

\item The metric perturbations perturb baryons and photons, and 
they form acoustic oscillations once the wavelength of the perturbations
becomes smaller than the horizon size.

\end{itemize}

Let us now see how quantum fluctuations are generated during inflation. In 
fact the mechanism by which the quantum fluctuations of the inflaton 
field are produced during an  
inflationary epoch is not peculiar to the inflaton field itself, rather 
it is generic to any scalar field evolving in an accelerated background.   
As we shall see, the inflaton field is peculiar in that it dominates the 
energy density of the Universe, thus possibly producing also metric 
perturbations. 

In the following, we shall describe in a quantitative way how 
the quantum fluctuations of a generic scalar field evolve 
during an inflationary stage. For more details we refer the reader to the
classical works~\cite{LindeLett,KL,Lindebook,Lidseyetal,MFB} and 
to some recent reviews on the 
subject~\cite{LythLiddlebook,Tonilectures}. 

Let us first consider 
the case of a scalar field $\chi$ 
with an effective potential $V(\chi)$ in a 
pure de Sitter stage, during which $H$ is constant. Notice that 
$\chi$ is a scalar field different from the inflaton -- or the inflatons -- 
that are driving the accelerated expansion.

\subsection{Quantum fluctuations of a generic scalar field during a de Sitter
stage}
\label{QFdeSitter}

Let us first consider the case of a scalar field $\chi$ 
with an effective potential $V(\chi)$ in a 
pure de Sitter stage, during which $H$ is constant. Notice that 
$\chi$ is a scalar field different from the inflaton -- or the inflatons -- 
that are driving the accelerated expansion.

We first split the scalar field $\chi(\tau, \bf x)$ as 
\be
\chi(\tau, \bf x)=\chi(\tau)+\delta \chi(\tau, \bf x)\, ,
\ee 
where $\chi(\tau)$ is the homogeneous classical value of the scalar field 
and $\delta \chi$ are its fluctuations and $\tau$ is the conformal time, 
related to the cosmic time $t$ through $d\tau= dt/a(t)$. 
The scalar field $\chi$ is quantized by implementing the standard 
technique of second quantization. To proceed we first make the following
field redefinition 
\begin{equation}
\label{redifinition}
\widetilde{\delta\chi}=a \delta\chi\, .
\end{equation}
Introducing the creation and annihilation operators 
$a_{\bf k}$ and $a_{\bf k}^{\dagger}$ 
we promote $\widetilde{\delta \chi}$ to an operator which can 
be decomposed as 
\begin{equation} 
\label{dec}
\widetilde{\delta\chi}(\tau,{\bf x})
=\int \frac{d^3{\bf k}}{(2 \pi)^{3/2}} \left[
u_k(\tau) a_{\bf k} e^{i{\bf{k \cdot x}}}+u^{*}_k(\tau) a^{\dag}_{\bf k}
e^{-i{\bf{k \cdot x}}} \right]\, .
\end{equation}

The creation and annihilation operators for $\widetilde{\delta\chi}$ 
(not for $\delta\chi$)
satisfy the standard commutation relations
\be
\label{commutators}
[a_{\bf k},a_{\bf k'}]=0,\quad [a_{\bf k},a^{\dag}_{\bf k'}]=
\delta^{(3)}({\bf k -k'})\, ,
\ee
and the modes $u_{k}(\tau)$ are normalized so that they satisfy the condition
\be \label{norm}
u^*_k u_k^{\prime}-u_k u^{* \prime}_k=-i,
\ee
deriving from the usual canonical commutation relations between the operators 
$\widetilde{\delta \chi}$ and its conjugate momentum $\Pi= 
{\widetilde{\delta \chi}}^{\prime}$. Here primes denote derivatives 
with respect to the conformal time $\tau$ (not $t$).\\  
The evolution equation for the scalar field $\chi(\tau, \bf x)$ is 
given by the Klein--Gordon equation 
\be
\label{KGX}
\square \chi = \frac{\partial V}{\partial \chi} \, ,
\ee
where $\square$ is the D'Alembert operator defined in 
Eq.~(\ref{def:Dalembert}).
The Klein--Gordon equation gives in an unperturbed FRW Universe 
\be
\chi^{\prime\prime} +2 \H \chi^\prime = 
- a^2 \frac{\partial V}{\partial \chi}\, ,
\ee
where $\H\equiv a^\prime/a$ is the Hubble expansion rate in conformal time. 
Now, we perturb the scalar field but neglect
the metric perturbations in the Klein--Gordon equation (\ref{KGX}), 
the eigenfunctions $u_k(\tau)$ obey the equation of motion
\be 
\label{KGsempl}
u_k^{\prime \prime}+\left( k^2-\frac{a{''}}{a}+m_{\chi}^2 a^2 \right)u_k=0\, ,
\ee
where $m_{\chi}^2=\partial^2V/{\partial \chi}^2$ is the effective mass 
of the scalar field.
The modes $u_k(\tau)$ 
at very short distances must reproduce the form for the ordinary
flat space--time quantum field theory. Thus, well within the horizon, in the 
limit $k/aH \rightarrow \infty$, the modes should approach plane waves of 
the form 
\be \label{smallscales}
u_k(\tau) \rightarrow \frac{1}{\sqrt{2 k}}e^{-ik\tau}.
\ee 
Eq.~(\ref{KGsempl}) has an exact solution in the case of a de Sitter
stage. Before recovering it, let us study the limiting behaviour of 
Eq.~(\ref{KGsempl}) on sub-horizon and superhorizon scales. 
On sub-horizon scales $k^2 \gg a''/a$, the mass term is negligible so 
that Eq~(\ref{KGsempl}) reduces to 
\be
u_k^{\prime \prime}+k^2u_k=0\, ,
\ee  
whose solution is a plane wave
\be
u_k \propto e^{-i k \tau}\, .
\ee
Thus fluctuations with wavelength within the cosmological horizon 
oscillate as in Eq.~(\ref{smallscales}). 
This is what we expect since in the ultraviolet  
limit, \emph{i.e.} for wavelengths much smaller than the horizon scales, 
we are approximating the space--time as flat. 
On superhorizon scales $k^2 \ll a''/a$, Eq.~(\ref{KGsempl}) reduces to
\be 
\label{bigscales}
u_k^{\prime \prime}-\left(\frac{a{''}}{a}-m_{\chi}^2 a^2 \right)u_k=0
\ee
Just for simplicity let us see what happens in the case of a massless 
scalar field ($m^2_\chi=0$). There are two solutions of 
Eq.~(\ref{bigscales}), a growing and a decaying mode:
\be
\label{Sbigscales}
u_k=B_{+}(k)a+B_{-}(k)a^{-2}\, .
\ee  
We can fix the amplitude of the growing mode, $B_{+}$, by matching the 
(absolute value
of the) solution~(\ref{Sbigscales}) to the plane wave 
solution~(\ref{smallscales}) when the fluctuation with wavenumber $k$ 
leaves the horizon $(k=aH)$
\be
|B_{+}(k)|=\frac{1}{a\sqrt{2k}}=\frac{H}{\sqrt{2k^3}}\, ,
\ee
so that the quantum fluctuations of the original scalar field $\chi$ on 
superhorizon scales are constant,
\be
|\delta \chi_{k}| = \frac{|u_{k}|}{a} = \frac{H}{\sqrt{2k^3}}\, .
\ee

Now, let us derive the exact solution without any matching tricks.
The exact solution to Eq.~(\ref{KGsempl}) introduces some corrections due 
to a non--vanishing mass of the scalar field. 
In a de Sitter stage, as $a=-(H\tau)^{-1}$
\be
\frac{a^{\prime \prime}}{a}-m^{2}_{\chi} a^2 = \frac{2}{\tau^2} \left( 
1-\frac{1}{2}\frac{m^2_{\chi}}{H^2} \right)\, ,
\ee 
so that Eq.~(\ref{KGsempl}) can be recast in the form 
\be \
\label{Bessel}
u_k^{\prime \prime}+\left( k^2 -\frac{\nu_{\chi}^2-\frac{1}{4}}{\tau^2} 
\right)u_k=0 \, , 
\ee
where 
\be
\nu_{\chi}^2=\left( \frac{9}{4}-\frac{m_{\chi}^2}{H^2} \right)\, .
\ee
When the mass $m^2_\chi$ is constant in time,
Eq.~(\ref{Bessel}) is a Bessel equation whose general solution 
for {\it real} $\nu_\chi$ reads
\be
\label{SOL}
u_{k}(\tau)=\sqrt{-\tau}\left[c_1(k)\,H_{\nu_\chi}^{(1)}(-k\tau)+
c_2(k)\,H_{\nu_\chi}^{(2)}(-k\tau)\right]\, ,
\ee
where $H_{\nu_\chi}^{(1)}$ and $H_{\nu_\chi}^{(2)}$ are the Hankel
functions of 
first and second kind, respectively. This result actually coincides with the 
solution found in the work by Bunch and Davies~\cite{BunchDavies}
for a free massive scalar field in de Sitter space--time.
If we impose that in the ultraviolet regime $k\gg aH$  $(-k\tau\gg 1$)
the solution matches the plane--wave solution $e^{-ik\tau}/\sqrt{2k}$
that we expect in flat space--time, and knowing that
$$
H_{\nu_\chi}^{(1)}(x\gg 1)\sim \sqrt{\frac{2}{\pi x}}\,e^{i\left(x-
\frac{\pi}{2}\nu_\chi-\frac{\pi}{4}\right)}\,\,\,\, ,
H_{\nu_\chi}^{(2)}(x\gg 1)\sim \sqrt{\frac{2}{\pi x}}\,
e^{-i\left(x-
\frac{\pi}{2}\nu_\chi-\frac{\pi}{4}\right)},
$$ 
we set $c_2(k)=0$ and $c_1(k)=
\frac{\sqrt{\pi}}{2}\,e^{i\left(\nu_\chi+\frac{1}{2}\right)\frac{\pi}{2}}$, 
which also satisfy the normalization condition~(\ref{norm}).
The exact solution becomes
\be
u_{k}(\tau)=\frac{\sqrt{\pi}}{2}\,
e^{i\left(\nu_\chi+\frac{1}{2}\right)\frac{\pi}{2}}\,
\sqrt{-\tau}\,H_{\nu_\chi}^{(1)}(-k\tau).
\label{exactm}
\ee
We are particularly interested in the asymptotic behaviour of the 
solution when the mode is
well outside the horizon. 
On superhorizon scales, since $H_{\nu_\chi}^{(1)}(x\ll 1)\sim
\sqrt{2/\pi}\, e^{-i\frac{\pi}{2}}\,2^{\nu_\chi-\frac{3}{2}}\,
(\Gamma(\nu_\chi)/\Gamma(3/2))\, x^{-\nu_\chi}$, 
the fluctuation (\ref{exactm}) becomes
\be
u_{k}(\tau)=e^{i\left(\nu_\chi-\frac{1}{2}\right)\frac{\pi}{2}}
2^{\left(\nu_\chi-\frac{3}{2}\right)}\frac{\Gamma(\nu_\chi)}{\Gamma(3/2)}
\frac{1}{\sqrt{2k}}\,(-k\tau)^{\frac{1}{2}-\nu_\chi}.
\ee
Thus we find that on superhorizon scales, the fluctuation of the scalar 
field $\delta\chi_k \equiv u_k/a$ with a non--vanishing mass
is not exactly constant, but it acquires a dependence upon time
\be
\label{exactSH}
\left|\delta\chi_{k}\right|=2^{(\nu_{\chi}-3/2)} 
\frac{\Gamma(\nu_{\chi})}{
\Gamma(3/2)}
\frac{H}{\sqrt{2k^3}}
\left(\frac{k}{aH}\right)^{\frac{3}{2}-\nu_\chi}
\,\, \textrm{(on superhorizon scales)}
\ee
Notice that the solution~(\ref{exactSH}) is valid for values of the scalar 
field mass $m_{\chi} \leqslant 3/2 H$. If the scalar field is very light,
$m_{\chi} \ll 3/2 H$, we can introduce the parameter 
$\eta_\chi=(m_\chi^2/3H^2)$ in analogy with the slow--roll 
parameters $\epsilon$ and $\eta$ for the inflaton field, 
and make an expansion of the solution in Eq.~(\ref{exactSH}) to 
lowest order in $\eta_\chi=(m_\chi^2/3H^2)\ll 1$ to find
\be
\label{Sitterslowroll}
\left|\delta\chi_{k}\right|=
\frac{H}{\sqrt{2k^3}}
\left(\frac{k}{aH}\right)^{\frac{3}{2}-\nu_\chi}\, ,
\ee
with 
\be
\frac{3}{2}-\nu_\chi\simeq \eta_\chi\, .
\label{nu}
\ee  
Eq.~(\ref{Sitterslowroll}) shows a crucial result. When the scalar 
field $\chi$ is light, its quantum fluctuations, first generated on 
subhorizon scales, are gravitationally amplified and stretched to
superhorizon scales because of the accelerated 
expansion of the Universe during inflation~\cite{LindeLett,Lindebook}.
\subsubsection{The power--spectrum}
A useful quantity to characterize the properties of a 
perturbation field is the \emph{power--spectrum}. 
For a given random field $f(t,{\bf x})$ which 
can be expanded in Fourier space (since we work in flat space) as \footnote{
The alternative Fourier--transform definition
$f(t, {\bf x})=\int\,\frac{d^3{\bf k}}{(2\pi)^3}\,e^{i{\bf k}
\cdot{\bf x}}\, f_{{\bf k}}(t)$ is also used in this review.} 
\be
f(t, {\bf x})=\int\,\frac{d^3{\bf k}}{(2\pi)^{3/2}}\,e^{i{\bf k}
\cdot{\bf x}}\, f_{{\bf k}}(t) \, ,
\ee
the (``dimensionless'') power--spectrum ${\mathcal P}_{f}(k)$ 
can be defined through
\be
\label{def:PS}
\langle f_{{\bf k}_1}f^{*}_{{\bf k}_2} \rangle
\equiv \frac{2\pi^2}{k^3}\,
{\mathcal P}_{f}(k) \, \delta^{(3)}\left({\bf k}_1-{\bf k}_2\right),
\ee
where the angled brackets denote ensemble averages. The power--spectrum 
measures the amplitude of the fluctuations at a give scale $k$; 
indeed from the definition~(\ref{def:PS}) 
the mean square value of $f(t,{\bf x})$ in 
real space is
\be
\langle f^2(t, {\bf x}) \rangle=\int\,\frac{dk}{k}\,
{\mathcal P}_{f}(k)\, .
\ee
Thus, according to our definition the power--spectrum,
${\mathcal P}_{f}(k)$ is the contribution to the variance per unit 
logarithmic interval in the wavenumber $k$. This is standard notation  
in the literature for the inflationary power--spectrum. However, 
another definition of power--spectrum, given by the quantity 
$P_f(k)$ related to ${\mathcal P}_{f}(k)$ by the relation 
$P_f(k) = 2 \pi^2 {\mathcal P}_{f}(k)/k^3$,
or $\langle f_{{\bf k}_1}f^{*}_{{\bf k}_2} \rangle
= P_{f}(k)\delta^{(3)}\left({\bf k}_1-{\bf k}_2\right)$, 
will be also used in this review. 

To describe the slope of the power--spectrum it is standard practice 
to define a \emph{spectral index} $n_f(k)$, through 
\be
n_f(k)-1 \equiv \frac{d \ln \mathcal{P}_f}{d \ln k}\, .
\ee 

In the case of a scalar field $\chi$ the power--spectrum 
$\mathcal P_{\delta \chi}(k)$ can be evaluated by combining 
Eqs.~(\ref{redifinition}), (\ref{dec}) and~(\ref{commutators})
\be
\langle \delta \chi_{\bf k_1} \delta \chi^{*}_{\bf k_2} \rangle= 
\frac{|u_k|^2}{a^2} \delta^{(3)}({\bf k_1}-{\bf k_2})\, ,
\ee
yielding 
\be
\label{PSgeneral}
\mathcal P_{\delta \chi}(k)=\frac{k^3}{2 \pi^2} |\delta \chi_k|^2\, ,
\ee
where, as usual, $\delta \chi_k \equiv u_k /a$.

The expression in Eq.~(\ref{PSgeneral}) is completely general. In the case 
of a de Sitter phase and a very light scalar field $\chi$, with 
$m_\chi \ll 3/2 H$ we find from Eq.~(\ref{Sitterslowroll}) 
that the power--spectrum on superhorizon scales is given by 
\be
\mathcal P_{\delta \chi}(k)=\left( \frac{H}{2 \pi} \right)^2 
\left( \frac{k}{aH}\right)^{3-2\nu_\chi}\, ,
\ee
where $\nu _\chi$ is given by Eq.~(\ref{nu}). Thus in this case 
the dependence on time is tiny, and the spectral index slightly 
deviates from unity 
\be 
n_{\delta \chi} -1=3-2\nu_ \chi=2\eta_\chi.
\ee
A useful expression to keep in mind is that of a massless free scalar 
field in de Sitter space. In this case from Eq.~(\ref{exactm}) with 
$\nu_\chi=3/2$ 
we obtain 
\be
\label{exactmasless}
\delta \chi_{k}=(-H \tau) \left( 1-\frac{i}{k \tau} \right) 
\frac{e^{-ik\tau}}{\sqrt{2 k}}\, .
\ee
The corresponding two--point correlation function for the Fourier modes 
is 
\bea
\langle \delta \chi({\bf k_1}) \delta^{*} \chi({\bf k_2})  \rangle& =& 
\delta^{(3)}({\bf k_1}-{\bf k_2}) \, \frac{H^2 \tau^2}{2k_1} 
\left( 1+\frac{1}{k^2 \tau^2} 
\right)\\
&\approx& \delta^{(3)}({\bf k_1}-{\bf k_2})\,
 \frac{H^2}{2 k^3_1}\, \, \, \, \, \, \,     
(\textrm{for} \, \, \, k_1 \tau \ll 1)\, ,
\label{2pointMFSF}
\eea
with a power--spectrum which, 
on superhorizon scales, is given by 
\be
{\mathcal{P}}_{\delta \chi}(k) = \left( \frac{H}{2 \pi} \right)^2\, ,
\ee
which is exactly scale invariant.

We conclude this section with an important remark. Fluctuations of the 
scalar field can be generated on superhorizon scales as in 
Eq.~(\ref{exactSH}) 
only if the scalar field is light. In fact it can be shown that for very 
massive scalar fields $m_{\chi} \gg 3/2H$ (when 
$\nu_{\chi}$ in Eq.~(\ref{nu}) becomes imaginary) the 
fluctuations of the scalar field remain in the vacuum state and 
do not produce perturbations on cosmologically relevant scales.
Indeed, the amplitude of the power--spectrum is damped
exponentially as $e^{-2 m^2_{\chi}/H^2}$ and the spectral index
is equal to 4 \cite{prz}.

\subsection{Quantum fluctuations of a generic scalar field in a quasi--de 
Sitter stage}
\label{QFqDeSitter}
So far, we have analyzed the time evolution and compuited the 
spectrum of the quantum fluctuations of a generic scalar field $\chi$ 
assuming that the scale--factor evolves like in a pure de Sitter expansion, 
$a(\tau)=-1/(H\tau)$. However, 
during Inflation the Hubble rate is not exactly constant, 
but changes with time
as $\dot H=-\epsilon H^2$ (quasi--de Sitter expansion).
In this subsection, we will solve for the perturbations
in a quasi--de Sitter expansion. According to the conclusions of the 
previous section we 
consider a scalar field $\chi$ with a very small effective mass,
$\eta_\chi=(m^2_\chi/3 H^2) \ll  1$, and we proceed by making an expansion to 
lowest order in $\eta_\chi$ and the inflationary parameter 
$|\epsilon |\ll1 $.
Thus from the definition of the conformal time  
\be
a(\tau) \simeq -\frac{1}{H}\frac{1}{\tau(1-\epsilon)}\, .
\ee    
and 
\be
\frac{a''}{a}=a^2 H^2 \left( 2+\frac{\dot{H}}{H^2} \right) \simeq 
\frac{2}{\tau^2} \left( 1+\frac{3}{2} \epsilon \right)\, .
\ee
In this way we obtain again the Bessel equation (\ref{Bessel}) where now
$\nu_\chi$ is given by
\be
\nu_\chi \simeq \frac{3}{2}+\epsilon-\eta_\chi\, , 
\ee
to lowest order in $\eta_\chi$ and $\epsilon$. 
Notice that the time derivatives of the slow--roll parameters are next--order
in the slow--roll parameters themselves, $\dot{\epsilon},\dot{\eta}\sim 
{\mathcal O}(\epsilon^2, \eta^2)$, 
we can safely treat $\nu_\chi$ as a constant to our order 
of approximation. Thus the solution is given by 
Eq.~(\ref{exactm}) with the new expression of $\nu_\chi$.  
On large scales and to lowest order in the slow--roll parameters we find
\be
\label{qdesitterslowroll}
\left|\delta\chi_{k}\right|=
\frac{H}{\sqrt{2k^3}}
\left(\frac{k}{aH}\right)^{\frac{3}{2}-\nu_\chi}\, .
\ee
Notice that the quasi--de Sitter expansion yields a correction of order 
$\epsilon$ in comparison with Eq.~(\ref{Sitterslowroll}).  
Since on superhorizon scales from Eq.~(\ref{qdesitterslowroll}) 
\be
\delta\chi_{\bf k}\simeq \frac{H}{\sqrt{2k^3}}
\left(\frac{k}{aH}\right)^{\eta_\chi-\epsilon}\simeq
\frac{H}{\sqrt{2k^3}}\left[1+\left(\eta_\chi-\epsilon\right)
{\rm ln}\,\left(\frac{k}{aH}\right)\right],
\ee
we get 
\begin{equation}
\label{e}
\left|\delta\dot{\chi}_{\bf k}\right|\simeq \left|
H\,  \eta_\chi 
\,\delta\chi_{\bf k}\right|\ll \left|H\,\delta\chi_{\bf k}\right|,
\end{equation}
which shows that the fluctuations are (nearly) 
frozen on superhorizon scales. Therefore, 
a way to characterize the perturbations is to compute 
their power--spectrum on scales larger than the horizon, where one finds
\be
{\mathcal P}_{\delta \chi}(k) \simeq \left( \frac{H}{2 \pi} \right)^2 
\left(\frac{k}{aH} \right)^{3-2\nu_\chi}\, .
\ee
Let us conclude this subsection with a comment. Indeed the spectrum of the 
fluctuations of the scalar field $\chi$ in a quasi--de Sitter stage can 
be also obtained ignoring the variation of the Hubble rate and at the 
end replace $H$ by its value, $H_k$, at the time when the 
fluctuation of wavenumber $k$ leaves the horizon. The fact that 
the fluctuations get frozen on superhorizon scales guarantees that we get the 
exact result. From Eqs.~(\ref{Sitterslowroll}) and~(\ref{nu})
the power--spectrum obtained with this approach would read
\be
{\mathcal P}_{\delta\chi}(k)=\left(\frac{H_k}{2\pi}\right)^2
\left(\frac{k}{aH}\right)^{3-2\nu_\chi}\, ,
\label{bb}
\ee
with $3-2\nu_\chi \simeq 2 \eta_\chi$.
In fact by using the relation 
\be \label{relation}
H \simeq H_k \left[ 1+\frac{\dot{H}}{H^2}  \Big |_{k=aH} \ln \left( 
\frac{aH}{k}\right) \right] \simeq H_k {\left( \frac{k}{aH} \right)}
^{\epsilon}\, ,
\ee
we reproduce our previous findings.

\subsection{Correlation functions of a self interacting scalar field}
\label{3pointfunctions}
The two--point correlation function or its Fourier transform, 
the power--spectrum, corresponds to the magnitude of a given cosmological 
perturbation $f(t, {\bf x})$. 
If such a quantity is \emph{Gaussian distributed} then the 
power--spectrum is all that is needed in order to completely characterize it  
from a statistical point of view. In fact, in such a case if we consider 
higher--order correlation functions we find that all the odd
correlation functions vanish, while the even correlation functions can be 
simply expressed in terms of the two--point function. Another way 
to say that is to introduce the {\it connected} part of the correlation 
functions, defined as the part of the expectation value 
$\langle f(t, {\bf x_1}) f(t, {\bf x_2}) \cdots f(t, {\bf x_1}) f(t, {\bf x_N})
\rangle$ which cannot be expressed in terms of expectation values of lower 
order. For a zero--mean random field, the second and third--order
connected correlation functions coincide with the 
correlation functions themselves, while at fourth--order, 
for example, one can write 
\bea
&&\langle f(t, {\bf x_1}) f(t, {\bf x_2}) f(t, {\bf x_3}) 
f(t, {\bf x_4}) \rangle = \nonumber \\
&&\langle f(t, {\bf x_1}) f(t, {\bf x_2}) \rangle 
\langle f(t, {\bf x_3}) f(t, {\bf x_4}) \rangle
+
\langle f(t, {\bf x_1}) f(t, {\bf x_3}) \rangle 
\langle f(t, {\bf x_2}) f(t, {\bf x_4}) \rangle \nonumber \\
&& + 
\langle f(t, {\bf x_1}) f(t, {\bf x_4}) \rangle 
\langle f(t, {\bf x_2}) f(t, {\bf x_3}) \rangle
+
\langle f(t, {\bf x_1}) f(t, {\bf x_2}) f(t, {\bf x_3}) 
f(t, {\bf x_4}) \rangle_c \, , \nonumber \\
\eea
where $\langle \cdot \rangle_c$ denotes the connected part.  
Following the same hierarchical expansion one can express correlations 
functions of higher order in a similar manner. 
For the case of a Gaussian distributed 
perturbation $f(t, {\bf x})$ all the connected parts for $N >2$ are zero.
In particular it follows that the three--point function, or its Fourier 
transform, the \emph{bispectrum} represents the lowest--order statistics 
able to distinguish non--Gaussian from Gaussian perturbations. Therefore 
a large fraction of this review will focus on the study of the bispectrum of 
the cosmological perturbations produced in different cosmological scenarios.

As an instructive example we start in this section 
by considering the bispectrum of a scalar field 
$\chi(t, \bf{x})$ during a de Sitter stage. Such a 
computation can be performed by using the 
techniques of quantum field theory in curved space--time. 
In the context of inflationary cosmologies these techniques have been used in 
Refs.~\cite{AllenGW,FRS}. However, only recently some 
critical aspects of this approach have been clarified and a systematic 
formalism has been developed~\cite{Maldacena} (see also 
Ref.~\cite{BUzan1} for a critical investigation on the bases of such 
a calculation). Therefore we will now summarize how to calculate higher order 
correlation functions for a scalar field in a de Sitter space--time 
following mainly Refs.~\cite{Maldacena,BUzan1}.

Higher--order correlation functions are generated as soon as the scalar field 
has some interaction with itself (or other fields). This amounts to saying 
that the potential for $\chi$ contains some 
terms beyond the quadratic mass term, 
so that the interaction part of the potential can be written as 
\bea
V_{int}(\chi)=\frac{V^{(3)}}{3!} \left( \delta \chi \right)^3+
\frac{V^{(4)}}{4!} \left( \delta \chi \right)^4 + \cdots \, ,
\eea   
where $V^{(k)}$ is the \emph{k}--th derivative of the potential. 
The scalar field is quantized as in Eq.~(\ref{dec}) 
in terms of the eigenmodes $u_k(\tau)$.
We want to calculate the correlation function for $N$--points 
$\langle \delta \chi(1) \delta \chi(2) \cdots \delta \chi(N)\rangle$.  
The $N$--point correlation functions can be in fact expressed perturbatively 
in terms of those of the free scalar field which have been 
computed in Section~\ref{QFdeSitter}. To completely take into account the 
effects of the interaction terms, the underlying idea is that it is 
necessary to calculate expectation values for the actual vacuum state, that 
is to say the interacting vacuum state, not just the free vacuum state 
$|0 \rangle$ defined by the requirement that $a_{\bf k} |0\rangle=0$ 
for all $\bf{k}$\footnote{In Ref.~\cite{lps,Martin:1999fa,ganng} 
the non--Gaussian signatures on the CMB arising from
inflationary models with non--vacuum initial states for cosmological 
perturbations have been addressed}. 
Such expectation values are defined in the following way 
by using the interaction picture~\cite{Maldacena,BUzan1}
\bea
\label{ExpectationValue}
\langle \widetilde{\delta \chi}_{\bf{k_1}} 
\widetilde{\delta \chi}_{\bf{k_2}} \cdots
\widetilde{\delta \chi}_{\bf{k_N}}  \rangle \equiv 
\langle 0|U^{-1}(\tau_0,\tau) \widetilde{\delta \chi}_{\bf{k_1}} 
\widetilde{\delta \chi}_{\bf{k_2}} \cdots \widetilde{\delta \chi}_{\bf{k_N}} 
U(\tau_0,\tau) |0 \rangle \, ,
\eea 
where $U(\tau_0,\tau)$ is the time evolution operator defined as 
\be
U(\tau_0,\tau)=\exp\left( -i \int_{\tau_0}^\tau d \tau' 
H_{int}(\tau') \right)\, .
\ee
We have already moved to Fourier space where the calculations are easier. 
Here $\tau_0$ is some early time at which the interactions of the field are 
supposed to switch on, while 
$H_{int}=\frac{V^{(3)}}{3!} \left( \delta \chi \right)^3+
\frac{V^{(4)}}{4!} \left( \delta \chi \right)^3+ \cdots$ 
is the Hamiltonian in the interaction picture.  
Thus, as it has been pointed out in Ref.~\cite{Maldacena} the quantity in 
Eq.~(\ref{ExpectationValue}) doe not correspond to a scattering amplitude, 
where the initial 
(at $t \rightarrow -\infty$)
and the final (at $t \rightarrow +\infty$) 
states are considered as free states. Moreover the expectation value 
defined in Eq.~(\ref{ExpectationValue}) is free of the critical divergences 
that occur when calculating the correlation functions on the free vacuum 
state in a de Sitter space--time, as explained in Ref.~\cite{BUzan1} (see also 
the references therein). \\
To first order in $H_{int}$ the evolution operator can be expanded as 
\be
U(\tau_0,\tau)=I-i\int_{\tau_0}^{\tau} d \tau' H_{int}(\tau')\, ,
\ee
and it follows that the connected part of the N--point correlation function is 
given by
\be
\label{connectedpart}
\langle \widetilde{\delta \chi}_{\bf{k_1}} 
\widetilde{\delta \chi}_{\bf{k_2}} \cdots
\widetilde{\delta \chi}_{\bf{k_N}}  \rangle=
-i \int_{\tau_0}^{\tau} d \tau' \, \,  \langle 0| 
[ \langle \widetilde{\delta \chi}_{k_1} \widetilde{\delta \chi}_{k_2} \cdots
\widetilde{\delta \chi}_{k_N}  \rangle, H_{int}(\tau') ] 0\rangle\, .
\ee
The final result can be expressed 
in terms of the Green function~\cite{BUzan1}
\be
G(k, \tau, \tau')=\frac{1}{2k}\left(1-\frac{i}{k\tau} \right)\left(1+
\frac{i}{k\tau'} \right) \exp[ik(\tau'-\tau)]\, ,
\ee
which can be obtained from the definition 
$\langle 0|\widetilde{\delta \chi}(\tau,{\bf{k}}) 
\widetilde{\delta \chi}(\tau',{\bf{k'}}) |0\rangle = 
\delta^{(3)} ({\bf k}+{\bf k'}) G(k,\tau,\tau')$ by employing perturbatively 
the solution of the free massless scalar field~(\ref{exactmasless}).

Let us now consider a specific example where the scalar field potential 
contains a cubic interaction term 
$(\lambda/3!)\chi^3$~\cite{FRS,BUzan1} so that we can write   
\be
H_{int}=\frac{\lambda}{3!} \delta \chi^3\, ,
\ee      
where $\lambda$ is a coupling constant.
From Eq.~(\ref{connectedpart}) it then follows
\bea
\label{Greens}
& &\langle \widetilde{\delta \chi}_{\bf{k_1}} 
\widetilde{\delta \chi}_{\bf{k_2}}
\widetilde{\delta \chi}_{\bf{k_3}}  \rangle=
-i \lambda \delta^{(3)}({\bf k_1}+{\bf k_2}+{\bf k_3}) \times \nonumber \\ 
& & \int_{-\infty}^{\tau} \frac{-\d \tau'}{H\tau'} [G(k_1,\tau,\tau')
G(k_2,\tau,\tau')G(k_3,\tau,\tau')
-G^{*}(k_1,\tau,\tau')G^{*}(k_2,\tau,\tau')G^{*}(k_3,\tau,\tau')]\, . 
\nonumber \\
\eea 
Here $\tau$ corresponds to the conformal time at the end of inflation.\\
Such an integral depends on some combinations of the norms of the 
wavevectors like $\pi_1=\sum_i k_i$, $\pi_2 =\sum_{i < j} k_i$, 
$\pi_3= \sum_{i<j<k} k_i$.
Actually it can be performed and expressed in the large scale limit, 
$k_i\tau \ll 1$ for all $i$, as~\cite{BUzan1}
\bea
\label{3pointchi}
\langle {\delta \chi}_{\bf{k_1}} 
{\delta \chi}_{\bf{k_2}} 
{\delta \chi}_{\bf{k_3}}  \rangle=
- \frac{\lambda H^2}{12} \frac{\delta^{(3)}(\sum_i {\bf k_i})}{\prod k_i^3} 
\left[ -\sum_i k_i^3 \Big( \gamma+\zeta_3(k_i)+\log[-k_t \tau] \Big) \right]
\, , \nonumber \\
\eea  
where we have switched to the field $\delta \chi=\widetilde{\delta \chi}/a$.
In formula~(\ref{3pointchi}) $k_T=k_1+k_2+k_3$, 
$\gamma$ is the Euler constant $\gamma \approx0.577$ and 
$\zeta_3(k_i)$ is a function which can be expressed in terms 
of the combinations $\pi_i$. As it has been shown in detail 
in Ref.~\cite{BUzan1} the function $\zeta (k_i)$ weakly depends on the 
wavevectors and it is always of the order of unity.\footnote{
The precise expression of $\zeta_3(k_i)$ is given by the appropriate limit 
$k_4 \rightarrow 0$ in the expression
\be
\zeta_4(k_i)=\frac{-\pi_1^4+2\pi_1^2\pi_2+\pi_1\pi_3-3\pi_4}
{\pi_1(\pi_1^3-3\pi_1\pi_2+3\pi_3)}\, ,
\ee
where $\pi_4=\sum_{i<j<k<l} k_i$. $\zeta_4(k_i)$ is the corresponding 
quantity that appears in the connected part of the correlation functions for a 
potential $\frac{\lambda}{4!} \chi^4$ 
(see Ref.~\cite{BUzan1} for more details).
}\\
Notice that it is standard use to express this result as the sum of 
products of the two--point correlation function on large scales given 
in Eq.~(\ref{2pointMFSF})  
\be
\label{3pointF}
\langle {\delta \chi}_{\bf{k_1}} 
{\delta \chi}_{\bf{k_2}} 
{\delta \chi}_{\bf{k_3}}  \rangle=\nu_{3}(k_i) \sum_i \prod_{J \neq i} 
\frac{H^2}{2k_j^3}\, ,
\ee
where
\be
\label{nuova}
\nu_3(k_i)=\frac{\lambda}{3H^2} 
\Big[ \gamma+\zeta(k_i)+ \log [-k_T\tau] \Big] \, .
\ee
The term $\log[- k_T \tau]$ corresponds to $N_{k_T}=\log 
\left( a_{end}/a_{k_T} \right)$ 
which is the number of e--foldings from the time the scale corresponding to 
$k_T$ leaves the horizon during inflation and the end of inflation. 
Typically $N_{k_T} \approx 60$ for observable cosmological large scales 
and thus it dominates over the other terms which are typically of 
the order of unity so that one can approximate 
\be
\label{expr:nu}
\nu_3(k_i) \approx -\frac{\lambda N_{k_T}}{3H^2}\, .
\ee
This result, first found in Refs.~\cite{AllenGW,FRS}, 
has actually a very transparent physical interpretation~\cite{BUzan1}. 
The first two terms in Eq.~(\ref{nuova}) can be interpreted as  
genuine quantum effects of 
the scalar field modes on scales smaller than the cosmological horizon  
which leave a characteristic (scale--dependent) imprint at the time of horizon 
crossing.\footnote{
Following Ref.~\cite{Maldacena} we just recall here how to perform 
the integrals like~(\ref{Greens}). One 
can split them as integrals over the region outside the horizon, the region 
around horizon crossing and the region much smaller than the horizon. 
Moreover in order to take automatically into account that we are considering 
expectation values on the interacting vacuum one has to deform the 
$\tau$ integration contour so that it has some evolution in Euclidean time. 
This is achieved by the change $\tau \rightarrow 
\tau +i\varepsilon |\tau|$, for 
large $|\tau|$. In this way since on subhorizon scales the fields oscillate 
rapidly, the integration over the region deep inside the horizon 
does not give any contribution.}
On the other hand after a few e--foldings after the modes 
leaves the horizon, the evolution of the field can be described in a 
classical way and it just corresponds to the term proportional to the number 
of e--foldings $N_{k_T}$. \\
In fact this can be shown by solving the Klein--Gordon equation for the 
scalar field $\chi$ in a de Sitter background in the large--scale limit
\emph{up to second order} in the perturbations. If we expand the scalar 
field as 
\be
\chi(\tau, {\bf x})=\chi_0(\tau)+ \delta \chi (\tau, {\bf x})= 
\chi_0(\tau) +\delta^{(1)} \chi(\tau, {\bf x}) 
+\frac{1}{2} \delta^{(2)} \chi(\tau, {\bf x})\, ,
\ee 
where we split the scalar field perturbation into a 
first and a second 
order part, the evolution equation for $\delta \chi^{(2)}$ on large scales 
reads in cosmic time
(see Ref.~\cite{noi})
\be
\label{KG2ndorder}
\ddot{\delta^{(2)} \chi}+3H \dot{\delta^{(2)} \chi}+
\frac{\partial^2 V}{\partial \chi^2} \delta^{(2)} \chi=
- \frac{\partial^3 V}{\partial \chi^3} \left( \delta^{(1)} \chi \right)^2\, . 
\ee
In a slow--roll approximation Eq.~(\ref{KG2ndorder}) becomes 
\be
3 H \dot{\delta^{(2)} \chi} \approx 
- \frac{\partial^3 V}{\partial \chi^3} \left( \delta^{(1)} \chi \right)^2
\, ,  
\ee 
whose solution is 
\be
\label{classical}
\delta^{(2)} \chi =-\frac{\lambda}{3 H^2} N_k 
\left( \delta^{(1)} \chi \right)^2+\delta^{(2)} \chi (t_k)\, ,
\ee
where $N_k=\int_{t_k}^{t_{end}} H dt=H \Delta t$ is the number of e--folds 
between the end of inflation and the time $t_k$ the scale of wavenumber $k$ 
leaves the horizon during inflation and we have used the fact that 
$V=(\lambda/3!) \chi^3$.  
The integration constant $\delta^{(2)} \chi (t_k)$ is the value of 
the field at horizon-crossing and  
corresponds to the terms coming from 
quantum effects on subhorizon scales. However as it is evident from the 
result in Eq.~(\ref{classical}) these terms will be subdominant with respect 
to that corresponding to the scalar field dynamics 
occurring once the mode leaves the horizon.
\footnote{Notice that such a result, obtained computing second--order 
perturbations is completely equivalent to solving the equation of motion 
for the scalar field outside the horizon using a perturbative expansion 
in the coupling $\lambda$, as done in Ref.~\cite{BUzan1}. 
The equation of motion is 
\be
\ddot{\chi}+3H\dot{\chi}= -\frac{\partial V}{\partial \chi}\, , 
\ee 
neglecting spatial gradients. This is equivalent to consider 
$\chi$ in this equation as a filtered field on scales that leave the horizon 
at a given time $t_k$~\cite{BUzan1}. At zeroth order in $\lambda$ 
the solution 
$\chi_{(0)}$ is a Gaussian field which remains constant. The first order 
correction then satisfies
\be
\ddot{\chi}_{(1)}+3H \dot{\chi}_{(1)}=
-\frac{\partial V}{\partial \chi}(\chi_{(0)})
\, ,
\ee
which for a potential $V(\chi)= \lambda\chi^3/3!$ gives 
\be
\label{calssical}
\chi_{(1)}=\chi_{(1)}(t_k)-\frac{\lambda}{2} \left( 
\chi_{(0)}\right)^2 \frac{t-t_k}{3 H}=
\chi_{(1)}(t_k)-\frac{\lambda}{2} \left( \chi_{(0)}\right)^2 
\frac{N_k}{3 H^2}\, .
\ee
}

We conclude this section with some further comments. Such calculations are 
performed without taking into account any perturbation of the metric, 
such as the gravitational potential. Notice that if the scalar 
field $\chi$ is the inflaton field then according to these 
results the magnitude of the non--linearity parameter $f_{\rm NL}$ 
for the gravitational potential should be proportional to $\lambda$ and then 
in terms of the slow--roll parameters 
$f_{\rm NL} \simeq {\mathcal O}(\xi^2)$, where 
$\xi^2=M_{\rm P}^2 \,  (V^{(1)}V^{(3)}/V^{(2)})$ which is second--order 
in the slow--roll parameters. However, it has been shown in 
Refs.~\cite{Getal,gan} (see also Ref.~\cite{barcol}) 
that, when accounting also for the non--linearities in 
the metric perturbations the level of non--Gaussianity 
turns out to be $f_{\rm NL} \simeq {\mathcal O}(\epsilon, \eta)$. In fact the 
main contribution to the non--Gaussianity in single--field 
models of slow--roll 
inflation comes from the non--linear gravitational perturbations, rather than 
the inflaton self--interactions. The authors of 
Ref.~\cite{Getal} have used 
the so--called stochastic approach to inflation~\cite{StaroStoc}(for a more 
recent approach, see Ref.~\cite{mmr}). 
Such a result has been also recently obtained in 
a more rigorous way in Refs.~\cite{noi,Maldacena} by studying the 
perturbations in the metric and in the inflaton field up to second order in 
deviations from the homogeneous background until the end of inflation. 
These results show a general principle holding for single--field models of 
slow--roll inflation. In order to have a period of inflation the inflaton 
potential must be very flat (\emph{i.e} $\epsilon,|\eta| \ll 1$), therefore 
the self--interaction terms in the inflaton potential and the gravitational 
coupling  must very small and then non--linearities are suppressed too.  
On the other hand if the scalar field $\chi$ is different from the inflaton 
and it gives a negligible contribution to the total energy density, its 
self--interactions are not constrained by any inflationary slow--roll 
condition, and thus sizeable non--Gaussianities can be generated. This is the 
scenario which has been proposed for example in Ref.~\cite{AllenGW}. However, 
in this case the perturbations produced have a very little impact on the 
total energy density perturbation since the energy density of the scalar 
field is subdominant (the so--called isocurvature perturbations), and this 
kind of scenario is not in accordance with present observational data.
There is one more interesting possibility for the self--interactions of
a scalar field to play an important role in producing non--Gaussian 
signatures. If the subdominant scalar field $\chi$ is {\emph coupled} to 
the inflaton field then it is possible that non--Gaussianities intrinsic 
in the scalar field $\chi$ are transferred to the inflaton sector, 
as it was first proposed in Ref.~\cite{BMT3}. We will come back to this 
scenario in detail in Sec.~\ref{multiplefields}. 

The intrinsic non--linearities of the scalar fields 
present during inflation are only a possible source of non--Gaussianity. 
Indeed, even if the fluctuations of the scalar field $\chi$ are Gaussian
distributed, it is possible that their energy density perturbations have some 
non--linearity. This is the 
case, for instance, of a scalar field $\chi$ different from the 
inflaton with a quadratic potential $V(\chi) \propto \chi^2$ leading to a 
vacuum expectation value $\langle \chi \rangle=0$, as it has been 
proposed in Ref.~\cite{LM} (see also~\cite{musc}) 
In this case the energy density perturbations are not 
given by the usual linear contribution $\delta \rho_\chi \propto 
\chi \delta \chi$ but will be non--Gaussian with $\delta \rho_\chi \propto 
\delta \chi^2$. Actually such a quadratic contribution to the primordial 
energy density perturbations is the key feature of the non--Gaussianities 
in the curvaton scenario~\cite{LW2}.\\ 
The gravitational dynamics itself introduces important non--linearities, 
which will contribute to the final non--Gaussianity in the large--scale CMB 
anisotropies. In fact as it has been shown in Ref.~\cite{BMR2} 
it is just because 
of the non--linear gravitational dynamics that the tiny non--Gaussianity 
produced during inflation gets amplified in the post--inflationary evolution.


\subsection{Metric perturbations and the energy--momentum tensor}
\label{MetricEMTensor}

In the previous sections we have shown how perturbations in a generic scalar 
field $\chi$ are generated on superhorizon scales during an inflationary 
period. This is the first step to understand the production and 
evolution of the cosmological perturbations in the different scenarios 
we are going to consider. In the standard single--field inflation, as well 
as in the curvaton and inhomogeneous reheating scenarios cosmological 
perturbations can be traced back initially to fluctuations of scalar fields; 
then they evolve in the radiation--dominated phase and, subsequently, in 
the matter and dark energy--dominated phases.

Let us first focus on the generation of the perturbations of scalar 
fields and make some remarks.  
As we have already emphasized in the discussion of the previous 
section, so far we have neglected the perturbations in the metric tensor 
around the homogeneous FRW background. 
In the case of the inflaton field taking into account the metric 
perturbations is of primary importance. The reason is very simple. The 
inflaton field drives the accelerated inflationary expansion, 
which means that it dominates the energy density of the 
Universe at that time. Thus any perturbation in 
the inflaton field 
$\delta\varphi$ implies  a perturbation of the energy--momentum tensor 
$\delta T_{\mu\nu}$, and a perturbation in the energy--momentum 
tensor implies, through Einstein's equations of motion 
$G_{\mu\nu}= 8\pi G_{\rm N} T_{\mu\nu}$, a perturbation of the metric. On the 
other hand perturbations in the metric affect the evolution of 
the inflaton fluctuations $\delta \varphi$ through the perturbed 
Klein--Gordon equation. We thus conclude that in the standard scenario of 
inflation perturbations of the inflaton field and perturbations of the metric 
are tightly coupled to each other and have to be studied together. 
Indeed this is the correct way to proceed. 
A very general scenario for the generation of the cosmological perturbations 
is one where other scalar fields $\chi_I$ are present besides the inflaton. 
This could be the case of inflation driven by several scalar fields whose 
contribution to the total energy density is comparable 
(multi--field inflation), or the case where 
the extra scalar fields are subdominant~\cite{Linde1,Starobinsky,moma,Polarstarobinsky,iso1,iso2,KL,MollMatOrtLucchin,LM,borga,langlois:1999,gw,BMT1,BMT2}. 
In such a general scenario a consistent way to study the production of 
cosmological fluctuations is to perturb both the 
scalar fields and the metric. The metric 
perturbations will then have a feedback also on the evolution of the 
subdominant scalar fields. Moreover, in such a general picture the different  
scalar fields can interact with one another through a generic potential 
$V(\varphi, \chi_I)$, while 
we have neglected such interactions for the scalar field $\chi$ so far.

During the radiation/matter--dominated 
eras, as we see again from the Einstein equations, a consistent study of the 
cosmological perturbations must take into account       
perturbations both in the energy momentum--tensor and 
in the metric tensor. 
We shall see that the relation between the energy--momentum 
perturbations and the metric is also justified in light of gauge issues. 
In fact we shall introduce some 
gauge--invariant quantities that mix both matter and metric perturbations. 
This will be essential in order to study the evolution of metric 
perturbations during the different stages, from  the early period of 
inflation/reheating to the subsequent radiation and matter dominated epochs.
As pointed out in the Introduction, in order to keep track of the 
non--Gaussianity of the cosmological perturbations  
throughout these different stages we perform our analysis \emph {up to 
second order} in the perturbations. In particular we will focus on some 
quantities which are gauge--invariant up to second order, and which allow 
us to follow the evolution of the metric perturbations 
(the gravitational potentials) taking into account the different 
second--order contributions to the non--linearities of the perturbations.\\

\section
{\bf Cosmological perturbations at first and second order}

In order to study the perturbed Einstein's equations, 
we first write down the perturbations on a spatially flat 
FRW background following the formalism of 
Refs.~\cite{BMMS,MMB,noi}. We shall first consider the fluctuations of the 
metric, and then the fluctuations of the energy--momentum tensor. 
Hereafter greek indices will be taken to run from $0$ to $3$, while latin 
indices, labelling spatial coordinates, will run from $1$ to $3$. If not 
otherwise specified we will work with conformal time $\tau$, and primes will 
denote differentiation with respect to $\tau$.  

\subsection{The metric tensor}
\label{metrictensor}
The components of a spatially flat FRW 
metric perturbed up to second order can be written as
\begin{eqnarray} \label{metric1}
g_{00}&=&-a^2(\tau)\left( 1+2 \phi^{(1)}+\phi^{(2)} \right)\, ,\nonumber\\
g_{0i}&=&a^2(\tau)\left( \hat{\omega}_i^{(1)}+\frac{1}{2} 
\hat{\omega}_i^{(2)} \right)
\, ,  \nonumber\\g_{ij}&=&a^2(\tau)\left[
(1 -2 \psi^{(1)} - \psi^{(2)})\delta_{ij}+
\left( \hat{\chi}^{(1)}_{ij}+\frac{1}{2}\hat{\chi}^{(2)}_{ij} \right)\right] 
\,.
\end{eqnarray}
The functions $\phi^{(r)}, \hat{\omega}_i^{(r)}, 
\psi^{(r)}$ and $\hat{\chi}^{(r)}_{ij}$, where $(r)=(1),(2)$, stand for the 
$r$th--order perturbations of the metric. 
Notice that such an expansion could a 
priori include terms of arbitrary order~\cite{MMB}, 
but for our purposes the first and second--order terms are sufficient. 
It is standard use to split the perturbations into the so--called scalar, 
vector 
and tensor parts according to their transformation properties with respect to 
the $3$-dimensional space with metric $\delta_{ij}$, where scalar parts are 
related to a scalar potential, vector parts to transverse (divergence--free) 
vectors and tensor parts to transverse trace--free tensors. Thus in our case
\begin{equation}
\hat{\omega}_i^{(r)}=\partial_i\omega^{(r)}+\omega_i^{(r)}\, ,
\end{equation}
\begin{equation}
\hat{\chi}^{(r)}_{ij}=D_{ij}\chi^{(r)}+\partial_i\chi^{(r)}_j
+\partial_j\chi^{(r)}_i
+\chi^{(r)}_{ij}\, ,
\end{equation}
where $\omega_i$ and $\chi_i$ are transverse vectors, {\it i.e.} 
$\partial^i\omega^{(r)}_i=\partial^i\chi^{(r)}_i=0$, $\chi^{(r)}_{ij}$ 
is a symmetric transverse and trace--free tensor,  {\it i.e.} 
$\partial^i\chi^{(r)}_{ij}=0$, $\chi^{i(r)}_{~i}=0$) and $D_{ij}=
\partial_i \partial_j - (1/3) \, \, 
\delta_{ij}\, \nabla^2$ is a trace--free operator.
\footnote{Here and in the following latin indices 
are raised and lowered using $\delta^{ij}$ and $\delta_{ij}$, respectively.}

Let us recall that the reason why such a splitting has been introduced 
\cite{Bardeen1,KodaSasa}
is that, \emph{at least} in linear theory, these different modes are decoupled
from each other in the perturbed evolution equations, so that they can 
be studied separately. As we shall see throughout 
the following sections this property 
does not hold anymore beyond the linear regime where 
second--order perturbations   
\emph{are} coupled -- sourced -- by first--order perturbations.

For our purposes the metric in Eq.~(\ref{metric1}) 
can be simplified. The fact that 
first--order vector perturbations have decreasing amplitudes and that 
are not generated in the presence of scalar fields, 
allows us to conclude that they can be safely disregarded.
Moreover, the first--order 
tensor part gives a negligible contribution to second--order
perturbations.
Thus, in the following we will neglect  
$\omega^{(1)}_i$, $\chi^{(1)}_{i}$ and $\chi^{(1)}_{ij}$.
However the same reasoning does not apply to second--order 
perturbations.  
Since in the non--linear case scalar, vector and tensor modes are dynamically 
coupled, the second--order vector and tensor 
contributions are generated by first--order scalar perturbations 
even if they were initially zero~\cite{MMB}. 
Thus we have to take them into account and we shall use the metric
\bea \label{metric2}
g_{00}&=&-a^2(\tau)\left( 1+2 \phi^{(1)}+\phi^{(2)} \right)\, ,\nonumber\\
g_{0i}&=&a^2(\tau)\left( \partial_i\omega^{(1)}+\frac{1}{2}\, 
\partial_i\omega^{(2)}+\frac{1}{2}\, \omega_i^{(2)} \right)
\, ,  \nonumber\\g_{ij}&=&a^2(\tau)\left[
\left( 1 -2 \psi^{(1)} - \psi^{(2)} \right)\delta_{ij}+
D_{ij}\left( \chi^{(1)} +\frac{1}{2} \chi^{(2)} \right)\right.\nonumber\\
&+&\left.\frac{1}{2}\left( \partial_i\chi^{(2)}_j
+\partial_j\chi^{(2)}_i
+\chi^{(2)}_{ij}\right)\right]
\,.
\eea 
The contravariant metric tensor is obtained by requiring (up to second order)
that $g_{\mu\nu}g^{\nu\lambda}=\delta_\mu^\lambda$ and is given by
\bea \label{cont}
g^{00}&=&-a^{-2}(\tau)\left( 1-2 \phi^{(1)}-\phi^{(2)} +4\left(
\phi^{(1)}\right)^2-\partial^i\omega^{(1)}\partial_i\omega^{(1)}  
\right)\, ,\nonumber\\
g^{0i}&=&a^{-2}(\tau)\left[ \partial^i\omega^{(1)}+\frac{1}{2}
\left( \partial^i\omega^{(2)}+\omega^{i(2)} \right) +2 \left( \psi^{(1)}
-\phi^{(1)} \right) \partial^i\omega^{(1)}\right. \nonumber\\
&-&\left. \partial^i\omega^{(1)}D^i\,_k \chi^{(1)} \right]
\, ,  \nonumber\\
g^{ij}&=&a^{-2}(\tau) \left[
\left( 1+2 \psi^{(1)} +\psi^{(2)}+4 \left( \psi^{(1)} \right)^2 
\right) \delta^{ij}-
D^{ij}\left( \chi^{(1)} +\frac{1}{2} \chi^{(2)} \right)\right.\nonumber \\
&-& \frac{1}{2}\left( \partial^i\chi^{j(2)}
+\partial^j\chi^{i(2)}
+\chi^{ij(2)} \right)-\partial^i\omega^{(1)}\partial^j\omega^{(1)}\nonumber\\
&-&\left.
4 \psi^{(1)}D^{ij}\chi^{(1)}+D^{ik}\chi^{(1)}D^j_{~k} \chi^{(1)} \right]\,.
\eea
Using $g_{\mu\nu}$ and $g^{\mu \nu}$ one can calculate the 
connection coefficients $\Gamma^{\alpha}_{\beta \gamma}$ and the Einstein 
tensor components $G^{\mu}_{\nu}$ up to second order in the metric 
fluctuations. We report their complete expressions in Appendix A; they can 
be also found in Ref.~\cite{noi}.

Let us conclude this section by noting that in the following we 
we will often adopt the {\it Poisson
gauge} \cite{Poisson1,Poisson2} which is defined
by the condition $\omega^{(r)}=\chi^{(r)}=\chi^{(r)}_{i}=0$. Then, one scalar
degree of freedom is eliminated from $g_{0i}$ and one scalar
and two vector degrees of freedom from $g_{ij}$. This gauge generalizes the
so--called longitudinal gauge to include vector and tensor modes.
\subsection{The energy--momentum tensor} 
\label{energy--momentum}
In this section we shall consider a 
fluid characterized by the energy--momentum tensor
\begin{equation}
\label{EM}
T^\mu_{~ ~\nu}= \left(\rho+P\right)u^\mu u_\nu + P \delta^{\mu}_{~\nu}\,,
\end{equation}
where $\rho$ is the energy density, $P$ the pressure, and $u^\mu$ is the 
fluid four--velocity subject to the constraint
$g^{\mu\nu}u_\mu u_\nu=-1$. Notice that we do not include any anisotropic 
stress term in our energy--momentum tensor, i.e. we make the perfect fluid
hypothesis, since in the different scenarios we are going to discuss 
anisotropic stresses are not present, as we only deal with scalar fields, 
matter and radiation components.
Indeed, we devote a specific section to the energy--momentum tensor 
of a scalar field, given its importance for the standard scenario 
of inflation. Here, we also
restrict ourselves to the case where the equation of state of the fluid 
$w=P/\rho$ is constant, with $w=1/3$ for a radiation fluid and $w=0$ for 
collisionless matter (dust).   
We now expand the basic matter variables $u^{\mu}$, $\rho$ and $P$ up to 
second order in the perturbations around the homogeneous background.
For the velocity we write
\begin{equation}
u^\mu=\frac{1}{a}\left(\delta^{\mu}_0+v^{\mu}_{(1)}+\frac{1}{2}v^{\mu}_{(2)}
\right)\, .
\end{equation}
From the normalization condition we obtain 
\bea
v^{0}_{(1)}&=&- \phi^{(1)},\nonumber\\
v^{0}_{(2)}&=& -\phi^{(2)} +3\left(\phi^{(1)}\right)^2 
+ 2\,  \partial_i \omega^{(1)} v^{(1)}
+ v_i^{(1)}v^i_{(1)} \, .
\eea
Notice that the velocity perturbation $v^{i}_{(r)}$ also 
splits into a scalar (irrotational) and a vector (solenoidal) part, as
\be
v^{i}_{(r)}=\partial^{i} v_{(r)} + v^{i}_{(r){\mathcal S}}\, , 
\ee
with $\partial_i v^{i}_{(r){\mathcal S}}=0$. According to what 
we said in the previous section
we can neglect the linear vector velocity perturbation.
\footnote{Notice, however, that in the following expression for the 
perturbed energy--momentum tensor we are still completely general by 
including also linear vector and tensor perturbation modes.} 

Using the metric of Eq.~(\ref{metric2}) we find for 
$u_{\mu}=g_{\mu \nu} u^{\nu}$   

\bea
u_{0}&=&a\left(-1-\phi^{(1)}-\frac{1}{2}\phi^{(2)}+\frac{1}{2}
\left(\phi^{(1)}\right)^2-\frac{1}{2}v_i^{(1)}
v^i_{(1)}\right)\,,\nonumber\\
u_{i}&=&a\left(v_i^{(1)}+\partial_i \omega^{(1)}
+\frac{1}{2}v_i^{(2)}+\frac{1}{2}\omega_i^{(2)}
-\phi^{(1)}\partial_i \omega^{(1)}
-2\psi^{(1)}v_i^{(1)} \right. \\ \nonumber
&+& \left. D_{ij}\chi^{(1)}v^{j}_{(1)} \right)\, .
\eea

The energy density $\rho$ 
can be split into a homogeneous background $\rho_0(\tau)$ 
and a perturbation $\delta \rho (\tau, x^{i})$ as follows

\begin{equation}
\label{rho}
\rho(\tau, x^i)=\rho_0(\tau)+\delta\rho(\tau, x^i)=\rho_0(\tau)+
\delta^{(1)}\rho(\tau, x^i)+\frac{1}{2}\delta^{(2)}
\rho(\tau, x^i)\, ,
\end{equation}
where the perturbation has been expanded into a first and a second--order
part. The same decomposition can be adopted for the
pressure $P$, where in our case $\delta P =w \delta \rho$.

Using the expression (\ref{rho}) for the energy density and the 
expressions for the velocity into Eq.~(\ref{EM}) we calculate
$T^\mu_{~\nu}$ up to second order and we find 
\bea 
T^\mu_{~\nu}=T^{\mu(0)}_{~\nu}+\deu T^{\mu}_{~\nu}+\ded
T^{\mu}_{~\nu} \, ,
\eea
where $T^{\mu(0)}_{~\nu}$ corresponds to the background value, and
\bea 
\label{scalT_00}
T^{0(0)}_{~0}+\deu T^{0}_{~0}&=& -\rho_0-\delta^{(1)}\rho\, , \\ 
\label{00}
\ded T^0_{~0}&=& -\frac{1}{2}\delta^{(2)}\rho-\left(1+w\right)\rho_0
v_i^{(1)}v^i_{(1)}-\left(1+w\right)\rho_0 \, \partial_i \omega^{(1)} 
v^i_{(1)}\, ,  \\
&& \nonumber \\
&& \nonumber \\
\label{scalT_0i}
T^{i(0)}_{~0}+\deu T^{i}_{~0}&=& -\left(1+w\right)\rho_0 v^i_{(1)}\, , \\
\label{0i}
\ded {T^i_{~0}}&=& -\left(1+w\right)\rho_0\left[
\frac{1}{2}v^i_{(2)}+
\left(\psi^{(1)}+
\frac{\delta^{(1)}\rho}{\rho_0}\right)v^i_{(1)}\right] \, ,\\
&& \nonumber \\
&& \nonumber \\
T^{i(0)}_{~j}+\deu T^{i}_{~j}&=& w\rho_0 \left(1+
\frac{\delta^{(1)}\rho}{\rho_0}\right)
\delta^{i}_{~j}\,, \\
\label{ij}
\ded{T^i_{~j}}&=& 
\frac{1}{2}w\, \delta^{(2)}\rho \, \delta^{i}_{~j}+ 
\left(1+w\right)\rho_0v^i_{(1)}\left(v_j^{(1)}+\partial_j \omega^{(1)} 
\right)\, .
\eea
A comment is in order here. As it can be seen from Eq.~(\ref{metric2}) 
and Eqs.~(\ref{00}), (\ref{0i}) and (\ref{ij}) 
the second--order perturbations always contain two different contributions, 
quantities which are intrinsically of second order, 
and quantities which are given by the product of two first--order 
perturbations. As a consequence, when considering the 
Einstein equations to second order in Section~\ref{2ndEE}, 
first--order perturbations behave as a source for the 
intrinsically second--order fluctuations. This 
is an important issue which was pointed out in different 
works on second--order perturbation 
theory~\cite{Tomita1,Tomita2,MPS1,MPS2,MMB} 
and it plays a central role in deriving our main results on non--Gaussianity of
cosmological perturbations.

\subsection{Gauge dependence at second order}

In the previous sections we have defined the perturbation 
$\delta T$ in a given quantity $T$
considering the difference between the physical value of $T$  
(the perturbed one) and the background unperturbed value $T_{0}$, 
and in the specific we have then expanded such a perturbation in a 
first and a second--order part. 
However the theory of perturbations in General Relativity intrinsically 
encodes 
a certain degree of arbitrariness in performing such a comparison between the 
physical and the reference background quantities. This is because we  
consider perturbations of the space--time itself on which a given quantity is  
defined. Thus we deal with two different space--times, the real physical 
space--time
and the unperturbed background (which in our case is the FRW space--time), 
where $T$ and $T_0$ are defined respectively. 
In order to compare the value of $T$ 
with the reference value $T_0$ it is necessary to establish a map, that is 
a one--to--one correspondence, between the physical and the background 
space--times.
Such a map is a gauge choice, and a change of the map is a gauge 
transformation. From the point of view of a set of coordinates this means a 
change of the coordinates. However the gauge choice is not unique, since 
General Relativity {\emph is} a theory based on the 
freedom of changing locally the system of coordinates. 
Therefore the value of the perturbation in the generic 
quantity $T$ depends on the gauge, or in other words, the perturbation 
$\delta T$ 
will transform after a change of coordinates thus acquiring different values, 
which nonetheless are on the same footing. This is the issue of 
gauge--dependence, which holds at any order of perturbations.  
There are two options to avoid such an ambiguity: identify combinations of 
perturbations which are gauge--invariant quantities, that is to say 
quantities which are independent of the gauge transformation, or choosing a 
given gauge and perform the calculations in that 
gauge. The second option could introduce pure gauge--modes, which have no 
physical meaning, and must be eliminated from the solutions.   
The gauge--dependence has been widely studied within linear perturbation theory
in different papers~\cite{sachs,SW,Bardeen1,EB,stewart} 
and discussed specifically in the
context of cosmological perturbations in some reviews~\cite{KodaSasa,MFB}. 
Only recently the gauge issue has been addressed in a systematic way beyond 
the linear 
regime~\cite{BMMS,SonegoBruni,MMB} giving a full description of gauge 
transformations at second order. 

It is not the purpose of this review to describe in detail
gauge--transformations for second--order perturbations, and we refer the 
reader to detailed papers on the subject~\cite{BMMS,SonegoBruni}. 
We want indeed to focus on some gauge--invariant quantities which play a major 
role when study the evolution of second--order perturbations 
and which allow us 
to determine accurately  the resulting non--Gaussianity.  
Thus in the following we just give some of the transformation rules that 
cosmological perturbations around a flat FRW background obey to up to second 
order. 
They can be useful to check some of the gauge--invariant quantities we shall 
introduce. 
\subsubsection{Gauge transformations}
Let us consider an infinitesimal coordinate transformation up to second order
\begin{equation}
\label{GT}
 \tilde{x}^\mu(\lambda)=x^\mu-\xi_{(1)}^\mu-\frac{1}{2}\,
\left({\xi_{(1)}^\mu}_{,\nu}\xi_{(1)}^\nu
+\xi_{(2)}^\mu\right)\, ,
\end{equation}
where $\xi^{(\mu)}_{(r)}(\tau, x^i)$ are vector fields defining the gauge 
transformation, being regarded as quantities of the same order as the 
perturbation variables. Specifying their time and space components, 
one can write 
\begin{equation}
\xi_{(r)}^0=\alpha_{(r)}\;,
\end{equation}
and
\begin{equation}
\xi_{(r)}^i=\partial^i\beta_{(r)}+d_{(r)}^{i}\;,
\end{equation}
where we have split the space component into a scalar and a vector part 
with $\partial_i d^{(r)i}=0$. From a practical point of view fixing a gauge is 
equivalent to fixing a coordinate system. In particular the 
function $\xi_{(r)}^0$ selects constant-$\tau$ hypersurfaces, {\emph i.e.} a 
time--slicing, while $\xi_{(r)}^i$ selects the spatial coordinates within 
those hypersurfaces. 

If we now expand a generic tensor $T(\tau,x^i)$ defined in the real physical 
word up to second order as 
\be
T(\tau,x^i)=T_0+\delta T(\tau,x^i)=T_0+\delta^{(1)} T(\tau,x^i)
+\frac{1}{2} \delta^{(2)} T(\tau,x^i)\, ,
\ee
where $T_0$ is the background value, then its perturbations transform as 
\cite{MMB}
\begin{eqnarray}
\label{eq:t1}
\widetilde{\delta^{(1)} T} &=&\delta^{(1)} T +\pounds_{\xi_{(1)}} T_{0}\, ,\\
\label{eq:t2} 
\widetilde{\delta^{(2)} T}& =&\delta^{(2)} T +2\pounds_{\xi_{(1)}} 
\delta^{(1)} T
+\pounds^{2}_{\xi_{(1)}} T_{0} +\pounds_{\xi_{(2)}}
T_{0} \, ,
\end{eqnarray}
where $\pounds_{\xi_{(r)}}$ is the Lie derivative along the vector 
$\xi^{\mu}_{(r)}$.

Thus, for example, the energy density perturbation transforms at 
first order as
\be
\label{eq:mut}
\widetilde{\delta^{(1)} \rho}=\delta^{(1)} \rho 
+\rho_0^\prime\alpha_{(1)}\, ,
\ee 
and at second order as~\cite{MMB}
\bea
\label{eq:mut2}
\widetilde{\delta^{(2)} \rho}&=&\delta^{(2)} \rho +\rho_0^\prime\alpha_{(2)} +
\alpha_{(1)}\left(\rho_0^{\prime\prime}\alpha_{(1)}
+\rho_0^\prime\alpha_{(1)}^\prime+2\delta^{(1)} \rho^\prime\right) \nonumber \\
&+&\xi^i_{(1)}\left( \rho^\prime_0\alpha^{(1)}_{,i} 
+  2\delta\rho_{,i} \right) \, .
\eea
By transforming the metric tensor perturbations $\delta^{(1)} g_{\mu \nu}$ 
and $\delta^{(2)} g_{\mu \nu}$ in the metric~(\ref{metric2})
according to Eq.~(\ref{eq:t1}) 
and Eq.~(\ref{eq:t2}) one finds that the metric perturbation $\psi=\psi^{(1)}
+\frac{1}{2} \psi^{(2)}$ transforms at first order as 
\begin{equation}
\label{eq:psi1}
\widetilde{\psi^{(1)}}=\psi^{(1)}-
\frac{1}{3}\,\nabla^2\beta_{(1)}-\frac{a'}{a}\,\alpha_{(1)}\, ,
\end{equation}
and at second order as~\cite{MMB}

\begin{eqnarray}\label{eq:phi2}
\widetilde{\psi^{(2)}} & = & \psi^{(2)}
+\alpha_{(1)}\left[
2\left(\psi^{(1)\prime}+
2\frac{a'}{a}\psi^{(1)}\right)
-\left(\frac{a''}{a}+\frac{a^{\prime 2}}{a^2}\right)\alpha_{(1)}
-\frac{a'}{a} \alpha_{(1)}^{\prime}\right]
\nonumber \\
&+& \xi^i_{(1)}\left( 2\psi^{(1)}_{,i}-
\frac{a'}{a}\alpha^{(1)}_{,i}\right)
-\frac{1}{3}\left(-4\psi^{(1)}+\alpha_{(1)}\partial_0
+\xi^i_{(1)}\partial_i
+4\frac{a'}{a}\alpha_{(1)}\right)\nabla^2\beta_{(1)} \nonumber\\
&-& \frac{1}{3}\left(2\omega^{,i}_{(1)}-\alpha^{,i}_{(1)}
+\xi^{i\prime}_{(1)}\right)\alpha^{(1)}_{,i} 
-\frac{1}{3}\left(2\chi^{(1)}_{ij}+\xi^{(1)}_{i,j}+
\xi^{(1)}_{j,i}\right)
\xi^{j,i}_{(1)}
-\frac{a'}{a}\alpha_{(2)} \nonumber \\
&-&\frac{1}{3}\nabla^2\beta_{(2)}\, .
\end{eqnarray}

\subsection{Second--order gauge--invariant perturbations}
In linear theory a gauge--invariant treatment of cosmological 
perturbations was introduced by Bardeen in his seminal work~\cite{Bardeen1}. 
As far as non--linear perturbations are concerned the first results were 
found in Refs.~\cite{Salopek1,Salopek2,AfshordiBrand},  
using a gradient--expansion technique (or long--wavelength approximation), 
and in Refs.~\cite{noi,Maldacena}, using a second--order perturbative 
approach. In these works a 
generalization of the so--called curvature perturbation to linear order 
was found in the context of single--field models of inflation, and its 
constancy in time was proved. Refs.~\cite{Salopek1,Salopek2,noi,Maldacena}
focused on the study of non--Gaussianity of cosmological perturbations.
These were the first papers to fully account for -- at least during the 
inflationary epoch -- the different second--order effects both in the 
inflaton field and in the metric perturbations.   

However no gauge--invariant theory for non--linear perturbations in the 
context of cosmological perturbations has been 
built up until some very recent papers on the 
subject~\cite{lw,mw,rigo2,HNP2,rigo3}. 
Specifically the authors of 
Refs.~\cite{lw,mw} use a second--order perturbative approach, while in
Ref.~\cite{rigo2,rigo3} a long--wavelength approximation is employed. 
Also, we refer the reader to Refs.~\cite{BruniSop,nak}, where the 
issue of gauge--invariance at second and higher 
order has been addressed from a broader point of view. 

We now give the expressions of some quantities which are gauge--invariant 
up to second order in the perturbations and which we shall 
use to follow the non--Gaussianity of cosmological perturbations from an early 
period of inflation through reheating and deep into the radiation/matter 
dominated epochs. In particular we give particular relevance to the 
gauge--invariant definition of curvature perturbations.

\subsection{The curvature perturbation on spatial slices of uniform density} 
\label{def:zeta}

At linear order the intrinsic spatial curvature on hypersurfaces on constant
conformal  time $\tau$ and for a flat Universe is given 
by~\cite{Bardeen1,KodaSasa}
\be
^{(3)}R=\frac{4}{a^2}\nabla^2\, \hat{\psi}^{(1)}\, ,
\ee
where for simplicity of notation\footnote{
Notice that our notation is different from that of 
Refs.~\cite{MFB,malik,mw} for the presence of $D_{ij}$ in the 
metric~(\ref{metric1}), 
while it is closer to the one used in Refs.~\cite{Bardeen1,KodaSasa}. 
As far as the first--order perturbations are concerned 
the metric perturbations 
$\psi$ and $E$ of Refs.~\cite{MFB,malik} are 
given in our notation as $\psi=\psi^{(1)}+(1/6)\, \nabla^2 \chi^{(1)}$ and 
$E=\chi^{(1)}/2$, respectively. The same is true at second order for the 
perturbation variables $\psi_2$ and $E_2$ of Ref.~\cite{mw}, which in terms 
of our quantities are given by $\psi_2=\psi^{(2)}+(1/6)\, \nabla^2 \chi^{(2)}$
and $E_2=\chi^{(2)}/2$. 
However, no difference appears 
in the calculations when using the Poisson gauge or the spatially flat gauge,
or when considering the perturbation evolution on large scales.
}
we have indicated 
\be
\label{notation}
\hat{\psi}^{(1)}=\psi^{(1)}+\frac{1}{6} \nabla^2 \chi^{(1)}\, .
\ee
The quantity $\hat{\psi}^{(1)}$ 
is usually referred to as the {\it curvature perturbation}.
However the curvature perturbation 
$\psi^{(1)}$ is {\it not}  gauge invariant, but is defined only on a 
given slicing. In fact, under a transformation on constant time hypersurfaces
$\tau \rightarrow \tau+\alpha_{(1)}$ (change of the slicing in 
Eq.~(\ref{eq:psi1}))
\be
\hat{\psi}^{(1)} \rightarrow \hat{\psi}^{(1)}-\H \alpha_{(1)}\, ,
\ee
where we have used Eq.~(\ref{eq:psi1}) and the transformation 
$\widetilde{\chi^{(1)}}=\chi^{(1)}+2 \beta_{(1)}$~\cite{MMB}.
If we consider the {\it slicing of uniform energy density} which 
is defined to be the slicing where there is no perturbation in the 
energy density, $\delta\rho=0$, from Eq.~(\ref{eq:mut}) we have 
$\alpha_{(1)}=\delta^{(1)} \rho/\rho_0'$ and the curvature perturbation 
$\hat{\psi}^{(1)}$ 
on uniform density perturbation slices -- usually indicated by 
$- \zeta^{(1)}$ is given by
\be
\label{zeta1}
-\zeta^{(1)} \equiv \widetilde{\hat{\psi}^{(1)}}|_{\rho}=\hat{\psi}^{(1)}+\H 
\frac{\delta^{(1)} \rho}{\rho_0'}\, .
\ee
This quantity is gauge--invariant and it is a clear example of how to find a 
gauge--invariant quantity by selecting in an unambiguous way a proper 
time slicing. It was first introduced in 
Refs.~\cite{BardeenSteinTur,Bardeen3} as a conserved quantity 
on large scales for purely adiabatic perturbations. 

Notice that such a combination can be regarded also as 
the density perturbation 
on uniform curvature slices, where $\psi^{(1)}=\chi^{(1)}=0$, 
the so--called \emph{ 
spatially flat gauge}~\cite{KodaSasa}. 
The energy density $\rho$ here has to be regarded as the total energy 
density.  
If the matter content of a system is made of several fluids it 
is possible to define similarly the curvature perturbations 
associated with each 
individual energy density components $\rho_i$, which to linear order are given 
by~\cite{LUW,wmu} 
\bea
\label{zetai}
\zeta^{(1)}_{i}
 &=& - \hat{\psi}^{(1)} - \H\left(\frac{\delta^{(1)}\rho_{i}}
{\rho_{i}'}\right)\, .
\eea
Here and in the following, if not specified, we drop the subscript `0' 
referring to the background quantities for simplicity of notation.
Notice that the total curvature perturbation in Eq.~(\ref{zeta1}) is given in 
terms of the 
individual curvature perturbations as
\be
\label{zsum}
\zeta^{(1)}=\sum_i \frac{\rho'_i}{\rho} \zeta^{(1)}_i\, .
\ee

We now come to the generalization at second order 
of the gauge--invariant curvature perturbation of Eq.~(\ref{zeta1}). 
As in Eq.~(\ref{metric2}) we can 
define the gravitational potential $\psi$ up to second order as 
$\psi=\psi^{(1)}+\frac{1}{2} \psi^{(2)}$ and we expand the energy density as 
in 
Eq.~(\ref{rho}). The authors of Refs.~\cite{lw,mw}
\footnote{The reader is also referred to Refs.~\cite{noi,Maldacena,rigo1} 
for related quantities and definitions.} have shown that 
the second--order curvature perturbation on uniform (total) density 
hypersurfaces is given by the quantity (up to a gradient term)
\bea
\label{qqq}
-\zeta^{(2)}&=&\widetilde{\hat{\psi}^{(2)}}|_{\rho} \nonumber \\
&=&\hat{\psi}^{(2)}+\H\frac{\delta^{(2)}\rho}{\rho^\prime}
-2\H\frac{\delta^{(1)}\rho^\prime}{\rho^\prime}
\frac{\delta^{(1)}\rho}{\rho^\prime}
-2\frac{\delta^{(1)}\rho}{\rho^\prime}\left(\hat{\psi}^{(1)\prime}
+2\H\hat{\psi}^{(1)}\right) \nonumber \\
&+&\left(\frac{\delta^{(1)}\rho}{\rho^\prime}
\right)^2 \left(\H \frac{\rho^{\prime\prime}}{\rho^\prime}-
\H^\prime-2\H^2\right) \, ,
\eea
where as in Eq.~(\ref{notation}) we have used the shorthand notation 
$\hat{\psi}^{(2)}=\psi^{(2)}+\frac{1}{6} \nabla^2 \chi^{(2)}$. 
As explained in Refs.~\cite{lw,mw} the quantity $\zeta^{(2)}$ is 
gauge--invariant,
being constructed on a well--defined time slicing corresponding to spatial 
hypersurfaces where $\delta^{(1)} \rho=\delta^{(2)} \rho=0$. 
In a similar manner to linear order let us introduce the gauge--invariant 
curvature perturbations $\zeta^{(2)}_i$ at second order relative to 
a particular component. These quantities will be given by the same formula 
as Eq.~(\ref{qqq}) relatively to each energy density $\rho_i$  
\bea
\label{zeta2singole}
-\zeta^{(2)}_i&=&\hat{\psi}^{(2)}+\H\frac{\delta^{(2)}\rho_i}
{\rho_{i}^\prime}
-2\H\frac{\delta^{(1)}\rho_{i}^\prime}{\rho_{i}^\prime}
\frac{\delta^{(1)}\rho_i}{\rho_{i}^\prime}
-2\frac{\delta^{(1)}\rho_i}{\rho_{i}^\prime}\left(\hat{\psi}^{(1)\prime}
+2\H \hat{\psi}^{(1)}\right) \nonumber \\
&+&\left(\frac{\delta^{(1)}\rho_i}{\rho_{i}^\prime}
\right)^2 \left(\H \frac{\rho_{i}^{\prime\prime}}{\rho_{i}^\prime}-
\H^\prime-2\H^2\right) \, .
\eea   

\subsection{Adiabatic and entropy perturbations}
The gauge--invariant curvature perturbations introduced in the previous 
sections 
are usually adopted to characterize the so--called adiabatic perturbations. 
In fact adiabatic perturbations are such that a net perturbation in the total 
energy density and -- via the Einstein equations -- in the intrinsic 
spatial curvature are produced. However, as we have seen, neither the energy 
density nor the curvature perturbations are gauge--invariant, 
hence the utility of using the variable $\zeta=\zeta^{(1)}+\frac{1}{2} 
\zeta^{(2)}$ to define such perturbations. 
Thus the notion of adiabaticity applies  when the properties of a fluid, 
in the physical perturbed space--time,  can be described \emph{uniquely}  
in terms of its energy density $\rho$. For example, the pressure perturbation
will be adiabatic if the pressure is a unique function of the energy density 
$P=P(\rho)$ (see Ref.~\cite{lw} for an exhaustive discussion on this point).

On the other hand, by the same token, to define 
a non--adiabatic (or entropy) perturbation of a given quantity $X$ it is 
necessary to ``extract'' that part of the perturbation which does not depend 
on the energy density. A very general prescription to do that is to consider 
the value of the perturbation 
$\delta X=\delta^{(1)} X+\frac{1}{2} \delta^{(2)} X$ on the hypersurfaces of 
uniform energy density

\be
\label{gendef}
\delta X_{\rm nad} \equiv \widetilde{\delta X|_\rho} \, ,
\ee  
since this quantity will vanish for adiabatic perturbations when $X=X(\rho)$.
Specifically the non--adiabatic pressure perturbation will be given by the 
pressure perturbation on slices of uniform energy density 
$\widetilde{\delta P|_\rho}$. Being specified in a non--ambiguous slicing, 
the entropy perturbations 
defined in Eq.~(\ref{gendef}) turn out to be gauge--invariant. 
Notice that such a definition holds true both 
when considering quantities on the uniform total energy density hypersurfaces 
and when considering hypersurfaces of uniform energy density relative to each 
individual component when more than one fluid is present.  

Before moving to the explicit expressions for the first and 
second--order adiabatic and entropy perturbations an important remark is 
in order. In general the perturbations will not be exclusively 
of adiabatic or of entropy type, but both perturbation modes 
will be present. Indeed, as we will see in the next section the 
non--adiabatic pressure perturbation $\widetilde{\delta P|_\rho}$ sources 
the total curvature perturbation $\zeta$ on large scales. Such a coupling 
is the mechanism responsible for the generation of cosmological perturbations 
in the curvaton and in the inhomogeneous reheating scenarios, contrary to the 
standard single--field inflationary scenario where only adiabatic 
perturbations are involved.
\subsubsection{Adiabatic and entropy perturbations at first order}
\label{AI1st}
At first order the non--adiabatic pressure perturbation is given 
by~\cite{Bardeen1,KodaSasa}
\be
\delta^{(1)} P_{\rm nad} \equiv \widetilde{\delta ^{(1)}P|_\rho}=
\delta ^{(1)}P -c_s^2 \delta^{(1)} \rho\, , 
\ee
where $c_s^2=P_0'/\rho_0'$ is the adiabatic sound speed of the fluid. 
As a check of what we said above, notice that indeed this quantity is 
gauge--invariant.  

It can be shown that in the presence of more than one fluid the total 
non--adiabatic pressure perturbation can be split into two parts

\be
\label{splitPnad}
\delta^{(1)} P_{\rm nad}=\delta^{(1)} P_{\rm int}+\delta^{(1)} P_{\rm rel}\, .
\ee
The first part is given by the sum of the intrinsic entropy perturbation 
of each fluid

\be
\label{intrins1}
\delta^{(1)} P_{\rm int}= \sum_i\delta^{(1)} P_{{\rm intr},i}\, ,
\ee
where 
\be
\label{intrinsicentropy1}
\delta^{(1)} P_{{\rm intr},i}=
\delta ^{(1)}P_i-c_i^2 \delta^{(1)} \rho_i
\ee
is the intrinsic non--adiabatic pressure perturbation of that fluid 
(which is a gauge--invariant quantity) with
$c_i^2={p_i}'/{\rho_i}'$ the adiabatic 
sound speed of the individual fluid.   
The second part
is given by the relative entropy perturbation between different 
fluids~\cite{wmu}
\be
\delta^{(1)} P_{\rm rel}=\frac{1}{6 \H \rho'} \sum_{ij} 
\rho_{i}' \rho_{j}' \left( c^2_{i} -c^2_{i} \right) 
\mathcal{S}^{(1)}_{ij}\, , 
\ee
where $\mathcal{S}_{ij}$ is the relative energy density 
perturbation whose gauge--invariant definition is expressed in terms of 
the curvature perturbations $\zeta^{(1)}_i$ of Eq.~(\ref{zetai}) 
as~\cite{WMLL,wmu}

\be
\label{Srelative1}
\mathcal{S}^{(1)}_{ij}=
3 \left( \zeta^{(1)}_i-\zeta^{(1)}_j\right)\, .
\ee  
Notice that for fluids with no intrinsic entropy perturbations, the pressure 
perturbation will be adiabatic if the relative entropy perturbations vanish
\be
\label{adcondition1}
\zeta^{(1)}_i=\zeta^{(1)}_j\, .
\ee  
In such a case this is the condition to have \emph{pure} adiabatic 
perturbations. 
As a consequence, from Eq.~(\ref{zsum}) we see that the total curvature is 
equally shared 
by the different components $\zeta^{(1)}=\zeta^{(1)}_i$.

On the other hand a \emph{pure} isocurvature perturbation is such that 
the individual components compensate with each other in order to leave the 
curvature 
perturbation unperturbed. This is the reason why these are also referred to as 
\emph{isocurvature} perturbations. 

In Refs.~\cite{MalikWands,WMLL} it has been shown how to derive the evolution 
equation for the curvature perturbation $\zeta^{(1)}$ simply from the 
continuity equation for the energy density, 
without making any use of Einstein's equations.  
The result is that even on large scales the curvature perturbation can evolve 
being sourced by the non--adiabatic pressure of the system according 
to~\cite{GBW,lrreview,MalikWands,WMLL}
\be
\label{MAIN1}
\zeta^{(1)'}=-\frac{\H}{\rho+P} \delta^{(1)} P_{\rm nad}\, .
\ee
For purely adiabatic perturbations the curvature perturbation is 
conserved on large scales, thus making $\zeta^{(1)}$ the proper quantity to 
characterize the amplitude of adiabatic perturbations.
Eq.~(\ref{MAIN1}) shows in particular that the notion of isocurvature 
perturbation is valid only at some initial 
epoch. \footnote{On the other hand, as far as the evolution of the entropy 
perturbation 
itself is concerned, it has been shown that the non--adiabatic part of a 
perturbation is 
sourced on large scales only by other entropy perturbations, and that there 
is no source term
coming from the overall curvature perturbation~\cite{LUW}. If we indicate 
generically an 
entropy perturbation as ${\bf {\mathcal S}}$ then its equation of motion on 
large scales reads
\bea
{\bf {\mathcal S}}'=\beta \H {\bf {\mathcal S}}\, , \nonumber 
\eea
where $\beta$ is a time--dependent function 
which depends on the particular system under study. This result has also
been obtained on very general grounds within the 
``separate Universe approach'' of Ref.~\cite{WMLL}.}    
Indeed the fact that the non--adiabatic pressure perturbation sources 
the curvature perturbation on large scales was already known 
in the literature~\cite{Bardeen1,MFB,Mollerach} 
However, it was only recently that this issue has received renewed 
attention, being applied   
in the context of the curvaton scenario as an alternative way to produce 
adiabatic density perturbation starting from an initial entropy mode. 

\subsubsection{Adiabatic and entropy perturbations at second order}
\label{AI2nd}

Up to second order in the perturbations it has been shown that the 
gauge--invariant non--adiabatic pressure perturbation is given by~\cite{mw}

\begin{eqnarray}
\label{Pnad2}
\delta^{(2)} P_{\rm nad} \equiv \widetilde{\delta^{(2)} P|_\rho}&=&
\delta^{(2)}P-\frac{P'}{\rho'}\delta^{(2)}\rho
+P'\left[2\left(\frac{\delta^{(1)'}\rho}{\rho'}-
\frac{\delta^{(1)'}P}{P'}\right)\frac{\delta^{(1)}\rho}{\rho'}\right.
\nonumber\\
&+&\left.\left(\frac{P''}{P}-\frac{\rho''}{\rho}\right)
\left(\frac{\delta^{(1)}\rho}{\rho'}\right)^2\right]\, .
\end{eqnarray}

In a similar manner as in Eq.~(\ref{adcondition1}), for a set of fluids 
(with no intrinsic entropy perturbations) we can define pure adiabatic 
perturbations as those obeying the gauge--invariant condition
\be
\label{adcondition2}
\zeta^{(2)}_i=\zeta^{(2)}_j\, .
\ee

Also at second order the curvature perturbation $\zeta^{(2)}$ on large 
scales evolves due 
to the non--adiabatic pressure perturbation, as shown in Ref.~\cite{mw}
\bea
\label{MAIN2}
\zeta^{(2)'}&=&- \frac{\H}{\rho+P} \delta^{(2)}P_{\rm nad}-\frac{2}{\rho+P}
\left[ \delta^{(1)}P_{\rm nad} -2(\rho+P) \zeta^{(1)} \right] 
\zeta^{(1)'} \, ,
\eea
where the first--order curvature perturbation obeys Eq.~(\ref{MAIN1}). 

The issue of the conservation of the curvature perturbation at second order 
(and beyond) for adiabatic perturbations has also been addressed in 
Ref.~\cite{lw}, while in Ref.~\cite{rigo2,rigo3} the 
evolution for the curvature perturbation has been obtained in the 
context of the long--wavelength approximation.   

\section{\bf Evolution of cosmological perturbations up to second order} 

Let us now consider the evolution of cosmological 
perturbations on large scales up to second order. Our aim is to follow 
the non--linearity in the perturbations from an early period of inflation 
through the different post--inflationary stages till today.
This will enable us to give a definite prediction for the level of 
non--Gaussianity in different scenarios for the generation of the 
cosmological perturbations, namely the standard single field 
inflation, the curvaton and the inhomogeneous reheating scenarios. 
In each of these the cosmological evolution can be divided into three main 
stages:
\begin{enumerate}
\item
A primordial epoch of accelerated expansion, when 
cosmological perturbations are produced on large scales 
from quantum fluctuations of a scalar field, which can be 
different from the inflaton, as in the curvaton and 
in the inhomogeneous reheating models. 
\vskip 0.5cm
\item
An epoch when the perturbations in the energy density of the scalar fields 
are transferred to a radiation fluid during the reheating stage. During 
this stage the inflaton field (and the curvaton field, if present) 
oscillates around the minimum of its potential behaving as non--relativistic 
matter and then it decays into light particles (the radiation fluid).
\vskip 0.5cm
\item
After the reheating stage an overall adiabatic perturbation is generated 
and the Universe enters into a ``post--inflationary'' 
phase dominated by radiation and, subsequently, matter (and dark energy).
\end{enumerate}
We shall follow the evolution of the different second--order effects 
throughout these phases using the gauge--invariant curvature perturbations 
introduced 
previously~\cite{BMR2,BMR3,BMR4}. This is a highly efficient method for 
different reasons. As we will see from the continuity equation for the 
energy--momentum tensor, the overall curvature perturbation evolves 
on large scales due to a non--adiabatic pressure perturbation which can 
be expressed in terms of the individual curvature perturbations; thus we 
will be able to connect the different evolutionary stages. Moreover, by using 
gauge--invariant quantities we can pass from one gauge to 
another, in order to simplify some calculations, in a straightforward way.
Finally, as stressed Sec.~\ref{MetricEMTensor}, the curvature perturbations 
are a combination of the gravitational potential 
$\psi=\psi^{(1)}+\frac{1}{2} \psi^{(2)}$ and of the energy density 
perturbations 
$\delta \rho=\delta^{(1)} \rho+\frac{1}{2} \delta^{(2)} \rho$, which is
very useful to obtain the final second--order contributions to the 
gravitational potentials from the different stages.  
When calculating such contributions during the radiation/matter 
dominated phases we use the Einstein equations in order to relate 
the energy density fluctuations and the gravitational potential.  
As we shall see the prototype of this procedure is given by the standard 
scenario of single field inflation, where the curvature perturbation is 
in fact conserved on large scales, since in this case perturbations remain 
adiabatic.    
Our last step will be to define how our results must be compared to the 
observations, to search for possible non--Gaussian signatures in the 
CMB temperature anisotropy on large scales. 
To this aim, in Sec.~\ref{DT/T} we will determine  
how the non--linearities in the gravitational potentials translate into 
non--linearities of the CMB temperature fluctuations on large
angular scales~\cite{BMR4}.  

In the following we derive the perturbed Einstein equations and the
energy--momentum continuity equations up to second order, for a Universe 
filled by multiple interacting fluids, consisting of a (oscillating) scalar 
field and a radiation fluid, and for a Universe which is radiation/matter 
dominated. This is all what we need in order to study the three different 
cosmological scenarios, apart from a 
detailed analysis of the generation \emph{during} 
inflation of second--order cosmological perturbations from the inflaton 
fluctuations. This analysis will be made  separately in 
Sec.~\ref{NGInfl}. 
We now strictly follow Refs.~\cite{BMR2,BMR3,BMR4} which are the 
first works to systematically address the 
evolution of the second--order primordial cosmological perturbations 
in the different scenarios for perturbation generation. 

\subsection{First--order Einstein equations} 

Our starting point are the 
perturbed Einstein equations $\delta G^\mu_{~\nu}= \kappa^2\, 
\delta T^\mu_{~\nu}$. Here $\kappa^2\equiv 8\pi \,G_{\rm N}$. 
As the matter content we take the generic fluid defined by the 
energy--momentum tensor given 
in Sec.~\ref{energy--momentum}. The detailed expressions for the Einstein 
tensor components $\delta G^\mu_{~\nu}$ from the metric in 
Eq.~(\ref{metric2}) are contained in Appendix A, and from 
there one can read the Einstein equations in different gauges. 
Here we only report those equations which we shall use to derive our main 
results.
Specifically, in the Poisson gauge defined in Sec.~\ref{metrictensor}
the first--order the $(0-0)-$ and the 
$(i-0)-$components of Einstein equations are 
\bea
\label{00Poisson}
\frac{1}{a^2}
\Bigg[ 6\,\H^2 \phi^{(1)} \,+6\,\H{\psi^{(1)}}^{\prime} -2
\nabla^2\psi^{(1)}\Bigg]
&=& -\kappa^2\delta^{(1)}\rho\, , \\
\label{0iPoisson}
\frac{2}{a^2}\left(\H \partial^i \phi^{(1)} 
+\partial^i
{\psi^{(1)}}^{\prime}\right)&=&\,-\kappa^2\left(1+w\right)\rho_0 v^i_{(1)}\, , 
\eea
where $w \equiv P/\rho$ is the equation of state of the fluid. 

In the Poisson gauge at first--order the non--diagonal 
part of the $(i-j)$-component of Einstein equations, gives 
\be
\psi^{(1)}=\phi^{(1)}\, ,     
\ee
and, on superhorizon scales, Eq.~(\ref{00Poisson}) gives 

\begin{equation}
\label{00Plargescales}
\psi^{(1)}=-\frac{1}{2}\frac{\delta^{(1)}\rho}{\rho_0}=
\frac{3(1+w)}{2}\H \frac{\delta^{(1)}\rho}{\rho^\prime}\, . 
\end{equation} 
where in the last step we have used the background continuity equation 
$\rho'=-3\H \rho \left( 1+w \right)$.

Using the spatially flat gauge $\psi^{(1)}=\chi^{(1)}=0$, from the $(0-0)-$ 
component of Einstein equation we get a similar result for the gravitational 
potential $\phi^{(1)}$\footnote{From this section onward we do not use 
different symbols for the perturbations evaluated in different gauges,
rather we will specify every time the gauge we are using.} 

\be
\label{00flatlargescales}
\phi^{(1)}=-\frac{1}{2}\frac{\delta^{(1)}\rho}{\rho_0}\, . 
\ee 
Notice that Eqs.~(\ref{00Poisson}),~(\ref{0iPoisson}),~(\ref{00Plargescales}) 
and~(\ref{00flatlargescales}) indeed 
hold also when referring to the total energy density $\rho$ 
and the total velocity perturbation and equation of state $w$ in the case of a 
multiple component system.

\subsection{Second--order Einstein equations}
\label{2ndEE}
Let us consider the Einstein equations perturbed at second--order 
$\delta^{(2)}G^\mu_{~\nu}=\kappa^2 \, \delta^{(2)}T^\mu_{~\nu}$. 
The second--order expression for the Einstein tensor 
$\delta^{(2)}G^\mu_{~\nu}$ can be found in any gauge in Appendix A. 

We first consider the Einstein equations in the Poisson gauge which 
will be used in particular to express the non--linearities in the 
gravitational potential 
$\phi^{(2)}$ during the radiation/matter dominated phases.

\begin{itemize}
\item
The $(0-0)-$component of Einstein equations leads to 

\bea
\label{00Poisson2}
&&3\H^2\phi^{(2)}+3\H\psi^{(2)\prime}-\nabla^2\psi^{(2)}-12\H^2
\left(\psi^{(1)}\right)^2-3\left(\nabla\psi^{(1)}\right)^2\nonumber\\
&&-8\psi^{(1)}\nabla^2\psi^{(1)}-3\left(\psi^{(1)\prime}\right)^2=
\kappa^2a^2\ded T^0_{~0}\, .
\eea

\item
At second order the gravitational potentials $\phi^{(2)}$ and $\psi^{(2)}$ 
differ even in the Poisson gauge for the presence of source terms which are 
quadratic in the first--order perturbations. 
In fact it is possible to find the second 
order equivalent of the linear constraint $\psi^{(1)}=\phi^{(1)}$ using the 
traceless part of the $(i-j)$- components of Einstein equations. 
One finds the following constraint relating the gravitational potentials  
$\psi^{(2)}$ and $\phi^{(2)}$\footnote{Such a 
constraint has first been derived for a Universe filled by a scalar field in 
Ref.~\cite{noi}.}  
\cite{BMR2}. 

\begin{eqnarray}
\label{constraint}
\psi^{(2)} - \phi^{(2)} & = & - 4 \left(\psi^{(1)}\right)^2 
- \nabla^{-2} \left(2 \partial^i \psi^{(1)} \partial_i
\psi^{(1)} + 3 \left(1+w\right) {\cal H}^2 v_{(1)}^i 
v_{(1)i} \right) \nonumber \\
& + & 3 \nabla^{-4} \partial_i \partial^j 
\left(2 \partial^i \psi^{(1)} \partial_j
\psi^{(1)} + 3 \left(1+w\right) {\cal H}^2 v_{(1)}^i 
v_{(1)j} \right) \; .
\end{eqnarray}
In particular in the case of a matter--dominated phase when $w=0$ the linear 
gravitational potential $\psi^{(1)}=\phi^{(1)}$ is constant in time and, 
using Eq.~(\ref{0iPoisson}), the constraint~(\ref{constraint}) reads
\bea
\label{constraintbis}
\psi^{(2)}-\phi^{(2)}&=& - \frac{2}{3}\left( \psi^{(1)} \right)^2
+ \frac{10}{3}\nabla^{-2}\left(\psi^{(1)}\nabla^2\psi^{(1)}\right) \nonumber\\
&-&10\,\nabla^{-4}\left(\partial^i\partial_j\left(\psi^{(1)}\partial_i 
\partial^j \psi^{(1)}\right)\right)\, .
\eea 
\end{itemize}

If we use the spatially flat gauge $\psi^{(1)}=\chi^{(1)}=0$ 
and $\psi^{(2)}=\chi^{(2)}=0$ we obtain 
for the $(0-0)$- component of Einstein equation on large scales 

\be
\label{00flat2largescales}
\phi^{(2)}=-\frac{1}{2}\frac{\delta^{(2)}\rho}{\rho_0}+
4\left( \phi^{(1)} \right)^2\, .
\ee

\subsection{Energy--momentum tensor conservation}

We now derive the time--evolution on large scales of the gauge--invariant 
curvature perturbations $\zeta_i$ introduced in Sec.~\ref{def:zeta}. 
Indeed the equations of motion for these quantities are a direct consequence 
of the energy continuity equation.  
In particular, we will focus on 
a system composed by a scalar field oscillating around the minimum of 
its potential and a radiation fluid having in mind the physical case of 
reheating. 
We will then describe the evolution of the curvature perturbations in the 
subsequent radiation/matter dominated phases.     

Let us consider the system composed by the oscillating scalar field
$\varphi$
and the radiation fluid. 
Averaged over several oscillations the 
effective equation of state of the scalar field $\varphi$ 
is $w_\varphi=\langle P_\varphi/\rho_\varphi 
\rangle = 0$, where $P_\varphi$ and $\rho_\varphi$  are  
the scalar field pressure and energy density respectively. The scalar field 
is thus equivalent to a fluid of non--relativistic particles~\cite{Turner}. 
Moreover it is supposed to decay into radiation
(light particles) with a decay rate $\Gamma$. We can thus describe this 
system as a pressureless and a radiation fluid which interact via
energy transfer triggered by the decay rate $\Gamma$. We follow 
the gauge--invariant approach developed in Ref.~\cite{wmu} to study 
cosmological perturbations at first--order for the general 
case of an arbitrary number of interacting fluids and we shall 
extend the analysis to second--order in the perturbations. 
Indeed the system under study
encompasses the dynamics of the three main mechanisms for the generation of 
the primordial cosmological density perturbations on large scales, namely 
the standard scenario of single field inflation~\cite{guth81,lrreview}, 
the curvaton scenario~\cite{curvaton1,LW2,curvaton3,LUW}, and 
the recently introduced scenario of ``inhomogeneous reheating'' 
\cite{gamma1,gamma2,gamma3,MatRio,allahverdi}.
Each component has energy--momentum tensor 
$T^{\mu\nu}_{(\varphi)}$ and $T^{\mu\nu}_{(\gamma)}$. The total energy 
momentum $T^{\mu\nu}=T^{\mu\nu}_{(\varphi)}+T^{\mu\nu}_{(\gamma)}$ is 
covariantly conserved 

\be
\label{conservtot}
T^{\mu\nu}_{\, \, ;\mu} = 0\, ,
\ee
but allowing for an interaction between the two fluids~\cite{KodaSasa}

\begin{eqnarray}
 \label{Qvector}
T^{\mu\nu}_{(\varphi);\mu}&=&Q^\nu_{(\varphi)}\,, \nonumber\\
T^{\mu\nu}_{(\gamma);\mu}&=&Q^\nu_{(\gamma)}\,, 
\end{eqnarray}
where $Q^\nu_{(\varphi)}$ and $Q^\nu_{(\gamma)}$ are
the generic energy--momentum transfer coefficients for
the scalar field and radiation sector respectively and are subject 
to the constraint 

\begin{equation}
\label{Qconstraint}
Q^\nu_{(\varphi)}+Q^\nu_{(\gamma)}=0 \, ,
\end{equation} 
derived from Eq.~(\ref{conservtot}). 
The energy--momentum transfer $Q^\nu_{(\varphi)}$ and 
$Q^\nu_{(\gamma)}$ can be decomposed for convenience as~\cite{KodaSasa}
\bea
\label{splitQ}
Q^\nu_{(\varphi)}&=&\hat{Q}_{\varphi}u^{\nu}+f_{(\varphi)}^{\nu}
\, , \nonumber \\
Q^\nu_{(\gamma)}&=&\hat{Q}_{\gamma}u^{\nu}+f_{(\gamma)}^{\nu}\, ,
\eea
where the $f^{\nu}$'s are 
required to be orthogonal to the total velocity of the fluid  $u^{\nu}$.
The energy continuity equations for the scalar field and radiation 
can be obtained from $u_{\nu} T^{\mu\nu}_{(\varphi);\mu} =
u_{\nu}Q^{\nu}_{(\varphi)}$ and $u_{\nu} T^{\mu\nu}_{(\gamma);\mu} =
u_{\nu}Q^{\nu}_{(\gamma)}$ and hence from Eq.~(\ref{splitQ})
\bea
\label{conseq}
u_{\nu} T^{\mu\nu}_{(\varphi);\mu}&=& \hat{Q}_{\varphi} \, ,\nonumber \\
u_{\nu} T^{\mu\nu}_{(\gamma);\mu}&=& \hat{Q}_{\gamma}\, .
\eea  
In the case of an oscillating scalar field decaying into radiation 
the energy transfer coefficient $\hat{Q}_\varphi$ is given 
by~\cite{Qreh}
\bea 
\label{Qreh}
\hat{Q}_\varphi&=&-\Gamma \rho_\varphi \nonumber \\
\hat{Q}_\gamma&=&\Gamma \rho_\varphi \ , 
\eea
where $\Gamma$ is the decay rate of the scalar field into radiation.

\subsubsection{Background equations}

The evolution of our spatially flat FRW background 
Universe is governed by the Friedmann constraint equation
\begin{eqnarray}
\label{Friedmann}
\H^2 &=& \frac{8\pi G_{\rm N}}{3}\rho a^2 \,,
\end{eqnarray}
and by the energy continuity equation derived from Eq.~(\ref{conservtot})
\begin{equation}
\label{continuity}
\rho'=-3\H\left( \rho+P\right)\,,
\end{equation}
where $\rho$ and $P$ are the total energy density and pressure of the system.
The total energy density
and the total pressure are related to the energy density and
pressure of the scalar field and radiation by

\begin{eqnarray}
\rho &=&\rho_\varphi+\rho_\gamma \,, \nonumber\\
P&=& P_\varphi+P_\gamma \,, 
\end{eqnarray}
where $P_\gamma$ is the radiation pressure. The energy continuity equations 
for the energy density of the scalar field $\rho_\varphi$ and radiation 
$\rho_\gamma$ in the background are 
\begin{eqnarray}
\label{rhophi'}
\rho_{\varphi}'
&=&-3\H\left(\rho_{\phi}+P_{\phi}\right) +a Q_{\varphi}\, , \\
\label{rhog'}
\rho_{\gamma}'
&=&-4\H\left(\rho_{\gamma}+P_{\gamma}\right) +a Q_{\gamma}\, ,
\end{eqnarray}
where $Q_{\varphi}$ and $Q_{\gamma}$ indicate the background values of 
the transfer coefficients $\hat{Q}_{\varphi}$ and $\hat{Q}_{\gamma}$, 
respectively. 

\subsection{Evolution of first--order curvature perturbations on large scales}

The curvature perturbations $\zeta^{(1)}_i$ associated with the 
energy density of the scalar field and the radiation fluid are

\begin{eqnarray}
\label{zeta1phi}
\zeta^{(1)}_{\varphi}
 &=& - \hat{\psi}^{(1)} - \H\left(\frac{\delta^{(1)}\rho_{\varphi}}
{\rho_{\varphi}'}\right)\, , \\
\label{zeta1g}
\zeta^{(1)}_{\gamma}
 &=& - \hat{\psi}^{(1)} - \H\left(\frac{\delta^{(1)}\rho_{\gamma}}
{\rho_{\gamma}'}\right)\, .
\end{eqnarray}   
Notice that the total curvature perturbation $\zeta^{(1)}$ can be 
expressed as a 
weighted sum of the single curvature perturbations of the scalar field 
and radiation fluid as~\cite{WMLL,wmu}
\be
\label{zetatot1}
\zeta^{(1)}= f \zeta^{(1)}_{\varphi}+ (1-f)\zeta^{(1)}_\gamma \, .
\ee
where
\be
\label{def:f}
f=\frac{\rho_\varphi'}{\rho'} \, ,\quad
1-f=\frac{\rho_\gamma'}{\rho'}\, 
\ee  
define the contribution of the scalar field and radiation to the total 
curvature perturbation $\zeta^{(1)}$, respectively.
We now perturb at first order the continuity equations 
Eqs.~(\ref{conseq}) for the scalar field and  
radiation energy densities,  
including the energy transfer. To this aim we first 
expand the transfer coefficients 
$\hat{Q}_{\varphi}$ and $\hat{Q}_{\gamma}$ up to first order 
in the perturbations around the homogeneous background as 
\begin{eqnarray}
\hat{Q}_{\varphi}&=& Q_{\varphi}+\delta^{(1)}Q_{\varphi}\, , \\
\hat{Q}_{\gamma}&=& Q_{\gamma}+\delta^{(1)}Q_{\gamma }\, .
\end{eqnarray}
Eqs.~(\ref{conseq}) give -- on wavelengths larger than the 
horizon scale --

\begin{eqnarray} 
\label{pertenergyexact1}
&&{\delta^{(1)}\rho'}_{\varphi}+3\H\left( \delta^{(1)}\rho_{\varphi}
+\delta^{(1)}P_{\varphi} \right)
- 3 \left( \rho_{\varphi}+P_{\varphi} \right)\psi^{(1)'}\nonumber \\ 
&&= a \, Q_{\varphi}\phi^{(1)}+a\,  \delta^{(1)} Q_{\varphi}\, , \\
&& \nonumber \\
\label{pertenergyexact2}
&&{\delta^{(1)}\rho'}_{\gamma}+3\H\left( 
\delta^{(1)}\rho_{\gamma} +\delta^{(1)}P_{\gamma} \right)
- 3 \left( \rho_{\gamma}+P_{\gamma} \right) \psi^{(1)'}\nonumber \\
&&= a\, Q_{\gamma}\phi^{(1)}+a\, \delta^{(1)}Q_{\gamma}\, .
\end{eqnarray}
Notice that the oscillating scalar field and radiation have fixed equations 
of state with $\delta^{(1)}P_\varphi = 0$ and $\delta^{(1)}P_\gamma =  
\delta^{(1)}\rho_\gamma/3$. This corresponds to vanishing intrinsic 
non--adiabatic pressure perturbations, as defined in 
Eq.~(\ref{intrinsicentropy1}).

Before proceeding further let us make a cautionary remark.
In Eqs.~(\ref{pertenergyexact1})--(\ref{pertenergyexact2}) 
and in the following, as in Ref.~\cite{BMR3}, in the long--wavelength limit 
we are neglecting gradient terms which, upon integration over time, 
may give rise to non--local operators when evaluating second--order 
perturbations. However, these gradient terms 
will not affect statistical quantities in momentum--space, such as
the gravitational potential bispectrum on large scales, as 
discussed in details in Sec.~\ref{Phinonline}. 

Following the procedure of Ref.~\cite{wmu} we can 
rewrite Eqs.~(\ref{pertenergyexact1}) and~(\ref{pertenergyexact2})
in terms of the gauge--invariant curvature
perturbations $\zeta^{(1)}_\varphi$ and $\zeta^{(1)}_\gamma$  
\begin{eqnarray}
\label{eq:zeta1phi}
\zeta^{(1)'}_\varphi&=& 
\frac{a\H}{\rho_\varphi'}\bigg[\delta^{(1)}Q_\varphi-
\frac{Q_\varphi'}{\rho_\varphi'} \delta^{(1)} \rho_\varphi 
+Q_\varphi \frac{\rho'}{2\rho} 
\bigg( \frac{\delta^{(1)}\rho_\varphi}{\rho_\varphi'} - 
\frac{\delta^{(1)}\rho}{\rho'} \bigg) \bigg]\, ,\\
&& \nonumber \\
\label{eq:zeta1g}  
\zeta^{(1)'}_\gamma&=& 
\frac{a\H}{\rho_\gamma'}\left[\delta^{(1)}Q_\gamma-
\frac{Q_\gamma'}{\rho_\gamma'} \delta^{(1)} \rho_\gamma+ 
Q_\gamma \frac{\rho'}{2\rho} 
\left(\frac{\delta^{(1)}\rho_\gamma}{\rho_\gamma'} -
\frac{\delta^{(1)}\rho}{\rho'} \right) \right] \, ,
\end{eqnarray}
where we have used the perturbed $(0-0)$-component
of Einstein equations for superhorizon wavelengths 
$\psi^{(1)'}+\H \phi^{(1)}=-\frac{\H}{2}\frac{\delta^{(1)}\rho}{\rho}$ 
(see Appendix A).
Notice that from the constraint in 
Eq.~(\ref{Qconstraint}) the perturbed energy 
transfer coefficients obey  
\be
\delta^{(1)}Q_{\gamma}=-\delta^{(1)}Q_{\varphi}\, .
\ee

\subsubsection{Perturbations in the decay rate}

If the energy transfer coefficients $\hat{Q}_\varphi$ and 
$\hat{Q}_\gamma$ are given in terms 
of the decay rate $\Gamma$ as in Eq.~(\ref{Qreh}), the first order 
perturbation 
are respectively
\begin{eqnarray}
\label{Qgamma1}
\delta^{(1)}Q_\varphi&=&-\Gamma \delta^{(1)}\rho_\varphi-\delta^{(1)}\Gamma 
\, \rho_\varphi \, , \\
\label{Qgamma2}
\delta^{(1)}Q_\gamma &=&\Gamma \delta^{(1)}\rho_\varphi+\delta^{(1)}\Gamma 
\, \rho_\varphi \, ,
\end{eqnarray}   
where notice in particular that we have allowed for a perturbation in the 
decay rate $\Gamma$, 
\be
\Gamma(\tau, {\bf x})=\Gamma(\tau)+\delta^{(1)}\Gamma(\tau, {\bf x})\, .
\ee
Perturbations in the inflaton decay rate are indeed the key feature 
of the ``inhomogeneous reheating'' 
scenario~\cite{gamma1,gamma2,gamma3,allahverdi}.
In fact from now on we shall consider the background value $\Gamma$ of the 
decay rate as constant in time, 
$\Gamma \approx \Gamma_{*}$ as this is the case for the
standard case of inflation and the inhomogeneous reheating mechanism.
In such a case $\delta^{(1)}\Gamma$ is automatically gauge--invariant.
\footnote{
The authors of Ref.~\cite{MatRio} have introduced 
a gauge--invariant generalization at first order 
in the case of $\Gamma'\neq 0$ which reads 
$\delta \Gamma ^{(1)}_{\rm GI}=\delta^{(1)} \Gamma 
-\Gamma'\frac{\delta \rho_\varphi}{\rho_\varphi'}$. Indeed such a 
time--variation can have interesting effects on the overall curvature 
perturbation evolution $\zeta^{(1)\prime}$. See Ref.~\cite{MatRio} 
for more details.}
Plugging the expressions ~(\ref{Qgamma1})--(\ref{Qgamma2}) 
into Eqs.~(\ref{eq:zeta1phi})--(\ref{eq:zeta1g}), and using 
Eq.~(\ref{zetatot1}), we find that the first order 
curvature perturbations for the scalar field and radiation obey on large 
scales~\cite{BMR4}
\begin{eqnarray}
\label{zeta1phi'}
\zeta^{(1)'}_\varphi&=&\frac{a \Gamma}{2} \frac{\rho_\varphi}{\rho_\varphi'} 
\frac{\rho'}{\rho} \left( \zeta^{(1)} -\zeta^{(1)}_\varphi 
\right)+a \H \frac{\rho_\varphi}{\rho_\varphi'} \delta^{(1)} \Gamma \, ,\\
\label{zeta1g'}
\zeta^{(1)'}_\gamma&=&-\frac{a}{\rho_\gamma'} \left[ \Gamma \rho' 
\frac{\rho_\varphi'}{\rho_\gamma'}\left(1-\frac{\rho_\varphi}{2\rho} 
\right)\left( \zeta^{(1)} -\zeta^{(1)}_\varphi 
\right)
+\H\rho_\varphi \delta^{(1)}\Gamma 
\right]\, . \nonumber \\
&&
\end{eqnarray}
From Eq.~(\ref{zetatot1}) it is thus possible to find the equation of 
motion for the total curvature perturbation $\zeta^{(1)}$ using the evolution 
of the individual curvature perturbations in Eqs.~(\ref{zeta1phi'}) 
and~(\ref{zeta1g'})
\begin{eqnarray}
\label{zeta1'}
\zeta^{(1)'}&=&f'\left( \zeta^{(1)}_\varphi- \zeta^{(1)}_\gamma \right)
+f \zeta^{(1)'}_\varphi+(1-f)\zeta^{(1)'}_\gamma \nonumber \\
&=& -\H f \left( \zeta^{(1)}-\zeta^{(1)}_\varphi \right)\, .
\end{eqnarray}
Notice that Eq.~(\ref{zeta1'}) can be rewritten as

\begin{eqnarray}
\label{zeta1'S}
\zeta^{(1)'}= \H f (1-f)\left( \zeta^{(1)}_\varphi-\zeta^{(1)}_\gamma \right) 
=\frac{\H}{3}f(1-f) {\mathcal S}_{\varphi \gamma}\, ,
\end{eqnarray}
which explicitly shows that, {\emph in general}, the total curvature 
perturbation 
can evolve on large scales due to a non--adiabatic pressure given by 
the relative entropy perturbation ${\mathcal S}_{\varphi \gamma}$ 
defined in Eq.~(\ref{Srelative1}). In fact by comparison with 
Eq.~(\ref{MAIN1}) the 
expression for the non--adiabatic pressure perturbation at first order reads
\be
\delta^{(1)}P_{\rm nad}=-\frac{(3 \rho_\varphi+4\rho_\gamma)}{3} f (1-f) 
 \left( \zeta^{(1)}_\varphi-\zeta^{(1)}_\gamma \right)\, .
\ee       
From Eqs.~(\ref{zeta1phi'}) and~(\ref{zeta1g'}) one can also obtain an 
equation of motion for the relative entropy perturbation, being 
${\mathcal S}_{\varphi \gamma}=3
\left( \zeta^{(1)}_\varphi-\zeta^{(1)}_\gamma \right)$, and use 
Eq.~(\ref{zeta1'S}) to close the system of equations. 
However notice that 
during the decay of the scalar field into the radiation fluid, 
$\rho_{\gamma}'$ may vanish and 
Eq.~(\ref{zeta1g'}) for $\zeta^{(1)}_\gamma$, and hence, the evolution 
equation 
for ${\mathcal S}_{\varphi \gamma}$ become singular. 
Therefore it is convenient to close the system of equations by using 
the two first--order 
Eqs.~(\ref{zeta1phi'}) and~(\ref{zeta1'}) for the evolution of 
$\zeta^{(1)}_{\varphi}$ and $\zeta^{(1)}$.

In the scenarios for the generation of cosmological perturbations that we 
are going to study in detail, an adiabatic 
perturbation is produced after the ``reheating phase''  and thus the 
total curvature perturbation $\zeta$ is conserved on large scales 
during the radiation and 
matter--dominated phases, as it is evident from Eqs.~(\ref{MAIN1}) 
and~(\ref{MAIN2}) 
for a vanishing non--adiabatic pressure perturbation. In particular  from the 
definition of the curvature perturbation at linear order 
$\zeta^{(1)}=-\hat{\psi}^{(1)}-\delta^{(1)} \rho/\rho^\prime$ 
and using Eq.~(\ref{00Plargescales}) in the Poisson gauge we determine 
   
\be
\label{relPsizeta}
\psi^{(1)}=-\frac{3(1+w)}{5+3 w}\, \zeta^{(1)}\, .
\ee

Such a relation is very useful to relate the gravitational potential 
$\psi^{(1)}$ 
during either the radiation or the matter dominated epoch 
to the gauge--invariant curvature perturbation $\zeta^{(1)}$ at the end of the 
``reheating'' phase. In fact, as we will see, 
in the case of standard single field 
inflation the perturbations are always adiabatic through the different phases 
and thus the curvature perturbation $\zeta^{(1)}$ remains always constant on 
superhorizon scales, so that we can write $\zeta^{(1)} = \zeta^{(1)}_I$, where 
the subscript ``$I$'' means that $\zeta^{(1)}$ is evaluated during the 
inflationary stage. 
On the other hand in the curvaton and in the inhomogeneous reheating scenarios 
the curvature perturbation $\zeta^{(1)}$ initially evolve on large scales due 
to a non--vanishing entropy perturbation, and thus the value of $\zeta^{(1)}$ 
during the radiation and matter dominated phase will be determined by 
the curvature perturbation produced at the end of the ``reheating'' phase.

\subsection{Evolution of second--order curvature perturbations on large scales}

We now generalize to second order in the density perturbations
the results of the previous section. In particular we obtain an equation of 
motion on large scales for the individual second--order curvature 
perturbations which include also the energy transfer between the scalar field 
and the radiation component~\cite{BMR4}.

Since the curvature perturbations $\zeta^{(1)}_i$ and $\zeta^{(2)}_i$  
are gauge--invariant, we choose to 
work in the spatially flat gauge $\psi^{(1)}=\chi^{(1)}=0$ and 
$\psi^{(2)}=\chi^{(2)}=0$ if not otherwise specified. 
Note that from Eqs.~(\ref{zeta1phi}) and (\ref{zeta1g})
$\zeta^{(1)}_\varphi$ and $\zeta^{(1)}_\gamma$
are thus given by
\begin{eqnarray}
\label{zeta1phiflat}
\zeta^{(1)}_{\varphi}
 &=& - \H\left(\frac{\delta^{(1)}\rho_{\varphi}}
{\rho_{\varphi}'}\right)\, , \\
\label{zeta1gflat}
\zeta^{(1)}_{\gamma}
 &=& - \H\left(\frac{\delta^{(1)}\rho_{\gamma}}
{\rho_{\gamma}'}\right)\, .
\end{eqnarray} 
Eqs.~(\ref{zeta1phiflat})--(\ref{zeta1gflat}) and the 
energy continuity equations at first order, Eqs.~(\ref{pertenergyexact1})-
(\ref{pertenergyexact2}), in the spatially flat gauge 
$\psi^{(1)}=\chi^{(1)}=0$ yield 
\bea
\frac{\delta^{(1)} \rho'}{\rho'}&=&3f\zeta^{(1)}_\varphi+4(1-f) 
\zeta^{(1)}_\gamma\, , \\
\H \frac{\delta^{(1)} \rho}{\rho'}&=&- f \zeta^{(1)}_\varphi- (1-f) 
\zeta^{(1)}_\gamma\, .
\eea
We can thus rewrite the total second--order 
curvature perturbation $\zeta^{(2)}$ in Eq.~(\ref{qqq}) as
\begin{eqnarray}
\zeta^{(2)}&=&-\H \frac{\delta^{(2)} \rho}{\rho'} \nonumber \\
&-&\left[ f \zeta^{(1)}_\varphi+(1-f) \zeta^{(1)}_\gamma\right] 
\left[ f^2 \zeta^{(1)}_\varphi+(1-f)(2+f) \zeta^{(1)}_\gamma\right] 
\, , \nonumber
\\
&&
\end{eqnarray}
where we have used the background continuity 
Eqs.~(\ref{rhophi'})--(\ref{rhog'})
to find $\H \frac{\rho''}{\rho'}-\H'-2\H^2=-\H^2 (6-f)$.

Following the same procedure, the individual curvature perturbations for 
the scalar field and the radiation fluid as  
defined in Eq.~(\ref{zeta2singole}) are given by~\cite{BMR4}
\bea
\label{zeta2phiflat}
\zeta^{(2)}_\varphi&=&-\H \frac{\delta^{(2)}\rho_\varphi}{\rho_\varphi'}
+[2-3(1+w_\varphi)]\left( \zeta^{(1)}_\varphi \right)^2 -2
\left( a \frac{Q_\varphi \phi^{(1)}}{\rho_\varphi'}+
a \frac{\delta^{(1)} Q_\varphi}{\rho_\varphi'} \right) \zeta^{(1)}_\varphi
\nonumber \\
&-& 
\left[ a \frac{Q_\varphi'}{\H \rho_\varphi'}-\frac{a}{2} 
\frac{Q_\varphi}{\H \rho_\varphi'} \frac{\rho'}{\rho} \right]
\left( \zeta^{(1)}_\varphi \right)^2 \, , \\
&& \nonumber \\
\label{zeta2gflat}
\zeta^{(2)}_\gamma&=&-\H \frac{\delta^{(2)}\rho_\gamma}{\rho_\gamma'}
+[2-3(1+w_\gamma)]\left( \zeta^{(1)}_\gamma \right)^2 
-2 \left( a \frac{Q_\gamma \phi^{(1)}}{\rho_\gamma'}+
a \frac{\delta^{(1)} Q_\gamma}{\rho_\gamma'} \right)
\zeta^{(1)}_\gamma
\nonumber \\
&-&  
\left[ a \frac{Q_\gamma'}{\H \rho_\gamma'}-\frac{a}{2} 
\frac{Q_\gamma}{\H \rho_\gamma'} \frac{\rho'}{\rho} \right]
 \left( \zeta^{(1)}_\gamma \right)^2 \, ,
\eea
where $w_\gamma=1/3$ is the radiation equation of state.
Using  Eqs.~(\ref{zeta2phiflat}) and~(\ref{zeta2gflat}) to 
express the perturbation of the total energy density $\delta^{(2)}\rho$
one obtains the following expression for the total curvature 
perturbation $\zeta^{(2)}$~\cite{BMR4}
\bea
\label{main0}
\zeta^{(2)}&=&f \zeta^{(2)}_\varphi+(1-f) \zeta^{(2)}_\gamma 
+f(1-f)(1+f)\left( \zeta^{(1)}_\varphi
-\zeta^{(1)}_\gamma\right)^2 \nonumber \\
&+&2
\left( a \frac{Q_\varphi \phi^{(1)}}{\rho'}+
a \frac{\delta^{(1)} Q_\varphi}{\rho'} \right)
\left[\zeta^{(1)}_\varphi- \zeta^{(1)}_{\gamma} \right] \nonumber \\
&+& \left( a \frac{Q_\varphi'}{\H \rho'}-\frac{a}{2} 
\frac{Q_\varphi}{\H\rho} \right)
\left[ \left( \zeta^{(1)}_\varphi \right)^2-
\left( \zeta^{(1)}_\gamma \right)^2 \right]   \nonumber \\
&& 
\eea 
Expressing the (0-0)-component of Einstein equations (\ref{00flatlargescales}) 
in the spatially flat gauge at first--order $\psi^{(1)}=\chi^{(1)}=0$ 
in terms of the total curvature $\zeta^{(1)}$
\be
\label{00first}
\phi^{(1)}=-\frac{1}{2} \frac{\delta^{(1)}\rho}{\rho}=\frac{1}{2} 
\frac{\rho'}{\H \rho} \zeta^{(1)}\, ,
\ee
and using the explicit expressions for the first--order perturbed coefficients 
in terms of the decay rate $\Gamma$, Eqs.~(\ref{Qgamma1})--(\ref{Qgamma2}), we 
finally obtain~\cite{BMR4}
\bea
\label{main1}
\zeta^{(2)}&=&f \zeta^{(2)}_\varphi+(1-f) \zeta^{(2)}_\gamma
+f(1-f)(1+f)\left( \zeta^{(1)}_\varphi-\zeta^{(1)}_\gamma\right)^2 
\nonumber \\
&+&\frac{a\ \Gamma}{\H} f \left( \zeta^{(1)}_\varphi-
\zeta^{(1)}_\gamma \right)^2 -2 a \delta^{(1)} 
\Gamma \frac{\rho_\varphi}{\rho'} 
\left( \zeta^{(1)}_\varphi-\zeta^{(1)}_\gamma \right) \nonumber \\
&+&\frac{a\Gamma}{\H} (1-2f) \frac{\rho_\varphi}{2\rho}
\left( \zeta^{(1)}_\varphi-\zeta^{(1)}_\gamma \right)^2\, .
\eea 
Eq.~(\ref{main1}) is an important result. It generalizes 
to second--order in the perturbations 
the weighted sum in Eq.~(\ref{zetatot1}) and extends the 
expression found in Ref.~\cite{BMR3} in the particular case of 
the curvaton scenario, under the sudden--decay approximation, 
where the energy transfer was neglected. Similarly to linear order such 
an expression will be useful to describe the large--scale evolution of 
$\zeta^{(2)}$ sourced by a non--adiabatic pressure perturbation through the 
evolution of the density perturbations in the scalar field and radiation.

As already mentioned in the previous section, the expressions for
the second--order perturbations have been found here by 
using the long--wavelength limit for the first--order 
perturbations. Indeed second--order quantities expressed in terms of 
first--order perturbations will depend also 
on the short-wavelength behaviour of the first--order perturbations, as it 
is evident going to momentum space. Thus, for example, even if 
$\zeta^{(1)}_i$ are constant on large scales at linear order, 
it would not be strictly correct to consider the second--order 
part of $\zeta$ depending on $\zeta^{(1)}_i$ as constant. 
However our procedure is fully justified when applied to the
evaluation of the bispectrum on superhorizon scales as we shall discuss in 
Sec.~\ref{Phinonline}.

Let us now give the equations of motion on large scales 
for the individual second--order curvature perturbations 
$\zeta^{(2)}_\varphi$ and $\zeta^{(2)}_\gamma$. 
The energy transfer coefficients $\hat{Q}_\varphi$ and $\hat{Q}_\gamma$ in 
Eqs.~(\ref{splitQ})
perturbed at second order around the homogeneous backgrounds are given by
\begin{eqnarray}
\label{Q2phi}
\hat{Q}_{\varphi}&=& Q_{\varphi}+\delta^{(1)}Q_{\varphi}
+\frac{1}{2}\delta^{(2)}Q_\varphi\, , \\
\label{Q2g}
\hat{Q}_{\gamma}&=& Q_{\gamma}+\delta^{(1)}Q_{\gamma }\, ,
+\frac{1}{2}\delta^{(2)}Q_\gamma \, .
\end{eqnarray}
Note that from Eq.~(\ref{Qconstraint}) it follows that
$\delta^{(2)}Q_\gamma=-\delta^{(2)}Q_\gamma$. 
Thus the energy continuity equations (\ref{conseq}) perturbed at second order 
give on large scales~\cite{BMR4}
\bea
\label{cont2phi}
&{\delta^{(2)}\rho_\varphi}'&+3\H \left(
\delta^{(2)}\rho_\varphi+\delta^{(2)}P_\varphi \right)
-3(\rho_\varphi+P_\varphi) \psi^{(2)'} \nonumber \\
&&-6\psi^{(1)'}
\left[ \delta^{(1)}\rho_\varphi+\delta^{(1)}P_\varphi+
2 (\rho_\varphi+P_\varphi ) \psi^{(1)}\right] = \nonumber \\
&&a\,  \delta^{(2)}Q_\varphi+a\,  Q_\varphi \phi^{(2)}
-a Q_\varphi \left( \phi^{(1)} \right)^2+2a  \phi^{(1)}\delta^{(1)}Q_\varphi
\, ,
\eea
\bea
\label{cont2g}
&{\delta^{(2)}\rho_\gamma}'&+3\H \left(
\delta^{(2)}\rho_\gamma+\delta^{(2)}P_\gamma \right)
-3(\rho_\gamma+P_\gamma) \psi^{(2)'} \nonumber \\ 
&&-6\psi^{(1)'}\left[ \delta^{(1)}\rho_\gamma+\delta^{(1)}P_\gamma+
2 (\rho_\gamma+P_\gamma ) \psi^{(1)}\right] = \nonumber \\
&& a\,  \delta^{(2)}Q_\gamma+a\, Q_\gamma \phi^{(2)}
-a Q_\gamma \left( \phi^{(1)} \right)^2+2a  
\phi^{(1)}\delta^{(1)}Q_\gamma 
\, ,
\eea
where $\phi^{(2)}$ is the second--order perturbation in the gravitational 
potential $\phi=\phi^{(1)}+\frac{1}{2}\phi^{(2)}$.  
Note that Eqs.~(\ref{cont2phi}) and~(\ref{cont2g}) hold true 
in a generic gauge. 
We can now recast these equations in terms of the gauge--invariant 
curvature perturbations  $\zeta^{(2)}_\varphi$ and $\zeta^{(2)}_\gamma$ in 
a straightforward way by choosing the spatially flat gauge 
$\psi^{(r)}=\chi^{(r)}=0$. 

The (0-0)-component 
of Einstein equations in the spatially flat gauge at first order     
is given by Eq.~(\ref{00first}), and at second order on large scales 
it reads 
\be
\label{00second}
\phi^{(2)}=-\frac{1}{2}\frac{\delta^{(2)}\rho}{\rho}+4\left( \phi^{(1)} 
\right)^2
\, .
\ee
Using Eqs.~(\ref{00first}) and~(\ref{00second}) 
with the expressions~(\ref{zeta2phiflat})--(\ref{zeta2gflat}) 
we find from the energy continuity equations (\ref{cont2phi})--(\ref{cont2g}) 
that the individual second--order curvature perturbations obey on 
large scales~\cite{BMR4}
\bea
\label{eq:zeta2phi}
\zeta^{(2)'}_\varphi&=&
-\frac{a \H}{\rho'}\left[\left( \delta^{(2)}Q_\varphi
-\frac{Q_\varphi'}{\rho_\varphi'}\delta^{(2)}\rho_\varphi \right) 
+Q_\varphi\frac{\rho'}{2\rho}
\left(\frac{\delta^{(2)}\rho_\varphi}{\rho_\varphi'}
-\frac{\delta^{(2)}\rho}{\rho'} \right)
\right] \nonumber \\
&-&3a Q_\varphi \frac{\H}{\rho_\varphi'} \left({\phi^{(1)}}\right)^2
-2a \frac{\H}{\rho_\varphi'}\delta^{(1)}Q_\varphi \phi^{(1)}
-2\zeta^{(1)}_\varphi\zeta^{(1)'}_\varphi\, \nonumber \\
&-&2\left[ \zeta^{(1)}_\varphi
\left( a \frac{Q_\varphi \phi^{(1)}}{\rho_\varphi'}+
a \frac{\delta^{(1)} Q_\varphi}{\rho_\varphi'} \right) \right]^{\prime}
- \left[ \left(\zeta^{(1)}_\varphi\right)^2 
\left( a \frac{Q_\varphi'}{\H \rho_\varphi'}-\frac{a}{2} 
\frac{Q_\varphi}{\H \rho_\varphi'} \frac{\rho'}{\rho} \right) \right]^{\prime} 
\nonumber \\
&&   
\eea
and 
\bea
\label{eq:zeta2g}
\zeta^{(2)'}_\gamma &=&
-\frac{a \H}{\rho'}\left[\left( \delta^{(2)}Q_\gamma
-\frac{Q_\gamma'}{\rho_\gamma'}\delta^{(2)}\rho_\gamma \right) 
+Q_\gamma\frac{\rho'}{2\rho}\left(\frac{\delta^{(2)}\rho_\gamma}{\rho_\gamma'}
-\frac{\delta^{(2)}\rho}{\rho'} \right)
\right] \nonumber \\
&-&3a Q_\gamma \frac{\H}{\rho_\gamma'} \left({\phi^{(1)}}\right)^2
-2a \frac{\H}{\rho_\gamma'}\delta^{(1)}Q_\gamma \phi^{(1)}
-4\zeta^{(1)}_\gamma\zeta^{(1)'}_\gamma\nonumber \\
&-&2\left[ \zeta^{(1)}_\gamma
\left( a \frac{Q_\gamma \phi^{(1)}}{\rho_\gamma'}+
a \frac{\delta^{(1)} Q_\gamma}{\rho_\gamma'} \right) \right]^{\prime}
- \left[ \left(\zeta^{(1)}_\gamma\right)^2 
\left( a \frac{Q_\gamma'}{\H \rho_\gamma'}-\frac{a}{2} 
\frac{Q_\gamma}{\H \rho_\gamma'} \frac{\rho'}{\rho} \right) 
\right]^{\prime}\, ,
\nonumber \\
&& 
\eea
where we have used the fact that $w_{\varphi}=0$ and $w_\gamma=1/3$.

Eqs.~(\ref{eq:zeta2phi}) and (\ref{eq:zeta2g}) 
allow to follow the time--evolution of the gauge--invariant
curvature perturbations at second order. 
     

The results contained in the previous section can now be used to study 
the evolution of the second--order curvature perturbations during the 
reheating phase after a period of standard single field inflation, and 
in the alternative scenarios for the generation of the primordial adiabatic 
perturbations which have been recently proposed, namely 
the curvaton scenario~\cite{curvaton1,LW2,curvaton3} 
and the inhomogeneous (or ``modulated'') 
reheating~\cite{gamma1,gamma2,gamma3,MatRio,allahverdi}.
In fact, in each of these scenarios a scalar field oscillates around the 
minimum of its potential and eventually decays into radiation.
The evolution at second order of the curvature perturbations is 
necessary in order to follow the non--linearity of the
cosmological perturbations and thus to accurately compute the level of 
non--Gaussianity, including all the relevant second--order effects.
We shall now consider in detail these contributions in the three 
mentioned scenarios. 

\section{\bf The standard scenario}
\label{standardInfl}

The {\it standard scenario} is associated
to one--single field models of inflation, and the observed density
perturbations are due to fluctuations of the inflaton field itself.
When inflation ends, the inflaton oscillates about the minimum of its
potential and decays, thereby reheating the Universe. 
The initial inflaton fluctuations are adiabatic on large scales and  
are transferred 
to the radiation fluid during reheating.
In such a standard scenario the inflaton decay rate has no spatial 
fluctuations.

\subsection{The first--order curvature perturbation}

During inflation the inflaton field dominates the energy density of the 
Universe and therefore the energy density perturbations produced by the 
inflaton quantum fluctuations generate an adiabatic curvature perturbation.
Let us consider the inflaton field $\varphi(\tau, \bf x)$ with a potential 
$V(\varphi)$ and minimally coupled to gravity. 
The evolution equation for the inflaton field is the Klein--Gordon equation
\be
\label{KG}
\square \varphi= \frac{\partial V}{\partial \varphi}\, .
\ee 
Perturbing Eq.~(\ref{KG}) at linear order we obtain that the inflaton 
fluctuations obey 
\bea
\label{KG1}
{\deu\varphi}^{\prime\prime}&+&2\,\H {\deu \varphi}^{\prime}
- \nabla^2 \deu \varphi+ a^2 \deu \varphi 
\,\frac{\partial^2 V}{\partial \varphi^2}\,a^2
+2\,\phi^{(1)}\,\frac{\partial V}{\partial \varphi} \nonumber \\
&-&{\varphi_0}^{\prime}\left[ {\phi^{(1)}}^{\prime}+3\,{\psi^{(1)}}^{\prime}
+\nabla^2 {\omega^{(1)}} \right]=0\, .
\eea
A straightforward way to calculate the curvature perturbation generated on 
large scales is to solve the Klein--Gordon equation 
in the spatially flat gauge 
defined by the requirement $\psi^{(1)}=0$ and $\chi^{(1)}=0$. In fact in this 
gauge the perturbations of the scalar field correspond to the Sasaki--Mukhanov 
gauge--invariant variables~\cite{v1,v2}
\be
\label{MSvar}
Q_\varphi=\delta^{(1)} \varphi+\frac{\varphi'}{\H} {\hat{\psi}}^{(1)}\, .
\ee
As usual we introduce the field $\widetilde{Q}_{\varphi}=a Q_{\varphi}$.
The Klein--Gordon equation in the spatially flat gauge now reads (in 
Fourier space)~\cite{TN}
\bea
\widetilde{Q}_{\varphi}^{\prime \prime}
+\left( k^2-\frac{a{''}}{a}+{\mathcal M}^2_{\varphi}a^2 
\right)\tilde{Q}_{\varphi}=0\, .
\eea
where
\be
{\mathcal M}^2_{\varphi}=V_{\varphi \varphi}
-\frac{8 \pi G_{\rm N}}{a^3} \left(\frac{a^3}
{H} \dot{\varphi}^2 \right)^{\displaystyle{\cdot}}
\ee
is an effective mass of the inflaton field in this gauge, which to
lowest order in the slow--roll parameters is given by
\be
\frac{{\mathcal M}^2_{\varphi}}{H^2}= 3\eta-6\epsilon \, ,
\ee
where 
$\epsilon=(1/ 16 \pi G_{\rm N}) \left( V_\varphi/V \right)^2$ and 
$\eta=(1/8 \pi G_{\rm N}) \left( V_{\varphi\varphi} /V \right)$ 
are the inflaton slow--roll parameters. 
This equation has the same form as Eq.~(\ref{KGsempl}) and thus
we can just follow the same procedure described in detail 
in Section~\ref{QFdeSitter}, simply replacing
$m^2_{\chi}$ with ${\mathcal M}^2_{\varphi}$.
The equation of motion for $\widetilde{Q}_{\varphi}$ or for the corresponding 
eigenvalues $u_k(\tau)$ thus becomes
\be 
\label{BesselInfl}
u_k^{\prime \prime}+\left( k^2 -\frac{\nu_{\varphi}^2-\frac{1}{4}}{\tau^2}
\right)u_k=0 \, ,
\ee
with
\be
\nu_{\varphi} \simeq 
\frac{3}{2}+3 \epsilon - \eta\, .
\ee
From Section~\ref{QFdeSitter} 
we conclude that on superhorizon scales and to lowest order 
in the slow--roll parameters the inflaton fluctuations are
\be
|Q_\varphi (k)|=\frac{H}{\sqrt{2 k^3}} \left( \frac{k}{a H} \right)^
{\frac{3}{2} -\nu_\varphi}\, .
\ee  

In order to calculate the curvature perturbation on large scales we can 
consider the curvature perturbation on comoving hypersurfaces, which in the 
case of a single scalar field reads~\cite{CCURV1,CCURV2,Lidseyetal,lrreview}
\be
\label{RInfl}
{\mathcal R}^{(1)}=\hat{\psi}^{(1)}+\frac{\H}{\varphi'} \delta^{(1)} 
\varphi\, . 
\ee
Notice that the comoving curvature perturbation ${\mathcal R}^{(1)}$
and the uniform energy density curvature perturbation $\zeta^{(1)}$ are 
simply related by (see for example Ref.~\cite{gw})
\be
-\zeta^{(1)}= {\mathcal R}^{(1)}+\frac{2 \rho}{9(\rho+p)} 
\left( \frac{k}{aH} \right)^2 \phi^{(1)}
\ee   
where here $\phi^{(1)}$ is the gravitational potential in the longitudinal 
gauge. Therefore on large scales ${\mathcal R}^{(1)} \simeq - \zeta^{(1)}$.
From Eq.~(\ref{MSvar}) it is evident that 
\be
\label{RQ}
{\mathcal R}^{(1)}=\frac{\H}{\varphi'} Q_\varphi\, .
\ee
Thus we obtain the power--spectrum of the curvature perturbation 
on large scales 
\be
\label{Pinflation}
{\mathcal P}_{\mathcal R}=\left( \frac{H^2}{2 \pi \dot{\varphi} }\right)^2 
\left( \frac{k}{aH} \right)^{3-2 \nu_\varphi} 
\simeq \left( \frac{H^2}{2 \pi \dot{\varphi}} \right)_*^2 \, , 
\ee
where the asterisk stands for the epoch a given perturbation mode leaves 
the horizon during inflation.
From Eq.~(\ref{Pinflation}) one
immediately reads the spectral index of the curvature 
perturbation to lowest order in the slow--roll parameters
\be
n_{\mathcal R}-1 \equiv  \frac{d \ln {\mathcal P}_{\mathcal R}}{d \ln k} =
{3-2 \nu_\varphi}=-6\epsilon+2\eta\, .
\ee
Notice that from our results one can check that during inflation  
the curvature perturbation mode is constant on superhorizon scales 
${\mathcal R^{(1)}}' \simeq -\zeta^{(1)'} \simeq 0$ 
(from which the last equality in Eq.~(\ref{Pinflation}) follows). 

This is a well--known result: the  
curvature mode is the quantity which  
allows to connect observable perturbations to primordial perturbations 
produced during inflation~\cite{BardeenSteinTur,LythLiddle,lrreview}. 
This result comes from the fact that in  
single--field slow--roll models of inflation the 
intrinsic entropy perturbation of the inflaton field is negligible 
on large scales~\cite{lrreview,Bassettetal,gw,Bartoloetal}. 
We will now show 
that the above result also holds during the reheating phase 
on large scales.

\subsection{Reheating after inflation} 

When inflation ends,
the inflaton oscillates about the minimum of its
potential and decays into radiation, thereby reheating the Universe.
In such a standard scenario the inflaton decay rate has no spatial
fluctuations.
Eq.~(\ref{zeta1'}) and Eq.~(\ref{zeta1phi'}) with $\delta^{(1)}
\Gamma=0 $ now read
\bea
\label{zeta1'g}
\zeta^{(1)'}&=& -\H f \left( \zeta^{(1)}-\zeta^{(1)}_\varphi \right)\, , \\
\label{zeta1phi'g}
\zeta^{(1)'}_\varphi&=&\frac{a \Gamma}{2} \frac{\rho_\varphi}{\rho_\varphi'}
\frac{\rho'}{\rho} \left( \zeta^{(1)} -\zeta^{(1)}_\varphi
\right)\, .
\eea
At the beginning of the reheating phase, after the end of inflation,
the total curvature perturbation is initially given by
the curvature perturbation of the inflaton fluctuations
$\zeta^{(1)}_{\rm in}=\zeta^{(1)}_{\varphi,{\rm in}}$. Therefore
Eqs.~(\ref{zeta1'g}) and (\ref{zeta1phi'g}) show that during the
reheating phase $\zeta^{(1)}=\zeta^{(1)}_\varphi=
\zeta^{(1)}_{\varphi,{\rm in}}$ is a fixed--point of the time--evolution.
Such a result has been obtained in this way 
at first order in Ref.~\cite{MatRio} 
(see also Refs.~\cite{KodaHama,Qreh}) and extended to
second--order in the perturbations in Ref.~\cite{BMR2}, under the 
sudden--decay approximation.

\subsection{The second--order curvature perturbation and non--Gaussianity
during inflation}
\label{NGInfl}

A complete analysis of the perturbations produced during single--field 
slow--roll inflation up to second order has been performed in 
Ref.~\cite{noi}.  
Such an analysis fully accounts 
for the inflaton self--interactions as well as for the second--order 
fluctuations of the background metric. Moreover it also provides a 
gauge--invariant expression for the second--order 
comoving curvature perturbation thus allowing to calculate the bispectrum of 
such a quantity during inflation. The results of Ref.~\cite{noi}  
agree with those of Ref.~\cite{Maldacena}, 
where the three--point function for the curvature perturbation 
is calculated using a different procedure. In 
Ref.~\cite{Maldacena} the starting point is the Lagrangian and one 
evaluates the cubic contributions 
to the curvature perturbations. In fact 
Refs.~\cite{noi,Maldacena} represent a step 
forward in the computation of the non--linearities produced during inflation. 
Before then, the problem of calculating the bispectrum of perturbations
produced during inflation had been addressed by either looking at the effect 
of inflaton self--interactions (which necessarily generate non--linearities in
its quantum fluctuations) in a fixed de Sitter background~\cite{FRS}, or 
using the so--called stochastic approach to inflation \cite{Getal}
\footnote{See also Refs.~\cite{wk,GM}, and Ref.~\cite{mmr}, for a more recent 
analysis.}, where back--reaction effects of field fluctuations on the 
background metric are partially taken into account. An intriguing result of 
the stochastic approach -- which is indeed confirmed by the 
second--order analyses of Refs.~\cite{noi,Maldacena} -- 
is that the dominant source 
of non--Gaussianity actually comes from non--linear gravitational 
perturbations, rather than by inflaton self--interactions. 

Before going into details, let us give here an 
estimate of the size of the non--Gaussianity that we expect to be produced 
during inflation. 

The inflaton field can be split 
into a homogeneous background $\varphi_0(\tau)$ 
and a perturbation $\delta \varphi (\tau, x^{i})$ as
\be \label{scalarfield}
\varphi(\tau, x^i)=\varphi_0(\tau)+\delta\varphi(\tau, x^i)=\varphi_0(\tau)+
\delta^{(1)}\varphi(\tau, x^i)+\frac{1}{2}\delta^{(2)}
\varphi(\tau, x^i)\, ,
\ee
where the perturbation has been expanded into a first and a second--order
part, respectively. First of all notice that, 
at first order in the perturbations, using Eq.~(\ref{0iPoisson}) 
for a scalar field in the longitudinal gauge 
$(\dot{\psi}^{(1)}+H \psi^{(1)}=\frac{\kappa^2}{2} \dot{\varphi_0} 
\deu \varphi)$ and the perturbed Klein--Gordon equation (\ref{KG1L}) one 
obtains $\psi^{(1)}=\epsilon \, H 
\deu \varphi/\dot{\varphi_0}$ to lowest order 
in the slow--roll parameters and on large scales. 
On the other hand, from the definition of the comoving curvature 
perturbation at first order, Eq.~(\ref{RInfl}), 
it follows that $\R^{(1)}= H \deu \varphi/\dot{
\varphi_0}$ to lowest order in the slow--roll 
parameters and on large scales, and hence under these approximations 
$\psi^{(1)} = \epsilon \R^{(1)}$. 
Let us consider the perturbed Klein--Gordon equation at second order 
(in the Poisson  gauge) on large scales 
(see Eq.~(\ref{KG2L}) in Appendix~\ref{B}) 
\bea
\label{KG2}
&\ddot{\delta^{(2)} \varphi}&+3 H \dot{\delta^{(2)} \varphi} 
+ 2 \frac{\partial V}{\partial \varphi} \phi^{(2)} 
- \dot{\varphi_{0}} \dot{\phi^{(2)}}-3 \dot{\varphi_{0}} \dot{\psi^{(2)}}     
-8 \dot{\varphi_{0}} \psi^{(1)} \dot{\psi^{(1)}} 
- 8 \dot{\psi^{(1)}} \dot{\delta^{(1)} \varphi} \nonumber \\ 
&-&\frac{8}{a^2} 
\psi^{(1)} \partial_i \partial^{i} \delta^{(1)} \varphi = 
- 4 \frac{\partial^{(2)} V}{\partial \varphi ^2} 
\psi^{(1)} \delta^{(1)} \varphi 
- \frac{\partial^{2} V}{\partial \varphi ^2} \delta^{(2)} \varphi 
- \frac{\partial^{3} V}{\partial \varphi ^3} 
\left( \delta^{(1)} \varphi \right)^2\, .\nonumber \\
\eea
Now in order to give our estimate we consider a second--order curvature 
perturbation $\R^{(2)} \sim H \delta^{(2)} \varphi/\dot{\varphi_0}$. For 
simplicity, let us just focus on the first source term on the 
R.H.S. of Eq.~(\ref{KG2}). Under a slow--roll approximation from 
Eq.~(\ref{KG2}) we see that (up to numerical coefficients of order unity) 
\be
\label{Req} 
\dot{\R}^{(2)} \sim \left( \frac{\partial^2 V}{\partial \varphi^2} \right) 
\left( \frac{\psi^{(1)}}{H}\right) 
\left(  H \frac{\delta^{(1)} \varphi}{\dot{\varphi_0}}\right)
\sim H \eta \, \epsilon \left( \R^{(1)} \right)^2\, , 
\ee      
where in the last step we have used  
$\psi^{(1)}= \epsilon \R^{(1)}$ 
with $\R^{(1)} \sim H \delta^{(1)} \varphi/ \dot{\varphi_0 }$ and the 
definition of the slow--roll parameters. Recalling that 
the time derivatives of the slow--roll parameters are next order 
in the parameters,  
$\dot{\epsilon},\dot{\eta}= \mathcal{O}(\epsilon^2, \eta^2)$, from 
Eq.~(\ref{Req}) we obtain 
\be
\R^{(2)} \sim \mathcal{O} (\epsilon, \eta) \left(\R^{(1)} \right)^2\, . 
\ee      
From this simple calculation we therefore see that the non--Gaussianity level 
in the standard scenario is of the order of the slow--roll parameters 
$\epsilon$ and $\eta$
\be
{\rm NG} \sim \mathcal{O}(\epsilon, \eta), 
\ee  
in qualitative agreement with the predictions of Refs.~\cite{Getal,gan}, 
within the stochastic inflation approach. 

Let us now turn to the exact results on the level of non--Gaussianity by 
summarizing some of the findings of Ref.~\cite{noi}. 
It is possible to extend at second--order the gauge--invariant large--scale 
comoving curvature perturbation ${\mathcal R}^{(1)}$ defined in 
Eq.~(\ref{RInfl}) by introducing a quantity 
${\mathcal R}={\mathcal R}^{(1)}+\frac{1}{2}{\mathcal R}^{(2)}$ defined as~
\cite{noi}
\be 
\label{R}
{\mathcal R} = {\mathcal R}^{(1)}+\frac{1}{2} 
\left[ {\mathcal{H}}\frac{\delta^{(2)}
\varphi}
{{\varphi'_0}}+ \psi^{(2)} \right]
+ \frac{1}{2} \frac{ \left( {\psi^{(1)}}^{\prime}+2{\mathcal{H}}\psi^{(1)}
+{\mathcal{H}} {\deu\varphi}^{\prime}/ \varphi'_0 \right)^2}{{\mathcal{H}}'
+2{\mathcal{H}}^2-{\mathcal{H}}\, \varphi''_0/\varphi'_0}
\ee
Such a quantity is gauge--invariant with respect to an infinitesimal 
second--order shift of the time coordinate, 
$\tau\rightarrow 
\tau-\xi^0_{(1)}+\frac{1}{2} \left({\xi^{0'}_{(1)}}\xi^0_{(1)}
-\xi^0_{(2)}\right)$.\\
By solving the Einstein equations during inflation 
in the longitudinal (Poisson) gauge on large scales, and by performing 
an expansion to lowest order in the slow--roll parameters 
the second--order curvature perturbation is determined in terms of 
its first--order counterpart. The result is~\cite{noi} 
\be
\label{final:RInfla}
{\mathcal R}^{(2)}= 
\left(\eta-3\epsilon\right) \left( {\mathcal R}^{(1)} \right)^2
+{\mathcal I}\, ,
\ee
where

\bea
\label{integralsI}
{\mathcal I}&=&-\frac{2}{\epsilon}\,\int \frac{1}{a^2} \psi^{(1)} 
\La \psi^{(1)} dt -\frac{4}{\epsilon}\,
\int  \frac{1}{a^2} 
\left( \partial_i \psi^{(1)} \partial^i \psi^{(1)}  
\right) dt\nonumber\\
&-&\frac{4}{\epsilon}\,\int\,\left(\ddot{\psi}^{(1)}\right)^2 dt
+\left(\epsilon-\eta\right)\triangle^{-1}\partial_i R^{(1)}
\partial^i R^{(1)}
\eea  
contains also terms which are
${\mathcal O}(\epsilon,\eta)$. Notice that the integrals in 
Eq.~(\ref{integralsI}) give rise to 
non--local operators which are not necessarily suppressed 
on large scales being of the form 
$\nabla ^{-2}[\nabla(\cdot)\nabla(\cdot)]$ or 
$\nabla ^{-2}[(\cdot) \La(\cdot)]$. 
This is due to the fact that a given perturbation mode during inflation 
first is subhorizon, where it oscillates, and then at a given epoch 
it leaves the horizon.  

The total comoving curvature perturbation thus receives a
contribution which is quadratic in ${\mathcal R}^{(1)}$ and it  
will then have a non--Gaussian $(\chi^2)$ component.
We conclude that during inflation a tiny intrinsic non--linearity 
is produced, being the slow--roll parameters $\epsilon, |\eta| \ll 1$. 
This does not come as a surprise, indeed, and it has a very transparent 
interpretation. Since the inflaton field is driving inflation, its potential 
must be very flat, with very small slow--roll parameters. This amounts to 
saying that the interaction terms in the inflaton potential must be 
suppressed, hence also the non--linearities eventually producing 
non--Gaussian features. Alternative mechanisms to generate a higher level 
of non--Gaussian adiabatic perturbations in the inflationary framework 
could be the presence of some features in the inflaton 
potential~\cite{Salopek1,KBHP,Salopek,wk} in the part corresponding 
to the last$~\sim 60$ e--foldings,   
or the presence of more than a single scalar field during 
inflation~\cite{BMT3}. 
In both cases the restrictions coming from the slow--roll 
conditions can be avoided. We shall come back later to these alternative 
scenarios. 
In the case of the standard single--field models of inflation in order to 
characterize the primordial non--Gaussianity we can expand ${\mathcal R}$ in 
Fourier space as 
\begin{eqnarray}
\label{Rmomspace}
{\mathcal R}({\bf k}) &=& {\mathcal R}^{(1)}({\bf k})+ 
\frac{1}{(2\pi)^3}
\int\, d^3 k_1\,d^3 k_2\, \delta^{(3)}\left({\bf k}_1+{\bf k}_2-{\bf k}\right)
\nonumber\\
&\times&
f^{\mathcal R}_{\rm NL}\left({\bf k}_1,{\bf k}_2\right)
{\mathcal R}^{(1)}({\bf k}_1){\mathcal R}^{(1)}({\bf k}_2)\, ,
\end{eqnarray}
where we have introduced a 
momentum--dependent non--linearity parameter \linebreak
$f^{\mathcal R}_{\rm NL}\left({\bf k}_1,{\bf k}_2\right)$, which from 
Eq.~(\ref{final:RInfla}) reads 
\be
\label{fR}
f^{\mathcal R}_{\rm NL}\left({\bf k}_1,{\bf k}_2\right)=
\frac{1}{2}(\eta-3\epsilon)+
I\left({\bf k}_1,{\bf k}_2\right)\, ,
\ee
where $I\left({\bf k}_1,{\bf k}_2\right)$ is directly related to the 
function ${\mathcal I}$ and is of first order in the slow--roll parameters.
Thus the level of non--Gaussianity generated during inflation is 
typically $f^{\mathcal R}_{\rm NL} \sim {\mathcal O}(10^{-1} 
\div 10^{-2})$. 
Eq.~(\ref{fR}) can also be recast in the form 
\be
f^{\mathcal R}_{\rm NL}\left({\bf k}_1,{\bf k}_2\right)
=\frac{1}{4}\left( n_{\mathcal R}-1 \right) 
+I\left({\bf k}_1,{\bf k}_2\right) \;,
\ee
where we have made use of the expression of the spectral index 
$n_{\mathcal R}-1=-6 \epsilon+2\eta$ in terms of the slow--roll parameters. 
Notice however that the result in Eq.~(\ref{fR}) refers only to the 
non--Gaussianity generated \emph{during inflation}\footnote{The generalization
of the calculation of the non--linearity parameter during inflation 
to two--field models of inflation has been recently presented in
Ref.~\cite{ev}.}. 

In order to determine the 
level of the non--Gaussianity which can be actually compared with observations 
it is necessary to consider the subsequent evolution of the 
(gravitational potential) perturbations after inflation ends, 
through reheating  and  the radiation/matter dominated 
epochs. Usually in the literature the matching has been performed 
by simply extending the linear relation on large scales between the 
gravitational potential and the curvature perturbation in the matter 
dominated epoch $\phi^{(1)}=- (3/5) {\mathcal R}^{(1)}$ to second order
~\cite{FRS,Getal,noi,Maldacena}.
We warn the reader that such a procedure is indeed not correct. In fact one 
has to take into account also after inflation a fully second--order 
relativistic analysis of the cosmological perturbations. As we shall 
consider in the next sections, in the case of the 
standard scenario of inflation
the matching between the inflationary epoch and the 
radiation/matter dominate phases, where observable quantities 
are defined, is achieved 
by exploiting the conservation on large scales of the curvature perturbation 
$\zeta$ up to second order. As a result, as 
it has been shown in Ref.~\cite{BMR2}, the small initial non--Gaussianity 
generated during single--field inflation is actually largely 
\emph{enhanced} by the second--order 
gravitational dynamics in the post--inflationary phases.   
Such an enhancement produces a non--linearity parameter in the CMB temperature 
anisotropy on large scales which is $f_{\rm NL} \sim {\mathcal O}(1)$.
Nonetheless, the results contained in Refs.~\cite{noi,Maldacena} are useful 
in that they allow to determine the initial conditions on the non--Gaussianity 
produced during slow--roll inflation. 

\subsubsection{Inflaton effective Lagrangian}

In this paragraph we want to mention the possibility discussed in 
Ref.~\cite{Creminelli} that the non--Gaussianity 
produced during inflation might receive additional contributions from 
some high-energy corrections which can modify the minimal inflaton 
Lagrangian given in Eq.~(\ref{inflatonaction}). Such corrections can arise 
if inflation takes place at relatively high energies, and they can be 
parametrized by an effective inflaton Lagrangian in 
which one integrates out degrees of freedom with momenta larger than some 
scale $M$ corresponding to the scale of the new phyics. This is realized by  
including some higher order operators suppressed by the appropriate power of 
$M$. Such operators must not spoil the flatness of the potential in order to 
have an inflationary phase, and, as argued in Ref.~\cite{Creminelli}, the best 
candidates are operators 
that just modify the kinetic part of the action such as those 
of the form $\left( \nabla \varphi \right)^4$ with a scale $M$ 
which can be taken as low as $\dot{\varphi}^2$. Therefore a possible 
effective action reads
\be
\label{action:higher operators}
S=\int d^4x \sqrt{-g} \left[-\frac{1}{2}\left(\nabla \varphi \right)^2
-V(\varphi)+\frac{1}{8 M^4} \left(\nabla \varphi \right)^2 
\left(\nabla \varphi \right)^2 + \cdots \right]
\ee
Notice that when $M^2$ tends to 
$\dot{\varphi}$ then 
such an effective description ceases to make sense, because one should keep 
track of all the higher terms in the action.   
The higher dimension operators represent 
additional self--interaction terms which will produce some 
non--Gaussianities during inflation. 
In fact starting from the action~(\ref{action:higher operators}), 
following Ref.~\cite{Maldacena}, it is possible to calculate
the contributions to the three--point function for the curvature perturbation 
$\R$ coming from the higher dimension operators to lowest order in 
$\dot{\varphi}^2/M^4$. In Ref.~\cite{Creminelli} 
it has been found that the typical magnitude of such contributions is
\footnote{The exact expression also contains a scale--dependent part of the 
same magnitude, whose precise form however differs from the one 
obtained in the standard case, see Ref.~\cite{Creminelli} for more details.}   
\be
\label{fRhigheroperators}
f^{\R}_{\rm NL} \sim \frac{\dot{\varphi}^2}{M^4}\, .
\ee
We see that the net effect of the introduction of a new scale 
$M$ is that the slow--roll parameters in the standard result~(\ref{fR}) are 
now replaced by a new order parameter which does not have to be restricted 
by slwo-roll conditions. However the lower limit allowed for $M^4$ is 
$\dot{\varphi}^2$ and therefore also in this case   
\be
f^{\R}_{\rm NL} \lesssim 1\, .
\ee   

We must stress here again the important point we discussed 
at the end of the previous section. As for Eq.~(\ref{fR}), also the 
estimate~(\ref{fRhigheroperators}) refers to the level of non--Gaussianity 
only during the inflationary epoch. In order to make a full and sensible 
comparison with the observations it is necessary to perform a fully 
second--order analysis of the evolution of the pertubations after inflation 
ends, through the radiation and matter dominated epochs. 
As we shall see, the non--linearities arising from the post--inflationary 
dynamics will anyway enhance the observable non--Gaussianity level in the CMB 
temperature anisotropy in such a way as to hide the initial imprint such as 
that in~(\ref{fRhigheroperators}). 

\subsection{Reheating after inflation}

The first step to follow the evolution of non--linearities on large scales 
after inflation is to analyze how the curvature perturbation $\zeta$ 
evolves on large scales during reheating.\\
In Ref.~\cite{BMR2} it was shown that also at second order the 
curvature perturbation $\zeta^{(2)}$ remains constant during inflation, 
under the inflaton sudden--decay approximation.  
Under such an approximation the individual energy density 
perturbations (and hence the corresponding curvature perturbations) are 
separately conserved until the decay of the scalar field, which amounts 
to saying that in the equations for the curvature perturbations 
Eqs.~(\ref{eq:zeta2phi}) and~(\ref{eq:zeta2g}) one can drop the energy 
transfer triggered by the decay rate $a \Gamma/\H \ll 1$. 
Going beyond the sudden--decay approximation, the first order 
results $\zeta^{(1)}_\varphi=\zeta^{(1)}$ in Eq.~(\ref{main1}) yield
\be
\label{zeta2reheating}
\zeta^{(2)}=f \zeta^{(2)}_\varphi+(1-f) \zeta^{(2)}_\gamma\, .
\ee
The equation of motion for $\zeta^{(2)}$ on large scales is obtained
by differentiating this expression and by using 
Eqs.~(\ref{eq:zeta2phi}) and~(\ref{eq:zeta2g}), with 
$\delta^{(2)} \Gamma=0$ and  $\zeta^{(1)}_\varphi=\zeta^{(1)}$; it reads
\be
\label{zeta2'reheating}
\zeta^{(2)'}= -\H f \left( \zeta^{(2)}-\zeta^{(2)}_\varphi \right)\, .
\ee
In the same way as at first order 
from Eqs.~(\ref{zeta2reheating}) and~(\ref{zeta2'reheating}) it follows 
that the second--order curvature perturbation $\zeta^{(2)}$ remains constant 
on large scales during the reheating phase, 
being given at the end of inflation by the curvature perturbation in the 
inflaton field $\zeta^{(2)}_{\rm in}=\zeta^{(2)}_{\varphi, {\rm in}}$.

\subsection{Post--inflationary evolution of the second--order curvature 
perturbation}

The superhorizon--scale evolution of the primordial non--linearity generated 
during inflation during the radiation and matter dominated 
phases has been studied in Ref~\cite{BMR2}. 
Following their approach, we consider the energy--momentum tensor for a
perfect fluid with constant but otherwise generic equation of state, as
defined in Sec.~\ref{energy--momentum}.\\
We will explicitly show that during the radiation and matter dominated 
epochs the second--order curvature perturbation $\zeta^{(2)}$ is conserved 
on large scales. From now on we shall adopt the Poisson gauge. 
Our starting point is the energy continuity equation at 
second--order  
\bea
\label{conservation}
\delta^{(2)}\rho^\prime&+&3\H\left(1+w\right)\delta^{(2)}\rho
-3\left(1+w\right)\rho_0\psi^{(2)^\prime}-
6(1+w)\psi^{(1)^\prime}\left[\delta^{(1)}\rho
+2\rho_0\psi^{(1)}\right]\nonumber\\
&=&-2(1+w)\rho_0\left(v_i^{(1)}v^i_{(1)}\right)^\prime-
2(1+w)(1-3w)\H\rho_0 v_i^{(1)}v^i_{(1)}\nonumber\\
&+&4(1+w)\rho_0\partial_i\psi^{(1)}v^i_{(1)}+
2\frac{\rho_0}{\H^2}\left(\psi^{(1)}\nabla^2\psi^{(1)\prime}-
\psi^{(1)\prime}\nabla^2\psi^{(1)}\right)\, ,
\eea
where we have also used the  $(0-i)$- component of Einstein equation (see 
Appendix A). 
This equation can be rewritten in a more suitable form
\bea
\label{qq}
&& \left[\psi^{(2)}+\H\frac{\delta^{(2)}\rho}{\rho_0^\prime}
+(1+3w)\H^2\left(\frac{\delta\rho^{(1)}}{\rho_0^\prime}\right)^2-4
\H\left(\frac{\delta\rho^{(1)}}{\rho_0^\prime}\right)\psi^{(1)}
\right]^\prime= \nonumber \\
&& \frac{2}{3}\left(v_i^{(1)}v^i_{(1)}\right)^\prime 
+ \frac{2}{3}(1-3w)\H
v_i^{(1)}v^i_{(1)} -\frac{4}{3}\partial_i\psi^{(1)}v^i_{(1)} + 
\frac{16}{27\left(1+w\right)^2\H}  \psi^{(1)} \nabla^2 \psi^{(1)}
\nonumber\\
&&
- \frac{2}{3\left(1+w\right)\H^2}\left[\left(1- 
\frac{8}{9\left(1+w\right)}\right) 
\psi^{(1)} \nabla^2\psi^{(1)\prime}
- \left(1 - \frac{4\left(1+3w\right)}{9\left(1+w\right)}\right)
\psi^{(1)\prime} \nabla^2\psi^{(1)} \right] 
\nonumber\\
&&
+ \frac{8\left(1+3w\right)}{27\left(1+w\right)^2 \H^3} 
\left[\frac{\left(\nabla^2 \psi^{(1)}\right)^2}{3}
- \psi^{(1)\prime} \nabla^2 \psi^{(1)\prime} 
+ \frac{\nabla^2 \psi^{(1)\prime} \nabla^2 \psi^{(1)}}{3\H}
\right] \, ,
\eea
where the argument on the L.H.S. can be further simplified to 
\bea
\label{3}
\psi^{(2)}+\H\frac{\delta^{(2)}\rho}{\rho_0^\prime}
-\left(5+3w\right)\H^2 \left(\frac{\delta^{(1)}\rho}{\rho_0^\prime}
\right)^2=\psi^{(2)}+\H\frac{\delta^{(2)}\rho}{\rho_0^\prime}
-\frac{4}{5+3w}\left(\zeta_I^{(1)}\right)^2\, , \nonumber \\
\eea
and the final form has been obtained 
employing Eqs.~(\ref{00Plargescales}) and (\ref{relPsizeta}).\\
Notice that the quantity in Eq.~(\ref{3}) is in fact the curvature 
perturbation defined in Eq.~(\ref{qqq}) in the case of the generic fluid 
with constant equation of state that we are considering here. 
From Eqs.~(\ref{qq}) and (\ref{3}) we find

\bea
\label{three}
\psi^{(2)}+\H\frac{\delta^{(2)}\rho}{\rho_0^\prime}
-\left(5+3w\right)\H^2 \left(\frac{\delta^{(1)}\rho}{\rho_0^\prime}
\right)^2={\mathcal C} +\frac{2}{3} \left(v_i^{(1)}v^i_{(1)}\right)+
\int^\tau \,d\tau^\prime\, {\mathcal S}(\tau^\prime) \, ,  \nonumber \\
\eea
where ${\mathcal C}$ is a constant in time, 
${\mathcal C}^\prime=0$, on large scales and

\bea
\label{def:S}
{\mathcal S}&=&
\frac{2}{3}(1-3w)\H
v_i^{(1)}v^i_{(1)} -\frac{4}{3}\partial_i\psi^{(1)}v^i_{(1)} + 
\frac{16}{27\left(1+w\right)^2\H}  \psi^{(1)} \nabla^2 \psi^{(1)}
\nonumber\\
&-&
\frac{2}{3\left(1+w\right)\H^2}\left[\left(1- 
\frac{8}{9\left(1+w\right)}\right) 
\psi^{(1)} \nabla^2\psi^{(1)\prime}
- \left(1 - \frac{4\left(1+3w\right)}{9\left(1+w\right)}\right)
\psi^{(1)\prime} \nabla^2\psi^{(1)} \right] 
\nonumber\\
&+&
\frac{8\left(1+3w\right)}{27\left(1+w\right)^2 \H^3} 
\left[\frac{\left(\nabla^2 \psi^{(1)}\right)^2}{3}
- \psi^{(1)\prime} \nabla^2 \psi^{(1)\prime} 
+ \frac{\nabla^2 \psi^{(1)\prime} \nabla^2 \psi^{(1)}}{3\H}
\right] \, .
\eea

We are interested in the determination of the non--linearities 
after the inflationary stage. We have seen in the previous section that 
also during the reheating phase the curvature perturbation $\zeta^{(2)}$ is 
conserved. Therefore we are allowed to fix the
constant  ${\mathcal C}$ by matching the conserved
quantity on large scales at the end of
inflation ($\tau=\tau_I$)
\begin{equation}
\label{four}
{\mathcal C}= \psi_I^{(2)}
+\H_I\frac{\delta^{(2)}\rho_I}{\rho_{0I}^\prime}-
2\left(\zeta_I^{(1)}\right)^2\, ,
\end{equation}
where we have used the fact that during inflation $w_I\simeq -1$ and
we have disregarded gradient terms which turn out to be negligible  
for the computation of the large--scale bispectrum.

In fact the inflationary quantity  $\left(\psi_I^{(2)}
+\H_I\frac{\delta^{(2)}\rho_I}{\rho_{0I}^\prime}\right)$ 
has been computed in Refs.~\cite{noi,Maldacena} 
\begin{equation}
\label{fivenuova}
\psi_I^{(2)} +\H_I\frac{\delta^{(2)}\rho_I}{\rho_{0I}^\prime}
\simeq \left( \eta - 3 \epsilon \right) \left( \zeta_I^{(1)} \right)^2  
+ {\mathcal O}(\epsilon,\eta)\, \left({\rm non-local}\,\,{\rm terms}\right)
\, ,
\end{equation}
in terms of the slow--roll parameters $\epsilon=1-\H_I^\prime/\H_I^2$ and 
$\eta = 1+\epsilon -\left(\varphi^{\prime\prime}/\H_I\varphi^\prime\right)$ 
where $\H_I$ is the Hubble parameter during inflation and 
$\varphi$ is the inflaton field~\cite{lrreview}. 
Since during inflation the slow--roll parameters are tiny, we can safely
disregard the intrinsically second--order 
terms originated from the inflationary epoch.
Thus from Eq.~(\ref{three}) and Eq.~(\ref{four}) we obtain a relation 
between the gravitational potential $\psi^{(2)}$ and the energy density 
perturbation $\delta^{(2)} \rho$ during the radiation/matter dominated epochs
\bea
\label{relPsirhoInfl}
\psi^{(2)}-\frac{1}{3(1+w)} \frac{\delta^{(2)} \rho}{\rho_0}=
-\frac{2}{3}\,  \frac{5+3w}{1+w} \left( \psi^{(1)} \right)^2+\frac{2}{3} 
\left( v_i^{(1)} v^i_{(1)} \right) + \int_{\tau_I}^{\tau} \S(\tau') 
d\tau' \, , \nonumber \\ 
\nonumber \\ 
\eea
where we have made use of Eq.~(\ref{00Plargescales}) and 
Eq.~(\ref{relPsizeta}), with
$\zeta^{(1)}=\zeta^{(1)}_I$ since the curvature perturbation is 
conserved on large scales. 

\section{\bf The curvaton scenario}   

Let us now consider the so--called curvaton 
mechanism~\cite{curvaton1,LW2,curvaton3} to generate an initially 
adiabatic perturbation deep in the radiation era, as an alternative to the 
standard inflationary picture. In fact in the curvaton scenario the 
cosmological perturbations are produced from fluctuations of a scalar field 
$\sigma$ (different from the inflaton) during a period of inflation, 
in the case where the perturbations from the inflaton field are considered 
to be negligible. 
The scalar field is subdominant during 
inflation and thus its fluctuations are initially of isocurvature type. 
Therefore a curvature perturbation is sourced on large scales according to 
Eq.~(\ref{MAIN1}) 
and Eq.~(\ref{MAIN2}). 
The curvature perturbation will become relevant when the energy density of the 
curvaton field is a significant fraction of the total energy. This happens 
after the end of inflation when the   
curvaton field begins to oscillate around the minimum of its potential 
once its mass has dropped below the Hubble rate, behaving 
like non--relativistic matter. Finally, 
well before primordial nucleosynthesis, 
the curvaton field is supposed to completely 
decay into thermalised radiation thus generating a final adiabatic 
perturbation.\footnote{In the curvaton scenario it is indeed possible that 
some residual isocurvature perturbations survive after the curvaton 
decay. This could be 
the case for example if the curvaton field decays when subdominant into a 
component of Cold Dark Matter (CDM) which does not thermalize with the 
existing radiation. This is 
due to the fact that an isocurvature perturbation is present initially, 
while in the standard scenario of inflation it is not possible since 
the perturbations initially are adiabatic. 
If this is the case, non--Gaussianity in the isocurvature perturbations 
are expected as well. 
We refer the reader to Refs.~\cite{LW2,LUW,Guptaetal} for more details. 
Here we will just consider the 
simplest setting of the curvaton scenario where only adiabatic 
perturbations are left 
after the curvaton decay.} From this epoch onwards the ``standard'' 
radiation dominated  
phase takes place.     

\subsection{Generating the curvature perturbation at linear order}

During inflation the curvaton field $\sigma$ is supposed to give a 
negligible contribution to the energy density and to be an almost free 
scalar field,  with a small effective mass 
$m^2_\sigma=|\partial^2 V/\partial \sigma^2| \ll H_I^2$ \cite{LW2,LUW}, where
$H_I=\dot{a}/a$ is the Hubble rate during inflation.

The unperturbed curvaton field satisfies the equation of motion
\be
\label{backg}
\sigma''+2 \H \sigma'+a^2 \frac{\partial V}{\partial \sigma}=0\, .
\ee
It is also usually assumed that the curvaton field is very weakly coupled to 
the scalar fields driving inflation and that the curvature perturbation 
from the inflaton fluctuations is negligible
\cite{LW2,LUW}. 
Notice that these are just the conditions under which we worked 
in Sec.~\ref{QFqDeSitter} when calculating the spectrum of perturbations 
generated by the quantum fluctuations of a generic light scalar field during 
inflation.
Thus, if we expand the curvaton field  up to first order 
in the 
perturbations around the homogeneous background as 
\be
\sigma(\tau,\bf x) = \sigma(\tau)+\delta^{(1)}\sigma\, ,  
\ee
the linear perturbations satisfy on large scales the equation
\be
\label{sigma1}
\delta^{(1)} \sigma ''+2 \H \delta^{(1)} \sigma' +a^2 
\frac{\partial^2 V}{\partial \sigma^2}\, \delta^{(1)} \sigma=0\, .
\ee
The fluctuations 
$\delta \sigma$ on superhorizon scales  will be Gaussian 
distributed and, from the results of Sec.~\ref{QFqDeSitter}, they will have 
a nearly scale--invariant spectrum -- see Eq.~(\ref{bb}) --
\be
\mathcal{P}_{\delta\sigma} (k) \approx \frac{H_*^2}{4\pi^2} 
\label{pinf}
\,,
\ee
where the subscript $*$ denotes the epoch of horizon exit $k=aH$.     
Once inflation is over the inflaton energy density will be converted into 
radiation ($\gamma$) and the curvaton field will remain approximately 
constant until $H^2 \sim m_\sigma^2$. At this epoch the curvaton field begins 
to oscillate around the minimum of its potential which can be safely 
approximated by the quadratic term $V \approx m_\sigma^2 \sigma^2/2$.
During this stage the energy density of the curvaton field just scales as 
non--relativistic matter $\rho_\sigma \propto a^{-3}$~\cite{Turner}.
The energy density in the oscillating field is
\be
\label{energyoscill}
\rho_\sigma(\tau,{\bf x}) \approx
m_\sigma^2 \sigma^2(\tau,{\bf x})
\,,
\label{rhosigma}
\ee
and it can be expanded into a homogeneous background 
$\rho_\sigma(\tau)$ and a first--order perturbation $\delta^{(1)} 
\rho_\sigma$ as 
\be
\label{rhocurv1}
\rho_\sigma(\tau,{\bf x})=\rho_\sigma(\tau)+\delta^{(1)} 
\rho_\sigma(\tau,{\bf x})=
m_\sigma^2 \sigma+2 m_\sigma^2\, \sigma \, \delta^{(1)} \sigma
\, . 
\ee
As it follows from Eqs.~(\ref{backg}) and (\ref{sigma1}) for a 
quadratic potential the ratio $\delta^{(1)} \sigma/\sigma$ remains 
constant and     
the resulting relative energy density perturbation is
\be
\label{relrhocurv}
\frac{\delta^{(1)}\rho_\sigma}{\rho_\sigma}=2 \left(
\frac{\delta^{(1)} \sigma}{\sigma} \right)_* \, ,
\ee
where the $*$ stands for the value at  horizon crossing.

Perturbations in the energy density of the 
curvaton field produce in fact a primordial density 
perturbation well after the end of inflation.     
The primordial adiabatic density perturbation is associated with a 
perturbation in the spatial curvature $\psi$ and it is characterized 
in a gauge--invariant manner by the curvature perturbation $\zeta$ on 
hypersurfaces of uniform total density $\rho$, introduced in 
Sec.~\ref{def:zeta}. 
At linear order $\zeta$ is 
defined by Eq.~(\ref{zeta1}) and on large scales its evolution is 
sourced by the non--adiabatic pressure perturbation 
$\delta^{(1)} P_{\rm nad}=\delta^{(1)} P - c_s^2\delta^{(1)}\rho$, 
obeying the equation of motion~(\ref{MAIN1}).  
In the curvaton scenario the curvature perturbation is generated well after 
the end of inflation during the oscillations of the curvaton field because 
the pressure of the mixture of matter (curvaton) and radiation produced by 
the inflaton decay is not adiabatic. A convenient way to study this 
mechanism is 
to consider the curvature perturbations $\zeta_i$ associated with each 
individual energy density components defined in Eq.~(\ref{zetai}).
In fact the weighted sum in Eq.~(\ref{zetatot1}) 
during the oscillations of the curvaton field can be written as 
\cite{lw,LUW}  
\be
\label{zetasum}
\zeta^{(1)}=(1-f)\zeta^{(1)}_\gamma+f\zeta^{(1)}_\sigma \, ,
\ee
with the quantity $f$  
defining the relative contribution of the curvaton 
field to the total curvature perturbation is now given by   
\begin{equation}
\label{deff}
f = \frac{3\rho_\sigma}{4\rho_\gamma +3\rho_\sigma}\, . 
\end{equation}
according to Eq.~(\ref{def:f}).

From now on we shall work under the {\emph approximation of  
sudden decay} of the curvaton field. 
Under this approximation the curvaton and 
the radiation components $\rho_\sigma$ and $\rho_\gamma$ satisfy 
separately the energy conservation equations  
\bea
\label{conseqs}
\rho_\gamma'=-4 \H \rho_\gamma\, ,\nonumber \\
\rho_\sigma'=-3 \H \rho_\sigma \, ,
\eea
and the curvature perturbations $\zeta^{(1)}_i$ remains constant 
on superhorizon scales until 
the decay of the curvaton, as it follows from 
Eqs.~(\ref{zeta1phi'})--(\ref{zeta1g'}) in the limit $a \Gamma/\H \ll 1$.

 Therefore from Eq.~(\ref{zetasum}) it follows 
that the first-oder curvature perturbation evolves on large scales as
\be
\label{eqzetasudden}
\zeta^{(1)'}=f'(\zeta^{(1)}_\sigma-\zeta^{(1)}_\gamma)
=\H f(1-f)(\zeta^{(1)}_\sigma-\zeta^{(1)}_\gamma)\, ,
\ee 
where we have used the conservation of the curvature perturbations. 
By comparison with Eq.~(\ref{MAIN1}) one obtains the expression for the
non--adiabatic pressure perturbation at first order~\cite{LW2,LUW}
\be
\label{pressurepert}
\delta^{(1)} P_{\rm nad}=\rho_\sigma (1-f)(\zeta^{(1)}_\gamma-
\zeta^{(1)}_\sigma)\, .
\ee

Since in the curvaton scenario 
it is supposed that the curvature perturbation 
in the radiation produced at the end of inflation is negligible 
\be
\label{zetaradiation}
\zeta^{(1)}_\gamma=-\hat{\psi}^{(1)}+\frac{1}{4}\frac{\delta^{(1)} \rho_\gamma}
{\rho_\gamma}=0 \, .
\ee
Similarly the value of $\zeta^{(1)}_\sigma$ is fixed by 
the fluctuations of  the curvaton during inflation
\be
\label{zetasigma}
\zeta^{(1)}_\sigma=-\hat{\psi}^{(1)}+\frac{1}{3}
\frac{\delta^{(1)}\rho_\sigma}{\rho_\sigma}= 
\zeta^{(1)}_{\sigma I}\, ,
\ee 
where $I$ stands for the value of the 
fluctuations during inflation.  
From Eq.~(\ref{zetasum}) the total curvature perturbation 
during the curvaton oscillations is given by  
\be
\label{zetaoscill}
\zeta^{(1)}=f \zeta^{(1)}_\sigma \, . 
\ee
As it is clear from Eq.~(\ref{zetaoscill}) initially, 
when the curvaton energy density is subdominant, the 
density perturbation in the curvaton field $\zeta^{(1)}_\sigma$ gives a 
negligible contribution to the total curvature perturbation, 
thus corresponding to an isocurvature (or entropy) perturbation. 
On the other hand during the oscillations $\rho_\sigma \propto a^{-3}$ 
increases with respect to the energy density of radiation 
$\rho_\gamma\propto a^{-4}$, and the perturbations in the curvaton field 
are then converted into the curvature perturbation.     
Well after the decay of the curvaton, 
during the conventional radiation and 
matter dominated eras, the total curvature perturbation  
will remain constant on superhorizon scales at a value which, 
in the sudden--decay approximation, is fixed by Eq.~(\ref{zetaoscill}) at 
the epoch of curvaton decay
\be
\label{atcurvdecay}
\zeta^{(1)}=f_D\, \zeta^{(1)}_\sigma \, ,
\ee
where $D$ stands for the epoch of the curvaton decay.

Going beyond the sudden--decay approximation it is possible to introduce a 
transfer parameter $r$ defined as~\cite{LUW,wmu}
\be
\label{rnuova}
\zeta^{(1)}=r\zeta^{(1)}_\sigma \, ,
\ee  
where $\zeta^{(1)}$ is evaluated well after the epoch of the curvaton 
decay and $\zeta^{(1)}_\sigma$ is evaluated well before this epoch.
The numerical study of the coupled perturbation equations has been performed 
in Ref.~\cite{wmu} showing that the sudden--decay approximation is exact when 
the curvaton dominates the energy density before it decays $(r=1)$, while 
in the opposite case   
\be
r\approx \left( \frac{\rho_\sigma}{\rho} \right)_D.
\ee

\subsection{Second--order curvature perturbation from the curvaton 
fluctuations}

As we have shown in Sec.~\ref{standardInfl} in the standard scenario
where the generation of cosmological perturbations is induced by 
fluctuations of a single inflaton field (and there is no curvaton) 
the evolution of the perturbations is purely adiabatic, and the total 
curvature perturbation $\zeta^{(2)}$ is indeed conserved. 
Thus, following~Ref.~\cite{BMR2}, we have used   
the conserved quantity $\zeta^{(2)}$      
to follow the evolution  on large scales of 
the primordial non--linearity in the cosmological perturbations from a period
of  
inflation to the matter dominated era. On the contrary in 
the curvaton and inhomogeneous reheating scenarios the total 
curvature perturbation $\zeta^{(2)}$
evolves on large scales due to a non--adiabatic pressure. 
In the present scenario the conversion of the 
curvaton isocurvature perturbations into a final curvature perturbation 
at the epoch of the curvaton decay can be followed through the sum 
(\ref{zetasum}) of the 
individual curvature perturbations weighted by the ratio $f$ of 
Eq.~(\ref{deff}).  

Let us now extend such a result at second order in the perturbations.
Since the quantities $\zeta^{(1)}_i$ and $\zeta^{(2)}_i$  
are gauge--invariant, we choose to 
work in the spatially flat gauge $\psi^{(r)}=\chi^{(r)}=0$ 
if not otherwise specified.
Note that from Eqs.~(\ref{relrhocurv}) and (\ref{zetasigma})
the value of $\zeta^{(1)}_\sigma$ is thus given by
\be
\label{zetasigmainfl}
\zeta^{(1)}_\sigma=\frac{1}{3} \frac{\delta^{(1)}\rho_\sigma}{\rho_\sigma}=
\frac{2}{3} \frac{\delta^{(1)} \sigma}{\sigma}=
\frac{2}{3} \left( \frac{\delta^{(1)} \sigma}{\sigma} \right)_* \, ,
\ee 
where we have used the fact that $\zeta^{(1)}_\sigma$ 
(or equivalently $\delta^{(1)} \sigma/ \sigma$) remains constant, while
from Eq.~(\ref{zetaradiation}) in the spatially flat gauge
\be
\label{zetaroscill}
\zeta^{(1)}_\gamma=\frac{1}{4}
\frac{\delta^{(1)} \rho_\gamma}{\rho_\gamma} \, .
\ee
During the oscillations of the scalar field Eq.~(\ref{main1})
with $\delta^{(1)} \Gamma =0$ reduces to 
\bea
\label{main}
\zeta^{(2)}&=&f \zeta^{(2)}_\sigma+(1-f) \zeta^{(2)}_\gamma
+f(1-f)(1+f)\left( \zeta^{(1)}_\sigma-\zeta^{(1)}_\gamma\right)^2\, ,
\eea 
where we have used the sudden--decay limit $a \Gamma/\H \ll1$ and within such 
an approximation $f$ is given by Eq.~(\ref{deff}). Similarly from 
Eqs~(\ref{zeta2phiflat})--(\ref{zeta2gflat}) the expression of the 
individual curvature perturbations in the spatially flat-gauge now read 

\bea
\label{zetass}
\zeta^{(2)}_\sigma&=&\frac{1}{3}\frac{\delta^{(2)} \rho_\sigma}{\rho_\sigma}-
\left( \zeta^{(1)}_\sigma \right)^2 \, , \\ 
\label{zetagammass}
\zeta^{(2)}_\gamma&=&\frac{1}{4} \frac{\delta^{(2)} 
\rho_\gamma}{\rho_\gamma}- 2
\left( \zeta^{(1)}_\gamma\right)^2\, .
\eea    
Such quantities are gauge--invariant and, in the sudden--decay approximation 
they are separately conserved until the curvaton decay.
 
Therefore from Eq.~(\ref{main}) it follows that $\zeta^{(2)}$ evolves 
according to Eq.~\cite{BMR3}
\be
\label{evolzeta2}
\zeta^{(2)'}=f'\left( \zeta^{(2)}_\sigma-\zeta^{(2)}_\gamma\right)
+f' (1-3f^2) \left( \zeta^{(1)}_\sigma-\zeta^{(1)}_\gamma\right)^2\, .
\ee
Note that Eq.~(\ref{evolzeta2}) can be rewritten as Eq.~(\ref{MAIN2}) 
derived in Ref.~\cite{mw}  
with  $\delta^{(1)} P_{\rm nad}$ given by Eq.~(\ref{pressurepert}) and
\bea
\delta^{(2)}P_{\rm nad} &=&\rho_{\sigma}(1-f)
\Big[ \left( \zeta^{(2)}_\gamma-\zeta^{(2)}_\sigma \right) 
+ (f^2+6f-1) \nonumber \\
&\times& \left( \zeta^{(1)}_\sigma-\zeta^{(1)}_\gamma \right )^2 
+ 4\zeta^{(1)}_\gamma \left( \zeta^{(1)}_\sigma-\zeta^{(1)}_\gamma
\right) \Big]\, ,
\eea
is the gauge--invariant non--adiabatic pressure perturbation 
on uniform density hypersurfaces on large scales which, as one can 
easily check, coincides with the generic expression in Eq.~(\ref{Pnad2}) 
which has been provided in Ref.~\cite{mw}.

The second--order curvature perturbation in the standard 
radiation or matter eras remain constant on superhorizon scales and, 
in the sudden--decay approximation, it is thus given by 
the quantity in Eq.~(\ref{main}) evaluated at the epoch of the curvaton 
decay
\be
\label{zeta2decay}
\zeta^{(2)}=f_D \zeta^{(2)}_\sigma+f_D \left( 1-f^2_D \right) \left( 
\zeta^{(1)}_\sigma \right)^2 \, ,
\ee 
where we have used the curvaton hypothesis that the curvature perturbation
in the radiation produced at the end of inflation is negligible so that 
$\zeta^{(1)}_\gamma\approx 0$ and 
$\zeta^{(2)}_\gamma\approx 0$.
The curvature perturbation $\zeta^{(1)}_\sigma$ is given 
by Eq.~(\ref{zetasigmainfl}), while
$\zeta^{(2)}_\sigma$ in Eq.~(\ref{zetass}) is obtained by   
expanding the energy density of the curvaton field, Eq.~(\ref{energyoscill}),
up to second order in the curvaton fluctuations
\bea
\rho_\sigma(\bfx,t)&=&\rho_\sigma(\tau)+
\delta^{(1)}\rho_\sigma(\tau, x^i)+\frac{1}{2}\delta^{(2)}
\rho_\sigma(\tau, x^i)\nonumber \\
&=&m_\sigma^2 \sigma+
2 m_\sigma^2\, \sigma \, \delta^{(1)} \sigma+m^2_\sigma \left( 
\delta^{(1)} \sigma \right)^2\, .
\eea
It follows that
\be
\frac{\delta^{(2)} \rho_\sigma}{\rho_\sigma}=\frac{1}{2} 
\left( \frac{\delta^{(1)}\rho_\sigma}{\rho_\sigma} \right)^2
=\frac{9}{2} \left( \zeta^{(1)}_\sigma \right)^2\, ,
\ee
where we have used Eq.~(\ref{zetasigmainfl}), and hence from 
Eq.~(\ref{zetass}) we obtain
\be
\label{zeta2infl}
\zeta^{(2)}_\sigma =\frac{1}{2} \left(\zeta^{(1)}_\sigma \right)^2=
\frac{1}{2} \left(\zeta^{(1)}_\sigma \right)_I^2\, ,
\ee
where we have emphasized that also $\zeta^{(2)}_\sigma$ is a conserved quantity
whose value is determined by the curvaton fluctuations during inflation. 
Plugging Eq.~(\ref{zeta2infl}) into Eq.~(\ref{zeta2decay}) the 
curvature perturbation during the standard radiation or matter dominated 
eras turns out to be~\cite{BMR3}
\be
\label{zetafinal}
\zeta^{(2)}=f_D \left( \frac{3}{2}-f_D^2 \right) 
\left( \zeta^{(1)}_\sigma \right)^2\, .
\ee

From now on we switch from the 
spatially flat gauge $\psi=\chi=0$ 
to the Poisson gauge defined in Sec.~\ref{metrictensor}.
Such a procedure is possible since the curvature perturbations 
$\zeta^{(2)}_i$ 
are gauge--invariant quantities. In particular this is evident from 
the expression found in Eq.~(\ref{zetafinal}).
In fact we  are interested in the non--linearities produced in the 
gravitational potentials in the Poisson gauge.  
By doing so we are in the position to obtain a relation 
between the gravitational 
potential $\psi^{(2)}$ and the energy density $\delta^{(2)} \rho$ in the 
radiation/matter dominated epochs.

From Eq.~(\ref{qqq}) we find that 
during the matter dominated era 
\bea
\label{zetam}
\zeta^{(2)}&=&-\psi^{(2)}+\frac{1}{3} \frac{\delta^{(2)}\rho}{\rho}+
\frac{5}{9} \left( \frac{\delta^{(1)}\rho}{\rho} \right)^2 \nonumber\\
&=&-\psi^{(2)}+\frac{1}{3} \frac{\delta^{(2)}\rho}{\rho}+\frac{20}{9}
\left( \psi^{(1)} \right)^2\, ,
\eea
where in  the last step we have used the first--order 
solution~(\ref{00Plargescales}) on 
large scales in the Poisson gauge. 
On the other hand the curvature perturbation in the 
radiation/matter dominated 
eras remains constant at a value which is fixed by  Eq.~(\ref{zetafinal}).
Thus Eq.~(\ref{zetam}) combined with Eq.~(\ref{zetafinal}),
yields~\cite{BMR3}
\be
\label{relpsirho}
\psi^{(2)}-\frac{1}{3} \frac{\delta^{(2)} \rho}{\rho}=\frac{1}{9}
\left[ 20-\frac{75}{2 f_D}+25 f_D \right] \left( \psi^{(1)} \right)^2\, ,
\ee
where we have used 
\be
f_D \zeta^{(1)}_\sigma=-\frac{5}{3} \psi^{(1)}
\ee
from Eq.~(\ref{atcurvdecay}) 
and the usual linear relation between the curvature
perturbation and the gravitational potential 
$\zeta^{(1)}=-\frac{5}{3} \psi^{(1)}$
during the matter dominated era, see Eq.~(\ref{relPsizeta}).

\section{\bf The inhomogeneous reheating scenario: $\delta \Gamma \neq 0$}

Recently, another mechanism for the generation of cosmological
perturbations has been 
proposed~\cite{gamma1,gamma2,gamma3,MatRio,allahverdi}.  
It acts during the reheating
stage after inflation and it was dubbed the ``inhomogeneous reheating'' 
mechanism in Ref.~\cite{gamma3} and ``modulated reheating'' in 
Ref.~\cite{gamma2}. This mechanism works as follows. 
As in the curvaton scenario it is supposed that the perturbations coming 
from the inflaton fluctuations are negligible.
To reheat the Universe the inflaton has to couple to ordinary particles and 
has to decay into radiation with a decay rate $\Gamma$ which depends on 
the couplings of the inflaton field.    
In the standard scenario of inflation such a coupling is  
constant. In fact it may be determined by the vacuum 
expectation value of fields $\chi$'s of the underlying theory. It could be 
the case of supersymmetric theories or theories inspired by superstrings, 
as discussed in some details in Ref.~\cite{gamma1} and~\cite{gamma2}, 
respectively, with the scalar fields $\chi's$ 
being some scalar super--partner or the so--called moduli fields. 
If those fields are light during inflation fluctuations   
$\delta\chi\sim H/2\pi$,
where $H$ is the Hubble rate during inflation, are left imprinted
on superhorizon scales, as we have recalled in Sec.~\ref{QFqDeSitter}. 
These perturbations lead to spatial 
fluctuations in the decay rate $\Gamma$ 
of the inflaton field to ordinary 
matter
\be
\frac{\delta\Gamma}{\Gamma} \sim\frac{\delta\chi}{\chi}\, ,
\ee
thus producing fluctuations in the radiation and in the 
reheating temperature in different regions of the Universe. 
These fluctuations are of isocurvature type and will be converted into 
curvature fluctuations after reheating, once the 
thermalized radiation starts do dominate
the energy density.\footnote{Indeed the idea that the
total curvature perturbation may be affected on large scales
by entropy perturbations when there exists a scalar field affecting
particle masses or couplings constants controlling the reheating
process has first been suggested in Ref.~\cite{Qreh}.}

\subsection{Generating the curvature perturbation at linear order 
from decay rate fluctuations}

Using the cosmic time as time variable, 
the first order Eq.~(\ref{zeta1phi'}) for $\zeta^{(1)}_\varphi$ 
on large scales reads 

\begin{equation}
\label{zeta1phidot}
\dot{\zeta}^{(1)}_\varphi
= \frac{\Gamma}{2} \frac{\rho_\varphi}{\dot{\rho_\varphi}} 
\frac{\dot{\rho}}{\rho} \left( \zeta^{(1)} -\zeta^{(1)}_\varphi 
\right)+ H \frac{\rho_\varphi}{\dot{\rho_\varphi}} \delta^{(1)} 
\Gamma \,. 
\end{equation}

We shall now adopt a ``mixed sudden--decay approximation''. 
We shall treat the pressureless scalar 
field and radiation fluids as if they were not interacting until 
the decay of the inflaton, when $\Gamma \approx H$. 
Since at the beginning of the 
reheating phase the energy density in radiation is negligible this means 
that $f =\dot{\rho_\varphi}/\dot{\rho} \approx 1$ 
and there is indeed only a single fluid with, from 
Eq.~(\ref{zetatot1}), $\zeta^{(1)} \approx \zeta^{(1)}_\varphi $ 
and $\zeta^{(1)}_\gamma \approx 0$. In fact under such an approximation we  
can neglect all the terms proportional to the decay rate 
$\Gamma$, \emph{but} we allow for the spatial fluctuations 
of the decay rate. Thus  
the first order Eq.~(\ref{zeta1phidot}) reads
\be 
\label{zeta1phiapprox}
\dot{\zeta}^{(1)}_\varphi \simeq -\frac{1}{3} \delta^{(1)} \Gamma\, ,
\ee
where we have used $\dot{\rho_\varphi}=-3H\rho_\varphi$ in the 
sudden--decay approximation.
Integration over time yields
\be
\label{solution1}
\zeta^{(1)}_\varphi=-\frac{t}{3} \delta^{(1)}\Gamma =-\frac{2}{9} 
\frac{\delta^{(1)} \Gamma}{H}\simeq \zeta^{(1)}\, , 
\ee   
where we have used the fact that during the oscillations of the 
scalar field which dominates the energy density $H=2/3t$.
The inhomogeneous reheating mechanism produces at linear level 
a gravitational potential which after the 
reheating phase, in the radiation dominated epoch, is given by   
(in the longitudinal gauge) \cite{gamma1}
\be
\label{resultgamma}
\psi^{(1)}=\frac{1}{9} \frac{\delta^{(1)} \Gamma}{\Gamma_*}\, ,
\ee
where $\Gamma_*$ stands for the value of the background decay rate, which in 
this scenario is approximately constant, being determined by the very 
light scalar field(s) $\chi$. 
During the radiation dominated epoch the usual relation between the 
gravitational potential and the curvature perturbation in 
Eq.~(\ref{relPsizeta}) yields
\be 
\psi^{(1)}=-\frac{2}{3} \zeta^{(1)}\, ,
\ee 
and thus from Eq.~(\ref{solution1})
we can set the ratio $\Gamma_{*}/H_{D}=3/4$ 
at the time of the inflaton decay in order to reproduce the result 
~(\ref{resultgamma}) of Ref.~\cite{gamma1}. 
Therefore 
from Eq.~(\ref{solution1}) it follows that the value of $\zeta^{(1)}$
is~\cite{BMR4}
\be
\label{zetagamma}
\zeta^{(1)} \simeq -\frac{1}{6} \frac{\delta^{(1)} \Gamma}{\Gamma_*}\, .
\ee

\subsection{Second--order curvature perturbation 
from inhomogeneous reheating}

We now expand the decay rate as 
\be
\label{Gamma2}
\Gamma=\Gamma_{*}+\delta \Gamma=\Gamma_{*}+\delta^{(1)}\Gamma+\frac{1}{2}
\delta^{(2)}\Gamma\, ,
\ee
and perturbing the energy transfer coefficient 
$\hat{Q}_\varphi=-\Gamma \rho_\varphi$ up to second order  
it follows from Eqs.~(\ref{Q2phi}) and~(\ref{Gamma2})
\be
\label{Q2gamma}
\delta^{(2)}Q_\varphi=-\rho_\varphi \delta^{(2)} \Gamma -
\Gamma_{*} \delta^{(2)}\rho_\varphi -2\delta^{(1)}\Gamma 
\delta^{(1)}\rho_\varphi\, .
\ee
Plugging Eq.~(\ref{Q2gamma}) into Eq.~(\ref{eq:zeta2phi}), 
the equation of motion on large scales for the curvature perturbation 
$\zeta^{(2)}_\varphi$ allowing for  possible fluctuations of the decay rate 
$\delta^{(1)} \Gamma$ and $\delta^{(2)} \Gamma$ turns out to be~\cite{BMR4}
\bea
\label{zeta2phidot}
\dot{\zeta}^{(2)}_\varphi&=& 
\frac{H}{\dot{\rho_\varphi}} \left( \delta^{(2)} \Gamma \rho_\varphi
+2\delta^{(1)} \Gamma \delta^{(1)} \rho_\varphi \right)
-\frac{\Gamma_* \rho_\varphi}{2} \frac{\dot{\rho}}{\rho} 
\left(\frac{\delta^{(2)}\rho_\varphi}{\dot{\rho_\varphi}}
-\frac{\delta^{(2)}\rho}{\dot{\rho}} \right) \nonumber \\
&+& 3\Gamma_* \rho_\varphi \frac{H}{\dot{\rho_\varphi}} {\phi^{(1)}}^2
+ \frac{2H}{\dot{\rho_\varphi}} \left( \delta^{(1)}\Gamma \rho_\varphi 
+\delta^{(1)} \rho_\varphi \Gamma_* \right) \phi^{(1)}-2\zeta^{(1)}_\varphi
\dot{\zeta}^{(1)}_\varphi\nonumber \\
&+& 2\left[ \zeta^{(1)}_\varphi \left(\Gamma_* 
\frac{\rho_\varphi}{\dot{\rho_\varphi}}\phi^{(1)}+\delta^{(1)}\Gamma 
\frac{\rho_\varphi}{\dot{\rho_\varphi}} + 
\Gamma_* \frac{\delta^{(1)} \rho_\varphi}{\dot{\rho}_\varphi} 
\right) \right]^{\displaystyle{\cdot}}
+ \left[ {\zeta^{(1)}_\varphi}^2 \frac{\Gamma}{H} \left(1-
\frac{\rho_\varphi}{\dot{\rho_\varphi}} \frac{\dot{\rho}}{\rho} \right) 
\right]^{\displaystyle{\cdot}}\, , \nonumber \\
&&
\eea
where we have used the 
fact that the  decay rate $\Gamma$ in the scenario under consideration 
remains constant. We shall use the result previously found in 
Eq.~(\ref{zetagamma}) to solve this equation. 
In fact under the sudden--decay approximation and using Eq.~(\ref{00first})  
the second--order Eq.~(\ref{zeta2phidot}) simplifies to 
\be
\label{eq2approx}
\dot{\zeta}^{(2)}_\varphi\simeq -\frac{1}{3}\delta^{(2)} \Gamma
-\zeta^{(1)}_\varphi\delta^{(1)} 
\Gamma -2\left(\zeta^{(1)}_\varphi \dot{\zeta}^{(1)}_\varphi
\right)-\frac{2}{3} \left(  \frac{\delta^{(1)} 
\Gamma}{H}\zeta^{(1)}_\varphi \right)^{\cdot}\, .
\ee  
Notice that the 
fluctuations $\delta \Gamma=\delta^{(1)}\Gamma+\frac{1}{2}\delta^{(2)}
\Gamma$ indeed depend on the underlying physics for the coupling of the 
inflaton field to the other scalar field(s) $\chi$. Let us take for example
$\Gamma(t,{\bf x}) \propto \chi^2(t,{\bf x})$. If the scalar field $\chi$ is
very light, its homogeneous value can be treated as constant 
$\chi(t)\approx \chi_*$ and 
during inflation quantum fluctuations $\delta^{(1)} \chi$ 
around its homogeneous value $\chi_{*}$ are left imprinted on 
superhorizon scales. Therefore non--linear fluctuations $\left( 
\delta^{(1)} \chi \right)^2$ of the decay rate $\Gamma$ are produced as well
\be
\label{expandchi}
\Gamma(t,\bfx) \propto \chi^2(t,\bfx)=\chi^2_{*}+2\chi_* \delta^{(1)} \chi+
\left( \delta^{(1)} \chi \right)^2\, .
\ee
From Eqs.~(\ref{Gamma2}) and~(\ref{expandchi}) it follows
\bea
\frac{\delta^{(1)}\Gamma}{\Gamma_*}&=&2\frac{\delta^{(1)}\chi}{\chi_{*}}\, ,
\nonumber \\
\label{gammanonlin}
\frac{\delta^{(2)}\Gamma}{\Gamma_*}&=&2\left( 
\frac{\delta^{(1)}\chi}{\chi_{*}} \right)^2=\frac{1}{2} 
\left(\frac{\delta^{(1)}\Gamma}{\Gamma_*} \right)^2\, .
\eea
Using the first order solution 
$\zeta^{(1)}_\varphi=- (t/3) \delta^{(1)}\Gamma$ and 
Eq.~(\ref{gammanonlin}) in Eq.~(\ref{eq2approx}),
the evolution of $\zeta^{(2)}_\varphi$ on large scales is
\bea
\dot{\zeta}^{(2)}_\varphi &\simeq& -\frac{1}{6\Gamma_*}
\left( \delta^{(1)} \Gamma \right)^2
+\frac{1}{3}\left( \delta^{(1)} \Gamma \right)^2 t
-2\left(\zeta^{(1)}_\varphi \dot{\zeta}^{(1)}_\varphi
\right) 
-\frac{2}{3} \left(  \frac{\delta^{(1)} 
\Gamma}{H}\zeta^{(1)}_\varphi \right)^{\displaystyle{\cdot}}\, .
\eea 
Integration over time is straightforward and yields
\begin{eqnarray}
\label{zeta2expr}
\zeta^{(2)}_\varphi&=&-\frac{t}{6} \frac{\left( \delta^{(1)} \Gamma \right)^2}
{\Gamma_*}
+\frac{1}{6} \left( \delta^{(1)} \Gamma \right)^2 t^2
-\left({\zeta^{(1)}_\varphi}\right)^2-\frac{2}{3} \zeta^{(1)}_\varphi
\frac{\delta^{(1)}\Gamma}{H}\, .
\end{eqnarray}
Now recall that at the time of inflaton decay 
$\Gamma_*/H_{D}=3/4$, and since 
$H=2/3\,t$, it follows $t_{D}=1/ 2 \Gamma_*$.
Thus $\zeta^{(2)}_\varphi$ 
in Eq.~(\ref{zeta2expr}) evaluated at the time $t_D$ of inflaton decay is  
\be
\zeta^{(2)}_\varphi \simeq -\frac{1}{24}\left(
\frac{\delta^{(1)} \Gamma}{\Gamma_*}\right)^2-{\zeta^{(1)}_\varphi}^2-
\frac{1}{2}\zeta^{(1)}_\varphi \frac{\delta^{(1)} \Gamma}{\Gamma_*}\, .       
\ee 
Finally, using Eq.~(\ref{zetagamma}), we find
that the total curvature 
perturbation $\zeta^{(2)}$ in the sudden--decay approximation 
is given by~\cite{BMR4}  
\be
\label{risfin}
\zeta^{(2)} \simeq \zeta^{(2)}_\varphi\simeq \frac{1}{2} 
\left( \zeta^{(1)} \right)^2\, .
\ee
Eq.~(\ref{risfin}) gives the value at which the curvature perturbation  
remains constant   during the radiation and dominated phases.
Notice that our results are gauge--invariant, involving the curvature
 perturbations, as it is clear for example from Eq.~(\ref{risfin}).
Thus, as in the previous section, we can now 
switch from the spatially flat gauge to the 
Poisson gauge, to obtain a relation analogous to the ones in
Eq.~(\ref{relPsirhoInfl}) and Eq.~(\ref{relpsirho}) 
between the energy density perturbation 
$\delta^{(2)} \rho$ and the gravitational potential $\psi^{(2)}$ 
during the matter dominated epoch. 
By combining Eq.~(\ref{risfin}) with the expression~(\ref{zetam}) 
for the curvature perturbation $\zeta^{(2)}$ in the Poisson gauge 
during the matter dominated phase we find 
\be
\label{relpsirhoinhreh}
\psi^{(2)}-\frac{1}{3}\frac{\delta^{(2)}\rho}{\rho}=\frac{5}{6}
\left( \psi^{(1)} \right)^2\, .
\ee

\section{\bf Non--linearities in the gravitational potential}
\label{Phinonline}

Let us now 
focus on the calculation of the non--linearity in the gravitational potential 
$\phi=\phi^{(1)}+\frac{1}{2}\phi^{(2)}$ (or $\psi$) in the Poisson gauge.  
In fact with our results we can express the gravitational potential 
$\phi$ in momentum space as 
\begin{eqnarray}
\label{phimomspace}
\phi({\bf k}) &=& \phi^{(1)}({\bf k})+ 
\frac{1}{(2\pi)^3}
\int\, d^3 k_1\,d^3 k_2\, \delta^{(3)}\left({\bf k}_1+{\bf k}_2-{\bf k}\right)
\nonumber\\
&\times&
f^\phi_{\rm NL}\left({\bf k}_1,{\bf k}_2\right)
\phi^{(1)}({\bf k}_1)\phi^{(1)}({\bf k}_2)\, ,
\end{eqnarray}
where we have defined an effective ``momentum--dependent'' non--linearity 
parameter $f^\phi_{\rm NL}$. Here the linear lapse function
$\phi^{(1)}=\psi^{(1)}$ is a Gaussian random field. 
Notice that indeed a momentum--dependent function 
must be added to the R.H.S. of Eq.~(\ref{phimomspace}) in order to satisfy 
the requirement that $\langle \phi\rangle =0$.
From Eq.~(\ref{phimomspace}) it follows that the gravitational potential 
bispectrum reads
\begin{eqnarray}
\langle \phi({\bf k}_1) \phi({\bf k}_2) \phi({\bf k}_3)
\rangle&=&(2\pi)^3\,\delta^{(3)}\left({\bf k}_1+{\bf k}_2+{\bf k}_3\right)
\nonumber\\
&\times& \left[2\,f^\phi_{\rm NL}
\left({\bf k}_1,{\bf k}_2\right)\,
{P}_\phi(k_1){P}_\phi(k_2)+{\rm cyclic}\right]\, ,
\end{eqnarray}
where ${P}_\phi(k)$ is related to the dimensionless
power--spectrum of the gravitational potential as defined in Eq.~(\ref{def:PS})
by ${P}_\phi(k)={\mathcal P}_{\phi}(k)\, 2 \pi^2/k^3$.  

We want to make here an important remark. The non--linearity parameter 
$f_{\rm NL}^{\phi}$ defines  the non--Gaussianity in the gravitational 
potential, but it does not define the non--Gaussianity level of the CMB 
temperature fluctuations. In order to predict such an observable it is
necessary to make a further step, and determine how the perturbations 
in the gravitational potentials translate into second--order fluctuations 
of the CMB temperature. We will carry out this calculation in the next 
section. We now give the expression for the non--linearities in the 
gravitational potential $\phi$. In each of the scenarios considered, solving 
the evolution for the curvature perturbation $\zeta^{(2)}$ for 
each of the scenarios considered, we 
have obtained the relations~(\ref{relPsirhoInfl}), (\ref{relpsirho}), 
and~(\ref{relpsirhoinhreh}) 
between the gravitational potential $\psi^{(2)}$  
and the energy density $\delta^{(2)} \rho$  
in terms of the linear gravitational potential squared 
$\left( \phi^{(1)} \right)^2$. We can now close our system and fully 
determine the variables $\psi^{(2)}$, $\phi^{(2)}$ and $\delta^{(2)} \rho$ 
by using the (0-0)-component of Einstein equation (\ref{00Poisson2}) 
and the constraint (\ref{constraint}) 
relating the gravitational potentials $\phi^{(2)}$ and $\psi^{(2)}$.

\subsection{The standard scenario}

Combining Eq.~(\ref{relPsirhoInfl}) obtained from the conservation of 
$\zeta^{(2)}$ with Eqs.~(\ref{00Poisson2}) and~(\ref{constraint}) we 
single out an equation for the gravitational potential $\phi^{(2)}$ on 
large scales
\bea
\label{new}
\phi^{(2)\prime}+\frac{5+3 w}{2} \H \phi^{(2)} &=& (5+3 w) \H  \left( 
\psi^{(1)} \right)^2 +\frac{3}{2} \H (1+w) 
\left[
\nabla^{-2} \left(2 \partial^i \psi^{(1)} \partial_i
\psi^{(1)} \right. \right.\nonumber \\ 
&+& \left. \left. 3 \left(1+w\right) {\cal H}^2 v_{(1)}^i 
v_{(1)i} \right)
- 3 \nabla^{-4} \partial_i \partial^j 
\left(2 \partial^i \psi^{(1)} \partial_j
\psi^{(1)} \right. \right.\nonumber \\ 
&+& 3 \left. \left. \left(1+w\right) {\cal H}^2 v_{(1)}^i 
v_{(1)j} \right)\right]    
+ \frac{3}{2} \H (1+w) \int^{\tau}_{\tau_I} \S(\tau') d\tau' -\S_1^{\prime}
\nonumber \\
&+&\frac{1}{\H} 
\left( \nabla\psi^{(1)} \right)^2
+ \frac{8}{3\H} \psi^{(1)} \left( \nabla^2 \psi^{(1)} \right) 
+ \frac{\nabla ^2 \S_1}{3\H}+ 
\frac{1}{\H} \left( \psi^{(1)\prime} \right)^2 \, ,
\nonumber \\
\eea
where $\S_1$ denotes the R.H.S. of Eq~(\ref{constraint}). 

We want to integrate this equation from $\tau_I$ to a time $\tau$ in 
the matter--dominated epoch. The general solution is given by the solution of 
the homogeneous equation plus a particular solution
\bea
\phi^{(2)}& = &
\phi^{(2)}(\tau_I) \exp\left[ - \int_{\tau_I}^\tau \frac{5+3 w}{2} \H d\tau'
\right ]\nonumber \\
&+&\exp\left[-\int_{\tau_I}^\tau \frac{5+3 w}{2} \H d\tau'
\right ] \times \int_{\tau_I}^{\tau} 
\exp \left[ \int_{\tau_I}^{\tau'} \frac{5+3 w}{2} \H ds \right]  
b(\tau') d\tau'\, ,
\eea    
where $b(\tau)$ stands for the source term in the R.H.S of Eq.~(\ref{new}).

Notice that the homogeneous solution during both the radiation and the 
matter--dominated epoch decreases in time. 
Therefore we can neglect the homogeneous solution and 
focus on the contributions from the source term $b(\tau)$.
At a time $\tau$ in the matter--dominated epoch 
$ \exp [ - \int_{\tau_I}^\tau d\tau'\, \H\, (5+3 w)/2] 
\propto \tau^{-5}$.
Thus if we are interested in the 
gravitational potential 
$\phi^{(2)}$ during the matter dominated epoch the contributions in the 
particular solution coming from the 
radiation--dominated epoch can be considered negligible.   
Recalling that during the matter--dominated epoch the linear gravitational 
potential $\psi^{(1)}$ is constant in time, it turns out that 
\bea
\label{expr:phi2}
\phi^{(2)}& \simeq & 2 \left( \psi^{(1)} \right)^2+\frac{3}{5} 
\left[ \nabla^{-2} \left( \frac{10}{3} \partial^i \psi^{(1)} \partial_i
\psi^{(1)} \right) 
-3 \nabla^{-4} \partial_i \partial^j 
\left(\frac{10}{3}  \partial^i \psi^{(1)} \partial_j
\psi^{(1)} \right)\right] \nonumber \\
&+& \exp\left[-\int_{\tau_I}^\tau \frac{5+3 w}{2} \H d\tau'
\right ] \times \int_{\tau_I}^{\tau} 
\exp \left[ \int_{\tau_I}^{\tau'} \frac{5+3 w}{2} \H ds \right]  
  \Bigg \{ \frac{3}{2} \H (1+w)  \Bigg.   \nonumber \\
& & \Bigg. \times  \int^{\tau'}_{\tau_I} \S(s) ds 
+  \frac{1}{\H} 
\left( \nabla\psi^{(1)} \right)^2 
+\frac{8}{3\H} \psi^{(1)} \left( \nabla^2 \psi^{(1)} \right)
+\frac{\nabla^2 \S_1}{3 \H}  \Bigg \} d \tau'\, ,
\eea 
where we have used Eq.~(\ref{0iPoisson}) to express the first--order 
velocities 
in terms of the gravitational potential, and we have taken into account 
that during the matter--dominated epoch $\S_1'=0$. 

As the gravitational potential $\psi^{(1)}$ on superhorizon scales is
generated during inflation, it is clear that the origin of the non--linearity
traces back to the inflationary quantum fluctuations.

The gravitational potential will then have a non--Gaussian 
$(\chi^2)$-component. 
Going to momentum space, from Eq.~(\ref{expr:phi2}) 
we directly read the non--linearity parameter of the gravitational potential 
$\phi=\phi^{(1)}+\frac{1}{2}\phi^{(2)}$ for scales entering 
the horizon during the matter--dominated stage~\cite{BMR2,BMR4}
\be
\label{ee1}
f^\phi_{\rm NL}({\bf k}_1,{\bf k}_2) \simeq - \frac{1}{2} +
g({\bf k}_1,{\bf k}_2)
\;,
\ee
where
\be
\label{functiong}
g({\bf k}_1,{\bf k}_2) =  4 \frac{{\bf k}_1\cdot {\bf k}_2}{k^2}
-3 \frac{\left( {\bf k}_1 \cdot {\bf k}_2 \right)^2}{k^4}+\frac{3}{2} 
\frac{k_1^4+k_2^4}{k^4}
\, ,
\ee
with ${\bf k} ={\bf k}_1 +  {\bf k}_2$. Notice that in deriving 
Eq.~(\ref{ee1}) we have neglected the contribution from the last term in 
Eq.~(\ref{expr:phi2}), since as we explain in Sec.~\ref{Remarks} this term is 
fully negligible when evaluating the bispectrum of the gravitational 
potential on large scales. Moreover in the final bispectrum expression, 
the diverging terms arising 
from the infrared behaviour of $f^\phi_{\rm NL}({\bf k}_1,{\bf k}_2)$
are automatically regularized once the monopole term is subtracted from the
definition of $\phi$, by requiring that $\langle \phi \rangle$=0. 
The non--Gaussianity provided by expression (\ref{ee}) will add to
the known Newtonian and relativistic second--order contributions which are  
relevant on subhorizon scales (a simple example being the Rees--Sciama effect
\cite{rs}), whose complete and detailed analysis has been given in 
Refs.~\cite{birk,pyne,molmat}.  

From Eq.~(\ref{ee1}) we conclude that the tiny non--Gaussianity generated 
during the inflationary epoch driven by a single scalar field, discussed in 
Sec.~\ref{NGInfl}, gets enhanced in the post--inflationary evolution giving 
rise to a non--negligible signature of large--scale non--linearity in 
the gravitational potentials. Once again, inflation provides the key 
generating mechanism to produce superhorizon seeds, which are later amplified 
by gravity.   

Finally it is interesting to note that as long as we are interested 
in the gravitational potential bispectrum on large scales, it is possible to 
obtain the same result as in Eq.~(\ref{ee1}) using the following appoximate 
solution to the (0-0)-component of Einstein equations (\ref{00Poisson2}) 
on large scales 
\be
\label{00P2largescales}
\phi^{(2)}=-\frac{1}{2}\frac{\delta^{(2)}\rho}{\rho_0}+
4\left(\psi^{(1)}\right)^2\, ,
\ee 
combining it with the equation obtained from the conservation of 
$\zeta^{(2)}$  Eq.~(\ref{relPsirhoInfl}) and with the 
constraint Eq.~(\ref{constraintbis}), both evaluated at the matter dominated 
phase, in order to close the system of equations for the variables 
$\psi^{(2)}$, $\phi^{(2)}$ and $\delta^{(2)} \rho$. The 
same holds true also for the other scenarios. Therefore, in the following we 
shall use the relations~(\ref{relpsirho}) and~(\ref{relpsirhoinhreh}), 
between the gravitational potential $\psi^{(2)}$ 
and the energy density $\delta^{(2)} \rho$ in the matter dominated epoch, 
together with Eq.~(\ref{00P2largescales}) 
and the constraint Eq.~(\ref{constraintbis}), which relates 
the gravitational potentials $\phi^{(2)}$ and $\psi^{(2)}$ 
in the matter--dominated epoch. As shown for the standard scenario, 
the terms which are neglected with such an approximation give a 
negligible contribution to the large--scale bispectrum of the gravitational 
potential. 

\subsection{The curvaton scenario}

We use Eq.~(\ref{relpsirho}) obtained from the evolution 
of $\zeta^{(2)}$ in the sudden--decay approximation with  
Eqs.~(\ref{00P2largescales}) and~(\ref{constraintbis}) and we conclude 
that in the curvaton 
scenario during the matter dominated epoch~\cite{BMR3}
\begin{eqnarray}
\label{finalcurv}
\phi^{(2)}&=&\left[\frac{10}{3}+\frac{5}{3}f_D-\frac{5}{2 f_D}\right]
\left(\psi^{(1)}\right)^2 \nonumber \\
&-&2 \nabla^{-2}\left(\psi^{(1)}\nabla^2\psi^{(1)}\right) 
+6\,\nabla^{-2}\left(\partial^i\partial_j\left(\psi^{(1)}\partial_i
\partial^j
\psi^{(1)}\right)\right)\, ,
\end{eqnarray}
where $f_D$ is given by Eq.~(\ref{deff}) at the time of the curvaton decay 
and defines the fractional energy density of the curvaton field. 
From 
Eq.~(\ref{finalcurv}) we obtain the non--linearity 
parameter for the gravitational 
potential $\phi=\phi^{(1)}+\frac{1}{2}\phi^{(2)}$~\cite{BMR3}
\begin{eqnarray}
\label{ee}
f^\phi_{\rm NL}=\left[\frac{7}{6}+\frac{5}{6}r-\frac{5}{4 r}\right]+
g({\bf k}_1,{\bf k}_2) \, ,
\end{eqnarray}
where we have replaced $f_D$ with 
$r\approx \left(\rho_\sigma /\rho \right)_D$ 
to go beyond the sudden--decay approximation, and the 
function $g({\bf k}_1,{\bf k}_2)$ 
is the same as in Eq.~(\ref{functiong}).
As far as the momentum--independent part of the non--linearity 
parameter $f_{\rm NL}^{\phi}$ is concerned, we note that
in the limit $r\ll 1$ we obtain $f^\phi_{\rm NL}=
-\frac{5}{4 r}$ which reproduces the estimate provided in~\cite{LW2,LUW},
while, in the limit $r\simeq 1$, we obtain $f^\phi_{\rm NL}=\frac{3}{4}$
for $r\simeq 1$.
\footnote{Notice that the formula (36) in Ref.~\cite{LUW}
for the estimate of the
non--linearity parameter contains a sign misprint and
should read $f^\phi_{\rm NL}\simeq -\frac{5}{4 r}$, giving
$f^\phi_{\rm NL}\simeq -\frac{5}{4}$ for $r\simeq 1$.}
Such a difference is due to the fact that we have taken into account all
the relevant second--order gravitational effects. 

An important comment is in order here. As it is evident from Eq.~(\ref{ee}) 
the level of non--Gaussianity increases for decreasing values of the 
parameter $r$, that is to say with a lower efficiency for generating the 
density perturbations. This relation between the inefficiency and 
the non--Gaussianity is in fact quite a general feature, that has been pointed 
out in Refs.~\cite{k,Zaldarriaga}. It is due to the fact that, in order to 
keep the density fluctuations at the observed level, as we decrease the 
efficiency to generate perturbations the second--order terms  
become more and more relevant in comparison with the linear contributions, 
thus increasing the level of non--Gaussianity. 

\subsection{The inhomogeneous reheating scenario}
\label{Inho}

In the inhomogeneous reheating scenario where $\Gamma \propto \chi^2$ 
by combining Eq.~(\ref{relpsirhoinhreh}) 
 with
Eqs.~(\ref{00P2largescales}) and~(\ref{constraintbis}) we find 
during the matter dominated epoch
\begin{eqnarray}
\label{final}
\phi^{(2)}&=&\frac{5}{2}
\left(\psi^{(1)}\right)^2
-2 \nabla^{-2}\left(\psi^{(1)}\nabla^2\psi^{(1)}\right) 
+6\,\nabla^{-2}\left(\partial^i\partial_j\left(\psi^{(1)}\partial_i
\partial^j
\psi^{(1)}\right)\right)\, . \nonumber \\
&&
\end{eqnarray}
We then read the non--linearity parameter for the
gravitational potential $\phi=\phi^{(1)}+\frac{1}{2}\phi^{(2)}$~\cite{BMR4}
\be
\label{ee2}
f^\phi_{\rm NL}=\frac{3}{4} + g({\bf k}_1,{\bf k}_2) \, ,
\ee
with $g({\bf k}_1,{\bf k}_2)$ defined in Eq.~(\ref{functiong}).

We would like to remind that the result in Eq.~(\ref{ee2}) has been 
obtained under certain minimal conditions for the inhomogeneous reheating to 
take place. This includes the assumption that during inflation 
$\Gamma_* \ll H$, and  that the decay rate 
is completely determined by a scalar field $\chi$ 
as $\Gamma \propto \chi^2$. However, the curvature perturbation produced in 
the inhomogeneous reheating scenario does have a dependence on the ratio 
$\Gamma_*/H$, which one can parametrize as~\cite{Zaldarriaga}
\be
\label{alphacorr}
\zeta^{(1)}=- \alpha\,  \frac{\delta^{(1)} \Gamma}{\Gamma_*}\, ,
\ee
where $\alpha$ is positive and decreases as the ratio 
$\Gamma_*/H$ at the end of inflation increases, with $\alpha=1/6$ in the limit 
$\Gamma_*/H \rightarrow 0$, thus recovering Eq.~(\ref{zetagamma}). Moreover 
the scalar field $\chi$ might set actually only a decay channel into which 
the inflaton decays; in addition the decay rate could have 
another channel which does not fluctuate, so that 
\be 
\label{gamma+gamma1}
\Gamma = \Gamma_0+\Gamma_1 \left( \frac{\chi}{\chi_*} \right)^2\, ,
\ee
as considered in Refs.~\cite{gamma1,Zaldarriaga}. 
The resulting linear curvature perturbation reads 
\be
\label{zetaInhvar}
\zeta^{(1)}=- 2 \alpha\, \frac{\Gamma_1}{\Gamma_*} 
\frac{\delta^{(1)} \chi}{\chi_*}\, .
\ee
As argued in Ref.~\cite{Zaldarriaga}, Eq.~(\ref{zetaInhvar}) shows that the
efficiency to generate the density perturbations can actually be 
very small either when the decay rate is not much smaller than the Hubble rate 
during inflation, or because the scalar field $\chi$ controls only one of the 
channels in which the inflaton field decays. Therefore, according to the 
previous considerations one expects that due 
to these effects the level of non--Gaussianity can be in fact higher than in 
Eq.~(\ref{ee2}). Moreover, as discussed in Ref.~\cite{Zaldarriaga} 
the decay rate could depend by several scalar fields and one 
should also account for the possible presence of intrinsic 
non--Gaussianities in the scalar field(s) $\chi$,  
produced by self--interactions of the type described 
in Sec.~\ref{3pointfunctions}. In
our formalims this means that in expanding $\Gamma(t,{\bf x})$ as in 
Eq.~(\ref{expandchi}) there might be an additional non--linear term given 
by $\delta^{(2)} \chi$ sourced by the self--couplings of the scalar field 
$\chi$. As shown in Ref.~\cite{Zaldarriaga} all 
these ``variations on the 
theme'' should increase the non--Gaussianity at a level very 
close to the limits set by {\sl WMAP}.
\footnote{Notice however that the analysis in 
Ref.~\cite{Zaldarriaga} focuses on the curvature perturbation $\zeta$ 
and does not take into account all the second--order 
effects which contribute to the level of the non--Gaussianity in the 
gravitational potential and in the CMB anisotropies which is actually the 
observable quantity.}  

In fact following the same steps which lead to Eq.~(\ref{ee2}) we are able 
to extend the result to the more general case which includes the 
dependence on the $\alpha$ parameter and a decay rate as in 
Eq.~(\ref{gamma+gamma1}), while keeping track of the 
different second--order effects arising in the determination of the 
non--linearity 
parameter of the gravitational potential. Using Eq.~(\ref{alphacorr}) 
and~(\ref{gamma+gamma1}) we find
\be
\label{ee2bis}
f^\phi_{\rm NL}=\frac{3}{4}+I+g({\bf k}_1,{\bf k}_2) \, ,
\ee   
where
\be
\label{defI}
I=-\frac{5}{2}+\frac{5}{12} \frac{\bar{\Gamma}}{\alpha \Gamma_1}\, ,
\ee
$\bar{\Gamma}$ being the mean value of the decay rate. Thus the 
``minimal case''~(\ref{ee2}) is 
recovered for $\alpha=1/6$ and $\Gamma_1=\bar{\Gamma}$ ($I=0$).

\subsubsection{Some remarks on the large--scale limit}
\label{Remarks}

Let us now clarify here again our procedure 
in deriving the expression of second--order 
quantities or equations of motion. Indeed, when dealing with second--order 
perturbations which are expressed in terms of first--order quantities, 
also the short--wavelength behaviour of the first--order 
perturbations must be taken into account, as it becomes evident going to 
momentum space. The crucial point here is which is the final 
quantity one is interested in.
We are interested in calculating the bispectrum of 
the gravitational potential and of the temperature anisotropies on 
large scales as a measure of the non--Gaussianity of the 
cosmological perturbations on those scales.  
The bispectrum of these quantities is twice the kernel which appears when 
expressing these second--order quantities in Fourier space, 
that is to say, e.g., 
$f^\phi_{\rm NL}\left({\bf k}_1,{\bf k}_2\right)$ in Eq.~(\ref{phimomspace}). 
This means that, when calculating the bispectrum, we can evaluate the kernel 
in the long--wavelength limit, irrespective of the integration over the 
whole range of momenta. This is the reason why we have used the 
long--wavelength approximation in the equations of motion   
when deriving the expressions of second--order 
quantities in terms of first--order perturbations.
Thus, the final result for 
the bispectrum is not affected by our procedure. 

\subsection{Second--order temperature  fluctuations on large scales and
the correct definition of the measured $f_{\rm NL}$}
\label{DT/T}

In this subsection we provide the expression for the
second--order temperature fluctuations on large scales which will allow
the exact definition of the non--linearity parameter $f_{\rm NL}$.
From now on, we will adopt the {\it Poisson gauge} defined in 
Sec.~\ref{metrictensor}. 

The second--order expression for the temperature fluctuation
field in the Poisson gauge has been obtained in Ref.~\cite{molmat},
by implementing the general formalism introduced in Ref.~\cite{pyne}.
We are interested here in the large--scale limit of that
expression, which allows to unambiguously define the primordial
non--Gaussian contribution.
Keeping only the large--scale limit of the linear
and second--order terms in Eqs.~(2.27) and~(2.28) of Ref.~\cite{molmat},
(ses also Eqs.~{4.11) and~(4.12) of Ref.~\cite{pyne}), we obtain
\begin{equation}
\label{complete}
\frac{\Delta T}{T} = \phi^{(1)}_{\mathcal E} + \tau^{(1)}_{\mathcal E} +
\frac{1}{2}\left(\phi^{(2)}_{\mathcal E} + \tau^{(2)}_{\mathcal E}\right) -
\frac{1}{2} \left(\phi^{(1)}_{\mathcal E}\right)^2 +
\phi^{(1)}_{\mathcal E} \tau^{(1)}_{\mathcal E}
\;,
\end{equation}
where $\phi_{\mathcal E} = \phi^{(1)}_{\mathcal E} +
\frac{1}{2}\phi^{(2)}_{\mathcal E}$ is the lapse perturbations at emission 
on the last scattering surface and 
$\tau_{\mathcal E} = 
\tau^{(1)}_{\mathcal E} + \frac{1}{2} \tau^{(2)}_{\mathcal E}$
is the intrinsic fractional temperature fluctuation at emission 
\be
\tau_{\mathcal E} \equiv \left. \frac{\Delta T}{T}\right|_{\mathcal E}\, .
\ee
Let us recall that, at linear order $\phi^{(1)}=\psi^{(1)}$.
In Eq.~(\ref{complete}) we dropped all those
terms which represent  {\it integrated} contributions such as
Integrated Sachs--Wolfe, Rees--Sciama and more complicated second--order 
integrated effects~\cite{rs1,rs2,rs3,rs4,rs5}. A full account of these  
effects is indeed provided by the general expressions for non--linear 
temperature anisotropies given in Refs.~\cite{birk,pyne,molmat} and will not 
be reported here. The form of the CMB temperature bispectrum arising from 
some non--linear effects has been inferred recently in Ref.~\cite{cz} 
(see also Ref.~\cite{babich}), for a particular triangle configuration.   
Notice that for a $\Lambda {\rm CDM}$ cosmology the 
Integrated Sachs--Wolfe effect would 
also give a contribution on large scales. In order to compute this effect 
one should study the complete evolution of the gravitational 
potentials from last scattering till $\Lambda$ (dark energy) domination.

It is important here to stress that the non--linearity parameter 
$f_{\rm NL}$ as introduced e.g. in Refs.~\cite{ks,komatsu:phd}
singles out the large--scale part of the second--order CMB anisotropies.
One should be able to distinguish secondary integrated terms from the
large--scale effects thanks to their specific angular--scale dependence. 
For the very same reason, we disregarded gravitational--lensing and
Shapiro time--delay effects, Doppler terms and all those second--order 
effects which are characterized by a high--$\ell$ harmonic content.
We finally dropped contributions at the observer position,
which only modify the {\it monopole} term. 

To obtain the intrinsic anisotropy in the photon temperature, we
can expand the photon energy density $\rho_\gamma \propto T^4$
up to second order and write
\begin{equation}
\tau^{(1)}_{\mathcal E} = \frac{1}{4} \frac{\delta^{(1)}
\rho_\gamma}{\rho_{\gamma}}\bigg\vert_{\mathcal E}
\end{equation}
where $\rho_{\gamma}$ is the mean photon energy density, and
\begin{equation}
\tau^{(2)}_{\mathcal E} = \frac{1}{4} \frac{\delta^{(2)}
\rho_\gamma}{\rho_{\gamma}}\bigg\vert_{\mathcal E} -
3\left( \tau^{(1)}_{\mathcal E} \right)^2=
\frac{1}{4} \frac{\delta^{(2)}
\rho_\gamma}{\rho_{\gamma}}\bigg\vert_{\mathcal E} -
\frac{3}{16}
\left( \frac{\delta^{(1)}
\rho_\gamma}{\rho_{\gamma}}\bigg\vert_{\mathcal E}\right)^2
\;.
\end{equation}
Next, we need to relate the photon energy density fluctuation to the
lapse perturbation, which we can easily do by implementing the 
adiabaticity condition up to second order. At first order
the adiabaticity condition reads $\zeta^{(1)}_m=\zeta^{(1)}_\gamma$
and we obtain
\begin{equation}
\frac{\delta^{(1)} \rho_\gamma}{\rho_{\gamma}} = \frac{4}{3}
\frac{\delta^{(1)} \rho_m}{\rho_{m}}\, ,
\end{equation}
where $\rho_m$ is the average energy density of the matter component.
At second order the adiabaticity condition
imposes $\zeta^{(2)}_m=\zeta^{(2)}_\gamma$, as explained in Sec.~\ref{AI2nd}.
From Eq.~(\ref{zeta2singole}) applied to matter and radiation we
find

\begin{equation}
\frac{\delta^{(2)} \rho_\gamma}{\rho_{\gamma}} = \frac{4}{3}
\frac{\delta^{(2)} \rho_m}{\rho_{m}}+
\frac{4}{9}
\left(\frac{\delta^{(1)} \rho_m}{\rho_{m}}\right)^2 \;.
\end{equation}
In the large--scale limit, the energy constraints~(\ref{00Plargescales}) 
and~(\ref{00P2largescales}) in the matter dominated era, yields
\begin{equation}
\frac{\delta^{(1)} \rho_m}{\rho_{m}} = -2 \psi^{(1)}
\end{equation}
and
\begin{equation}
\frac{\delta^{(2)} \rho_m}{\rho_{m}} = -2 \phi^{(2)} +
8 \left( \psi^{(1)} \right)^2 \;.
\end{equation}
We finally obtain the fundamental relation
\begin{equation}
\label{deltaT/T}
\frac{\Delta T}{T} = \frac{1}{3}\left[ \psi^{(1)}_{\mathcal E} +
\frac{1}{2}\left(\phi^{(2)}_{\mathcal E} -
\frac{5}{3} \left(\psi^{(1)}_{\mathcal E} \right)^2 \right) \right]\;.
\end{equation}

From Eq.~(\ref{deltaT/T}), it is clear that the expression for the
second--order temperature fluctuations is {\it not} a simple
extension of  the first--order Sachs--Wolfe effect $\Delta T^{(1)}/T
= \psi^{(1)}_{\mathcal E}/3$ to second order since it receives
a correction provided by the term 
$-(5/3) \left(\psi^{(1)}_{\mathcal E} \right)^2$.

We can express the lapse function at second order as in 
Eq.~(\ref{phimomspace}),  
or equivalently as a general convolution (see, e.g., Ref.~\cite{noi}) 

\begin{equation}
\label{lll}
\phi = \phi^{(1)} + \frac{1}{2}\phi^{(2)} = \psi^{(1)}
+ f_{\rm NL}^\phi * \left(\psi^{(1)}\right)^2 \;,
\end{equation}
up to a constant offset.
In order to connect the inflationary predictions with the definition of 
$f_{\rm NL}$ which has become standard in the CMB-related literature 
(see, e.g., Ref.~\cite{ks}) we remind the standard Sachs--Wolfe formula
\be
\label{maber}
\frac{\Delta T}{T}({\bf {\hat n}},\tau_0) = - \frac{1}{3} 
\Phi({\bf {\hat n}}(\tau_0 - \tau_{\mathcal E})) \;, 
\ee
where $\Phi \equiv - \phi$ is Bardeen's potential 
\cite{Bardeen1}, which is conventionally expanded in a form 
analogous to Eq.~(\ref{lll}), namely 
\begin{equation}
\label{lll2}
\Phi = \Phi_{\rm L} + f_{\rm NL} * \left(\Phi_{\rm L}\right)^2 \;
\end{equation}
(up to a constant offset, which only affects the temperature monopole term),
where $\Phi_{\rm L} = - \phi^{(1)}$.

\subsubsection{Angular averaging in a perturbed Universe and 
the value of $f_{\rm NL}$}

One more non--linear effect that one should take into account
is provided by the angular averaging implicit in the definition
of observables such as the harmonic amplitudes of the CMB temperature
as defined by an observer. Restricting ourselves to the pure
Sachs--Wolfe effect, this amounts to performing an angular average
with the physical (perturbed) metric on null hypersurfaces 
at fixed radial distance from the receiver. One can easily show that
only the first--order correction to the metric gives a contribution
to second--order quantities like the bispectrum and the effect
can be accounted for by multiplying the angular differential element
$d\Omega$ by a conformal factor $(1-2\psi^{(1)})$. 
This operation implies shifting the value of 
\begin{equation}
\label{shift}
f_{\rm NL}^\phi\to
(f_{\rm NL}^\phi-1)\, .
\end{equation}
It is interesting at this point to consider the particular 
``squeezed'' configuration considered in Ref.~\cite{Maldacena},  
which consists in taking one of the wavenumbers to be much smaller 
than the other two in the bispectrum $\langle\phi({\bf k}_1)
\phi({\bf k}_2)\phi({\bf k}_3)\rangle$, for instance $k_3\ll k_1,k_2$. 
It is immediate to verify that in such a limit, 
and taking into account the form of the function (\ref{functiong}), 
the bispectrum vanishes in the case of a scale--invariant power-spectrum
within the standard scenario, where
cosmological perturbations are due to the inflaton field. This
is in full agreement with the general argument given in 
Ref.~\cite{Maldacena}, where it was shown that in the squeezed limit
the effect of the perturbation with the lowest momentum
is only to rescale the other momenta in the corresponding fluctuations. 
For perturbations generated during inflation, 
this amounts to saying that fluctuations leaving the horizon at much earlier
times act as a classical background for the evolution of the other modes. 

From Eqs.~(\ref{shift}), (\ref{deltaT/T}) and (\ref{lll2}) 
we can now immediately derive the {\it true} non--linearity parameter 
$f_{\rm NL}$ which is the quantity actually measurable by high-resolution CMB
experiments, after properly subtracting instrumental noise, foreground
contributions and small--scale second--order terms. We find
\begin{equation}
\label{ppp}
f_{\rm NL} = - f_{\rm NL}^\phi + \frac{5}{6} +1=
- f_{\rm NL}^\phi + \frac{11}{6}
\;.
\end{equation}
We warn the reader that this is the quantity which enters in the
determination of higher--order statistics (such as the bispectrum of the 
temperature anisotropies) and to which the phenomenological study 
performed e.g. in Ref.~\cite{ks} applies. 
A number of present and future CMB experiments, such as
{\sl WMAP} \cite{k} and {\sl Planck}, have enough resolution to either 
constrain or detect non--Gaussianity of CMB anisotropy data parametrized 
by $f_{\rm NL}$ with high precision~\cite{ks}. 

Notice that the CMB temperature bispectrum does not vanish in the so--called 
squeezed limit discussed in Ref.~\cite{Maldacena}, owing to the presence of 
second--order Sachs--Wolfe--like terms which give the extra term $5/6$ in 
Eq.~(\ref{ppp}). This statement applies to all scenarios of generation of 
cosmological perturbations and contrasts with the results in Ref.~\cite{cz}
(in the limit in which the two calculations can be compared), where the CMB 
bispectrum has been inferred directly from that of the gravitational potential
(so that second--order Sachs--Wolfe--like terms were not included) and there 
is no matching to the primordial non--Gaussianity.

In Refs.~\cite{BMR2,BMR4} the matching among different
cosmological eras has been obtained using the gauge--invariant
curvature perturbation $\zeta^{(2)}$ defined in Eq.~(\ref{qqq}).
If one goes to the uniform energy--density gauge defined
by $\delta\rho=0$ at any order, one recovers (at second order) the
Salopek--Bond curvature perturbation defined in Ref.~\cite{Salopek1} 
$\zeta_{\rm SB}$ through the metric 
$ds^2=a^2(\tau)\left[-d\tau^2+ e^{2\zeta_{\rm SB}}
d{\bf x}^2\right]$. Indeed, expanding $\zeta_{\rm SB}$ as
$\zeta_{\rm SB}=\zeta^{(1)}_{\rm SB}+\frac{1}{2}\zeta^{(2)}_{\rm SB}$
and comparing to the metric (\ref{metric2}), one immediately
finds (on super--horizon scales) 
\begin{eqnarray}
\zeta^{(1)} & = &-\left.\psi^{(1)}\right|_{\rho}=
\zeta^{(1)}_{\rm SB}\, ,\nonumber\\
\zeta^{(2)} &=&-\left.\psi^{(2)}\right|_{\rho}=
\zeta^{(2)}_{\rm SB}+2\left(
\zeta^{(1)}_{\rm SB}\right)^2\, .
\end{eqnarray}
The extra--term $2\left(\zeta^{(1)}_{\rm SB}\right)^2$ beatifully
matches the last term in the R.H.S. of Eq.~(\ref{four}) immediately
explaining why $\zeta^{(2)}_{\rm SB}={\mathcal O}\left(\epsilon,\eta\right)$
during inflation,
as found in Ref.~\cite{Maldacena}. Notice, however, that
the computation of the second--order temperature anisotropy in any gauge
requires the use of the full gauge--invariant quantity $\zeta^{(2)}$
in order to  properly account for terms proportional to  
$\left(\psi^{(1)}\right)^2$. This step seems to be missing in 
Ref.~\cite{cz}, where both the uniform energy--density
and the longitudinal gauges have been used. 

After showing how the large scales perturbations in the gravitational 
potentials produce corresponding fluctuations in the CMB temperatures, 
we are finally in the position to give the predictions for the level of the 
non--Gaussianity in the three scenarios considered so far.

\vskip1cm
{\bf A. Standard scenario}

Using Eq.~(\ref{ppp}) and Eq.~(\ref{ee1}), 
we conclude that in the standard scenario where
cosmological perturbations are generated by the inflaton field,
the value of the non--Gaussianity parameter is provided by
\be
\label{FINALE1}
f_{\rm NL}({\bf k}_1,{\bf k}_2) \simeq  \frac{7}{3} -
g({\bf k}_1,{\bf k}_2)
\,.
\ee

\vskip 1cm

{\bf  B. Curvaton scenario}

From expression (\ref{ppp}) and Eq.~(\ref{ee}), 
we find that the level of non--Gaussianity
in the curvaton scenario is given by
\begin{equation}
\label{FINALE2}
f_{\rm NL}= - \left[-\frac{2}{3}+\frac{5}{6}r-\frac{5}{4 r}\right]
- g({\bf k}_1,{\bf k}_2) \, ,
\end{equation}
where we recall that $r \approx (\rho_\sigma / \rho)_{decay}$ is 
the ratio of the 
curvaton energy density to the total energu density at the curvaton decay.

\vskip 1cm

{\bf  C. Inhomogeneous reheating scenario}

Using the expression~(\ref{ppp}) and Eq.~(\ref{ee2}), 
we find that the level of non--Gaussianity
in the inhomogeneous reheating scenario where $\Gamma\propto
\chi^2$ is provided by 

\be
\label{FINALE3}
f_{\rm NL}= \frac{13}{12} - g({\bf k}_1,{\bf k}_2) \, .
\ee
As explained in Sec.~\ref{Inho} we can relax some conditions and 
obtain an extension of Eq.~(\ref{FINALE3}) for a decay rate $\Gamma$ which is 
only partially controlled by a scalar field $\chi$ as in 
Eq.~(\ref{gamma+gamma1}).
Using (\ref{ppp}) and Eq.~(\ref{ee2bis}) we find  
\be
f_{\rm NL}= \frac{13}{12} - I - g({\bf k}_1,{\bf k}_2)\, ,
\ee
where $I=-5/2+(5/12) \, \bar{\Gamma}/(\alpha \Gamma_1)$, with the parameter 
$0 < \alpha <1/6$ and $\bar{\Gamma}/\Gamma_1 >1$.

\subsubsection{A comment on primordial non--Gaussianity and the 
post--inflationary evolution}

The expressions for the non--linearity parameter $f_{\rm NL}$ obtained 
in the previous section are the results of three physical processes. 
The first one is the generation of an intrinsic non--Gaussianity during a 
primordial epoch and is strictly dependent on the particular mechanism which 
gives rise to the cosmological perturbations. 
This contribution sets the initial conditions 
for the evolution of the second--order perturbations in the 
radiation--dominated epoch obtained using the conservation of the curvature 
fluctuation $\zeta^{(2)}$, namely 
Eqs.~(\ref{relPsirhoInfl}),~(\ref{relpsirho}) and~(\ref{relpsirhoinhreh}).
The initial contribution is then processed by 
the second--order gravitational dynamics in 
the post--inflationary evolution given by Eq.~(\ref{00P2largescales}) 
and the constraint Eq.~(\ref{constraintbis}). Finally, the non-linearities 
thus produced in the gravitational potential are transferred 
to the temperature 
anisotropies on large scales, where new second--order corrections arise (see 
Eq.~(\ref{deltaT/T})). 
In fact, it has been shown in Ref.~\cite{BMR5} that the initial 
contribution from the primordial epoch can be neatly disentangled 
from the other contributions coming from the post--inflationary evolution. 
The key point here is that 
the  gauge--invariant comoving curvature perturbation 
$\zeta^{(2)}$ remains {\it constant} on super--horizon scales after 
it has been generated and possible isocurvature perturbations are 
no longer present. Therefore, $\zeta^{(2)}$ 
provides us with all the necessary information about the
``primordial'' level of non--Gaussianity 
generated either during inflation, as in the standard scenario, 
or immediately after inflation, as in the curvaton scenario. 
Different scenarios are characterized by different values of 
$\zeta^{(2)}$, while the post--inflationary non--linear evolution induced by
gravity is common to all scenarios~\cite{BMR2,BMR3,BMR4}.~\footnote{
Once the initial conditions are set, one uses the same 
equations~(\ref{00P2largescales}),~~(\ref{constraintbis}) 
and~(\ref{deltaT/T}).}
For example, in standard single-field inflation $\zeta^{(2)}$ is 
generated during inflation and its value is given by $\zeta^{(2)}=2\left( 
\zeta^{(1)} \right)^2+{\mathcal O}
\left(n_\zeta-1\right) $~\cite{noi,BMR2} (as it can be seen also from 
Eqs.~(\ref{four}) and~(\ref{fivenuova})).    
Notice that such a disentanglement can be performed unambiguously only by 
expressing the temperature anisotropies in a gauge--invariant way and 
identifying the primordial content in the gauge--invariant curvature 
perturbation $\zeta^{(2)}$.~\footnote{The 
observable large--scale temperature anisotropies are then given by the 
different contributions, and in that respect we used the word ``primordial'' 
in Sec.~\ref{DT/T} to define the overall non--Gaussianity 
which survives on large scales.}      

In Ref.~\cite{BMR5} it has been shown that the general 
expression for the second--order temperature anisotropies given in  
Refs.~\cite{pyne} and ~\cite{molmat} is in fact gauge--invariant under a 
time--shift $\tau\rightarrow 
\tau-\alpha_{(1)}+\frac{1}{2} ({\alpha'_{(1)}}\alpha_{(1)}
-\alpha_{(2)})$ and that it is indeed possible to express the temperature 
anisotropies by properly defining some 
gauge--invariant metric and density perturbations. Notice that the 
gauge--invariance refers to the contributions to the temperature 
anisotropies on all scales. The definition of the gauge--invariant quantities 
proceeds by choosing the shifts $\alpha^{(r)}$ such that 
$\omega^{(r)}=0$\footnote{In fact one could easily extend such a procedure 
by including general coordinate transformations such that the expression 
of the gauge--invariant perturbations actually turns out to coincide 
with that of the corresponding quantities in the Poisson gauge.}.
For example the gauge--invariant definition of 
the gravitational potential $\phi^{(2)}$ reads
\begin{eqnarray}
\label{newgaugeinv}
\phi^{(2)}_{\rm GI}&=&\phi^{(2)}+\omega^{(1)}\left[2\left(
\psi^{(1)'}+2\frac{a'}{a}\psi^{(1)}\right)+\omega^{(1) \prime \prime} 
+ 5 \frac{a'}{a}\omega^{(1) '}+\left( {\mathcal H}'+2 {\mathcal H}^2 \right)
\omega^{(1)}\right]\nonumber \\
&+& 2\omega^{(1)'}\left(2\psi^{(1)}+\omega^{(1)'}\right)+
\frac{1}{a} \left( a\alpha^{(2)} \right)' \, , \nonumber \\
\end{eqnarray}
where 
\begin{eqnarray}
\label{newalpha2}
\alpha^{(2)}=\omega^{(2)}+\omega^{(1)}\omega^{(1)'}
+\nabla^{-2}\partial^i\left[-4\psi^{(1)}\partial_i\omega^{(1)}-2
\omega^{(1)'}\partial_i\omega^{(1)}\right] \, .
\end{eqnarray}

In terms of gauge--invariant quantities, 
the large--scale limit brings the same expression as in 
Eq.~(\ref{complete}) where now each quantity is given by the corresponding 
gauge--invariant definition~\cite{BMR5}. In the 
large--scale limit one again drops all those terms which represent integrated 
contributions and other second--order 
small--scale effects that can be distinguished from the large--scale part 
through their peculiar scale dependence. 
Taking the explicit expression of the conserved curvature perturbation 
$\zeta^{(2)}$ from Eq.~(\ref{qqq}) for a matter--dominated epoch, and using 
the $(0-0)$ component 
together with the traceless part of the $(i-j)$ Einstein equations at second 
order, one finds that on large scales the gauge--invariant expression for 
the temperature anisotropies reads~\cite{BMR5}
\begin{equation}
\label{mainnuova}
\frac{\Delta T^{(2)}_{\rm GI}}{T}=
\frac{1}{18} \left( \psi^{(1)}_{\rm GI} \right)^2 
-\frac{{\mathcal K}}{10}-\frac{1}{10} \left[ \zeta^{(2)}_{\rm GI}-
2 \left( \zeta^{(1)}_{\rm GI} \right)^2 \right]\, , 
\end{equation} 
where 
\be
\psi^{(1)}_{\rm GI}=\psi^{(1)}- {\mathcal H} \omega^{(1)}
\ee
is the gauge--invariant definition of the linear gravitational potential 
$\psi^{(1)}$, $\zeta^{(1)}_{\rm GI}$ and $\zeta^{(2)}_{\rm GI}$ are 
large--scale curvature perturbations $\zeta^{(1)}$ and $\zeta^{(2)}$ 
expressed in terms of our gauge--invariant quantities, {\it e.g.}  
$\zeta^{(1)}_{\rm GI}=-\psi^{(1)}_{\rm GI}-
{\mathcal H} (\delta^{(1)} \rho_{\rm GI}/\rho')$, 
and we have introduced a kernel  
\be
{\mathcal K}=
10 \nabla^{-4} \partial_i \partial^j 
 \left( \partial^i \psi^{(1)}_{\rm GI} \partial_j
\psi^{(1)}_{\rm GI} \right) -\nabla^{-2} 
\left( \frac{10}{3} \partial^i \psi^{(1)}_{\rm GI} 
\partial_i \psi^{(1)}_{\rm GI} \right)\, .
\ee 
Eq.~(\ref{mainnuova}) clearly shows that 
there are two contributions to the final non--linearity in the 
large--scale temperature anisotropies. 
The contribution $[\zeta^{(2)}_{\rm GI}-
2 ( \zeta^{(1)}_{\rm GI} )^2]$, 
comes from the ``primordial'' conditions set during or immediately 
after inflation. 
It is encoded in the curvature perturbation $\zeta$ which remains constant 
once it has been generated. 
The remaining part of Eq.~(\ref{mainnuova}) describes the post--inflationary
processing of the primordial non--Gaussian signal due to the non--linear 
gravitational dynamics. Thus, the expression in 
Eq.~(\ref{mainnuova}) allows to neatly disentangle the 
primordial contribution to non--Gaussianity 
from the one coming from the post--inflationary 
evolution. While the non--linear evolution after inflation is the same in 
each scenario, the primordial content will be different and depending  
on the particular mechanism generating the cosmological perturbations. 
We parametrize the primordial non--Gaussianity in terms of the 
conserved curvature perturbation (in the radiation or matter--dominated epochs)
\begin{equation}
\label{primordial}
\zeta^{(2)}=2a\left(\zeta^{(1)}\right)^2\, ,
\end{equation}    
where $a$ will depend on the physics of the given scenario. For example in the 
curvaton case $a=(3/4r)-r/2$, where 
$r \approx (\rho_\sigma/\rho)_{\rm D}$ is the relative   
curvaton contribution to the total energy density at the curvaton 
decay, as it follows from Eqs.~(\ref{rnuova}) and~(\ref{zetafinal}). 
In the minimal picture for the inhomogeneous reheating scenario
from Eq.~(\ref{risfin}) we find $a=1/4$. 
From Eq.~(\ref{mainnuova}) we can extract the non--linearity parameter 
$f_{\rm NL}$ which is usually adopted to parametrize in a phenomenological 
way the level of non--Gaussianity in the cosmological perturbations and 
has become the standard quantity to be observationally
constrained by CMB experiments~\cite{ks,k}.
Using the parametrization~(\ref{primordial}) and 
$\zeta^{(1)}=-\frac{5}{3} \psi^{(1)}_{\rm GI}$ during matter domination, 
from Eqs.~(\ref{maber}) and~(\ref{lll2}) we immediately read 
the non--linearity parameter in momentum space  
\begin{equation}
\label{f_NLnuova}
f_{\rm NL}({\bf k}_1,{\bf k}_2)
=-\left[ \frac{5}{3} \left(1-a \right) 
+\frac{1}{6}-\frac{3}{10} {\mathcal K} 
\right]+1\, 
\end{equation}
where ${\mathcal K}=10\, ({\bf k}_1 \cdot {\bf k}_3) 
({\bf k}_2 \cdot {\bf k}_3)/k^4 - (10/3) 
({\bf k}_1 \cdot {\bf k}_2)/k^2$ 
with ${\bf k}_3+{\bf k}_1+{\bf k_2}=0$ and $k=\left|
{\bf k}_3\right|$. In fact the formula
~(\ref{f_NLnuova}) already takes into account the 
additional non--linear effect 
entering in the angular three--point function of the CMB from the angular 
averaging performed with a perturbed line--element 
$(1-2 \psi^{(1)}_{\rm GI}) d \Omega$, implying a $+1$ shift 
in $f_{\rm NL}$. 
Notice that the procedure to get Eq.~(\ref{mainnuova}) and~(\ref{f_NLnuova}) 
is the same that we have 
used to compute the final values of the non--linearity parameter 
in the previous section, the only difference being that instead of  
determining the non--linearity in the gravitational potential $\phi^{(2)}$ 
and from that deducing $f_{\rm NL}$ through Eq.~(\ref{deltaT/T}) 
here we kept track of the curvature perturbation $\zeta^{(2)}$ in the final 
expression for the temperature anisotropies. 
In fact it is immediate to recover Eqs.~(\ref{FINALE1}),~(\ref{FINALE2}) 
and~(\ref{FINALE3}) taking 
into account that
\begin{equation}
{\mathcal K}=5-\frac{10}{3} g({\bf k}_1,{\bf k}_2)\, .
\end{equation}
In particular, within the standard scenario 
where the cosmological perturbations are due to the inflaton field, the 
primordial contribution to the non--Gaussianity is given by 
$a=1-\frac{1}{4} (n_{\zeta}-1)$~\cite{noi,BMR2} and the 
non--linearity parameter from inflation now reads 
\begin{equation}
f^{\rm inf}_{\rm NL}=-\frac{5}{12} (n_{\zeta}-1)  
+\frac{5}{6}+\frac{3}{10} {\mathcal K}\, . 
\end{equation} 
Therefore, the main contribution to non--Gaussianity comes from the 
post--inflationary evolution of the second--order cosmological perturbations 
which give rise to order--one coefficients, while the primordial 
contribution is proportional to 
$|n_{\zeta}-1| \ll 1$. This is true even in   
the ``squeezed'' limit first discussed by Maldacena~\cite{Maldacena}, 
where one of the wavenumbers is much smaller than the other two, 
\emph{e.g.} $k_1 \ll k_2,k_3$ and ${\mathcal K}\rightarrow 0$.

\section{\bf Other mechanisms generating non--Gaussian density perturbations}

In this section we describe in some detail some scenarios to generate 
non--Gaussianities in the observed cosmological perturbations which represent 
a plausible alternative to the mechanisms already discussed.
For these scenarios a complete analysis of the perturbation evolution and 
hence a precise determination of the level of the non--Gaussianity in the 
large--scale CMB anisotropies is still missing, but nevertheless they offer 
some general and interesting insight on the ways non--Gaussianities can be 
produced from an inflationary epoch at a higher level than predicted by 
the single--field models of slow--roll inflation. 

\subsection{Non--Gaussianities from multiple interacting scalar fields during 
inflation}
\label{multiplefields}

The standard models of inflation are based on the simple assumption that 
only the inflaton field is relevant both for the background evolution and 
for the produced density perturbations. However, 
especially on particle physics grounds, it is hard to believe that 
only one single scalar field $\varphi$ plays a role during the inflationary 
stage.
On the contrary it is quite  natural that during the inflationary dynamics 
several other scalar fields $\chi_I$ $(I=1,\dots,N)$ are present. 
The contribution to the total energy density of the extra scalar fields 
$\chi_I$ might or might not be negligible,  
compared to the one provided by the scalar field $\varphi$. 
If the latter is the case, then the model of inflation is called 
a multiple--field model~\cite{Linde1,Starobinsky,Polarstarobinsky,iso1,iso2}. 
However, as soon as 
one considers more than one scalar field, one must also consider the role of 
the isocurvature perturbations produced by the relative fluctuations of the 
scalar fields~\cite{iso1,iso2,KL} in addition to the usual adiabatic mode. 
It is well known that in such a 
framework non--Gaussian \emph{isocurvature} perturbations can be 
produced~\cite{AllenGW,MollMatOrtLucchin,Yamamotoetal,Salopek1,Salopek,YiV1,YiV2,YiV3,LM,BucherZhu,Peebles1,Peebles2,Peebles3}. 
The disadvantage of these scenarios is that 
in general the observed pattern of CMB anisotropies and the observations 
of LSS constrain the amount of isocurvature perturbations to contribute 
only a small fraction (see, e.g. Ref.~~\cite{spergel/etal:2003}). 
However, until very 
recently the adiabatic and the isocurvature perturbation modes had been 
considered as statistically independent, without taking into account that 
there can be a non--vanishing correlation between the curvature and the 
entropy modes. 
The physical origin of this correlation is actually due to the fact that the 
entropy mode on large scales can feed the adiabatic curvature perturbation 
as described by Eqs.~(\ref{MAIN1}) and~(\ref{MAIN2}). It is just such a    
cross-correlation produced during an inflationary epoch when several scalar 
fields are present that can introduce non--Gaussianity in the adiabatic mode 
too a it has been first suggested in Ref.~\cite{BMT3}. 

Before entering in some details let us here summarize the 
underlying idea of this mechanism. The starting point is the simple 
observation that it is quite natural to expect that 
the inflaton field is coupled to the extra scalar 
fields present during inflation. It has been shown that such a coupling gives 
rise to a new mechanism for generating quantum fluctuations in the scalar 
fields, which was dubbed the \emph{oscillation mechanism} 
in Ref.~\cite{BMT1}. In this case the quantum fluctuations of the scalar 
fields are not generated only because of gravitational amplification 
during the de Sitter epoch as described in Subsection~\ref{QFqDeSitter}, 
but also because --
due to the interaction terms -- the quantum fluctuations 
of a scalar field $\chi$ can 
oscillate  (evolve) into fluctuations of the scalar field $\varphi$ with 
a calculable probability, in a way similar to the 
phenomenon of neutrino oscillations. The probability of oscillation is 
resonantly amplified when the perturbations leave the horizon and the 
perturbations in the scalar field $\chi$ may disappear at horizon crossing 
giving rise to perturbations in the scalar field $\varphi$. Adiabatic 
and entropy perturbation are inevitably correlated at the end of 
inflation~\cite{BMT1}. 
The crucial observation is that -- 
since the degree of mixing is governed by the squared mass matrix of the 
scalar fields -- the oscillations can take place even if the energy 
density of the 
extra scalar fields is much smaller than the energy density of the inflaton 
field. This is an important point.  
Gaussian perturbations are usually
expected in inflationary models because the inflaton potential
is required to be very flat. This amounts to saying that the 
interaction terms in the inflaton potential are present, but
small and non--Gaussian features are suppressed since the non--linearities 
in the inflaton potential are suppressed too. On the other hand, nothing 
prevents the inflaton field from being coupled to another scalar degree 
of freedom 
whose energy density is much smaller than the one stored in the inflaton
field. This extra scalar field will not be constrained by slow--roll 
conditions and and it is natural
to expect that the self--interactions of such an extra field 
or the interaction terms with the inflaton field are sizeable, thereby
representing potential sources for non--Gaussianity. 
If during the inflationary epoch, oscillations between
the perturbation of the inflaton field and the perturbations of the 
other scalar degrees of freedom occur, the non--Gaussian features generated
in the system of the extra field are efficiently communicated to the inflaton
sector. As it has been shown in Ref.~\cite{BMT3} these 
non--Gaussianities can be left imprinted in the CMB anisotropies.

\subsubsection{The oscillation mechanism}

We now briefly describe the oscillation mechanism with an illustrative 
example. Consider two scalar fields, $\varphi$ and $\chi$ interacting 
through a generic potential $V(\varphi,\chi)$. We will dub $\varphi$
the inflaton field, even if this might be a misnomer as the two fields
might give a comparable contribution to the total energy density
of the Universe. In Fourier space the Klein--Gordon equations read 
\begin{eqnarray}
\label{m}
\delta\ddot{\varphi} + 3H\delta\dot{\varphi}
 + \frac{k^2}{a^2} \delta\varphi +  V_{\varphi\varphi}
\delta\varphi+V_{\varphi\chi}
\delta\chi&=&0  \nonumber\\
\delta\ddot{\chi} + 3H\delta\dot{\chi}
 + \frac{k^2}{a^2} \delta\chi +  V_{\chi\chi}
\delta\chi+V_{\chi\varphi}
\delta\varphi&=&0 ,
\end{eqnarray}
where we have used the notation $V_{\varphi\varphi} \equiv (\partial^2 
V/\partial\varphi\partial\varphi)$ and similarly for the other
derivatives. 

The interactions between the two scalar fields is manifest 
in that the squared mass matrix 

\be
{\mathcal M}^2=\left(
\begin{array}{cc}
V_{\varphi\varphi} & V_{\varphi\chi}\\
V_{\varphi\chi} & V_{\chi\chi}
\end{array}\right)
\ee

is in general non--diagonal. This introduces a mixing between the two 
scalar fields.
To estimate such a mixing one can diagonalize the system of 
equations (\ref{m}) by introducing 
a time--dependent $2\times 2$ 
unitary matrix ${\mathcal U}$ such that 
$ {\mathcal U}^\dagger {\mathcal M}^2 {\mathcal U}=
{\rm diag}\,(\omega_1^2,\omega_2^2)\equiv \omega^2$.
In the following we will assume that 
all  the entries of the squared mass matrix
${\mathcal M}^2$ are  real, so that 
 the unitary matrix ${\mathcal U}$ reduces to an orthogonal matrix
\be
\label{u}
{\mathcal U}=\left(\begin{array}{cc}
\cos\theta & -\sin\theta\\
\sin\theta & \cos\theta
\end{array}\right),
\ee
where $
\tan 2\theta=\frac{2\,V_{\chi\varphi}}{V_{\varphi\varphi}-V_{\chi\chi}}
$
and the mass eigenvalues are given by 
\be
\omega_{1,2}^2=\frac{1}{2}\left[\left(V_{\varphi\varphi}+
V_{\chi\chi}\right)\pm\sqrt{\left(V_{\varphi\varphi}-V_{\chi\chi}\right)^2
+4\,V_{\chi\varphi}^2}\right].
\ee

If for simplicity we work in the slow--roll approximation and assume 
that the entries of the squared mass matrix are constant in time, we
obtain for the states $\Psi=(\Psi_1,\Psi_2)^T=
{\mathcal U}^T(\varphi,\chi)^T$ 
\begin{eqnarray}
\label{pure}
\delta\wpsi_1^{\prime\prime}+\left(k^2-\frac{a^{\prime\prime}}{a}+
\omega_1^2 a^2\right)\delta\wpsi_1&=&0\, ,
\nonumber\\
\delta\wpsi_2^{\prime\prime}+\left(k^2-\frac{a^{\prime\prime}}{a}+
\omega_2^2 a^2\right)\delta\wpsi_2 &=&0\, .
\end{eqnarray}
These equations are exactly of the form of Eq.~(\ref{KGsempl}) and if we 
suppose a pure de Sitter phase the solutions 
for $\delta\wpsi_i$ are then given by Eq.~(\ref{exactm}), with 
$\nu_i^2=9/4-(\omega_i/H)^2$. The masss eigenvalues can be simply expressed 
in terms of the slow--roll parameters $\epsilon_I=(1/16 \pi G_{\rm N})\,  
\left(V_{\chi_I}/V\right)^2$ and $\eta_{IJ} = 
(1/8 \pi G_{\rm N} )\, \left(V_{\chi_I}V_{\chi_J}/V^2\right)$.  

Since at a given time $\tau$ -- the scalar perturbations
$\delta\varphi$ and $\delta\chi$ are a linear combination
of the scalar perturbations mass eigenstates $\delta\Psi_1$ and
$\delta\Psi_2$ 
\be
\label{combination}
\delta\varphi=\sum_{\ell=1,2}{\mathcal U}_{1\ell}\,\delta\Psi_\ell,\,\,\,\,
\delta\chi=\sum_{\ell=1,2}{\mathcal U}_{2\ell}\,\delta\Psi_\ell, ,
\ee
it is possible to calculate the  probability that a scalar perturbation 
$\delta\chi$ at the time
$\tau_0$ becomes a scalar perturbation in the ``inflaton'' field
$\delta\varphi $ at the time $\tau$ by computing 
$P\left[\delta\varphi(\tau_0)\rightarrow\delta\chi(\tau)\right]=
\left| \langle \delta\varphi(\tau_0) \delta\chi^*(\tau)
\rangle\right|^2$.   
As it has been shown in Ref.~\cite{BMT1} on subhorizon scales 
$k\simgt aH$  
\be
P\left[\delta\varphi(\tau_0)\rightarrow\delta\chi(\tau)\right]\simeq
0\,\,\,(k\gg aH)\, ,
\ee
but on superhorizon scales and in the limit $\omega^2_{1,2}\ll H^2$ 
the conversion probability is non--vanishing 
\be
\label{Prob}
P\left[\delta\varphi(\tau_0)\rightarrow\delta\chi(\tau)\right]\simeq 
\sin^2 2\theta\,\sin^2\left(\frac{\pi}{12}\frac{\Delta\omega^2}{H^2}
\right)\, .
\ee
Such a formula reminds   the well--known formula 
which describes the evolution
in time of the probability of oscillation between two neutrino
flavours (see, e.g. Ref.~\cite{bil}).
This result shows that at horizon crossing there is a mechanism of 
amplification for the perturbations of the scalar field $\varphi$ due to a 
conversion of the $\chi$ fluctuations into the inflaton perturbations. 
\footnote{
The phenomenon of resonant amplification is easily understood if 
one remembers that a given wavelength leaves the horizon when $k=aH$,
{\it i.e.} when $k^2=a^{\prime\prime}/a$ using the conformal time. As
long as
the wavelength is subhorizon, $k^2\gg a^{\prime\prime}/a$, the presence of
the mass terms in the equations of motion (\ref{pure}) is
completely negligible compared to the factor $(k^2-a^{\prime\prime}/a)$.
On the other hand, when
the wavelength leaves the
horizon the term $(k^2-a^{\prime\prime}/a)$ vanishes and the effect of the
mixing in the mass squared matrix is magnified, giving rise to the
resonant effect. Finally, when the wavelength is larger than the horizon,
$k^2\ll a^{\prime\prime}/a$, the term $(k^2-a^{\prime\prime}/a)$ starts
to dominate again over the mass terms and the oscillations get frozen.
} In fact such an analysis can be extended to fully account for metric 
perturbations, and to include time--dependent terms in the squared mass 
matrix~\cite{BMT1}. Still, the results are of the same form as 
in Eq.~(\ref{Prob}).

Two important remarks are in order. First of all, we wish to stress
that the oscillation mechanism operates even if the
energy of the inflaton field $\varphi$ is much larger than the energy stored in
 the other scalar field $\chi$. This is because what is crucial for
the oscillations to occur is the {\it relative} magnitude of the elements
of the squared mass matrix ${\mathcal M}^2$. Secondly, the magnitude of the
probability depends upon two quantities, $\sin^2 2\theta$ and 
$\Delta\omega^2/H^2$. Both can be readily expressed in terms
of the slow--roll parameters.
The first factor is not necessarily small, in fact it may
be even of order unity for maximal mixing. If expanded in 
terms of the slow--roll
parameters, it is ${\mathcal O}(\eta^0,\epsilon^0)$.
The second term is naturally
smaller than unity and is linear in the
slow--roll parameters. This reflects the fact that during inflation
only perturbations in scalar fields with masses smaller 
than the Hubble rate may be excited. However, $\Delta\omega^2/H^2$
is not necessarily much smaller than unity and the amplification of
the conversion probability at horizon crossing may be sizeable.

\subsubsection{Transfer of non--Gaussianities}

The oscillation mechanism is responsible for the transfer of 
non--Gaussianities from the isocurvature perturbation mode to the adiabatic 
mode. In order to see that, we can follow the elegant treatment of 
Ref.~\cite{gw} to study adiabatic and entropy perturbations 
in the case of multiple interacting scalar fields. 
The adiabatic and the entropy parts of the perturbations are expressed in 
terms of the original field fluctuations as 
\be \label{dA}
Q_A=(\cos \beta)Q_\varphi+(\sin \beta)Q_\chi \, ,
\ee
\be \label{ds}
\delta s=(\cos \beta)Q_\chi-(\sin \beta)Q_\varphi \label{s}\, ,
\ee
where
\be \label{angledef}
\cos \beta \equiv c_{\beta}=\frac{\dot{\varphi}}
{\sqrt{\dot{\varphi}^{2}+\dot{\chi}^{2}}}\, , 
\qquad  
\sin \beta \equiv s_{\beta}=\frac{\dot{\chi}}
{\sqrt{\dot{\varphi}^{2}+\dot{\chi}^{2}}}\, .
\ee

Here we have used the gauge--invariant Sasaki--Mukhanov variables 
\be
Q_I \equiv \delta \chi_I+\frac{\dot{\chi_I}}{H} \psi
\ee 
in order to take into account also the metric perturbations. For these fields
the evolution equations are like Eqs.~(\ref{m}), with the squared mass matrix 
given by ${\mathcal M}^2_{IJ}= V_{\varphi_I\varphi_J} - 1/M_{\rm P}^2 a^3\, 
\left( a^3/H\,\, \dot \varphi_I \dot \varphi_J \right)^{\cdot}\simeq
\frac{V}{M_{\rm P}^2}\left[ \eta_{IJ}-2\, (\pm \sqrt{\epsilon_{I}})
(\pm \sqrt{\epsilon_{J}})\right]$, and the mixing angle reads 
$\tan 2\theta=2\, 
{\mathcal M}^2_{\chi\varphi}/({\mathcal M}^2_{\varphi\varphi}-
{\mathcal M}^2_{\chi\chi})$. 

The cross-correlation between the adiabatic and the entropy perturbations 
is

\begin{equation}
\label{def:cross}
\langle Q_A({\bf k}) \delta s^*({\bf k}^\prime) 
\rangle \equiv {2\pi^2\over k^3}\, {\mathcal
C}_{Q_A\delta s} \, \delta^{(3)}({\bf k}-{\bf k'})\, ,
\end{equation}
in analogy with the definition~(\ref{def:PS}) for 
the power--spectrum of a given perturbation. 
Therefore, the origin of the cross-correlation is
due to a rather transparent physical behaviour in terms of the oscillation 
mechanism. During the inflationary epoch, the 
gauge invariant perturbations $Q_\varphi$ and $Q_\chi$
are generated with different wavelengths stretched by the
superluminal expansion of the scale--factor. Since the
squared mass matrix of $Q_\varphi$ and $Q_\chi$ is not diagonal, 
oscillations between the two quantities are expected.
As long as the wavelength remains subhorizon, 
$Q_\varphi$ and $Q_\chi$ evolve independently and may be 
considered good mass eigenstates. However, 
as soon as the wavelength leaves the horizon,  
an amplification in the probability of oscillation
between $Q_\varphi$ and $Q_\chi$ occurs: a non--vanishing correlation between
$Q_\varphi$ and $Q_\chi$ is created on superhorizon scales
because of the nondiagonal mass matrix
${\mathcal M}^2_{IJ}$.
Since the adiabatic and the isocurvature modes are a linear combination
of  $Q_\varphi$ and $Q_\chi$, at horizon crossing a non--vanishing
correlation between the adiabatic and the isocurvature modes is left 
imprinted in the spectrum in the form\footnote{
For simplicity we quote from Ref.~\cite{BMT1} the expression which 
neglects the time dependence of the square mass matrix of the fields 
$Q_\varphi$ and $Q_\chi$.
}
\begin{eqnarray}
\label{cor1}
a^2\langle Q_A(k) \delta s^*(k') \rangle &=&
\left(s_\beta c_\theta-c_\beta s_\theta\right)\left(
c_\beta c_\theta+s_\beta s_\theta\right)\left[
\left|Q_{\chi}\right|^2-\left|Q_{\varphi}\right|^2\right]\, .
\end{eqnarray}
In Ref.~{\cite{gw} it has been shown that the cross-correlation 
between the adiabatic and the entropy perturbations arise when the 
trajectories of the scalar fields in the background bend in the 
field space $(\varphi,\chi)$, which amounts to saying that 
$\dot{\beta} \neq 0$.
Actually such a bending is simply due to the 
interactions between the two scalar fields~\cite{BMT2,Wandsetal}. 
In particular if $\beta$ is constant no correlation is produced at the 
end of inflation. It could be the case where $\beta=0$, 
which corresponds to 
the case where the scalar field $\chi$ is approximately static, 
or when there is 
some kind of attractor solution with $\varphi \propto \chi$. Notice from 
Eqs.~(\ref{dA}),~(\ref{ds}) and~(\ref{angledef}) that in the 
former case the entropy perturbation is due enterely to the scalar 
field $\chi$ (in agreement with the results of Ref.~\cite{BMT1}).

During inflation the (comoving) curvature perturbation is given by~\cite{gw} 
\be
\label{Rmultiplefields} 
{\mathcal R}= \frac{H}{\dot{A}} Q_A, \quad , \qquad     
\dot{A}=(\cos \beta) \dot{\varphi}+(\sin \beta) \dot{\chi}\, ,  
\ee
while the entropy perturbation between the two scalar fields can be 
defined as 
\be
\label{Smultiplefields}
S_{\varphi\chi}=H \frac{\delta s}{\dot{A}}\, .
\ee 
In fact one needs to consider the evolution of the perturbations 
throughout 
the reheating stage, and after the end of inflation in order to link 
the curvature and the entropy perturbations to their corresponding 
quantities, defined in the large--scale limit deep in the radiation era, 
which are actually the quantities that can be constrained observationally. 
However as explained in Secs.~\ref{AI1st} and~\ref{AI2nd} the adiabatic 
perturbation is sourced on large scales by the entropy mode, while an entropy 
perturbation cannot be generated on large scales from an adiabatic 
perturbation. Therefore generically one can describe the 
time evolution of the curvature and entropy perturbation modes on large 
scales as
\begin{eqnarray}
\label{dR}
\dot{\R} = \alpha H \S \,, \,\,\,\,
\quad
\label{dS}
\dot{\S} = \beta H \S \,,
\end{eqnarray}
where $\alpha$ and $\beta$ are in general time--dependent
dimensionless functions.
The explicit form of the coupling between the curvature and
entropy perturbations will depend on the particular model under consideration:
it has been computed for the case of interacting scalar
fields~\cite{gw,Hwangfields,BMT2,Nibbelink,Wandsetal} and non--interacting
fluids~\cite{Hwangfluids}. In particular, as we already mentioned, 
in Ref.~\cite{gw} it has been shown 
that in the case of two scalar fields $\alpha=2\dot{\beta}/H$. 
Integrating Eqs.~(\ref{dR}) one can parametrize the evolution of 
the perturbations on large scales through some transfer functions 
$T(t_*,t)$ relating curvature and entropy perturbations 
generated when a given mode is stretched outside the 
Hubble scale during inflation ($k=aH$, denoted by an asterisk) to
curvature and entropy perturbations at some later 
time~\cite{amendola/etal:2001,Wandsetal}
\begin{equation}
\label{defT}
\left(
\begin{array}{c}
{\R} \\ {\S}
\end{array}
\right) = \left(
\begin{array}{cc}
1 & {T}_{\R\S} \\ 0 & {T}_{\S\S}
\end{array}
\right) \left(
\begin{array}{c}
\R \\\S
\end{array}
\right)_* \, ,
\end{equation}
where 
\begin{eqnarray}
 \label{TRS}
T_{\R\S}(t_*,t) &=& \int^t_{t_*} \alpha(t') T_{\S\S}(t_*,t') H(t')
dt' \,,
\nonumber\\
 \label{TSS}
T_{\S\S}(t_*,t) &=& \exp \left( \int^t_{t_*} \beta(t') H(t')
dt' \right) \,.
\end{eqnarray} 
For example, if the decay products of the reheating completely 
thermalize, then after the reheating process, in the radiation dominated era, 
$T_{\S\S}=0$. On the other hand if among the decay products a 
CDM species remains decoupled, than an isocurvature 
perturbation between this component and the photons will survive after 
inflation. The simplest possibility is that one of the scalar fields 
(or its decay products) is just identified with the CDM.

The gravitational potential $\phi$ (in the longitudinal gauge)
is indeed related to the curvature perturbation 
so that (at least at linear order) one can write for example at the 
beginning of the radiation epoch 
\be
\label{superp}
\phi \simeq \frac{2}{3} {\mathcal R}=\R_*+{T}_{\R\S} \S_*=
g_* (Q_{A*}+ {T}_{\R\S} \delta s_*) \, ,
\ee
where $g_*=(H/\dot{A})_*$ and we have used Eqs.~(\ref{Smultiplefields}) 
and~(\ref{Rmultiplefields}). 
Therefore we are now in the position to estimate, for example, 
the bispectrum of the gravitational potential. According to the 
considerations of Sec.~\ref{3pointfunctions} it is reasonable to 
consider that the dominant term in the bispectrum is given by the terms 
proportional to ${T}_{\R\S} \delta s|_*$, so that we want to estimate 
\be
\label{bisp1}
\langle {\phi}({\bf{k}}_1) {\phi}({\bf{k}}_2) {\phi}({\bf{k}}_3) 
\rangle \propto {T}^3_{\R\S} \langle \delta s_*({\bf{k}}_1)  
\delta s_*({\bf{k}}_2)  \delta s_*({\bf{k}}_3) \rangle\, .
\ee

Eq.~(\ref{bisp1}) shows that the correlation between the adiabatic and the 
entropy perturbations during inflation, 
parametrized by ${T}_{\R\S}$ actually sources the 
bispectrum of the gravitational potential (the adiabatic mode). Notice that 
this remains valid even if at the end of inflation only an adiabatic mode 
perturbation is left imprinted on very large scales deep in the radiation 
era, for example if all the decay products of the scalar fields thermalize 
after the 
reheating stage. If there is enough time during inflation for the 
conversion from isocurvature to adiabatic perturbations to occur 
then a non--Gaussian adiabatic 
perturbation mode is generated. On the other hand a 
residual isocurvature perturbation might 
survive the reheating stage, for example if the inflaton field decays into 
ordinary matter (the present day photons neutrinos and baryons) while the 
additional scalar field decays into decoupled dark matter, or it does not 
decay at all (like the case of an axion). In this case, as it has been shown 
in Ref.~\cite{BMT3}, the bispectrum of the CMB anisotropies receives 
two additional contributions, one from the intrinsic bispectrum of the 
isocurvature mode $\langle {\S}({\bf{k}}_1) 
{\S}({\bf{k}}_2) {\S}({\bf{k}}_3) 
\rangle$, and the other from the bispectrum of cross-correlation 
terms of the type 
$\langle {\phi}({\bf{k}}_1) {\phi}({\bf{k}}_2) {\S}({\bf{k}}_3) 
\rangle$, providing a characteristic signatures of these non--Gaussian 
inflationary perturbations.

In both cases one has to estimate terms like the one appearing in 
Eq.~(\ref{bisp1}) with a cubic combination of the transfer functions 
$T_{\S\S}$ and $T_{\R\S}$. 
We can do that in terms of the original scalar fields
by evaluating 
\bea
\label{bisp:originalfields}
&\widehat{T}^3& \langle \delta s_*({\bf{k}}_1)  
\delta s_*({\bf{k}}_2)  \delta s_*({\bf{k}}_3)=
\widehat{T}^3        \langle \delta s_*({\bf{k}}_1) 
\delta s_*({\bf{k}}_1)
\delta s_*({\bf{k}}_1) \rangle \nonumber\\  
&=&\widehat{T}^3 
\langle (c_\beta Q_{\chi1} - s_\beta Q_{\varphi1}) 
(c_\beta Q_{\chi2} - s_\beta Q_{\varphi2}) (c_\beta Q_{\chi3} 
- s_\beta Q_{\varphi3}) \rangle \\ \nonumber
& = & \widehat{T}^3\,  [c_\beta^3 \, \langle Q_{\chi1} Q_{\chi2} Q_{\chi3} 
\rangle - c_\beta^2 s_\beta \langle Q_{\chi1} Q_{\chi2} Q_{\varphi3} 
\rangle - c_\beta^2 s_\beta \langle Q_{\chi1} Q_{\varphi2} Q_{\chi3} \rangle 
\nonumber\\
& & + c_\beta s_\beta^2 \langle Q_{\chi1} Q_{\varphi2} Q_{\varphi3} \rangle - 
s_\beta c_\beta^2 \langle Q_{\varphi1} Q_{\chi2} Q_{\chi3} \rangle + 
s_\beta^2 c_\beta \langle Q_{\varphi1} Q_{\chi2} Q_{\varphi3} 
\rangle  \nonumber \\
& & + s_\beta^2 c_\beta \langle Q_{\varphi1} Q_{\varphi2} Q_{\chi3} 
\rangle - 
s_\beta^3 \langle Q_{\varphi1} Q_{\varphi2} Q_{\varphi3} \rangle ]\, 
\nonumber      
\eea
where, for example, $Q_{\varphi1}$ stands for 
$Q_{\varphi}({\bf{k}}_1)$ and we have used Eq.~(\ref{dA}), and for 
simplicity of notation we will omit the asterisk from 
now on. Also, we have used the notation $\widehat{T}$ to indicate that 
actually the proper analysis should be performed by extending these results to 
second order in the perturbations, and therefore $\widehat{T}$ should be 
considered as an effective transfer function accounting for the 
second--order effects.

Note that the bispectrum is a sum of different  
three--point correlation functions. The coefficients in front of each
correlation function involve  mixing angles which parametrize the amount of 
mixing between
the adiabatic and the isocurvature modes. If such mixing is sizeable,
all coefficients are of order unity and one expects that non--linearities
in the perturbation of the scalar field $\chi$ may be efficiently
transferred to the inflaton sector, thus generating large
non--Gaussian features. We shall now consider some specific examples which 
may help in understanding how such a mechanism acts during inflation. 

\subsubsection{Some worked examples}

One can  envisage different situations:

{\it i)} Inflation is driven by the inflaton field $\varphi$ and
there is another scalar field $\chi$ 
which does not interact with the inflaton and has a simple polynomial 
potential $V(\chi)\propto \chi^n$ leading to zero vacuum expectation value,
$\langle\chi\rangle =0$. 
In such a case, $\sin\beta=\sin \theta= 0$ and there 
is no mixing between the inflaton field and the $\chi$-field
as well as no cross-correlation between the adiabatic and isocurvature
modes. Non--vanishing non--Gaussianity will be present in the isocurvature
mode. This is indeed a known result~\cite{AllenGW,LM}. 
We stress here that in 
particular for a potential like  $V(\chi)\propto \chi^2$ the non--Gaussianity 
is generated in the \emph{energy density} of the scalar field $\chi$, while 
the scalar field is intrinsically Gaussian, as we discussed in  
Sec.~\ref{3pointfunctions}. Actually this is the scenario considered in 
Ref.~\cite{LM}. Notice that the same authors suggested that if 
the $\chi$--field decays late after 
inflation into the CMB photons, the non--Gaussianity in its energy density 
will then be transferred to the final adiabatic perturbations, which is 
at the basis of the curvaton mechanism for the generation of non--Gaussian 
adiabatic perturbations.

{\it ii)} Inflation is driven by two scalar fields $\varphi$ and $\chi$
with equal mass, $V=\frac{m^2}{2}(\varphi^2+\chi^2)$. In such a case the
mixing is maximal, $\beta=
\theta=\pi/4$. Nevertheless, the cross-correlation is again vanishing
\cite{gw,BMT1,BMT2} and the bispectrum gets contributions from 
adiabatic and isocurvature modes independently.
A term $\frac{\mu}{3!}\chi^3$ in the Lagrangian would be a source of 
non--Gaussianity and at the same time it would switch on a cross 
correlation between the adiabatic and the isocurvature modes, thus
 producing nonzero cross terms in the bispectrum of the type 
$\langle {\phi}({\bf{k}}_1) {\phi}({\bf{k}}_2) {\S}({\bf{k}}_3) 
\rangle$. However, 
these non--Gaussianities would be small because of the slow--roll
conditions. In fact in such a situation, since the two scalar fields have 
equal mass, their evolution mimicks a single--field 
slow--roll inflaton. 

{\it iii)} Let us now sketch a general way for the oscillation 
mechanism to be operative. Let us suppose that inflation 
is driven by an inflaton field $\varphi$ 
and there is another scalar field $\chi$ whose vacuum expectation value 
depends on the inflaton field and -- eventually -- on the Hubble parameter
$H$ and some other mass scale $\mu$, $\langle \chi\rangle=f(\varphi,H,\mu)$.
Under these circumstances, $\langle
\dot\chi\rangle=(\partial f/\partial \varphi) \,  \dot\varphi+
(\partial f/\partial H) \, \dot H$. As in illustrative case, let us 
restrict ourselves 
to the case in which  $(\partial f/\partial \varphi) \,  \dot\varphi$ is 
the dominant term and  we can approximate
$\langle\dot\chi\rangle = (\partial f/\partial \varphi) \dot\varphi$. 
We have therefore
$\tan\beta\simeq \partial f/\partial \varphi$ and $\dot\beta \simeq 
(\partial f/\partial \varphi)^\cdot / [1+ (\partial f/\partial \varphi)^2]$.
In such a case, the cross-correlation between the adiabatic and the 
isocurvature modes may be large and non--Gaussianity
may be efficiently transferred from one mode to the other.

An implementation of the transfer of non--Gaussianities from an 
isocurvature perturbation in the scalar field $\chi$ to the inflaton field 
$\varphi$ has been given in Refs.~\cite{BUzan2,BUzan3}. 
A key point to bear in mind is that some kind of coupling between the 
inflaton field and the extra scalar field is nedeed for such transfer to 
occur. On the other hand we must always require the scalar field $\chi$ 
to have an effective mass which is less than the Hubble rate during 
inflation in order for $\chi$ to develop non--negligible fluctuations. 
The two requirements seem 
to act somewhat in opposite directions~\cite{BUzan2,BUzan3}. However, a 
model where the transfer is efficient is the one with a potential of the 
form~\cite{BUzan2}
\be
\label{example1}
V(\varphi,\chi)=U(\varphi)+m^2_{\times}(\varphi-\varphi_0) 
\chi+\frac{\lambda}{n!} \chi^n
\, .
\ee
The transfer of non--linearities from the $\chi$ sector to the inflaton 
can be easily understood looking at the Klein--Gordon equation 
for the inflaton field fluctuations at second order. Such an equation 
will contain a source term deriving from the coupling, which will be of the 
form $V_{\varphi\chi} \delta^{(2)} \chi \sim m^2_{\times} \delta^{(2)} \chi$.
Therefore the scalar field $\chi$ will first develop some non--Gaussianites as 
described in 
Sec.~\ref{3pointfunctions} during inflation. Such non--Gaussianities
are of isocurvature type since the scalar field $\chi$ remains subdominant 
and for cubic self interactions 
they will be of the order $\delta^{(2)} \chi \sim 
\left( \delta^{(1)} \chi \right)^2 \lambda (t-t_k)/H$, where $t{_k}$ 
is the time of horizon crossing and $H$ is the Hubble 
parameter during inflation. Then, through the coupling 
the inflaton will acquire such non--linearities  
\be
\label{transferinKG}
\delta^{(2)} \varphi \sim \int dt \frac{m^2_{\times}}{H} \delta^{(2)} \chi 
\sim  \lambda \frac{m^2_{\times}}{H^2} \frac{N^2_k}{H^2} 
\left( \delta^{(1)} \chi \right)^2 \sim 
\lambda N^2_k  \frac{m^2_{\times}}{H^2}  \, .
\ee
The condition that the scalar field $\chi$ is subdominant and light enough  
that $\dot{\chi} < \dot{\varphi}$ implies that $N_k < \left( H/ m_{\times} 
\right)^2$ and it gives $\delta^{(2)} \chi < \lambda N_k/H^2$, 
which can be of the order of the Gaussian component $\delta^{(1)} \chi$.
In Ref.~\cite{BUzan2} it has been argued that only with 
a quartic self--coupling for the scalar field $\chi$ 
it is possible to develop non--Gaussianities on a long time scale  
without any severe fine tuning on the model.

Of course the potential in Eq.~(\ref{example1}) 
can be considered only as a toy 
model or as an effective potential at some stage of the inflationary dynamics.
In Ref.~\cite{BUzan3} some more realistic particle physics 
realizations of the transfer of non--Gaussianities have been discussed. 
An example is the hybrid--type model of inflation involving three scalar 
fields. Let us consider the potential~\cite{BUzan3}
\bea
V(\varphi,\chi,\sigma)=\frac{1}{2} m^2 \varphi^2
+\frac{\lambda}{4!} \chi^4+
\frac{\mu}{2} \left( \sigma^2-\sigma_0^2\right)^2
+\frac{g}{2} \sigma^2 \left(\varphi 
\cos \alpha+ \chi \sin \alpha \right)^2\, . \nonumber \\
\eea
Here $\varphi$ is the inflaton field, the second field $\chi$ is a light 
and subdominant scalar field with a quartic coupling and 
the third field $\sigma$ is coupled to the other two scalar fields so as to 
trigger the end of inflation by a phase transition. Here 
$\sigma_0$ is the final 
vacuum expectation value of $\sigma$ and $\alpha$ parametrizes the couplings 
of $\varphi$ and $\chi$ to $\sigma$. For large values of $\varphi$, 
the scalar field $\sigma$ is 
anchored to the origin. When $\varphi$ reaches the critical value 
$\varphi_{\rm end}$ 
for which 
the effective mass of $\sigma$ vanishes, then inflation ends and the fields 
roll down towards their true minima $\sigma= \pm \sigma_0$, $\varphi=0$ and 
$\chi=0$. The effective mass of $\sigma$ is $g (\varphi \cos \alpha+\chi 
\sin \alpha )^2-2\mu \sigma_0^2$ and the instability point is 
\be
\label{inst}
\varphi_{\rm end}=
\frac{\pm \sqrt{2 \mu/g} \sigma_0-\chi \sin \alpha}{\cos \alpha}
\, .
\ee 
In this model the transfer of non--Gaussianity proceeds as follows. 
For $\varphi > \varphi_{\rm end}$ the inflaton field $\varphi$ and the 
scalar field 
$\chi$ evolve independently. The scalar field $\varphi$ drives inflation, 
meanwhile $\chi$ develops non--Gaussianities. 
Notice in particular that $\chi$ is almost 
constant, $\dot{\chi} \approx 0$. From Eqs.~(\ref{dA}) and~(\ref{ds}) 
this means that the 
isocurvature perturbations are just given by the fluctuations of the 
subdominant scalar field $\chi$, while the fluctuations of the 
inflaton are curvature 
perturbations. In Eq.~(\ref{bisp:originalfields}) only the term
$\widehat{T}^3\,  c_\beta^3 \, \langle Q_{\chi1} Q_{\chi2} Q_{\chi3} 
\rangle$ survives. However at this stage there is no 
correlation between the two perturbations, 
the mixing angle $\beta$ defined in Eqs.~(\ref{angledef}) being constant and 
equal to zero. In fact since the transfer 
function which measures the degree of correlation $T_{\R\S}$ is proportional 
to $\dot{\beta}/H$, in Eq.~(\ref{bisp:originalfields}) the 
transfer of non--Gaussianites from $\chi$ 
to $\varphi$ is suppressed. However, because of the coupling between 
$\varphi$ and $\chi$ through their interactions to $\sigma$, near the critical 
point $\varphi_{\rm end}$ 
the trajectories in the field space $(\varphi,\chi)$ start bending, 
{\emph i.e.} $\dot{\beta} \neq 0$. 
Now in Eq.~(\ref{bisp:originalfields}) 
the transfer of non--Gaussianities from 
$\chi$ to the inflaton $\varphi$ is acting since 
$\widehat{T}^3\,  c_\beta^3 \, \langle Q_{\chi1} Q_{\chi2} Q_{\chi3} 
\rangle$ is non--vanishing and $c_\beta \neq 0$. 
In order to give an estimate of the non--Gaussianities, we recall that they 
depend on the coupling 
$\lambda$ and on the number of e--folds $N_k$, according to 
Eq.~(\ref{classical}). In fact to calculate the bispectra in 
Eq.~(\ref{bisp:originalfields}) one should first recast the scalar field 
fluctuations in terms of a combination of 
mass eigenstates $\delta \Psi_i$ with mixing angles $\theta$ as defined in 
Eq.~(\ref{combination}). This is due to the fact that the $Q_{\chi}$ 
fluctuations correspond to interaction eigenstates, and not to mass 
eigeinstate~\cite{BMT3}. In this way one can actually borrow the 
expression in Eqs.~(\ref{3pointF})--(\ref{expr:nu}). 
However notice that the term 
$\widehat{T}^3\,  c_\beta^3 \, \langle Q_{\chi1} Q_{\chi2} Q_{\chi3}\rangle$ 
will not
be given only by the expression~(\ref{3pointF})--(\ref{expr:nu}), 
but it will contain also contributions that might arise in the dynamics 
after the end of inflation, \emph{e.g.} 
depending on the details of the reheating stage. Apart from that, and 
most importantly, a precise determination of the level of 
non--Gaussianity produced by the transfer mechanism should require a full
second--order analysis of the perturbation evolution. 
Therefore in the transfer 
function $\widehat{T}^3$ we are hiding the effective 
non--Gaussianity which will follow from such an analysis. 

As noticed in Ref.~\cite{BUzan3} a general condition for this mechanism to 
work is that the terms responsible for the bending of the trajectories in 
field space must be 
different from the non--linear coupling term, otherwise 
there are not enough e--foldings for non--linearities to develop; 
moreover a kind of attractor trajectory is established which tends to suppress 
the $\chi$ fluctuations and the transfer of non--Gaussianities. In fact this 
situation corresponds to the first example we gave.
In Ref.~\cite{BUzan2,BUzan3} an analytical 
expression for the one--point Probability Distribution Function (PDF) for the 
isocurvature mode was also obtained. It reads~\cite{BUzan2,BUzan3} 
\be
\label{PDF}
P(\chi) d \chi=\sqrt{\frac{1}{2\pi} \left| 
\frac{1-\chi^2 \nu_3}{(1+\chi^2 \nu_3/3)^3} \right|} \exp \left[ 
-\frac{3 \chi^2}{(6+2\chi^2 \nu_3) \sigma^2_\chi} \right]\, .
\ee
As it can be seen from Eq.~(\ref{PDF}) the PDF depends only on two 
parameters, the variance of $\chi$ (which is proportional to $H$ 
during inflation) and the parameter $\nu_3$ which quantifies the amount of 
non--Gaussianity. Notice that this parameter just corresponds to 
the quantity in Eq.~(\ref{expr:nu}) for a quartic potential. 
In Ref.~\cite{BUzan2,BUzan3} the time evolution of 
the PDF was studied and compared with numerical results finding good 
agreement with the expression above. 
Finally, notice from Eq.~(\ref{superp}) 
that the total curvature perturbation will be given by a Gaussian 
perturbation $\R_{*}$, which  mainly corresponds to the curvature 
perturbations in the inflaton field $\varphi$, and by the non--Gaussian 
curvature perturbation $T_{\R\S} \S_{*}$,   
induced by the isocurvature mode of the scalar field $\chi$.  

\subsubsection{Some estimates of the non--linearity 
parameter $f^{\phi}_{\rm NL}$}

We can now give a general order--of--magnitude estimate 
of the non--linearity parameter $f^{\phi}_{\rm NL}$, in those inflationary 
models where the mixing between adiabatic and entropy perturbations is 
operative, as follows 
\be
\label{1estimate}
f^\phi_{\rm NL} \simeq 
\frac{\langle \phi^3 \rangle}{6 \langle \phi^2 \rangle^2} \, .
\ee 
To estimate the quantity on the R.H.S. of the last equation 
we look at the dominant contribution given in 
Eq.~(\ref{bisp1}). In fact we can extend Eq.~(\ref{superp}) to second order 
in the perturbations and parametrize the transfer of the entropy 
perturbations to the adiabatic mode ${\mathcal R}^{(2)}$ as\footnote{
By merely extending the first--order expression~(\ref{superp}) to second order 
in the perturbations, corrections of the form (first--order)$^2$ 
are not considered. However, we can account for them including them in the 
transfer functions we introduce. Moreover, these contributions 
are expected to give corrections of order unity to the 
overall non--linearity parameter, which 
on the other hand turns out to be generically much larger than unity, 
for the scenario we are considering. 
}
\be
\label{superp2}
{\mathcal R}^{(2)}={\mathcal R}^{(2)}_*+T^{(2)}_{\R \S} \, \S^{(2)}_* \, .
\ee
If we express the 
second--order entropy perturbation during inflation as 
$S^{(2)}_*=g_* \delta^{(2)} s_*$ from 
Eq.~(\ref{Smultiplefields}), then we can write 
\bea
\label{Transfer}
T^{(2)}_{\R \S} \, \S^{(2)}_* &=& g_*\,  T^{(2)}_{\R \S}\, \delta^{(2)} s_*  
\nonumber \\
&=&g_* \, {\widehat{T}}^{(2)}_{\R \S}\,  \nu_3({\delta s})\,  
\left( {\delta^{(1)} s} \right)^2\, ,  
\eea  
where $g_*=(H/\dot{A})_*$ and  
$\nu_3({\delta s})$ is the non--linearity parameter characterizing the 
non--Gaussianities in the entropy field $\delta s$, 
in a similar way to what discussed in Eqs.~(\ref{classical}) 
and~(\ref{expr:nu}). Here ${\widehat{T}}^{(2)}_{\R \S}$ should account for 
the transfer of these non--Gaussianities, additional  
second--order effects from the gravitational dynamics after inflation, and it 
should also account for second--order corrections of the form 
(first--order)$^2$. The latter contributions are expected to 
be ${\mathcal O}(1)$. On the other hand using Eq.~(\ref{superp}) we find
\be
\label{PS}
\langle \phi^2 \rangle=\frac{4}{9} g_*^2 \left[ 1
+( T^{(1)}_{\R \S} )^2 \right] \langle (\delta^{(1)} s_*)^2  \rangle  \, ,
\ee 
where we have used the fact that $\langle (\delta^{(1)} s_*)^2 \rangle 
\simeq \langle (Q^{(1)}_{A*})^2 \rangle$ and the approximation that 
the adiabatic and entropy fields are uncorrelated when they approach 
the horizon during inflation~\cite{Wandsetal}.

Thus, combining Eq.~(\ref{Transfer}) and Eq.~(\ref{PS}) we find
\be
\label{result1estimate}
f^\phi_{\rm NL} \simeq \frac{3}{2} \frac{\left( T^{(1)}_{\R \S} \right)^2
 \widehat{T}^{(2)}_{\R \S} }{\left[1+\left(T^{(1)}_{\R \S} \right)^2 
\right ]^2 } \, g^{-1}_* \, \nu_3(\delta s) \, . 
\ee
In fact the expression in Eq.~(\ref{result1estimate}) shows how the 
non--linearities 
in the entropy field are acquired by the gravitational potential.  

In order to give an order--of--magnitude estimate of $f_{\rm NL}^\phi$ 
we notice that it is possible to introduce a 
dimensionless measure of the correlation in 
terms of the correlation angle~\cite{Wandsetal}
\be
\label{cangle}
\cos \Delta \equiv \frac{\C_{\R \S}}{\P_{\R}^{1/2} \P_{\S}^{1/2}} \, ,
\ee
where $\C_{\R \S}=T^{(1)}_{\R}T^{(1)}_{\R \S} \P_{\R *}$ 
is the cross-correlation between adiabatic and entropy 
perturbations defined as in Eq.~(\ref{def:cross}), and $\P_{\R}
=[1+\left( T^{(1)}_{\R \S} \right)^2] \P_{\R *}$, $\P_{\S}=  
\left( T^{(1)}_{\S \S} \right)^2 \P_{\S *}$ are the linear adiabatic and entropy 
power--spectra~\cite{Wandsetal}. Such a correlation angle 
is indeed a measurable quantity~\cite{amendola/etal:2001,Wandsetal}.  
Moreover it 
is possible to argue on general grounds that the transfer 
$\widehat{T}^{(2)}_{\R \S}$ function can be written as  
$\widehat{T}^{(2)}_{\R \S} \simeq \alpha' N_k 
\left(T^{(1)}_{\S \S} \right)^2$, where $\alpha'$ 
is a model--dependent coefficient.
Then the non--linearity 
parameter in Eq.~(\ref{result1estimate}) reads
\be
\label{genericestimate}
f^\phi_{\rm NL} \simeq \frac{3}{2} \frac{\P_{\S}}{\P_{\R}} 
\cos^2 \Delta\,  
\alpha' N_k  \, g^{-1}_* \,  \nu_3(\delta s)\, . 
\ee
For $\nu_3(\delta s)$ we can use the expression~(\ref{expr:nu}) if we 
complitely identify the entropy field with the scalar field $\delta \chi$ 
with cubic self--interactions. Therefore we find
\be
\label{borrownu}
 g^{-1}_* \,  \nu_3(\delta s) = \sqrt{2} \sqrt{\epsilon} M_{\rm P}
\, \frac{\lambda N_k}{3 H^2}\, , 
\ee
where 
we have used the fact that $(\dot{A}/H)^2= 2 \epsilon M^2_{\rm P}$. 
In order for the generation and the 
transfer of non--Gaussianities to be effective, some constraints 
are to be satisfied. One of these requirements is that the effective 
mass of the entropy field is sufficiently small 
($< H$ during inflation), 
to generate the primordial entropy perturbations. We now turn back 
to the toy model defined by the potential in Eq.~(\ref{example1}) and, 
under some approximations, we specifically impose 
such constraints. First of all notice that we are able to  
recover an expression similar 
to Eqs.~(\ref{genericestimate})--(\ref{borrownu}) starting from the 
second--order curvature perturbation
\be
\zeta^{(2)}_{\varphi} \sim \frac{H}{\dot{\varphi}} \delta^{(2)} \varphi 
\sim \frac{H}{\dot{\varphi}} 
\lambda \frac{m^2_{\times}}{H^2} \frac{N^2_k}{3 H^2} 
\left( \delta^{(1)} \chi \right)^2\, ,
\ee
where we have used Eq.~(\ref{transferinKG}) in the case of cubic 
self--interactions for the scalar field $\chi$. If we now say that 
$\delta^{(1)} \chi \sim \delta^{(1)} \varphi$ then 
$\zeta^{(2)}_{\varphi} \sim (\dot{\varphi}/H) 
(\lambda N_k^2 m^2_{\times}/H^4) \left( \zeta^{(1)}_\varphi \right)^2$, 
from which we read the non--linearity parameter 
(up to factors of order unity)
\footnote{
We would like to warn the reader about some technical issues which
appear in deriving the formula in Eq.~(\ref{similarestimate}). 
It coincides with 
Eqs.~(\ref{genericestimate})--(\ref{borrownu}) in the limit where during 
inflation  $(\P_{\S}/\P_{\R}) 
\cos^2 \Delta \approx 1$ (the factor $m^2_{\times}/H^2$, being contained in 
the $\alpha'$ coefficient for this particulr model). In fact 
Eq.~(\ref{similarestimate}) has been derived under the approximation that 
the transfer of the entropy perturbation is extremely efficient and that the 
final curvature perturbation is essentilay determined by such a transfer. 
This is of course a limiting case. Related to that, notice that in using 
Eq.~(\ref{transferinKG}) we have assumed that the transfer is operative 
during all the $N_k$ e--folds from the time the mode leaves the horizon 
till the end of inflation. Actually the transfer could be efficient only for 
a shorter period of time. Therefore in the $\alpha'$ coefficient we should 
account for a fraction of the $N_k$ e--folds. However, as we explain later, 
our estimate of the non--linearity parameter $f^{\phi}_{\rm NL}$ is, for other 
reasons, quite a conservative one.  
}
\be
\label{similarestimate}
f^\phi_{\rm NL} \sim \frac{\dot{\varphi}}{H} \, \lambda \, 
\frac{N_k^2}{3 H^2} 
\frac{m^2_{\times}}{H^2} \sim \sqrt{2} \sqrt{\epsilon}\, M_{\rm P}\, 
\lambda \, \frac{N_k^2}{3 H^2} \frac{m^2_{\times}}{H^2}\, . 
\ee

We now relate the third derivative of the potential with respect to the 
scalar field $\chi$ $V_{\chi\chi\chi}= \lambda$ to its effective 
mass $V_{\chi \chi}$ as $V_{\chi\chi}=V_{\chi\chi\chi}\,  \chi$ so that
\be
f^\phi_{\rm NL} \sim \sqrt{2} \sqrt{\epsilon}\, 
\frac{M_{\rm P}}{H}\, 
\frac{V_{\chi \chi}}{3 H \chi} \, N_k^2  \frac{m^2_{\times}}{H^2}\, .
\ee
Notice that we expect $(\sqrt{2} \sqrt{\epsilon}\, 
M_{\rm P}/H)^{-1}$ to be 
at most of the order of the amplitude of the produced 
density perturbations $2 \pi \P^{1/2}_{\R} \sim 2.5 \times 10^{-4}$
~\cite{lrreview}. 
Therefore we conclude that 
\be
\label{estimateP}
f^\phi_{\rm NL} \sim 4 \times 10^3 \frac{V_{\chi \chi}}{3 H^2} \frac{H}{\chi}
 N_k^2\,   \frac{m^2_{\times}}{H^2}\, .
\ee
If we require that the scalar field $\chi$ is light enough,  
$V_{\chi \chi} \ll H^2$, to acquire some 
fluctuations $\delta \chi$, then we can estimate 
$ 4 \times 10^3 (V_{\chi \chi}/3 H^2) \sim 1$.  
This is actually a conservative estimate. In fact the mass 
of the scalar field $m^2_{\chi}=V_{\chi \chi}$ can be also of the order of 
the Hubble rate during inflation. The scalar field $\chi$ can be subdominant 
with respect to the (inflaton) field $\varphi$ and thus the stringent 
slow--roll conditions are widely relaxed.\footnote{In the case of
single--field slow--roll inflation the self--interactions of the 
inflaton field 
$\varphi$ produce a non--linearity parameter $f_{\rm NL} \sim \P^{-1/2}_\R 
(V_{\varphi \varphi \varphi}/H) N_k$. The slow--roll condition on the 
third derivative of the inflaton potential imposes
$\P^{-1/2}_\R  (V_{\varphi \varphi \varphi}/H) \ll 1$.}
It is also worth to notice that in the 
inhomogeneous reheating scenario, even if the scalar field(s) $\chi$ 
determining the inflaton decay rate do not need to satisfy slow--roll 
conditions, the non--Gaussianities induced by the self--interactions of $\chi$
are constrained from the requirement that 
$2/3 (V_{\chi \chi}/H^2) \sim 10^{-2}$~\cite{Zaldarriaga}, 
to satisfy the observational limits on the spectral index of density 
perturbations which can be traced back directly to the fluctuations 
$\delta \chi$ (see Eq.~(\ref{zetaInhvar})). 
This constraint does not apply to the case described in this subsection, 
since the fluctuations of the 
additional scalar field $\chi$ may give only a 
subdominat contribution to the total amplitude of the 
density perturbations. Another condition that we impose on the   
model is that $\dot{\chi} < \dot{\varphi}$ during inflation, 
in order to allow for non--Gaussianities in the $\chi$ sector to develop. 
This condition implies that $N_k < H^2/m^2_{\times}$. Therefore, with our 
conservative estimate from Eq.~(\ref{estimateP}) we find that
\be
\label{finalestimate}
f^\phi_{\rm NL} \lesssim \frac{H}{\chi} N_k \sim 60 \frac{H}{\chi}\, .
\ee
We see that if the $\chi$ field is not much larger than $H$ then significant 
non--Gaussianities can be produced very close to the limits set by {\sl WMAP}.

We conclude this section by introducing a simple parametrization for the 
non--Gaussianity generated in the gravitational potential by the transfer 
mechanism. Such a parametrization is nedeed for practical purposes when 
confronting with observations. Our results for the non--linearity 
parameter of the gravitational potential indicate that we expect the 
gravitational potential to be of the form
\be
\label{FNLadia}
\phi=\phi_1+f^\phi_{\rm NL} (\phi_2^2-\langle \phi_2^2 \rangle) 
+ {\mathcal O} (f^{\phi\,\, 2}_{\rm NL})\, ,
\ee
as it derives from Eqs.~(\ref{superp}),~(\ref{superp2}) and~(\ref{Transfer}).
Here $\phi_1$ and $\phi_2$ are zero--mean Gaussian fields with non--vanishing 
cross-correlation $\langle \phi_1 \phi_2 \rangle \neq 0$, with the field 
$\phi_2$ corresponding to that part of the gravitational potential 
induced by the evolution of the entropy pertubations.
Such a parametrization has been introduced in Refs.~\cite{FanBardeen,BMT3} 
and also envisaged in Ref.~\cite{k} as a possible extension of 
the formula~(\ref{phimomspace}), to look for specific non--Gaussian 
signatures from two--field inflationary models. If an isocurvature 
perturbation mode survives after inflation it is reasonable to parametrize 
its non--linearities as 
\be
\label{FNLiso}
\S=\S^{(1)}+f^{iso}_{\rm NL} (\S^{(1) 2} 
- \langle \S^{(1) 2} \rangle )\, .
\ee
Therefore, one can use also this 
parametrization to search for non--Gaussianities in primordial 
(correlated) adiabatic perturbations $\phi$ and 
isocurvature perturbations $\S$, as explained in detail in 
Ref.~\cite{KSW} (see also Ref.~\cite{BMT3} for more generic cases).
\vskip 2cm


\subsection{Non--Gaussianity in unconventional inflation set--ups}

While inflation driven by a scalar field with a very flat potential provides
an early de Sitter phase of the Universe and elegantly solves the horizon and
flatness problems, it is certainly worthwhile to look for alternatives
whose predictions might be discriminated with present and future
observations.Since a de Sitter phase of expansion obliterates the 
horizon and flatness problems, any alternative to
the standard slow--roll  inflation has to preserve this property. 
It is legitimate to ask whether the slow--roll picture is really
necessary or if slow--roll can be obtained in some unconventional way.
In the following, we review some possibilities 
which may predict a large amount of non--Gaussianity.

\subsubsection{Warm inflation}

Warm inflation~\cite{berera95,berera96}
is an alternative to the standard scenario of
supercooled inflation, where dissipative effects are assumed to play
a dynamical role during inflation so that 
radiation production occurs simultaneously with the inflationary expansion. 
The warm inflation picture is a comprehensive set of possible interactions 
between fields during inflation; no a priori assumptions  
about multi--field interactions, thus particle
production, during the inflationary epoch are made.  
As such, the warm inflation picture makes explicit that
the thermodynamic state of the Universe during inflation 
is a dynamical question. Supercooled inflation then emerges as
the limiting case in which interactions are negligible. 

The evolution of the (minimally--coupled) inflaton in warm inflation is 
described by the phenomonological equation 
\be
\label{wi}
\ddot \varphi + (3H + \Gamma) \dot \varphi - 
\frac{\nabla^2\varphi}{a^2} 
+ V^\prime(\varphi)=0 \;, 
\ee
where the dissipation rate $\Gamma$ may generally depend upon $\varphi$. 

The presence of radiation during inflation influences the seeds of density 
perturbations. It is therefore natural to ask whether in such a model  
the level of non--Gaussianity in the primordial perturbations 
might be sensibly different from that of the standard slow--roll scenario. 
This problem was analyzed in Ref.~\cite{guptaetal} (see also 
Ref.~\cite{gupta}), 
where the bispectrum of the gravitational field fluctuations 
was calculated through a simple generalization of the stochastic approach 
adopted in Ref.~\cite{Getal}. In analyzing the dynamics of 
inflaton fluctuations, 
metric fluctuations were disregarded for simplicity, as in Ref.~\cite{FRS}, 
so that the bispectrum is non--zero only due to the presence of inflaton 
self--interactions.  
Requiring the slow--roll condition $|\ddot{\varphi}| \ll
(3H+\Gamma)|\dot{\varphi}|$ and imposing a near--thermal--equilibrium, 
Markovian approximation, the equation of motion for the inflaton field 
emerges as
\be
\frac{d\varphi({\bf x},t)}{dt}=\frac{1}{\Gamma}
\left[e^{-2Ht}\nabla^2
\varphi({\bf x},t) - V'(\varphi({\bf x},t)) + \eta({\bf x},t) \right].
\label{lang}
\ee
Implementing the {\it fluctuation--dissipation} theorem determines 
the properties of the noise, which read  
\be
\langle \eta \rangle=0 \;, 
\label{noise1}
\end{equation}
\begin{equation}
\langle \eta({\bf k},t) \eta({\bf k}',t') \rangle=
2\Gamma T (2\pi)^3 \delta^{(3)}({\bf k}- {\bf k}')
\delta(t-t') \;, 
\label{noise2}
\ee
where $T$ is the temperature and 
${\bf k}$ and ${\bf k}^\prime$ denote physical momenta. 

By splitting as usual the inflaton field into a homogeneous background  
$\varphi_0(t)$ and a fluctuation field $\delta \varphi({\bf x},t)$, 
one can expand the equation of motion in powers of the fluctuations around 
the background. The bispectrum is then immediately obtained from the 
second--order contribution. The corresponding non--linearity strength 
$f_{\rm NL}^\phi$ in the strong-dissipative regime, $\Gamma/H \gg 1$, is found
\cite{guptaetal} 
\be 
\label{biswarm} 
f_{\rm NL}^\phi =  \frac{5}{6} \left( \frac{\dot \varphi_0}{H^2} \right)
\left[\ln\left(\frac{\Gamma}{H}\right)\frac{V^{\prime\prime\prime}}{\Gamma}
\right] \;,
\ee

Applying this formalism to the $\lambda \phi^4$ model, and imposing 
that the amplitude of density fluctuations matches the {\sl COBE} normalization
one finds $f_{\rm NL}^\phi \approx 3.7 \times 10^{-2}$. A similar analysis 
in the weak-dissipative limit, $\Gamma/H \ll 1$, leads to a value of 
$f_{\rm NL}^\phi$ smaller by an order of magnitude~\cite{gupta}.  

\subsubsection{Ghost inflation}

A new possibility of having a de Sitter phase in the Universe in a way 
differing from a cosmological constant has been proposed in 
Ref.~\cite{ghostinflation}. It can be thought of as arising from a 
derivatively coupled ``ghost'' scalar field $\varphi$ 
which ``condenses" in a background where it has non--zero velocity
\begin{equation}
\langle \dot{\varphi} \rangle = M^2 \, \, \to \langle \varphi \rangle =
M^2 t \;.
\end{equation}
Unlike other scalar fields, the velocity
$\dot{\varphi}$ does {\it not} redshift to zero as the Universe
expands, it stays constant, and indeed the energy momentum tensor
is identical to that of a cosmological constant. However, the
ghost condensate is {\it not} a cosmological constant, it is a
physical fluid with a physical fluctuation $\pi$ defined as
\begin{equation}
\varphi = M^2 t + \pi \;.
\end{equation}
The ghost condensate then gives an alternative way of realizing
a de Sitter phase in the Universe. Furthermore, it can be shown that
the symmetries of the theory
allow  to construct a systematic and reliable effective
Lagrangian for $\pi$ and gravity at energies lower than the ghost
cut-off $M$. Neglecting the interactions with gravity, the
effective Lagrangian for $\pi$ (around flat space) has the form~\cite{ghost}
\begin{equation}
\label{eq:piaction}
S = \int \!d^4x \; \left( \frac12 \dot{\pi}^2 - \frac{\alpha^2}{2 M^2}
(\nabla^2 \pi)^2 - \frac{\beta}{2 M^2} \dot{\pi} (\nabla \pi)^2 + 
\cdots \right)
\end{equation}
where $\alpha$ and $\beta$ are order one coefficients.
The Lagrangian is non--Lorentz invariant, as it should be
expected, since the background $\dot{\varphi} = M^2$ breaks Lorentz
invariance spontaneously (the $\pi$ field can be thought as the
Goldstone boson for this symmetry breaking). The low-energy
dispersion relation for $\pi$ is of the unusual form
\begin{equation}
\omega^2 = \alpha^2 \frac{k^4}{M^2} \;.
\end{equation}
The main motivation for such an approach is that 
coupling  this sector to gravity
leads to a variety of interesting modifications of gravity
in the infrared, including antigravity and oscillatory modulation of the
Newtonian potential at late times and large distances
\cite{ghost}.

As pointed out in Ref.~\cite{ghostinflation}, two are the important
differences here from ordinary slow--roll inflation. First, there is no
slow--roll. Even in the approximation where the potential is 
exactly flat, $\dot{\varphi} = M^2$ is non-zero.
The second important difference with standard slow--roll 
inflation concerns the size of
the fluctuations in $\varphi$ (or equivalently $\pi$). Since the
effective Lagrangian for $\pi$ is non--relativistic, and in
particular there are no $k^2$ spatial kinetic terms, the
fluctuations of $\pi$ are less suppressed than in a relativistic
theory. In a relativistic theory, the fact that scalar fields have
scaling dimension one tells that the size of the fluctuations of a
scalar field inside a region of size $R$ is given by $\sim 1/R$,
similarly, at a frequency $E$ it is given by $E$. In ordinary
inflation, the inflaton fluctuations freeze when they have a typical 
energy $E \sim H$, so that their typical size is $\delta\varphi \sim H$. 
We can determine the size of the fluctuations in the ghost inflation 
case by a simple scaling argument familiar
from power--counting for non--relativistic effective theories. 
Suppose one scales energies by a factor of $s$, $E \to s E$, or
alternatively $t \to s^{-1} t$. Clearly, because of the $\omega^2
\propto k^4$ dispersion relation, one has to scale $k$
differently, $k \to s^{1/2} k$ or $x \to s^{-1/2} x$. We then
determine the scaling dimension of $\pi$ by requiring the
quadratic action to be invariant, and one  finds that $\pi$ has
scaling dimension $1/4$
\begin{equation}
\pi \to s^{1/4} \pi \;.
\end{equation}
Now, $\pi$ has {\it mass} dimension one, so the fluctuations at
frequencies of order the cutoff $M$ is $\delta \pi_M \sim M$.
But the fact that $\pi$ has scaling dimension $1/4$ tells  that
the fluctuation at a lower energy $E$ is $\delta \pi_E  \sim (E
M^3)^{1/4}$. In particular, the size of ghost fluctuations that 
freeze, as usual, by Hubble friction when its frequency is of 
order $E \sim H$, is
\begin{equation}
\delta \pi_H \sim (H M^3)^{1/4} \;.
\end{equation}
Of course, for consistency of the effective theory, one has to require $H
\ll M$. The field $\varphi$ fluctuates and is stretched out until the Hubble
damping becomes important at frequency $E \sim H$;  this
does {\it not} correspond to $k \sim H$ but rather, from the
dispersion relation, $k \sim \sqrt{H M}$. The fluctuation $\delta \pi_H$ 
causes inflation to end at slightly different times in different places, and 
so one  has the estimate
\begin{equation} 
\frac{\delta \rho}{\rho} \sim H \delta t =
\frac{H \delta \pi_H}{\dot{\varphi}} \;. 
\end{equation} 
Notice that 
$\dot{\varphi} = M^2$, has nothing to do with slow--roll
parameters. Furthermore,  $\delta \pi_H$ is much larger. One then finds
\begin{equation} 
\label{eq:drho}
\frac{\delta \rho}{\rho} \sim
\left(\frac{H}{M}\right)^{5/4} \;,
\end{equation} 
which can be compared with the standard inflationary case
\begin{equation} 
\frac{\delta \rho}{\rho} \sim \frac{H}{M_{\rm P} \sqrt{\epsilon}} \;.
\end{equation} 
A detailed computation confirms the expectation (\ref{eq:drho}) 
and gives
\cite{ghostinflation}
the primordial curvature spectrum  
\begin{equation}
\label{eq:curspok} 
P_{\R}^{1/2} = \frac{1}{\sqrt{\pi}
\Gamma(1/4)} \frac{(H^5 M^3 \alpha^{-3})^{1/4}}{\dot\varphi} =
\frac1{\sqrt{\pi} \Gamma(1/4)} \left(\frac{H}{M}\right)^{5/4}
\alpha^{-3/4} \;. 
\end{equation} 

As for  the non--Gaussianity since the mass scales are
all much smaller than the
Planck scale, there is no effect coming from gravitational
interactions. The dominant effect comes from the trilinear
interaction in the $\pi$ effective Lagrangian
\begin{equation} 
\label{ll}
\frac{\beta}{M^2} \dot{\pi} (\nabla \pi)^2 \;.
\end{equation}
This leads to a non-zero three--point function for the 
density perturbations. To get an idea of the dimensionless size
of this effect, one  needs to find the dimensionless size of this
coupling at an energy of order $H$. Since one
knows the scaling dimension of $t,x$ and $\pi$, 
the coefficient of the operator (\ref{ll}) has scaling dimension $1/4$.
The dimensionless
size of this interaction at the cutoff $M$ is $\sim 1$; scaling it
down to an energy of order $H$ one gets an estimate for the non--Gaussianity
of the perturbations~\cite{ghostinflation}
\begin{equation} 
{\rm NG} \sim \left(\frac{H}{M}\right)^{1/4} \sim \left(\frac{\delta
\rho}{\rho}\right)^{1/5} \;.
\end{equation} 
These are much larger than in standard inflation, 
where one expects ${\rm NG} \sim \epsilon \cdot(\delta\rho/\rho)$ during 
inflation. A detailed calculation of the level of
non--Gaussianity has been performed in Ref.~\cite{ghostinflation} 
(see also Ref.~\cite{senat}) and shows that $f_{\rm NL}$ has a non--trivial 
momentum dependence. However, for an equilateral configuration, defined by 
setting $k_1 = k_2 = k_3$, one can define an ``effective'' $f_{\rm NL}$ 
\begin{equation}
\label{eq:eqfNL} 
f_{\rm NL}^{\rm eff} \simeq -140 \cdot \beta \cdot \alpha^{-8/5} \;.
\end{equation} 
This value can be much larger than unity depending upon the parameters
$\alpha$ and $\beta$. However, if the resulting $f_{\rm NL}^{\rm eff}$
turns out to be of order unity, one expects relevant corrections
coming from the post--de Sitter phase. 

\subsubsection{``D--cceleration'' mechanism of inflation}

In Ref.~\cite{saltman} an uncoventional mechanism for slow--roll inflation 
was introduced, motivated by the behavior of rolling
scalar fields in strongly interacting theories, 
analyzed using the AdS/CFT correspondence~\cite{CFT}. The key feature of
this model is that the inflaton field $\varphi$ is 
naturally slowed as it approaches a point where many light
degrees of freedom $\chi$ emerge, the slow--down 
arising from the virtual effects of the light particles. From the
stringy  perspective, this scenario 
translates to a probe $D$3--brane travelling 
down a five-dimensional warped throat
geometry. The ultraviolet end of the throat 
joins smoothly onto a compactification in the manner of Randall-Sundrum
\cite{randalls}, 
ensuring that gravity in four dimensions is dynamical while coupling the
field theory to sectors in the compactification which 
can generate corrections to the effective action for
$\varphi$. These corrections generically 
produce a nontrivial potential energy for $\varphi$ including a mass term
$m^2\varphi^2$ and a corresponding closing up of the throat in the infrared 
 region at a scale $\varphi_{IR}$. 

The dynamics of the probe $D$3--brane is captured by the 
Dirac--Born--Infeld (DBI) action
coupled to gravity,

\begin{equation}
S = \int d^4 x \sqrt{-g}\left( \frac{1}{2} M_{\rm P}^2 \R+
 L_{\rm eff} + \cdots \right)
\label{act1} 
\end{equation}
with
\begin{equation}
L_{\rm eff} = -\frac{1}{g_{s}} \left(\, f(\varphi)^{-1}
\sqrt{1+f(\varphi)g^{\mu\nu}\partial_\mu\varphi\partial_\nu\varphi}
+ V(\varphi)\right)
\label{act2}
\end{equation}
Here $g_s$ is the string coupling constant,  the function $f(\varphi)$ is the
(squared) warp factor of the AdS-like throat. For example, 
for a pure $AdS_5$ of radius $R$, it is simply
$f(\varphi)= \lambda/\varphi^4$ with $\lambda\equiv R^4/\alpha^{\prime\,2}$
and $\alpha^\prime$ is the string tension.
The non--analytic behavior of the square-root in (\ref{act2})
gives rise to a speed limit restricting how fast the
scalar field may roll. When $f(\varphi)= \lambda/\varphi^4$, 
the  speed limit is given by  $|\dot{\varphi}| \leq
\varphi^2/\sqrt{\lambda}$ and 
a useful measure of how close we are to the limit 
is given by 
\begin{equation} 
\gamma =
\frac{1}{\sqrt{1-f(\varphi)\dot{\varphi}^2}} \label{gamma}
\end{equation} 
which is analogous to the Lorentz contraction factor
defined in special relativity and grows without bound 
as the speed limit is approached.

The late time behaviour of the scale--factor 
$a$ and the scalar field $\varphi$ 
were determined in Ref.~\cite{saltman} for a variety of 
potentials $V(\varphi)$ and for $f(\varphi)=\lambda/\varphi^4$ using the 
Hamilton-Jacobi approach. This method elevates the scalar field $\varphi$ 
to the role of cosmological time, so that the Hubble parameter $H=\dot{a}/a$ 
is considered as a function $H=H(\varphi)$
determined in terms of the potential by,
\begin{equation}
V(\varphi)=3g_s\,M_{\rm P}^2\,H(\varphi)^2-\gamma(\varphi)/f(\varphi)
\label{V}
\end{equation}

where $\gamma(\varphi)$ is given by

\begin{equation}
\gamma(\varphi)=\left(1+4g_s^2\,M_{\rm P}^4\,
f(\varphi)\,H^\prime(\varphi)^2\right)^{1/2}
\label{gamma2}
\end{equation}

The evolution of $\varphi(t)$ is then fixed by the 
first--order Friedmann equation,

\begin{equation}
\dot{\varphi}=-2g_s\,M_{\rm P}^2\,\frac{H^\prime(\varphi)}{\gamma(\varphi)}
\label{phidot}
\end{equation}

As in the standard
inflationary scenarios, it is useful to introduce a slow--roll parameter

\begin{equation}
\epsilon =
\frac{2g_s\,M_{\rm P}^2}{\gamma}\left(\frac{H^\prime}{H}\right)^2 
\label{slowrollepsilon}
\end{equation}
which parameterises the
deviation from a pure de Sitter phase. In particular one has 
$\ddot{a}/a=H^2(1-\epsilon)$.

In Ref.~\cite{saltman} the first--order equations (\ref{V}) and (\ref{phidot})
were studied for a variety of potentials
$V(\varphi)$. For the massive scalar field case, 
$V(\varphi)\sim m^2\varphi^2$, the late time dynamics was shown
to be a power--law inflation given by $a(t)\to a_0t^{1/\epsilon}$ and 
\begin{equation}
 \varphi \rightarrow
\frac{\sqrt{\lambda}}{t},\ \ \ \ \ \ \ \ \ \ \ \gamma \rightarrow 
\sqrt{\frac{4 g_s}{3\lambda}}\,M_{\rm P} m\,t^2 ,\ \
\ \ \ \ \ \ \ \ \ \ \ H \rightarrow \frac{1}{\epsilon\,t}\, . 
\label{wehavethepower}
\end{equation} 
The coefficient of the
(time--dependent) Hubble parameter is given by the slow--roll parameter 
which is a constant when evaluated on this
background, 
\begin{equation} 
\frac{1}{\epsilon} = \frac{1}{3}\left( 1+
\sqrt{1+\frac{3m^2\lambda}{g_{s}M_{\rm P}^2}}\right) \approx
\sqrt{\frac{\lambda}{3g_{s}}}\frac{m}{M_{\rm P}} \label{epsilon}
\end{equation}
For $\epsilon < 1$, one obtains a phase of power--law inflation. 
The exponential de Sitter phase
can be thought of as the limit as $\epsilon\rightarrow 0$. 
Note that in contrast to usual single field
slow--roll inflation, the accelerated expansion occurs only if the mass of 
inflaton $m$ is suitably
large. This is one novel aspect of this model.

It is clear from the action (\ref{act2}) that expanding 
$\varphi$ in fluctuations
$\varphi\to\varphi+\delta\varphi$ involves expanding the square root in 
(\ref{act2}).
This produces powers of $\gamma$
accompanying the powers of the fluctuation $\delta\varphi$. 
The origin of the strong non--Gaussianities can be therefore understood 
rather simply.  The Lagrangian is proportional
to $\sqrt{1-v_p^2}$, where $v_p=\sqrt{\lambda}\dot\varphi/\varphi^2$ 
is the proper velocity of the brane probe 
whose position collective coordinate is the inflaton $\varphi$. The
inflationary solution involves a proper velocity 
approaching the speed of light as $\varphi$
approaches the origin (with both $\varphi$ and $\dot\varphi$ decreasing 
towards zero). Expanding the action in
fluctuations of $\varphi$ involves expanding the square root in the 
Lagrangian, which produces powers of $\gamma=1/\sqrt{1-v_p^2}$ accompanying 
powers of the fluctuations of the inflaton.  Since $\gamma$ is relatively
large, this produces a large contribution to non--Gaussianities in the model.
It is also easy to understand why linear perturbations with momentum $k$
freeze when they cross the sound horizon at $aH= k/\gamma$ rather than
at $aH=k$. The resulting power--spectrum reads~\cite{dbi}
$\P^\zeta_k\sim g_sH^4/\dot{\varphi}^2 \sim g_s/\epsilon^4\lambda$, 
where we have used the inflationary background (\ref{wehavethepower}); 
and the string coupling $g_s$ enters because the canonically normalised 
scalar field is ${\varphi}/{\sqrt{g_s}}$.

In the  large $\gamma$, small $\epsilon$ regime,
the leading terms in the interaction Hamiltonian are given by

\begin{equation} 
{\cal H}_{\rm
int}=-\frac{a^3\gamma^5}{g_s}\left[
\frac{1}{2\dot{\varphi}}(\delta\dot{\varphi})^3+\frac{k^2}{a^2\gamma^2}
\frac{1}{2\dot{\varphi}}\delta\dot{\varphi}(\delta\varphi)^2\right]\, .
\end{equation}
The computation of the three--point correlation function goes along the
lines described in Subsection~\ref{3pointfunctions} and gives 
an 
$f_{\rm NL}$ with a non--trivial momentum dependence~\cite{dbi}. However,  
for an equilateral
configuration defined by setting for $k_1 = k_2 = k_3$
one can define an ``effective'' $f_{\rm
NL}$ 

\begin{equation}
\label{eq:eqfNL2} 
f_{\rm NL}^{\rm eff} \simeq  -10^{-1}\,\gamma^2 \;,
\end{equation} 
which can be rather large. Again, if $\gamma \sim 20$ and 
$f_{\rm NL}^{\rm eff}$ is of order
unity, one needs  a detailed study of 
the post--inflationary evolution of the non--Gaussianities.

In Table~\ref{Table1} we summarize the level of primordial 
non--Gaussianity as generated in different cosmological scenarios. 
Actually for the 
first three cases the values displayed have been obtained by a complete 
study of the 
perturbation evolution from an early inflationary phase, through reheating 
till the 
radiation and the matter dominated epoch (to which the results are referred). 
For the other scenarios the values of 
$f_{\rm NL}$ actually do not refer to the non--Gaussianities generated in the 
CMB 
temperature anisotropies, rather to non--Gaussianities in the curvature 
perturbation $\zeta$ (and for the uncoventional inflationary scenarios they 
refer 
only to the inflationary phase). In this case when the values shown are 
$f_{\rm NL} \gg 1$, then 
they should be a good approximation for the actual observable quantity, but 
when they are of order unity one expects significant corrections from a 
detailed study of the post--inflationary evolution of the non--Gaussianites. 
The function $g({\bf k}_1,{\bf k}_2)$ is the same in the first three cases
and the 
infrared behaviour of $f_{\rm NL}({\bf k}_1,{\bf k}_2)$ is 
automatically regularized once the monopole term is subtracted 
by requiring that $\langle \Delta T/T \rangle$=0. In the curvaton scenario 
the parameter $r$ corresponds to the ratio of the 
curvaton energy density to the total 
energy density at the epoch of the curvaton decay. In the inhomogeneous 
reheating scenario we have accounted for a fraction of the total decay rate 
to be 
dependent on a light scalar field, $\Gamma_1/\bar{\Gamma}$, and for a ratio 
$\bar{\Gamma}/H$ at the end of inflation not much smaller than one, allowing  
the (positive) $\alpha$ parameter to be $ \alpha < 1/6$, as discussed in 
Sec.~\ref{Inho}.   
The ``minimal case'' correponds to an inflaton decay rate 
$\Gamma$ that is fully controlled by a light scalar field $\chi$, 
$\Gamma \propto 
\chi^2$,  and such that $\Gamma \ll H$ during inflation ($\alpha = 1/6$).
In the multiple--field case $V_{\chi\chi}/H^2$ 
represents the ratio of the mass of the additional self--interacting 
light scalar field $\chi$ and 
the Hubble parameter during inflation. The value shown is an
estimate obtained from extending some linear relations to second order in the 
pertubations, and some corrections are expected when the estimate shown is 
of the order one. Here we must make a remark. Notice that the 
resulting $f_{\rm NL}$ depends on the ``isocurvature fraction'' $\P_\S/\P_\R$, 
determining the relative amplitue of the 
isocurvature perturbation 
to the adiabatic one, and on the correlation between them.
Actually such a formula is valid not only if after inflation an 
isocurvature 
perturbation survives, but also if no entropy mode is left over. In 
the latter case 
one should interpret the combination $\left( \P_\S /\P_\R \right) \cos^2 
\Delta$ evaluated at the end of inflation. If an entropy mode survives for 
the data analysis 
one must also consider a corresponding non--linearity parameter 
$f^{iso}_{\rm NL}$ as shown in Eq.~(\ref{FNLiso}) and~(\ref{FNLadia}). 
In the warm scenario, $\Gamma$ indicates the inflaton dissipation rate. 
Finally, notice that in the ghost and D--cceleration scenarios, the values 
reported refer to an equilateral configuration of the wavenumbers, 
$k_1=k_2=k_3$.

\begin{sidewaystable}
\begin{center}
\begin{tabular}{l|cc}
\hline
\hline
\makebox[4cm]                   &
\makebox[7cm]{\large{$f_{\rm NL}({\bf k}_1,{\bf k}_2)$}} & 
\makebox[7cm]{\large{Comments}}   \\
\hline              
Single--field inflation &  $\frac{7}{3} - g({\bf k}_1,{\bf k}_2)$ & 
$g({\bf k}_1,{\bf k}_2)=
4 \frac{{\bf k}_1\cdot {\bf k}_2}{k^2}
-3 \frac{\left( {\bf k}_1 \cdot {\bf k}_2 \right)^2}{k^4}+\frac{3}{2} 
\frac{k_1^4+k_2^4}{k^4} $
\\
\\
Curvaton scenario & 
$- \left[-\frac{2}{3}+\frac{5}{6}r-\frac{5}{4 r}\right]
- g({\bf k}_1,{\bf k}_2)$ & $r \approx 
\left( \frac{\rho_{\sigma}}{\rho} \right)_{decay}$
\\ 
\\
Inhomogeneous reheating &  
$ \frac{13}{12} - I - g({\bf k}_1,{\bf k}_2)$ & $I=-\frac{5}{2}+\frac{5}{12} 
\frac{\bar{\Gamma}}{\alpha \Gamma_1}$\\
& & ``minimal case'' $I=0$ ($\alpha=\frac{1}{6}$, $\Gamma_1=\bar{\Gamma})$
\\
\\
Multiple scalar fields & 
$\lesssim \frac{\P_\S}{\P_\R} \cos^2 \Delta \left( 4 \cdot 10^3 
\cdot \frac{V_{\chi\chi}}{3H^2} \right) \cdot 60 \frac{H}{\chi}$ &  
order of magnitude estimate 
\\
& & 
of the absolute value
\\
\hline
``Uncoventional'' inflation\\set-ups\\
\hline
Warm inflation  &$-\frac{5}{6}\left(\frac{\dot\varphi_0}{H^2}\right)
\left[\ln\left(\frac{\Gamma}{H}\right)\frac{V^{\prime\prime\prime}}{\Gamma}
\right]$
& second--order corrections\\ & & not included\\   
Ghost inflation &  $-140 \cdot \beta \cdot \alpha^{-8/5}$ & 
post--inflationary corrections\\
& & not included\\
D--cceleration    &  $-10^{-1} \gamma^2$ & 
post--inflationary corrections\\
& & not included\\ 
\hline \hline
\end{tabular}
\end{center}
\caption[Table 1]{{\bf Predictions of the non--linearity 
parameter $f_{\rm NL}$ from 
different scenarios for the generation of cosmological perturbations}. 
In the inhomogeneous reheating scenario $\Gamma_1/\bar{\Gamma} 
\leqslant 1$ and $0 < \alpha \leqslant 1/6$.
In the multiple--field case $-1\leqslant \cos  \Delta \leqslant 1$ is defined 
in Eq.~(\ref{cangle}) measuring the correlation between the adiabatic and 
entropy perturbations, while $\P_\S/\P_\R$ is the isocurvature fraction. 
In ghost inflation the coefficients $\alpha$ and $\beta$ are typically 
$\sim {\mathcal{O}}(1)$. In the D--cceleration mechanism of 
inflation the coefficient 
$\gamma$ is expected to be $\gamma > 1$. For the multiple--field case and 
the unconventional inflation set--ups, the estimates can receive relevant 
corrections in the range 
$f_{\rm NL} \sim 1$ from 
the post--inflationary evolution of the perturbations. \label{Table1}
}
\end{sidewaystable} 
\newpage

\section{\bf Observational constraints on non--Gaussianity}

In this section, we discuss how to test observationally Gaussianity of 
primordial fluctuations and constrain the non--linearity 
parameter, $f_{\rm NL}$.
Currently, the best constraint comes from {\sl WMAP}'s measurements of
CMB anisotropy~\cite{k}; thus, we focus on 
testing Gaussianity of
the CMB. This section is organized as follows: 
In Sec.~\ref{sec:npointspectrum} we study the statistical properties of 
the angular $n$--point harmonic spectra for $n=2$ (power--spectrum), 
3 (bispectrum), and 4 (trispectrum).
In Sec.~\ref{sec:theory_bl} we make theoretical predictions for the 
CMB angular bispectrum from inflation.
In Sec.~\ref{sec:secondary} we calculate the secondary bispectrum
contribution from the Sunyaev--Zel'dovich effect and the weak-lensing 
effect, and foreground contribution from extragalactic 
radio and infrared astronomical sources.
In Sec.~\ref{sec:measure} we estimate how well one can 
constrain $f_{\rm NL}$ with observations, and
discuss how to distinguish between primordial, secondary, and
foreground bispectra.
In Sec.~\ref{sec:measure2} we present a practical way to determine
$f_{\rm NL}$ as well as the point--source contribution 
from nearly full--sky CMB experiments.
In Sec.~\ref{sec:obs} we review the observational constraints on $f_{\rm NL}$
from the {\sl WMAP} experiment using the method described here.

\subsection{Angular $n$--point harmonic spectrum on the sky}
\label{sec:npointspectrum}

As we have mentioned in the
introduction, the angular $n$--point correlation function,
\begin{equation}
 \label{eq:n_corr}
  \left<f(\hat{\mathbf{n}}_1)f(\hat{\mathbf{n}}_2)
   \dots f(\hat{\mathbf{n}}_n)\right>,
\end{equation}
is a simple statistic characterizing a clustering pattern of
fluctuations on the sky, $f(\hat{\mathbf{n}})$.
Here, the bracket denotes the ensemble average, and
Figure~\ref{fig:ensemble} sketches its meaning.
If the fluctuation is Gaussian, then the two--point correlation function
specifies all the statistical properties of $f(\hat{\mathbf{n}})$, for
the two--point correlation function is the only parameter in a Gaussian 
distribution. 
If it is not Gaussian, then we need higher--order correlation 
functions to determine the statistical properties.

Yet simple, one disadvantage of the angular correlation function
is that data points of the correlation function at different angular
scales are generally not independent of each other, but correlated: 
the two--point correlation at 1 degree is correlated with that at 2 degrees, 
and so on.
This property makes a detailed statistical analysis and interpretation 
of the data rather complicated.

Hence, one finds it more convenient to expand $f(\hat{\mathbf{n}})$
into spherical harmonics, the orthonormal basis on the sphere, as
\begin{equation}
 \label{eq:f_alm}
 f(\hat{\mathbf{n}}) 
 = \sum_{l=0}^\infty\sum_{m=-l}^{l} a_{lm}Y_{lm}(\hat{\mathbf{n}}),
\end{equation}
and then to consider the angular $n$--point harmonic spectrum, 
$\left<a_{l_1m_1}a_{l_2m_2}\dots a_{l_nm_n}\right>$.
While $a_{lm}$ for $m\neq 0$ is complex, reality of
$f(\hat{\mathbf{n}})$ gives $a_{l-m}=a_{lm}^*(-1)^m$, and thus the 
number of independent modes is not $4l+1$, but $2l+1$.

In particular, the angular two--, three--, and four--point harmonic spectra
are called the angular {\it power--spectrum}, {\it bispectrum}, and 
{\it trispectrum}, respectively. By isotropy the angular power--spectra 
at different angular scales, or at different $l$'s, are uncorrelated. 
Moreover, since the spherical harmonics are orthogonal for different $l$'s, 
they highlight characteristic structures on the sky at a given $l$.
In other words, even if the angular correlation function is featureless,
the angular spectrum may have a distinct structure, 
for inflation predicts a prominent peak in the angular power--spectrum, 
not in the angular correlation function.
In this section, we study statistical properties of the angular $n$--point 
harmonic spectra.

\subsubsection{Statistical isotropy of the Universe}
\label{sec:isotropy}

In reality, we cannot measure the ensemble average of the angular 
harmonic spectrum, but one realization such as 
$a_{l_1m_1}a_{l_2m_2}\dots a_{l_nm_n}$, which is so noisy that we want to 
average it somehow to reduce the noise.

We assume {\it statistical isotropy} of the Universe from which
it follows that our sky is isotropic and has no preferred direction.
Isotropy of the CMB justifies the assumption.
The assumption readily implies that one can average the spectrum over $m_i$
with an appropriate weight, as $m_i$ represent an azimuthal orientation 
on the sky.
The average over $m_i$ enables us to reduce the statistical error of 
the measured harmonic spectra.

How can we find the weight?
One finds it as a solution to statistical isotropy, or 
{\it rotational invariance} of the angular correlation function on the sky,
\begin{equation}
 \label{eq:rotinv}
  \left<Df(\hat{\mathbf{n}}_1)Df(\hat{\mathbf{n}}_2)
   \dots Df(\hat{\mathbf{n}}_n)\right>
  = \left<f(\hat{\mathbf{n}}_1)f(\hat{\mathbf{n}}_2)
     \dots f(\hat{\mathbf{n}}_n)\right>,
\end{equation}
where $D=D(\alpha,\beta,\gamma)$ is a rotation matrix for the Euler 
angles $\alpha$, $\beta$, and $\gamma$.
Figure~\ref{fig:rotinv} sketches the meaning of statistical isotropy. 
Substituting Eq.~(\ref{eq:f_alm}) for $f(\hat{\mathbf{n}})$ 
in Eq.~(\ref{eq:rotinv}), we then need 
rotation of the spherical harmonic, $DY_{lm}(\hat{\mathbf{n}})$.
It is formally represented by the rotation matrix element, 
$D_{m'm}^{(l)}(\alpha,\beta,\gamma)$, as~\cite{rotenberg:1959}
\begin{equation}
 \label{eq:rotYlm}
  DY_{lm}(\hat{\mathbf{n}})= 
  \sum_{m'=-l}^{l} D_{m'm}^{(l)} Y_{lm'}(\hat{\mathbf{n}}).
\end{equation}
The matrix element, $D_{m'm}^{(l)}=\left<l,m'\left|D\right|l,m\right>$,
describes finite rotation of an initial state whose 
orbital angular momentum is represented by $l$ and $m$ into
a final state represented by $l$ and $m'$.
Finally, we obtain the statistical isotropy condition on the 
angular $n$--point harmonic spectrum:
\begin{equation}
 \label{eq:condition}
  \left<a_{l_1m_1}a_{l_2m_2}\dots a_{l_nm_n}\right>
  = \sum_{{\rm all}~m'} \left<a_{l_1m'_1}a_{l_2m'_2}
			       \dots a_{l_nm'_n}\right>
  D_{m_1'm_1}^{(l_1)}D_{m_2'm_2}^{(l_2)}\dots D_{m'_nm_n}^{(l_n)}.
\end{equation}

\begin{figure}
 \begin{center}
  \leavevmode\epsfxsize=9cm \epsfbox{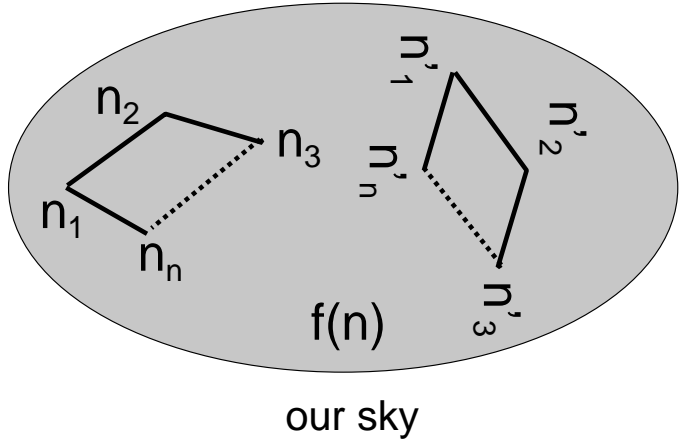}
 \end{center}
 \caption
 {\bf Statistical Isotropy of Angular Correlation Function} 
 \mycaption{A schematic view of statistical isotropy of the angular 
 correlation function.
 As long as its configuration is preserved, we can average
 $f(\hat{\mathbf{n}}_1)\dots f(\hat{\mathbf{n}}_n)$ 
 over all possible orientations and 
 positions on the sky.}
\label{fig:rotinv}
\end{figure}

Using this equation, in Ref.~\cite{hu:2001} appropriate 
weights for averaging the angular power--spectrum ($n=2$), 
bispectrum ($n=3$), and trispectrum ($n=4$), over azimuthal angles 
have been systematically evaluated. 
Some of those may be found more intuitively; however, this method
allows us to find the weight for any higher--order harmonic spectrum.
In the following sections, we derive rotationally invariant, 
azimuthally averaged harmonic spectra for $n=2$, 3, and 4, and 
study their statistical properties.

\subsubsection{Angular power--spectrum}\label{sec:powerspectrum}

The angular power--spectrum measures how much fluctuations exist on a 
given angular scale.
For example, the variance of $a_{lm}$ for $l\ge 1$,
$\left<a_{lm}a_{lm}^*\right>$, measures the amplitude of fluctuations
at a given $l$.

Generally speaking, the covariance matrix of $a_{lm}$, 
$\left<a_{l_1m_1}a_{l_2m_2}^*\right>$, is not necessarily diagonal.
It is, however, actually diagonal once we assume full sky coverage and 
rotational invariance of the angular two--point correlation function, 
as we will show in this section.
The variance of $a_{lm}$ thus describes the two--point correlation completely.

Rotational invariance (Eq.~(\ref{eq:condition})) requires 
\begin{equation}
 \label{eq:condition2}
  \left<a_{l_1m_1}a^*_{l_2m_2}\right>
  = \sum_{m_1'm_2'} \left<a_{l_1m'_1}a^*_{l_2m'_2}\right>
  D_{m_1'm_1}^{(l_1)}D_{m'_2m_2}^{(l_2)*}
\end{equation}
to be satisfied, where we have used the complex conjugate for simplifying 
calculations.
From this equation, we seek for a rotationally invariant representation 
of the angular power--spectrum. 
Suppose that the covariance matrix of $a_{lm}$ is diagonal, i.e., 
$  \left<a_{l_1m_1}a^*_{l_2m_2}\right>= 
  \left<C_{l_1}\right> \delta_{l_1l_2} \delta_{m_1m_2}$.
Equation~(\ref{eq:condition2}) then reduces to
\begin{equation}
 \nonumber
  \left<a_{l_1m_1}a^*_{l_2m_2}\right>
  =
  \left<C_{l_1}\right> \delta_{l_1l_2}
  \sum_{m_1'}D_{m_1'm_1}^{(l_1)}D_{m'_1m_2}^{(l_1)*}
  =
  \left<C_{l_1}\right> \delta_{l_1l_2} \delta_{m_1m_2}.
\end{equation}
Thus, we have proven that $\left<C_l\right>$ is rotationally invariant.
Rotational invariance implies that the covariance matrix is diagonal.

Observationally, the unbiased estimator of $\left<C_l\right>$ should be
\begin{eqnarray}
 \nonumber
  C_l&=& \frac1{2l+1}\sum_{m=-l}^{l} a_{lm}a_{lm}^*
     = \frac1{2l+1}\left(a_{l0}^2 + 2\sum_{m=1}^{l}a_{lm}a_{lm}^*\right)\\
 \label{eq:estimate2}
     &=& \frac1{2l+1}\left\{
		      a_{l0}^2 + 2\sum_{m=1}^{l}
     \left[\left(\Re{a_{lm}}\right)^2 + \left(\Im{a_{lm}}\right)^2\right]
     \right\}.
\end{eqnarray}
The second equality follows from $a_{l-m}=a_{lm}^*(-1)^m$, 
i.e., $a_{l-m}a_{l-m}^*=a_{lm}a_{lm}^*$, and hence
we average $2l+1$ independent samples for a given $l$.
It suggests that fractional statistical error of $C_l$ is 
reduced by $\sqrt{1/(2l+1)}$.
This property is the main motivation for considering the azimuthally
averaged harmonic spectrum.

We find it useful to define an azimuthally averaged
harmonic transform, $e_{l}(\hat{\mathbf n})$, as 
\begin{equation}
 \label{eq:el}
   e_{l}(\hat{\mathbf n})
   \equiv
   \sqrt{\frac{4\pi}{2l+1}}
   \sum_{m=-l}^{l} a_{lm} Y_{lm}(\hat{\mathbf n}),
\end{equation}
which is interpreted as a square-root of $C_l$ at a given position of 
the sky,
\begin{equation}
  \label{eq:cl}
   \int\frac{d^2\hat{\mathbf n}}{4\pi}~
   e^2_{l}(\hat{\mathbf n})
   = C_l.
\end{equation}
$e_{l}(\hat{\mathbf n})$ is particularly useful for measuring 
the angular bispectrum~\cite{rs4,komatsu/etal:2002}, 
trispectrum (Chapter 6 of Ref.~\cite{komatsu:phd}), and 
probably any higher--order harmonic spectra, because of being 
computationally very fast to calculate.
This is very important, as the new satellite experiments,
{\sl WMAP} and {\sl Planck}, have more than millions of pixels, for which
we will crucially need a fast algorithm for measuring these 
higher--order harmonic spectra.

We derive the covariance matrix of $C_l$,
$\left<C_lC_{l'}\right>-\left<C_l\right>\left<C_{l'}\right>$, with
 the four--point function, the trispectrum.
Starting with
\begin{equation}
 \label{eq:error2}
  \left<C_lC_{l'}\right>
  =
  \frac1{(2l+1)(2l'+1)}
   \sum_{mm'}\left<a_{lm}a_{lm}^*a_{l'm'}a_{l'm'}^*\right>,
\end{equation}
we obtain the power--spectrum covariance matrix
\begin{eqnarray}
 \nonumber
  \left<C_lC_{l'}\right> - \left<C_l\right>\left<C_{l'}\right>
  &=&
  \frac{2\left<C_l\right>^2}{2l+1}\delta_{ll'}
   + \frac1{(2l+1)(2l'+1)}
   \sum_{mm'}\left<a_{lm}a_{lm}^*a_{l'm'}a_{l'm'}^*\right>_{\rm c}\\
  \label{eq:error2*}
 &=&
  \frac{2\left<C_l\right>^2}{2l+1}\delta_{ll'}
   + \frac{(-1)^{l+l'}}{\sqrt{(2l+1)(2l'+1)}}
   \left<T^{ll}_{l'l'}(0)\right>_{\rm c},
\end{eqnarray}
where $\left<a_{lm}a_{lm}^*a_{l'm'}a_{l'm'}^*\right>_{\rm c}$
is the connected four--point harmonic spectrum, the connected trispectrum, 
which is exactly zero for a Gaussian field.
It follows from this equation that the covariance matrix of $C_l$ is 
exactly diagonal only when $a_{lm}$ is Gaussian.
$\left<T^{l_1l_2}_{l_3l_4}(L)\right>_{\rm c}$ is the ensemble average of
the angular averaged connected trispectrum, which we will 
define in Sec.~\ref{sec:trispectrum} (Eq.~(\ref{eq:tl})).

Unfortunately, we cannot measure the connected $T^{ll}_{l'l'}(0)$
directly from the angular trispectrum (see Sec.~\ref{sec:trispectrum}).
We will thus never be sure if the power--spectrum covariance is 
precisely diagonal, as long as we use the angular trispectrum.
We need other statistics able to pick up information on the 
connected $T^{ll}_{l'l'}(0)$, even though they are indirect.
Otherwise, we need a model for the connected trispectrum, and
use the model to constrain the connected $T^{ll}_{l'l'}(0)$ from 
the other trispectrum configurations.

There is no reason to assume the connected $T^{ll}_{l'l'}(0)$ is small.
It is produced on large angular scales, 
if the topology of the Universe is closed hyperbolic~\cite{inoue:2001}.
An analytic prediction for the 
connected trispectrum produced in a closed hyperbolic Universe
is derived in Appendix D of Ref.~\cite{komatsu:phd}.
On small angular scales, several authors have shown that 
the weak gravitational lensing effect produces a non--zero connected
trispectrum or four--point correlation 
function~\cite{bernardeau:1997,zaldarriaga/seljak:1999,zaldarriaga:2000};
in Ref.~\cite{hu:2001} the induced off-diagonal terms are found to be 
negligible compared with the diagonal terms out to $l\sim 2000$.

If the connected trispectrum is negligible, then we obtain
\begin{equation}
 \label{eq:error2**}
  \left<C_lC_{l'}\right> - \left<C_l\right>\left<C_{l'}\right>
  \approx
  \frac{2\left<C_l\right>^2}{2l+1}\delta_{ll'}.
\end{equation}
The fractional error of $C_l$ is thus proportional to 
$\sqrt{1/(2l+1)}$, as expected from our having $2l+1$ independent samples 
to average for a given $l$.
The exact form follows from $C_l$ being $\chi^2$ distributed
with $2l+1$ degrees of freedom when $a_{lm}$ is Gaussian.
If $a_{lm}$ is Gaussian, then its probability density distribution is
\begin{equation}
 \label{eq:probalm}
  P\left(a_{lm}\right)
  =
  \frac{\exp\left[-{a_{lm}^2}/(2\left<C_l\right>)\right]}
  {\sqrt{2\pi \left<C_l\right>}}.
\end{equation}
One can use this distribution to generate Gaussian random realizations
of $a_{lm}$ for a given $\left<C_l\right>$.
First,  calculate $\left<C_l\right>$ with the {\sf CMBFAST} code 
\cite{seljak/zaldarriaga:1996} for a set of cosmological parameters.
Next, generate a realization of $a_{lm}$,
$a_{lm}= \epsilon\left<C_l\right>^{1/2}$, where $\epsilon$ is a
Gaussian random variable with unit variance.

\subsubsection{Angular bispectrum}\label{sec:bispectrum}

The angular bispectrum consists of three harmonic transforms, 
$a_{l_1m_1}a_{l_2m_2}a_{l_3m_3}$.
For Gaussian $a_{lm}$, the expectation value is exactly zero.
By imposing statistical isotropy upon the angular three--point
correlation function, one finds that the angular averaged bispectrum, 
$B_{l_1l_2l_3}$, given by
\begin{equation}
  \label{eq:blll*}
  \left<a_{l_1m_1}a_{l_2m_2}a_{l_3m_3}\right>
  =
  \left<B_{l_1l_2l_3}\right>
  \left(\begin{array}{ccc}l_1&l_2&l_3\\m_1&m_2&m_3\end{array}\right)
\end{equation}
satisfies rotational invariance (Eq.~(\ref{eq:condition})).
Here, the matrix denotes the Wigner--$3j$ symbol
(see Appendix~\ref{app:wigner}).
Since $l_1$, $l_2$, and $l_3$ form a triangle, $B_{l_1l_2l_3}$ satisfies the 
triangle condition, $\left|l_i-l_j\right|\leq l_k \leq l_i+l_j$ 
for all permutations of indices. Parity invariance of 
the angular correlation function demands $l_1+l_2+l_3={\rm even}$.
Figure~\ref{fig:triangle} sketches a configuration of the angular bispectrum.

\begin{figure}
 \begin{center}
  \leavevmode\epsfxsize=5cm \epsfbox{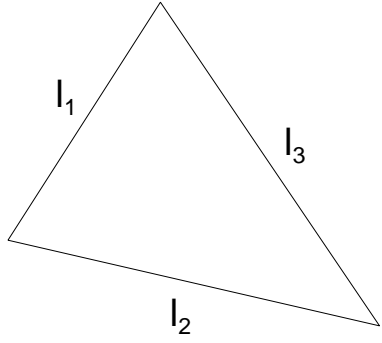}
 \end{center}
 \caption
 {\bf Angular Bispectrum Configuration}
\label{fig:triangle}
\end{figure}

The Wigner--$3j$ symbol, which describes the coupling of two angular momenta,
represents the azimuthal angle dependence of the angular bispectrum, 
since the bispectrum forms a triangle.
Suppose that two ``states'' with $(l_1,m_1)$ and $(l_2,m_2)$ angular momenta 
form a coupled state with $(l_3,m_3)$. 
They form a triangle whose orientation is represented by 
$m_1$, $m_2$, and $m_3$, with satisfying $m_1+m_2+m_3=0$. 
As we rotate the system, the Wigner--$3j$ symbol transforms $m$'s, 
yet preserving the configuration of the triangle.
Similarly, rotational invariance of the angular bispectrum demands
that the same triangle configuration gives the same amplitude of
the bispectrum regardless of its orientation, and thus 
the Wigner--$3j$ symbol describes the azimuthal angle dependence.

The proof of $\left<B_{l_1l_2l_3}\right>$ to be rotationally invariant 
is as follows.
Substituting Eq.~(\ref{eq:blll*}) for the statistical isotropy
condition (Eq.~(\ref{eq:condition})) for $n=3$, we obtain 
\begin{eqnarray}
 \nonumber
  & &
 \left<a_{l_1m_1}a_{l_2m_2}a_{l_3m_3}\right>\\
 \nonumber
  &=& 
  \sum_{{\rm all}~m'} \left<a_{l_1m_1'}a_{l_2m_2'}a_{l_3m_3'}\right>
  D_{m_1'm_1}^{(l_1)}D_{m_2'm_2}^{(l_2)}D_{m_3'm_3}^{(l_3)}\\
 \nonumber
 &=&
  \left<B_{l_1l_2l_3}\right>
 \sum_{{\rm all}~m'}
 \left(\begin{array}{ccc}l_1&l_2&l_3\\m_1'&m_2'&m_3'\end{array}\right)  \\
 \nonumber
 & &\times
 \sum_{LMM'}
 \left(\begin{array}{ccc}l_1&l_2&L\\m_1'&m_2'&M'\end{array}\right)
 \left(\begin{array}{ccc}l_1&l_2&L\\m_1&m_2&M\end{array}\right)
 (2L+1)
 D_{M'M}^{(L)*} D_{m_3'm_3}^{(l_3)}\\
 \nonumber
 &=&
  \left<B_{l_1l_2l_3}\right>
 \sum_{m_3'} \sum_{LMM'} \delta_{l_3L} \delta_{m_3'M'}
 \left(\begin{array}{ccc}l_1&l_2&L\\m_1&m_2&M\end{array}\right)
 D_{M'M}^{(L)*} D_{m_3'm_3}^{(l_3)}\\
 &=&
  \left<B_{l_1l_2l_3}\right>
 \left(\begin{array}{ccc}l_1&l_2&l_3\\m_1&m_2&m_3\end{array}\right).
\end{eqnarray}
In the second equality, we have reduced 
$D_{m_1'm_1}^{(l_1)}D_{m_2'm_2}^{(l_2)}$ to $D_{M'M}^{(L)*}$,
using Eq.~(\ref{eq:rotreduce}).
In the third equality, we have used the identity~\cite{rotenberg:1959},
\begin{equation}
 \label{eq:orthonormal}
 \sum_{m'_1m'_2}
  \left(\begin{array}{ccc}l_1&l_2&l_3\\m'_1&m'_2&m'_3\end{array}\right)
  \left(\begin{array}{ccc}l_1&l_2&L\\m'_1&m'_2&M'\end{array}\right)
  =
   \frac{\delta_{l_3L}\delta_{m'_3M'}}{2L+1}.
\end{equation}

To obtain the unbiased estimator of the angular averaged bispectrum,
$B_{l_1l_2l_3}$, we invert Eq.~(\ref{eq:blll*}) with the 
identity Eq.~(\ref{eq:orthonormal}), and obtain 
\begin{equation}
  \label{eq:best}
  B_{l_1l_2l_3}= \sum_{{\rm all}~m}
  \left(
  \begin{array}{ccc}
  l_1&l_2&l_3\\
  m_1&m_2&m_3
  \end{array}
  \right)
  a_{l_1m_1}a_{l_2m_2}a_{l_3m_3}.
\end{equation}
We can rewrite this expression into a more computationally useful form.
Using the azimuthally averaged harmonic transform, 
$e_{l}(\hat{\mathbf n})$ (Eq.~(\ref{eq:el})), and the identity 
\cite{rotenberg:1959},
\begin{eqnarray}
 \nonumber
  \left(\begin{array}{ccc}l_1&l_2&l_3\\m_1&m_2&m_3\end{array}\right)
  &=&
  \left(\begin{array}{ccc}l_1&l_2&l_3\\0&0&0\end{array}\right)^{-1}
  \sqrt{
  \frac{(4\pi)^3}{\left(2l_1+1\right)\left(2l_2+1\right)\left(2l_3+1\right)}
  }\\
 \label{eq:gaunt_spec}
 & &\times
  \int \frac{d^2\hat{\mathbf n}}{4\pi}~
  Y_{l_1m_1}(\hat{\mathbf n})
  Y_{l_2m_2}(\hat{\mathbf n})
  Y_{l_3m_3}(\hat{\mathbf n}),
\end{eqnarray}
we rewrite Eq.~(\ref{eq:best}) as
\begin{equation}
  B_{l_1l_2l_3}=
  \left(\begin{array}{ccc}l_1&l_2&l_3\\0&0&0\end{array}\right)^{-1}
  \int \frac{d^2\hat{\mathbf n}}{4\pi}~
  e_{l_1}(\hat{\mathbf n})
  e_{l_2}(\hat{\mathbf n})
  e_{l_3}(\hat{\mathbf n}).
\end{equation}
This expression is computationally efficient; we can quickly calculate 
$e_l(\hat{\mathbf n})$ with the spherical harmonic transform. 
Then, the average over the full sky, $\int d^2\hat{\mathbf n}/(4\pi)$,
is done by the sum over all pixels divided by the total number 
of pixels, $N^{-1}\sum_i^{N}$, if all the pixels have equal area. 
Note that the integral over $\hat{\mathbf n}$ must be done over the 
full sky even when a sky--cut is applied, as 
$e_{l}(\hat{\mathbf n})$ already encapsulates information on 
partial sky coverage through the $a_{lm}$, which may be measured 
on an incomplete sky.

We calculate the covariance matrix of $B_{l_1l_2l_3}$, provided that
non--Gaussianity is weak, $\left<B_{l_1l_2l_3}\right>\approx 0$.
Since the covariance matrix is a product of six $a_{lm}$'s,
we have $_6C_2\cdot{}_4C_2/3!= 15$ terms to evaluate, 
according to the Wick's theorem; however, using the identity
\cite{rotenberg:1959},
\begin{equation}
 \label{eq:helpidentity}
\sum_m (-1)^m 
 \left(\begin{array}{ccc}l&l&l'\\m&-m&0\end{array}\right)
 = \delta_{l'0},
\end{equation}
and assuming that none of the $l$'s zero, we find only $3!=6$ terms that do 
not include $\left<a_{l_im_i}a_{l_jm_j}\right>$ but 
include only $\left<a_{l_im_i}a_{l_jm_j}^*\right>$ non--vanishing.
Evaluating these 6 terms, we obtain 
\cite{luo:1994,heavens:1998,rs4,gangui/martin:2000}
\begin{eqnarray}
 \nonumber
  & &\left<B_{l_1l_2l_3}B_{l_1'l_2'l_3'}\right>\\
  \nonumber
  &=&
  \sum_{{\rm all}~mm'}
  \left(\begin{array}{ccc}l_1&l_2&l_3\\m_1&m_2&m_3\end{array}\right)
  \left(\begin{array}{ccc}l_1'&l_2'&l_3'\\m_1'&m_2'&m_3'\end{array}\right)
  \left<a_{l_1m_1}a_{l_2m_2}a_{l_3m_3}
   a_{l_1'm_1'}^*a_{l_2'm_2'}^*a_{l_3'm_3'}^*\right>\\
 \nonumber
 &=&
 \left<C_{l_1}\right>\left<C_{l_2}\right>\left<C_{l_3}\right>
 \left[
  \delta_{l_1l_2l_3}^{l_1'l_2'l_3'} + \delta_{l_1l_2l_3}^{l_3'l_1'l_2'}
  + \delta_{l_1l_2l_3}^{l_2'l_3'l_1'} 
  + (-1)^{l_1+l_2+l_3} \left(\delta_{l_1l_2l_3}^{l_1'l_3'l_2'}
  + \delta_{l_1l_2l_3}^{l_2'l_1'l_3'} + \delta_{l_1l_2l_3}^{l_3'l_2'l_1'}
  \right)\right],\\
 \label{eq:error3}
\end{eqnarray}
where $\delta_{l_1l_2l_3}^{l_1'l_2'l_3'}\equiv
\delta_{l_1l_1'}\delta_{l_2l_2'}\delta_{l_3l_3'}$, and so on.
Hence, the covariance matrix is diagonal in the weak 
non--Gaussian limit.
The diagonal terms for $l_i\neq 0$ and $l_1+l_2+l_3={\rm even}$ are
\begin{equation}
 \label{eq:variance3}
  \left<B_{l_1l_2l_3}^2\right>
  =
  \left<C_{l_1}\right>\left<C_{l_2}\right>\left<C_{l_3}\right>
  \left(1+ 2\delta_{l_1l_2}\delta_{l_2l_3} +
   \delta_{l_1l_2}+\delta_{l_2l_3}+\delta_{l_3l_1}\right).
\end{equation}
The variance is amplified by a factor of 2 or 6, when two or all 
$l$'s are the same, respectively.

One finds that Eq.~(\ref{eq:variance3}) becomes non--exact on 
the incomplete sky, where the variance distribution becomes more scattered.
Using simulated realizations of a Gaussian sky, the authors of ref.~\cite{k} 
measured the variance on the full sky as well as on the incomplete sky for 
three different Galactic sky--cuts, $20^\circ$, $25^\circ$, and $30^\circ$.
Figure~\ref{fig:variance_bare} plots the results; 
one finds that Eq.~(\ref{eq:variance3}) holds only approximately 
on the incomplete sky.

\begin{figure}
 \begin{center}
  \leavevmode\epsfxsize=9cm \epsfbox{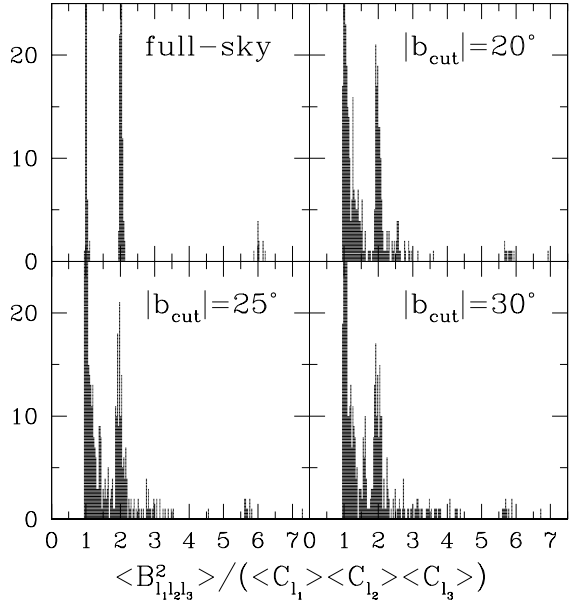}
 \end{center}
 \caption
 {\bf Variance of Angular Bispectrum} 
 \mycaption{Histograms of variance of the angular bispectrum
 for $l_1\leq l_2\leq l_3$  up to a maximum multipole of 20.
 There are 466 modes.
 These are derived from simulated realizations of a Gaussian sky.
 The top-left panel shows the case of full sky coverage, while
 the rest of panels show the cases of incomplete sky coverage.
 The top-right, bottom-left, and bottom-right panels use
 the $20^\circ$, $25^\circ$, and $30^\circ$ Galactic sky-cuts, respectively.}
\label{fig:variance_bare}
\end{figure}

\subsubsection{Angular trispectrum}\label{sec:trispectrum}

The angular trispectrum consists of four harmonic transforms, 
$a_{l_1m_1}a_{l_2m_2}a_{l_3m_3}a_{l_4m_4}$. In Ref.~\cite{hu:2001} 
a rotationally invariant solution for
the angular trispectrum was found as
\begin{equation}
 \label{eq:tl}
 \left<a_{l_1m_1}a_{l_2m_2}a_{l_3m_3}a_{l_4m_4}\right>
  =
  \sum_{LM}
  \left(\begin{array}{ccc}l_1&l_2&L\\m_1&m_2&-M\end{array}\right)
  \left(\begin{array}{ccc}l_3&l_4&L\\m_3&m_4&M\end{array}\right)
  (-1)^M \left<T^{l_1l_2}_{l_3l_4}(L)\right>.
\end{equation}
One can prove that this solution, $\left<T^{l_1l_2}_{l_3l_4}(L)\right>$, 
is rotationally invariant by similar calculations to those proving the 
angular bispectrum to be so.
By construction, $l_1$, $l_2$, and $L$ form one triangle,
while $l_3$, $l_4$, and $L$ form the other triangle in a quadrilateral
with sides $l_1$, $l_2$, $l_3$, and $l_4$.
$L$ represents a diagonal of the quadrilateral.
Figure~\ref{fig:quad} sketches a configuration of the angular 
trispectrum.
When we arrange $l_1$, $l_2$, $l_3$, and $l_4$ in order of 
$l_1\le l_2\le l_3\le l_4$, $L$ lies in 
$\max(l_2-l_1,l_4-l_3)\le L\le \min(l_1+l_2,l_3+l_4)$.
Parity invariance of the angular four--point correlation function 
demands $l_1+l_2+L=\mbox{even}$ and $l_3+l_4+L=\mbox{even}$.

\begin{figure}
 \begin{center}
  \leavevmode\epsfxsize=7cm \epsfbox{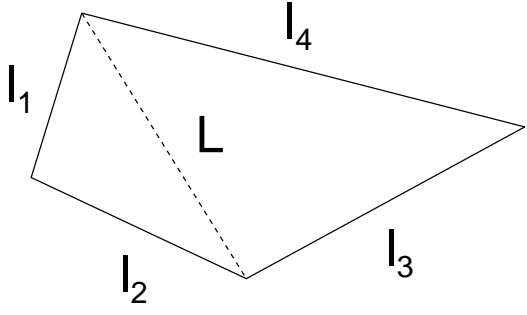}
 \end{center}
 \caption
 {\bf Angular Trispectrum Configuration}
\label{fig:quad}
\end{figure}

The angular trispectrum generically consists of two parts.
One is the disconnected part, the contribution from Gaussian fields,
which is given by the angular power--spectra~\cite{hu:2001},
\begin{eqnarray}
 \nonumber
  & &
  \left<T^{l_1l_2}_{l_3l_4}(L)\right>_{\rm disconnected}\\
 \nonumber
  &=&
  (-1)^{l_1+l_3}\sqrt{(2l_1+1)(2l_3+1)}\left<C_{l_1}\right>\left<C_{l_3}\right>
  \delta_{l_1l_2}\delta_{l_3l_4}\delta_{L0} \\
 \label{eq:disconnected*}
  & & 
  + (2L+1)\left<C_{l_1}\right>\left<C_{l_2}\right>\left[(-1)^{l_2+l_3+L}
		       \delta_{l_1l_3}\delta_{l_2l_4}
		     +\delta_{l_1l_4}\delta_{l_2l_3}\right].
\end{eqnarray}
For $l_1\le l_2\le l_3\le l_4$, the disconnected terms are 
non--zero only when $L=0$ or $l_1=l_2=l_3=l_4$.
We have numerically confirmed that the estimator given below 
(Eq.~(\ref{eq:tobs*})) accurately reproduces the disconnected terms
(Eq.~(\ref{eq:disconnected*})) on a simulated Gaussian sky.

The other is the connected part whose expectation value is 
exactly zero for Gaussian fields; thus, the connected part is sensitive 
to non--Gaussianity.
When none of the $l$'s are the same in $T^{l_1l_2}_{l_3l_4}(L)$,
one might expect the trispectrum to comprise the connected part only;
however, this is true only on the full sky.
The disconnected terms on the incomplete sky, which are often much bigger 
than the connected terms, leak the power to the other modes for which 
all $l$'s are different.
One should take this effect into account in the analysis.

Inverting Eq.~(\ref{eq:tl}), we obtain the unbiased estimator of 
$T^{l_1l_2}_{l_3l_4}(L)$ \cite{hu:2001},
\begin{eqnarray}
 \nonumber
  T^{l_1l_2}_{l_3l_4}(L)
  &=& (2L+1)\sum_{{\rm all}~m}\sum_M(-1)^M
  \left(\begin{array}{ccc}l_1&l_2&L\\m_1&m_2&M\end{array}\right)
  \left(\begin{array}{ccc}l_3&l_4&L\\m_3&m_4&-M\end{array}\right)\\
 \label{eq:test}
 & &\times
  a_{l_1m_1}a_{l_2m_2}a_{l_3m_3}a_{l_4m_4}.
\end{eqnarray}
Note that this expression includes both the connected and the
disconnected terms.

We find that this estimator has a special property for $L=0$
(which demands $l_1=l_2$ and $l_3=l_4$).
The trispectrum estimator for these configurations,
$T^{l_1l_1}_{l_3l_3}(0)$, reduces to a product of two power--spectrum 
estimators, $C_{l_1}C_{l_3}$,
\begin{eqnarray}
 \nonumber
  T^{l_1l_1}_{l_3l_3}(0)
  &=&
  \sum_{m_1m_3}
  \left(\begin{array}{ccc}l_1&l_1&0\\m_1&-m_1&0\end{array}\right)
  \left(\begin{array}{ccc}l_3&l_3&0\\m_3&-m_3&0\end{array}\right)
  a_{l_1m_1}a_{l_1-m_1}a_{l_3m_3}a_{l_3-m_3}\\
  &=& 
   \label{eq:special}
   (-1)^{l_1+l_3}\sqrt{(2l_1+1)(2l_3+1)}C_{l_1}C_{l_3},
\end{eqnarray}
where $C_l=(2l+1)^{-1}\sum_ma_{lm}a_{lm}^*$.
We have used the identity, Eq.~(\ref{eq:helpidentity}),
and $a_{l-m}=(-1)^ma_{lm}^*$ in the second equality.
From this equation, one may assume that $T^{l_1l_1}_{l_3l_3}(0)$ 
coincides with the disconnected
terms for $L=0$ (see Eq.~(\ref{eq:disconnected*})),
\begin{equation}
 \label{eq:unco}
  \left<T^{l_1l_1}_{l_3l_3}(0)\right>_{\rm disconnected}
  =
  (-1)^{l_1+l_3}\sqrt{(2l_1+1)(2l_3+1)}\left<C_{l_1}\right>\left<C_{l_3}\right>
  + 2\left<C_{l_1}\right>^2 \delta_{l_1l_3}.
\end{equation}
They are, however, different for non--Gaussian fields, 
because of the power--spectrum covariance, Eq.~(\ref{eq:error2*}).
By taking the ensemble average of $T^{l_1l_1}_{l_3l_3}(0)$, and 
substituting Eq.~(\ref{eq:error2*}) for
$\left<C_{l_1}C_{l_3}\right>$, we find a rather trivial result:
\begin{eqnarray}
 \nonumber
 \left<T^{l_1l_1}_{l_3l_3}(0)\right>
  &=&
  (-1)^{l_1+l_3}\sqrt{(2l_1+1)(2l_3+1)}\left<C_{l_1}C_{l_3}\right>\\
 &=&
  \nonumber
  (-1)^{l_1+l_3}\sqrt{(2l_1+1)(2l_3+1)}\left<C_{l_1}\right>\left<C_{l_3}\right>
  + 2\left<C_{l_1}\right>^2 \delta_{l_1l_3}
      +
      \left<T^{l_1l_1}_{l_3l_3}(0)\right>_{\rm c}\\   
  &=&
  \left<T^{l_1l_1}_{l_3l_3}(0)\right>_{\rm disconnected}
  +
  \left<T^{l_1l_1}_{l_3l_3}(0)\right>_{\rm c}.
\end{eqnarray}
Hence, $T^{l_1l_1}_{l_3l_3}(0)$ contains information not only 
on the disconnected trispectrum, but also of the connected one.

Unfortunately, we cannot measure the connected part of 
$T^{l_1l_1}_{l_3l_3}(0)$ directly
from the angular trispectrum because of the following reason.
To measure the connected terms, we have to subtract the
disconnected terms from the measured trispectrum first.
Since we are never able to measure the ensemble average of the 
disconnected terms (Eq.~(\ref{eq:unco})), 
we estimate them by using the estimated power--spectrum, $C_l$.
If we subtract the estimated disconnected terms,
$\propto C_{l_1}C_{l_3}$, from measured $T^{l_1l_1}_{l_3l_3}(0)$,
then it follows from Eq.~(\ref{eq:special}) that 
$T^{l_1l_1}_{l_3l_3}(0)$ vanishes {\it exactly}: 
$T^{l_1l_1}_{l_3l_3}(0)=0$; thus, 
$T^{l_1l_1}_{l_3l_3}(0)$ has no statistical power of measuring the 
connected terms.

For practical measurement of the angular trispectrum, we rewrite
the trispectrum estimator given by Eq.~(\ref{eq:test}) with
the azimuthally averaged harmonic transform, 
$e_{l}(\hat{\mathbf n})$ (Eq.~(\ref{eq:el})).
We find that the following form is particularly computationally efficient:
\begin{equation}
  \label{eq:tobs*}
   T^{l_1l_2}_{l_3l_4}(L)
   =
   \frac1{2L+1} \sum_{M=-L}^{L} t_{LM}^{l_1l_2*} t_{LM}^{l_3l_4},
\end{equation}
where $t_{LM}^{l_1l_2}$ is given by
\begin{equation}
 \label{eq:tLM}
  t_{LM}^{l_1l_2}
  \equiv
  \sqrt{\frac{2L+1}{4\pi}}  
  \left(
   \begin{array}{ccc}
    l_1 & l_2 & L \\ 0 & 0 & 0 
   \end{array}
 \right)^{-1} 
  \int d^2\hat{\mathbf n}
  \left[e_{l_1}(\hat{\mathbf n}) e_{l_2}(\hat{\mathbf n})\right]
  Y_{LM}^*(\hat{\mathbf n}).
\end{equation}
Since $t_{LM}^{l_1l_2}$ is the harmonic transform on the full sky, 
we can calculate it quickly. 
This method makes measurement of the angular trispectrum computationally
feasible even for the {\sl WMAP} data in which we have more than millions of
pixels; thus, the methods developed here can be applied not only to the 
{\sl COBE} DMR data, but also to the {\sl WMAP} data.

Let us calculate the covariance of the trispectrum in the weakly non--Gaussian
limit. Since the trispectrum covariance comprises eight $a_{lm}$'s,
the total number of terms is $_8C_2\cdot{}_6C_2\cdot{}_4C_2/4!=105$
according to the Wick's theorem.
The full calculation will be a nightmare, for we have to deal with
\begin{eqnarray}
 \nonumber
 & &\left<T^{l_1l_2}_{l_3l_4}(L)T^{l_1'l_2'}_{l_3'l_4'}(L')\right>\\
 \nonumber
 &=& (2L+1)(2L'+1)\sum_{{\rm all}~mm'}\sum_{MM'}(-1)^{M+M'}\\
 \nonumber
  & &\times
  \left(\begin{array}{ccc}l_1&l_2&L\\m_1&m_2&M\end{array}\right)
  \left(\begin{array}{ccc}l_3&l_4&L\\m_3&m_4&-M\end{array}\right)
  \left(\begin{array}{ccc}l_1'&l_2'&L'\\m_1'&m_2'&M'\end{array}\right)
  \left(\begin{array}{ccc}l_3'&l_4'&L'\\m_3'&m_4'&-M'\end{array}\right)\\
 \label{eq:error4}
  & &\times \left<
   a_{l_1m_1}a_{l_2m_2}a_{l_3m_3}a_{l_4m_4}
   a^*_{l_1'm_1'}a^*_{l_2'm_2'}a^*_{l_3'm_3'}a^*_{l_4'm_4'}
   \right>.
\end{eqnarray}
One can reduce this intricate expression to much a simpler
form for some particular configurations. 
For $L,L'\neq 0$ terms, thanks to the identity 
(\ref{eq:helpidentity}), only $4!=24$ terms that do not include 
$\left<a_{l_im_i}a_{l_jm_j}\right>$ but include only 
$\left<a_{l_im_i}a_{l_jm_j}^*\right>$ are non--vanishing.
For $L=L'=0$ terms, the triangle conditions in a quadrilateral
demand $l_1=l_2$ and $l_3=l_4$ (see Figure~\ref{fig:quad}).
As we have shown, these configurations have no statistical power of 
measuring the connected trispectrum of interest.
Hence, we evaluate $L,L'\neq 0$ terms in the following.

Evaluating the 24 $L,L'\neq 0$ terms is still a headache; however, for  
$l_1\leq l_2 < l_3\leq l_4$, we have only 8 terms left:
\begin{eqnarray}
 \nonumber
 & &\frac{\left<T^{l_1l_2}_{l_3l_4}(L)T^{l_1'l_2'}_{l_3'l_4'}(L')\right>}
 {(2L+1)\left<C_{l_1}\right>\left<C_{l_2}\right>\left<C_{l_3}\right>
  \left<C_{l_4}\right>}\\
 &=&
  \nonumber
  \delta_{LL'}
  \left[\delta^{l_1l_2l_3l_4}_{l_1'l_2'l_3'l_4'}
       +\delta^{l_1l_2l_3l_4}_{l_3'l_4'l_1'l_2'}
       + (-1)^{l_1+l_2+l_3+l_4}\left(
				\delta^{l_1l_2l_3l_4}_{l_2'l_1'l_4'l_3'}
				+\delta^{l_1l_2l_3l_4}_{l_4'l_3'l_2'l_1'}
                               \right)\right. \\
 \label{eq:error4*}
   & &\left. + (-1)^{l_1+l_2+L}\left(
		      \delta^{l_1l_2l_3l_4}_{l_2'l_1'l_3'l_4'}
		      +\delta^{l_1l_2l_3l_4}_{l_4'l_3'l_1'l_2'}
		    \right)
   + (-1)^{l_3+l_4+L}\left(
		    \delta^{l_1l_2l_3l_4}_{l_1'l_2'l_4'l_3'}
		    +\delta^{l_1l_2l_3l_4}_{l_3'l_4'l_2'l_1'}
		    \right)\right],
\end{eqnarray}
where $\delta_{l_1l_2l_3l_4}^{l_1'l_2'l_3'l_4'}\equiv
\delta_{l_1l_1'}\delta_{l_2l_2'}\delta_{l_3l_3'}\delta_{l_4l_4'}$, and so on.
Using parity invariance, $l_1+l_2+L={\rm even}$ and 
$l_3+l_4+L={\rm even}$, one finds the covariance matrix diagonal. 
Thus, the diagonal terms for $L\neq 0$ and $l_1\leq l_2<l_3\leq l_4$ are
simplified very much as
\begin{equation}
 \label{eq:variance4}
  {\left<\left[T^{l_1l_2}_{l_3l_4}(L)\right]^2\right>} 
  = {(2L+1)\left<C_{l_1}\right>\left<C_{l_2}\right>\left<C_{l_3}\right>
  \left<C_{l_4}\right>}
  \left(
   1 + \delta_{l_1l_2} + \delta_{l_3l_4} + \delta_{l_1l_2}\delta_{l_3l_4}
   \right).
\end{equation}
This result is strictly correct only on the full sky;
the incomplete sky makes the variance distribution much more scattered.
Figure~\ref{fig:variance_tl_bare_c1} plots the variance on the full sky 
as well as on the incomplete sky.

\begin{figure}
 \begin{center}
  \leavevmode\epsfxsize=9cm \epsfbox{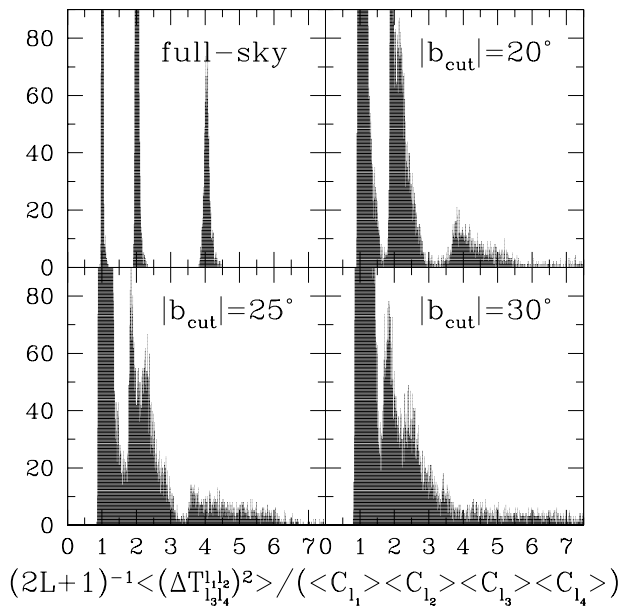}
 \end{center}
 \caption
 {\bf Variance of Angular Trispectrum I} 
 \mycaption{Histograms of variance of the angular trispectrum for
 $L\neq 0$ and $l_1\leq l_2<l_3\leq l_4$, for which the disconnected 
 terms vanish on the full sky.
 There are 16,554 modes, up to a maximum multipole of 20. 
 The meaning of the panels is the same as in Figure~\ref{fig:variance_bare}.}
\label{fig:variance_tl_bare_c1}
\end{figure}

For the rest of configurations for which the disconnected terms vanish, 
$L\neq 0$, $l_2=l_3$, and $l_1\neq l_4$, 
the covariance matrix is no longer diagonal in $L,L'$ \cite{hu:2001}.
Figure~\ref{fig:variance_tl_bare_c2} plots the numerically evaluated 
variance on the full sky as well as on the incomplete sky.
The variance divided by
${(2L+1)\left<C_{l_1}\right>\left<C_{l_2}\right>\left<C_{l_3}\right>
\left<C_{l_4}\right>}$ is no longer an integer, but
more scattered than that for $L\neq 0$ and $l_1\leq l_2<l_3\leq l_4$ 
even on the full sky.

\begin{figure}
 \begin{center}
  \leavevmode\epsfxsize=9cm \epsfbox{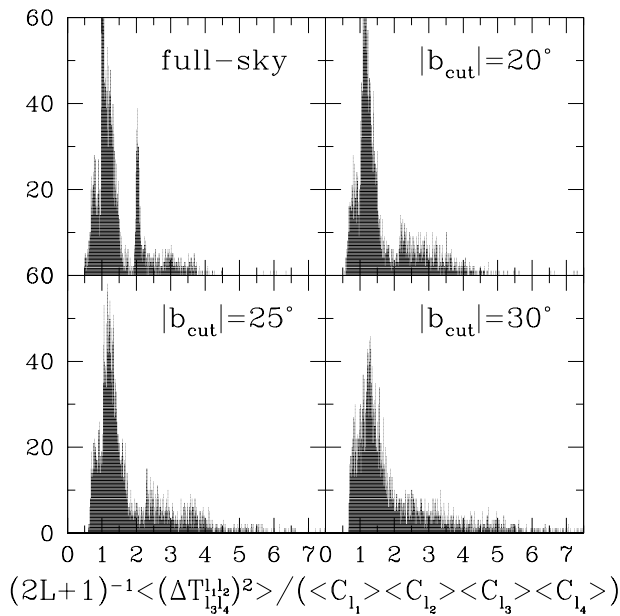}
 \end{center}
 \caption
 {\bf Variance of Angular Trispectrum II} 
 \mycaption{Histograms of variance of the angular trispectrum
 for $L\neq 0$, $l_2=l_3$, and $l_1\neq l_4$,
 for which the disconnected terms vanish on the full sky.
 There are 4,059 modes, up to a maximum multipole of 20.
 The meaning of the panels is the same as in Figure~\ref{fig:variance_bare}.}
\label{fig:variance_tl_bare_c2}
\end{figure}

\subsubsection{Power--spectrum and bispectrum on the incomplete sky}

Incomplete sky coverage destroys the orthonormality of the 
spherical harmonics on the sky.
The degree to which orthonormality is broken is often
characterized by the coupling integral~\cite{peebles:1980},
\begin{equation}
 \label{eq:coupling}
  W_{ll'mm'}
  \equiv
  \int
  d^2\hat{\mathbf n}~
  W(\hat{\mathbf n})
  Y_{lm}^*\left(\hat{\mathbf n}\right)
  Y_{l'm'}\left(\hat{\mathbf n}\right)
  =
  \int_{\Omega_{\rm obs}} 
  d^2\hat{\mathbf n}~
  Y_{lm}^*\left(\hat{\mathbf n}\right)
  Y_{l'm'}\left(\hat{\mathbf n}\right),
\end{equation}
where $W(\hat{\mathbf n})$ is zero in a cut region otherwise 1, and 
$\Omega_{\rm obs}$ denotes the solid angle of the observed sky.
When $W_{ll'mm'}\neq \delta_{ll'}\delta_{mm'}$,
the measured harmonic transform of the temperature 
anisotropy field, $a_{lm}$, becomes a 
{\it biased} estimator of the true harmonic transform, 
$a_{lm}^{\rm true}$, through
\begin{equation}
 \label{eq:alm_obs}
  a_{lm}=
  \sum_{l'=0}^\infty\sum_{m'=-l'}^{l'}a_{l'm'}^{\rm true} W_{ll'mm'}.
\end{equation}
Hence, we must correct our estimators of the power--spectrum 
and the bispectrum for the bias arising from incomplete sky coverage.

First, we derive a relationship between the angular power--spectrum
on the incomplete sky and that on the full sky. 
Taking the ensemble average of the estimator of the power--spectrum,
the pseudo-$C_l$ \cite{wandelt/hivon/gorski:2001},
$C_l=(2l+1)^{-1}\sum_m\left|a_{lm}\right|^2$,
we have
\begin{eqnarray}
 \nonumber
 \left<C_l\right>
  &=& \frac1{2l+1}\sum_{l'}
  C_{l'}^{\rm true}
  \sum_{mm'}
  \left|W_{ll'mm'}\right|^2\\
 \nonumber
  &\approx&
  \frac1{2l+1}
  C_{l}^{\rm true}
  \sum_{m}
  \sum_{l'm'}
  \int 
  {d^2\hat{\mathbf n}}~
  W(\hat{\mathbf n})
  Y_{lm}^*\left(\hat{\mathbf n}\right)
  Y_{l'm'}\left(\hat{\mathbf n}\right)
    \int 
  {d^2\hat{\mathbf m}}~
  W(\hat{\mathbf m})
  Y_{lm}\left(\hat{\mathbf m}\right)
  Y_{l'm'}^*\left(\hat{\mathbf m}\right)\\
 \nonumber
  &=&
  \frac1{2l+1}
  C_{l}^{\rm true}
  \sum_m
  \int 
  {d^2\hat{\mathbf n}}~
  W(\hat{\mathbf n})
  Y_{lm}^*\left(\hat{\mathbf n}\right)
  \int 
  {d^2\hat{\mathbf m}}~
  W(\hat{\mathbf m})
  Y_{lm}\left(\hat{\mathbf m}\right)
  \delta^{(2)}\left(\hat{\mathbf n}-\hat{\mathbf m}\right)\\
 \nonumber
  &=&
  C_{l}^{\rm true}
  \int
  \frac{d^2\hat{\mathbf n}}{4\pi}~
  W(\hat{\mathbf n})
  P_{l}\left(1\right)\\
 &=&
  C_{l}^{\rm true}\frac{\Omega_{\rm obs}}{4\pi}.
\end{eqnarray}
In the second equality, we have taken $C_{l'}^{\rm true}$ out of the 
summation over $l'$,
as $\left|W_{ll'mm'}\right|^2$ peaks very sharply at $l=l'$, and
$C_{l'}^{\rm true}$ varies much more slowly than 
$\left|W_{ll'mm'}\right|^2$ in $l'$.
This approximation is good for nearly full sky coverage.
In the third equality, we have used 
$\sum_{l'm'}Y_{l'm'}\left(\hat{\mathbf n}\right)
Y_{l'm'}^*\left(\hat{\mathbf m}\right)=
\delta^{(2)}\left(\hat{\mathbf n}-\hat{\mathbf m}\right)$.
In the fourth equality, we have used 
$\sum_m Y^*_{lm}\left(\hat{\mathbf n}\right)
Y_{lm}\left(\hat{\mathbf m}\right)= 
\frac{2l+1}{4\pi}P_l(\hat{\mathbf n}\cdot\hat{\mathbf m})$.
The result indicates that the bias amounts approximately to a 
fraction of the sky covered by observations.

Next, we derive a relationship between the angular bispectrum on the 
incomplete sky and that on the full sky. 
We begin with 
\begin{equation}
  \left<a_{l_1m_1}a_{l_2m_2}a_{l_3m_3}\right>
 = 
 \sum_{{\rm all}~l'm'}
 \left<a^{\rm true}_{l_1'm_1'}a^{\rm true}_{l_2'm_2'}
  a^{\rm true}_{l_3'm_3'}\right>
  W_{l_1l_1'm_1m_1'}W_{l_2l_2'm_2m_2'}W_{l_3l_3'm_3m_3'}.
\end{equation}
Rotational and parity invariance of the bispectrum 
implies that the bispectrum is given by
\begin{equation}
  \left<a_{l_1m_1}a_{l_2m_2}a_{l_3m_3}\right>
   =
   b_{l_1l_2l_3}
   \int
  d^2\hat{\mathbf n}~
  Y_{l_1m_1}^*\left(\hat{\mathbf n}\right)
  Y_{l_2m_2}^*\left(\hat{\mathbf n}\right)
  Y_{l_3m_3}^*\left(\hat{\mathbf n}\right),
\end{equation}
where $b_{l_1l_2l_3}$ is an arbitrary real symmetric function,
which is related to the angular averaged bispectrum, $B_{l_1l_2l_3}$.
When $b^{\rm true}_{l_1l_2l_3}$ varies much more slowly than 
the coupling integral, we obtain
\begin{eqnarray}
 \left<a_{l_1m_1}a_{l_2m_2}a_{l_3m_3}\right>
 &=&
  \nonumber
  \sum_{{\rm all}~l'}
  b_{l_1'l_2'l_3'}^{\rm true}
  \sum_{{\rm all}~m'}
  \int
  d^2\hat{\mathbf n}~
  Y_{l'_1m'_1}^*\left(\hat{\mathbf n}\right)
  Y_{l'_2m'_2}^*\left(\hat{\mathbf n}\right)
  Y_{l'_3m'_3}^*\left(\hat{\mathbf n}\right) \\
 \nonumber
 & &\times
 \int  d^2\hat{\mathbf n}_1~
  W(\hat{\mathbf n}_1)
  Y_{l_1'm_1'}\left(\hat{\mathbf n}_1\right)
  Y_{l_1m_1}^*\left(\hat{\mathbf n}_1\right) \\
 \nonumber
 & &\times
  \int  d^2\hat{\mathbf n}_2~
  W(\hat{\mathbf n}_2)
  Y_{l_2'm_2'}\left(\hat{\mathbf n}_2\right)
  Y_{l_2m_2}^*\left(\hat{\mathbf n}_2\right) \\
 \nonumber
 & &\times
  \int  d^2\hat{\mathbf n}_3~
  W(\hat{\mathbf n}_3)
  Y_{l_3'm_3'}\left(\hat{\mathbf n}_3\right)
  Y_{l_3m_3}^*\left(\hat{\mathbf n}_3\right)\\
 \label{eq:bl_obs}
  &\approx&
   b_{l_1l_2l_3}^{\rm true}
   \int  d^2\hat{\mathbf n}~
  W(\hat{\mathbf n})
  Y_{l_1m_1}^*\left(\hat{\mathbf n}\right)
  Y_{l_2m_2}^*\left(\hat{\mathbf n}\right)
  Y_{l_3m_3}^*\left(\hat{\mathbf n}\right).
\end{eqnarray}
Then, we calculate the angular averaged bispectrum, $B_{l_1l_2l_3}$
(Eq.~(\ref{eq:blll*})). 
By convolving Eq.~(\ref{eq:best}) with the Wigner--3$j$ symbol
and using the identity Eq.~(\ref{eq:gaunt_spec}),
we obtain
\begin{eqnarray}
 \nonumber
 \left<B_{l_1l_2l_3}\right>
 &\approx&
 b_{l_1l_2l_3}^{\rm true}
 \sqrt{\frac{4\pi}{(2l_1+1)(2l_2+1)(2l_3+1)}}
   \left(\begin{array}{ccc}l_1&l_2&l_3\\0&0&0\end{array}\right)^{-1}\\
 \nonumber
  & &\times
 \sum_{{\rm all}~m}
  \int
  d^2\hat{\mathbf m}~
  Y_{l_1m_1}\left(\hat{\mathbf m}\right)
  Y_{l_2m_2}\left(\hat{\mathbf m}\right)
  Y_{l_3m_3}\left(\hat{\mathbf m}\right) \\
 & &\times 
 \nonumber
  \int
  d^2\hat{\mathbf n}~
  W(\hat{\mathbf n})
  Y_{l_1m_1}^*\left(\hat{\mathbf n}\right)
  Y_{l_2m_2}^*\left(\hat{\mathbf n}\right)
  Y_{l_3m_3}^*\left(\hat{\mathbf n}\right)\\
 \nonumber
 &=&
 b_{l_1l_2l_3}^{\rm true}
 \sqrt{\frac{(2l_1+1)(2l_2+1)(2l_3+1)}{4\pi}}
   \left(\begin{array}{ccc}l_1&l_2&l_3\\0&0&0\end{array}\right)^{-1}\\
 \nonumber
  & &\times
  \int  \frac{d^2\hat{\mathbf m}}{4\pi}
  \int  \frac{d^2\hat{\mathbf n}}{4\pi}~
  W(\hat{\mathbf n})
  P_{l_1}\left(\hat{\mathbf m}\cdot\hat{\mathbf n}\right)
  P_{l_2}\left(\hat{\mathbf m}\cdot\hat{\mathbf n}\right)
  P_{l_3}\left(\hat{\mathbf m}\cdot\hat{\mathbf n}\right)\\
 \nonumber
 &=&
  b_{l_1l_2l_3}^{\rm true}
  \sqrt{\frac{(2l_1+1)(2l_2+1)(2l_3+1)}{4\pi}}
	\left(\begin{array}{ccc}l_1&l_2&l_3\\0&0&0\end{array}\right)
	\frac{\Omega_{\rm obs}}{4\pi}\\
 &=&
  \label{eq:biasbl}
  B_{l_1l_2l_3}^{\rm true}\frac{\Omega_{\rm obs}}{4\pi},
\end{eqnarray}
where we have used the identity,
\begin{equation}
 \int_{-1}^{1}\frac{dx}2~
  P_{l_1}(x)P_{l_2}(x)P_{l_3}(x)
  =
  \left(\begin{array}{ccc}l_1&l_2&l_3\\0&0&0\end{array}\right)^2.
\end{equation}
Thus, the bias for the angular bispectrum on the incomplete sky is also 
approximately given by a fraction of the sky covered by observations.

\subsection{Theoretical predictions for the CMB bispectrum from inflation}
\label{sec:theory_bl}

In this section, we derive analytical predictions for the angular bispectrum
from inflation.
We expand the observed CMB temperature fluctuation field, 
$\Delta T(\hat{\mathbf n})/T$, into the spherical harmonics,
\begin{equation}
  a_{lm}= \int d^2\hat{\mathbf n}\frac{\Delta T(\hat{\mathbf n})}{T}
  Y_{lm}^*(\hat{\mathbf n}),
\end{equation}
where the hats denote unit vectors. 
The CMB angular bispectrum is given by 
\begin{equation}
  \label{eq:blllmmm}
  B_{l_1l_2l_3}^{m_1m_2m_3}\equiv 
  \left<a_{l_1m_1}a_{l_2m_2}a_{l_3m_3}\right>,
\end{equation}
and the angular averaged bispectrum is (Eq.~(\ref{eq:best}))
\begin{equation}
  \label{eq:blll}
  B_{l_1l_2l_3}= \sum_{{\rm all}~m}
  \left(
  \begin{array}{ccc}
  l_1&l_2&l_3\\
  m_1&m_2&m_3
  \end{array}
  \right)
  B_{l_1l_2l_3}^{m_1m_2m_3}, 
\end{equation}
where the matrix is the Wigner--$3j$ symbol (see Appendix~\ref{app:wigner}).
The bispectrum, $B_{l_1l_2l_3}^{m_1m_2m_3}$,
satisfies the triangle conditions and parity invariance:
$m_1+m_2+m_3=0$, $l_1+l_2+l_3={\rm even}$, and 
$\left|l_i-l_j\right|\leq l_k \leq l_i+l_j$ for all permutations
of indices. 
It implies that $B_{l_1l_2l_3}^{m_1m_2m_3}$ consists of the Gaunt integral,  
${\mathcal G}_{l_1l_2l_3}^{m_1m_2m_3}$, defined by
\begin{eqnarray}
  \nonumber
  {\mathcal G}_{l_1l_2l_3}^{m_1m_2m_3}
  &\equiv&
  \int d^2\hat{\mathbf n}
  Y_{l_1m_1}(\hat{\mathbf n})
  Y_{l_2m_2}(\hat{\mathbf n})
  Y_{l_3m_3}(\hat{\mathbf n})\\
  \label{eq:gaunt}
  &=&\sqrt{
   \frac{\left(2l_1+1\right)\left(2l_2+1\right)\left(2l_3+1\right)}
        {4\pi}
        }
  \left(
  \begin{array}{ccc}
  l_1 & l_2 & l_3 \\ 0 & 0 & 0 
  \end{array}
  \right)
  \left(
  \begin{array}{ccc}
  l_1 & l_2 & l_3 \\ m_1 & m_2 & m_3 
  \end{array}
  \right).
\end{eqnarray}
${\mathcal G}_{l_1l_2l_3}^{m_1m_2m_3}$ is real, and satisfies 
all the conditions mentioned above.

Rotational invariance of the angular three--point correlation
function implies that $B_{l_1l_2l_3}$ is written as
\begin{equation}
  \label{eq:func}
  B_{l_1l_2l_3}^{m_1m_2m_3}
  ={\mathcal G}_{l_1l_2l_3}^{m_1m_2m_3}b_{l_1l_2l_3}, 
\end{equation}
where $b_{l_1l_2l_3}$ is an arbitrary real symmetric function 
of $l_1$, $l_2$, and $l_3$.
This form, Eq.~(\ref{eq:func}), is necessary and
sufficient to construct generic $B_{l_1l_2l_3}^{m_1m_2m_3}$ under 
rotational invariance; thus, we will use $b_{l_1l_2l_3}$ more frequently than
$B_{l_1l_2l_3}^{m_1m_2m_3}$ in this section, and call this function
the {\it reduced} bispectrum, as $b_{l_1l_2l_3}$ 
contains all physical information in $B_{l_1l_2l_3}^{m_1m_2m_3}$.
Since the reduced bispectrum does not contain the Wigner--$3j$ symbol,
which merely ensures the triangle conditions and parity invariance, 
it is easier to calculate physical properties of the bispectrum.

We calculate the angular averaged bispectrum, $B_{l_1l_2l_3}$,
by substituting Eq.~(\ref{eq:func}) into Eq.~(\ref{eq:blll}),
\begin{equation}
  \label{eq:wigner*}
  B_{l_1l_2l_3}
  =
  \sqrt{\frac{(2l_1+1)(2l_2+1)(2l_3+1)}{4\pi}}
  \left(
  \begin{array}{ccc}
  l_1&l_2&l_3\\
  0&0&0
  \end{array}
  \right)b_{l_1l_2l_3},
\end{equation}
where we have used the identity,
\begin{equation}
  \label{eq:wigner}
  \sum_{{\rm all}~m}
  \left(
  \begin{array}{ccc}
  l_1&l_2&l_3\\
  m_1&m_2&m_3
  \end{array}
  \right)
  {\mathcal G}_{l_1l_2l_3}^{m_1m_2m_3}
  =
  \sqrt{\frac{(2l_1+1)(2l_2+1)(2l_3+1)}{4\pi}}
  \left(
  \begin{array}{ccc}
  l_1&l_2&l_3\\
  0&0&0
  \end{array}
  \right).
\end{equation}

Alternatively, one can define the bispectrum in the flat--sky
approximation, 
\begin{equation}
 \label{eq:smallangle}
  \left<a({\mathbf l}_1)a({\mathbf l}_1)a({\mathbf l}_3)\right>
  =(2\pi)^2\delta^{(2)}\left({\mathbf l}_1+{\mathbf l}_2+{\mathbf l}_3\right)
  B({\mathbf l}_1,{\mathbf l}_2,{\mathbf l}_3),
\end{equation}
where ${\mathbf l}$ is a two--dimensional wave vector on the sky.
This definition of $B({\mathbf l}_1,{\mathbf l}_2,{\mathbf l}_3)$ 
reduces to Eq.~(\ref{eq:func}) with the correspondence
\begin{equation}
{\mathcal G}_{l_1l_2l_3}^{m_1m_2m_3}\rightarrow 
(2\pi)^2\delta^{(2)}\left({\mathbf l}_1+{\mathbf l}_2+{\mathbf l}_3\right) \;,
\end{equation}
in the flat--sky limit~\cite{hu:2000}. 
Thus, we have
\begin{equation}
 \label{eq:smallangle*}
  b_{l_1l_2l_3}\approx
  B({\mathbf l}_1,{\mathbf l}_2,{\mathbf l}_3)
  \qquad \mbox{(flat--sky approximation)}.
\end{equation}
This fact motivates our use of the reduced bispectrum, 
$b_{l_1l_2l_3}$, rather than the angular averaged bispectrum, $B_{l_1l_2l_3}$.
Note that $b_{l_1l_2l_3}$ is similar to $\hat{B}_{l_1l_2l_3}$ defined in
Ref.~\cite{magueijo:2000}; the relation is 
$b_{l_1l_2l_3}=\sqrt{4\pi}\hat{B}_{l_1l_2l_3}$.

If primordial fluctuations are adiabatic scalar fluctuations, then
\begin{equation}
  \label{eq:almphi}
  a_{lm}=4\pi(-i)^l
  \int\frac{d^3{\mathbf k}}{(2\pi)^3}\Phi({\mathbf k})g_{{\rm T}l}(k)
  Y_{lm}^*(\hat{\mathbf k}),
\end{equation}
where $\Phi({\mathbf k})$ is the primordial curvature perturbation
in Fourier space, and $g_{{\rm T}l}(k)$ is the radiation transfer function.
$a_{lm}$ takes over the non--Gaussianity, if any, from $\Phi({\mathbf k})$.
Although Eq.~(\ref{eq:almphi}) is valid only if the
Universe is flat, it is straightforward to extend this
to an arbitrary geometry.
We can calculate the isocurvature fluctuations similarly by
using the entropy perturbation and the proper transfer function.

As it has been shown in Sec.~\ref{Phinonline}, the primordial non--Gaussianity 
may be parameterized as a linear plus quadratic term in the gravitational 
potential in the general form of Eq.~(\ref{phimomspace}), where the 
non--linearity parameter $f_{\rm NL}$ appears as a kernel in Fourier space, 
rather than a constant. This gives rise to an angular modulation of the 
quadratic non--linearity, which might be used to search for specific 
signatures of inflationary non--Gaussianity in the CMB~\cite{LMR}.      
In this section, however, we restrict ourselves to the simplest weak 
non--linear coupling case, assuming that $f_{\rm NL}$ is merely 
a multiplicative constant, as done in data analyses so far. Hence we write
\begin{equation}
  \label{eq:modelreal}
  \Phi({\mathbf x})
 =\Phi_{\rm L}({\mathbf x})
 +f_{\rm NL}\left[
              \Phi^2_{\rm L}({\mathbf x})-
	      \left<\Phi^2_{\rm L}({\mathbf x})\right>
        \right],
\end{equation}
in real space, where $\Phi_{\rm L}({\mathbf x})$ denotes the linear Gaussian 
part of the perturbation, and $\left<\Phi({\mathbf x})\right>=0$ is guaranteed.

In Fourier space, we decompose $\Phi({\mathbf k})$ into two parts,
\begin{equation}
  \label{eq:model}
  \Phi({\mathbf k})=\Phi_{\rm L}({\mathbf k})+\Phi_{\rm NL}({\mathbf k}),
\end{equation}
and accordingly we have
\begin{equation}
  \label{eq:model*}
  a_{lm}=a_{lm}^{\rm L}+a_{lm}^{\rm NL},
\end{equation}
where $\Phi_{\rm NL}({\mathbf k})$ is a non--linear curvature
perturbation defined by
\begin{equation}
  \label{eq:nonlinear}
  \Phi_{\rm NL}({\mathbf k})\equiv 
  f_{\rm NL}
  \left[
  \int \frac{d^3{\mathbf p}}{(2\pi)^3}
  \Phi_{\rm L}({\mathbf k}+{\mathbf p})\Phi^*_{\rm L}({\mathbf p})
  -(2\pi)^3\delta^{(3)}({\mathbf k})\left<\Phi^2_{\rm L}({\mathbf x})\right>
  \right].
\end{equation}
One can immediately check that $\left<\Phi({\mathbf k})\right>=0$ is 
satisfied. In this model, a non--vanishing component of the 
$\Phi({\mathbf k})$--field bispectrum is
\begin{equation}
  \label{eq:phispec}
  \left<\Phi_{\rm L}({\mathbf k}_1)
	\Phi_{\rm L}({\mathbf k}_2)
	\Phi_{\rm NL}({\mathbf k}_3)\right>
  = (2\pi)^3\delta^{(3)}({\mathbf k}_1+{\mathbf k}_2+{\mathbf k}_3)
    \, \, 2 f_{\rm NL}P_\Phi(k_1)P_\Phi(k_2),
\end{equation}
where $P_\Phi(k)$ is Bardeen's potential 
linear power--spectrum given by 
\begin{equation}
 \left<\Phi_{\rm L}({\mathbf k}_1)\Phi_{\rm L}({\mathbf k}_2)\right>
  =(2\pi)^3P_\Phi(k_1)\delta^{(3)}({\mathbf k}_1+{\mathbf k}_2).
\end{equation}
We have also used 
\begin{equation}
 \left<\Phi_{\rm L}({\mathbf k}+{\mathbf p})\Phi^*_{\rm L}({\mathbf p})\right>
  =(2\pi)^3P_\Phi(p)\delta^{(3)}({\mathbf k}),
\end{equation}
and 
\begin{equation}
 \left<\Phi^2_{\rm L}({\mathbf x})\right>
  =(2\pi)^{-3}\int d^3{\mathbf k} P_\Phi(k).
\end{equation}

Substituting Eq.~(\ref{eq:almphi}) into Eq.~(\ref{eq:blllmmm}),
using Eq.~(\ref{eq:phispec}) for the $\Phi({\mathbf k})$--field
bispectrum, and then integrating over angles 
$\hat{\mathbf k}_1$, $\hat{\mathbf k}_3$, and $\hat{\mathbf k}_3$, we obtain 
the primordial CMB angular bispectrum,
\begin{eqnarray}
 B_{l_1l_2l_3}^{m_1m_2m_3}
  \nonumber
  &=& 
  \left<a_{l_1m_1}^{\rm L}a_{l_2m_2}^{\rm L}a_{l_3m_3}^{\rm NL}\right>
  + \left<a_{l_1m_1}^{\rm L}a_{l_2m_2}^{\rm NL}a_{l_3m_3}^{\rm L}\right>
  + \left<a_{l_1m_1}^{\rm NL}a_{l_2m_2}^{\rm L}a_{l_3m_3}^{\rm L}\right>\\
 \nonumber
  &=& 2{\mathcal G}_{l_1l_2l_3}^{m_1m_2m_3}
	\int_0^\infty r^2 dr 
    \left[
          b^{\rm L}_{l_1}(r)b^{\rm L}_{l_2}(r)b^{\rm NL}_{l_3}(r)+
	  b^{\rm L}_{l_1}(r)b^{\rm NL}_{l_2}(r)b^{\rm L}_{l_3}(r)\right.\\
  \label{eq:almspec}
  & &\left.+
	  b^{\rm NL}_{l_1}(r)b^{\rm L}_{l_2}(r)b^{\rm L}_{l_3}(r)
    \right],
\end{eqnarray}
where 
\begin{eqnarray}
  \label{eq:bLr}
  b^{\rm L}_{l}(r) &\equiv&
  \frac2{\pi}\int_0^\infty k^2 dk P_\Phi(k)g_{{\rm T}l}(k)j_l(kr),\\
  \label{eq:bNLr}
  b^{\rm NL}_{l}(r) &\equiv&
  \frac2{\pi}\int_0^\infty k^2 dk f_{\rm NL}g_{{\rm T}l}(k)j_l(kr).
\end{eqnarray}
Note that $b^{\rm L}_{l}(r)$ is dimensionless, 
while $b^{\rm NL}_{l}(r)$ has a dimension of $L^{-3}$.

One can immediately check that Eq.~(\ref{eq:func}) holds; thus, the reduced 
bispectrum, $b_{l_1l_2l_3}$ (Eq.~(\ref{eq:func})), for the primordial 
non--Gaussianity reads
\begin{eqnarray}
 \nonumber
  b_{l_1l_2l_3}^{\rm prim}
  &=& 2\int_0^\infty r^2 dr
    \left[
          b^{\rm L}_{l_1}(r)b^{\rm L}_{l_2}(r)b^{\rm NL}_{l_3}(r)+
	  b^{\rm L}_{l_1}(r)b^{\rm NL}_{l_2}(r)b^{\rm L}_{l_3}(r)\right.\\
 \label{eq:blprim}
  & &\left.+ b^{\rm NL}_{l_1}(r)b^{\rm L}_{l_2}(r)b^{\rm L}_{l_3}(r)
    \right].
\end{eqnarray}
We can fully specify $b_{l_1l_2l_3}^{\rm prim}$ by a single constant
parameter, $f_{\rm NL}$, as the CMB angular power--spectrum, $C_l$, 
will precisely measure cosmological parameters.
We stress again that this formula is valid only when the scale--dependence of
$f_{\rm NL}$ is weak, which is a good approximation if the
momentum--independent part of $f_{\rm NL}$ is larger than unity. 

One can calculate the primordial CMB bispectrum 
(Eqs.(\ref{eq:almspec})--(\ref{eq:blprim})) numerically as follows.
One computes the full radiation transfer function, $g_{{\rm T}l}(k)$,
with the {\sf CMBFAST} code~\cite{seljak/zaldarriaga:1996}, assuming
a single power--law spectrum, $P_\Phi(k)\propto k^{n-4}$,
for the primordial curvature fluctuations.
After doing the integration over $k$ (Eqs.(\ref{eq:bLr}) and (\ref{eq:bNLr})) 
with the same algorithm of {\sf CMBFAST},
one performs the integration over $r$ (Eq.~(\ref{eq:blprim})),
$r=c\left(\tau_0-\tau\right)$, where $\tau$ is the conformal time.
$\tau_0$ is the present--day value.
In our model, $c\tau_0=11.8\ {\rm Gpc}$, and 
the decoupling occurs at $c\tau_*=235\ {\rm Mpc}$ 
at which the differential visibility has a maximum.
Our $c\tau_0$ includes radiation effects on the expansion of the
Universe; otherwise, $c\tau_0=12.0\ {\rm Gpc}$.
Since most of the primordial signal is generated at $\tau_*$,
we choose the $r$ integration boundary as 
$c\left(\tau_0-2\tau_*\right)\leq r\leq c\left(\tau_0-0.1\tau_*\right)$.
We use a step-size of $0.1c\tau_*$, as we have found that a step size of  
$0.01c\tau_*$ gives very similar results.
As cosmological model, let us assume a scale--invariant 
Standard Cold Dark Matter (SCDM) model with $\Omega_{\rm m}=1$, 
$\Omega_\Lambda =0$, $\Omega_{\rm b}=0.05$, $h=0.5$, and $n=1$, and with 
power--spectrum $P_\Phi(k)$ normalized to {\sl COBE} \cite{bunn/white:1997}.
Although this model is almost excluded by current observations,
it is still useful to depict the basic effects of the transfer function 
on the bispectrum (see also Ref.~\cite{LMR}). 

Figure~\ref{fig:bl} shows $b_l^{\rm L}(r)$ (Eq.~(\ref{eq:bLr})) and 
$b_l^{\rm NL}(r)$ (Eq.~(\ref{eq:bNLr})) for several different values of $r$.
We find that $b^{\rm L}_l(r)$ and $C_l$ look very similar to each other in 
shape and amplitude at $l\simgt 100$, 
although the amplitude in the Sachs--Wolfe regime is different by a 
factor of $-3$. 
This is because $C_l\propto P_\Phi(k)g_{{\rm T}l}^2(k)$,
while $b_l^{\rm L}(r)\propto P_\Phi(k)g_{{\rm T}l}(k)$, 
where $g_{{\rm T}l}=-1/3$.
We also find that $b_l^{\rm L}(r)$ has a good phase coherence over  a
wide range of $r$, while the phase of $b_l^{\rm NL}(r)$ in the 
high--$l$ regime oscillates rapidly as a function of $r$. 
This strongly damps the integrated result (Eq.~(\ref{eq:almspec})) 
in the high-$l$ regime.
The main difference between $C_l$ and $b_l(r)$ is that
$b_l(r)$ changes the sign, while $C_l$ does not.

\begin{figure}
 \begin{center}
  \leavevmode\epsfxsize=9cm \epsfbox{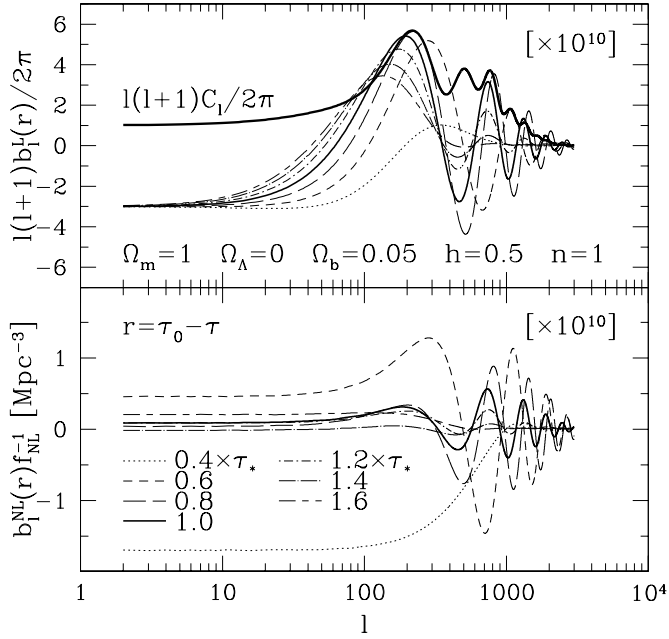}
 \end{center}
 \caption{{\bf Components of Primordial CMB Bispectrum}~\cite{ks}}
 \mycaption{
 This figure shows $b_l^{\rm L}(r)$ (Eq.~(\ref{eq:bLr})) and 
 $b_l^{\rm NL}(r)$ (Eq.~(\ref{eq:bNLr})), the two terms in our calculation of 
 the primordial CMB angular bispectrum, as a function of $r$. 
 Various lines in the top panel show 
 $\left[l(l+1)b_l^{\rm L}(r)/2\pi\right]\times 10^{10}$, where
 $r=c\left(\tau_0-\tau\right)$, at $\tau=0.4,0.6,0.8,1.0,1.2,1.4$, 
 and $1.6\times \tau_*$ (decoupling time);
 $\left[b_l^{\rm NL}(r)f^{-1}_{\rm NL}\right]\times 10^{10}$ 
 are shown in the bottom panel. 
 $\tau_0$ is the present-day conformal time.
 Note that $c\tau_0=11.8\ {\rm Gpc}$, and $c\tau_*=235\ {\rm Mpc}$ in our 
 cosmological model chosen here.
 The thickest solid line in the top panel is the CMB angular power--spectrum,
 $\left[l(l+1)C_l/2\pi\right]\times 10^{10}$.  
 $C_l$ is shown for comparison.}
\label{fig:bl}
\end{figure}

Looking at Figure~\ref{fig:bl}, we find
$l^2b_l^{\rm L}\sim 2\times 10^{-9}$ and
$b_l^{\rm NL}f^{-1}_{\rm NL}\sim 10^{-10}\ {\rm Mpc^{-3}}$.
As most of the signal is coming from the decoupling epoch, 
the volume element at $\tau_*$ is 
$r_*^2\Delta r_*\sim (10^4)^2\times 10^2\ {\rm Mpc^3}$;
thus, we can give an order--of--magnitude estimate of the primordial reduced
bispectrum (Eq.~(\ref{eq:blprim})) as 
\begin{equation}
  \label{eq:orderest}
  b_{lll}^{\rm prim}\sim 
  l^{-4}
  \left[2 r_*^2\Delta r_*\left(l^2b_l^{\rm L}\right)^2
   b_l^{\rm NL}\times 3\right]
  \sim l^{-4}\times 2\times 10^{-17}f_{\rm NL}. 
\end{equation}
Since $b_l^{\rm NL}f^{-1}_{\rm NL}\sim r_*^{-2}\delta(r-r_*)$ 
(see Eq.~(\ref{eq:deltadelta})),
$r_*^2\Delta r_* b_l^{\rm NL}f^{-1}_{\rm NL}\sim 1$.
This rough estimate agrees with the numerical result below 
(Figure~\ref{fig:bispectrum}).

Figure~\ref{fig:bispectrum} 
shows the integrated bispectrum (Eq.~(\ref{eq:almspec}))
divided by the Gaunt integral, ${\mathcal G}_{l_1l_2l_3}^{m_1m_2m_3}$,
which is the reduced bispectrum, $b_{l_1l_2l_3}^{\rm prim}$.
While the bispectrum is a 3--d function, we show different 1--d slices of the 
bispectrum in this figure.
We plot 
$$l_2(l_2+1)l_3(l_3+1)
\left<a_{l_1m_1}^{\rm NL}a_{l_2m_2}^{\rm L}a_{l_3m_3}^{\rm L}\right>
\left({\mathcal G}_{l_1l_2l_3}^{m_1m_2m_3}\right)^{-1}/(2\pi)^2$$ as a 
function of $l_3$ in the top panel, while we plot
$$l_1(l_1+1)l_2(l_2+1)
\left<a_{l_1m_1}^{\rm L}a_{l_2m_2}^{\rm L}a_{l_3m_3}^{\rm NL}\right>
\left({\mathcal G}_{l_1l_2l_3}^{m_1m_2m_3}\right)^{-1}/(2\pi)^2$$
in the bottom panel.
We have multiplied each $b_{l}^{\rm L}(r)$ which contains $P_\Phi(k)$
by $l(l+1)/(2\pi)$ so that the Sachs--Wolfe plateau at $l_3\simlt 10$ 
is easily seen.
We have chosen $l_1$ and $l_2$ so as $(l_1,l_2)=(9,11),(99,101),(199,201)$,
and $(499,501)$.
We find that the $(l_1,l_2)=(199,201)$ mode,  
the first acoustic peak mode, has the largest signal in this 
family of parameters.
The top panel has a prominent first acoustic peak, and strongly damped
oscillations in the high-$l$ regime; the bottom panel also has a first peak, 
but damps more slowly. The 
typical amplitude of the reduced bispectrum is
$l^4b^{\rm prim}_{lll}f^{-1}_{\rm NL}\sim 10^{-17}$, which agrees with
the order--of--magnitude estimate of Eq.~(\ref{eq:orderest}).

\begin{figure}
 \begin{center}
  \leavevmode\epsfxsize=9cm \epsfbox{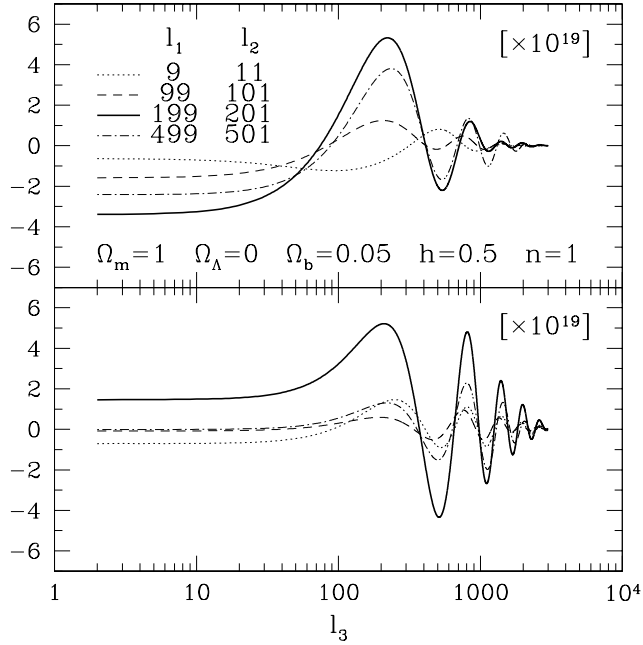}
 \end{center}
 \caption{{\bf Primordial CMB Bispectrum}~\cite{ks}}
 \mycaption{The primordial angular bispectrum (Eq.~(\ref{eq:almspec})), 
 divided by
 the Gaunt integral, ${\mathcal G}_{l_1l_2l_3}^{m_1m_2m_3}$ 
(Eq.~(\ref{eq:gaunt})).
 The bispectrum is plotted as a function of $l_3$ for 
 $(l_1,l_2)=$(9,11), (99,101), (199,201), and (499,501).
 Each panel plots a different 1-dimensional slice of the bispectrum. 
 The top panel plots 
 $l_2(l_2+1)l_3(l_3+1)\left<
 a_{l_1m_1}^{\rm NL} a_{l_2m_2}^{\rm L} 
a_{l_3m_3}^{\rm L}\right>f_{\rm NL}^{-1}
 \left({\mathcal G}_{l_1l_2l_3}^{m_1m_2m_3}\right)^{-1}/(2\pi)^2$,
 while the bottom panel plots
 $l_1(l_1+1)l_2(l_2+1)\left<
 a_{l_1m_1}^{\rm L} a_{l_2m_2}^{\rm L} 
a_{l_3m_3}^{\rm NL}\right>f_{\rm NL}^{-1}
 \left({\mathcal G}_{l_1l_2l_3}^{m_1m_2m_3}\right)^{-1}/(2\pi)^2$.
 Note that we have multiplied the bispectrum in each panel by a factor 
 of $10^{19}$.}
\label{fig:bispectrum}
\end{figure}

The formula in Eq.~(\ref{eq:blprim}) and numerical results agree with 
\cite{Getal} in the Sachs--Wolfe regime, where 
$g_{{\rm T}l}(k)\approx -j_l(kr_*)/3$, and
\begin{equation}
  \label{eq:SWapp}
  b_{l_1l_2l_3}^{\rm prim}
  \approx
  -6f_{\rm NL}
  \left(C_{l_1}^{\rm SW}C_{l_2}^{\rm SW}+
        C_{l_1}^{\rm SW}C_{l_3}^{\rm SW}+
        C_{l_2}^{\rm SW}C_{l_3}^{\rm SW}\right)
  \qquad
  \mbox{(SW approximation)}.
\end{equation}
Each term is of the same order as Eq.~(\ref{eq:blprim}).
Here, $C_l^{\rm SW}$ is the CMB angular power--spectrum in the Sachs--Wolfe
approximation,
\begin{equation}
  \label{eq:clsw}
  C_l^{\rm SW}
  \equiv
  \frac2{9\pi}\int_0^\infty k^2 dk P_\Phi(k)j^2_l(kr_*).
\end{equation}
In deriving Eq.~(\ref{eq:SWapp}) from Eq.~(\ref{eq:blprim}), 
we have approximated $b_l^{\rm NL}(r)$ (Eq.~(\ref{eq:bNLr})) with
\begin{equation}
  \label{eq:deltadelta}
  b_l^{\rm NL}(r)
  \approx
  \left(-\frac{f_{\rm NL}}3\right)
  \frac2{\pi}\int_0^\infty k^2 dk j_{l}(kr_*)j_l(kr)
  = -\frac{f_{\rm NL}}3 r_*^{-2}\delta(r-r_*).
\end{equation}

The Sachs--Wolfe approximation (Eq.~(\ref{eq:SWapp})) 
is valid only when $l_1$, $l_2$, and $l_3$ are all smaller than 
$\sim 10$, for which the authors of Ref.~\cite{Getal} give 
$\sim -6\times 10^{-20}$ in Figure~\ref{fig:bispectrum}.
We stress again that the Sachs--Wolfe approximation 
gives a qualitatively different result from our full calculation 
(Eq.~(\ref{eq:blprim})) at $l_i\simgt 10$. 
The full bispectrum changes sign, while the 
approximation never changes sign because of the use of $C_l^{\rm SW}$.
The acoustic oscillation and the sign--change 
are actually great advantages when we try
to separate the primordial bispectrum from various secondary bispectra.
We will analyze this point later.

As we have calculated the full bispectrum at all scales, it is now possible to
calculate the 3--point function in real space. Unlike the bispectrum, 
however, the form of the full 3--point function is fairly complicated; 
nevertheless, one can obtain a simple form for the skewness, $S_3$, given by
\begin{equation}
  S_3\equiv \left<\left(\frac{\Delta T(\hat{\mathbf n})}{T}\right)^3\right>,
\end{equation}
which is perhaps the simplest (but less powerful) statistic 
characterizing non--Gaussianity. 
We expand $S_3$ in terms of $B_{l_1l_2l_3}$ (Eq.~(\ref{eq:blll})),
or $b_{l_1l_2l_3}$ (Eq.~(\ref{eq:func})), as 
\begin{eqnarray}
  \nonumber
  S_3
  &=&
  \frac1{4\pi}\sum_{l_1l_2l_3}
  \sqrt{
  \frac{\left(2l_1+1\right)\left(2l_2+1\right)\left(2l_3+1\right)}
        {4\pi}
        }
  \left(
  \begin{array}{ccc}
  l_1 & l_2 & l_3 \\ 0 & 0 & 0 
  \end{array}
  \right)
   B_{l_1l_2l_3}
   W_{l_1}W_{l_2}W_{l_3}\\
 \nonumber
  &=&
  \frac1{2\pi^2}\sum_{2\leq l_1l_2l_3}
  \left(l_1+\frac12\right)\left(l_2+\frac12\right)\left(l_3+\frac12\right)
  \left(
  \begin{array}{ccc}
  l_1 & l_2 & l_3 \\ 0 & 0 & 0 
  \end{array}
  \right)^2\\
  \label{eq:skewness}
   & &\times
   b_{l_1l_2l_3}
   W_{l_1}W_{l_2}W_{l_3},
\end{eqnarray}
where $W_l$ is the experimental window function. 
We have used Eq.~(\ref{eq:wigner*}) to replace 
$B_{l_1l_2l_3}$ by the reduced bispectrum, $b_{l_1l_2l_3}$, 
in the last equality. 
Since $l=0$ and $1$ modes are not observable, we have excluded them 
from the summation.
Throughout this section, we consider a single-beam window function,
$W_l=e^{-l(l+1)/(2\sigma_{\rm b}^2)}$, where 
$\sigma_{\rm b}={\rm FWHM}/\sqrt{8\ln2}$.
Since $\left(\begin{array}{ccc}l_1&l_2&l_3\\0&0&0
\end{array}\right)^2 b_{l_1l_2l_3}$ is symmetric under permutation of
indices, we change the way of summation as
\begin{equation} 
  \label{eq:sumchange}
  \sum_{2\leq l_1l_2l_3}
  \longrightarrow
  6 \sum_{2\leq l_1\leq l_2\leq l_3}.
\end{equation}
This reduces the number of summations by a factor of $\simeq 6$.
We will use this convention henceforth.

The top panel of Figure~\ref{fig:skewness} plots $S_3(<l_3)$, 
which is $S_3$ summed up to a certain $l_3$, for FWHM beam sizes 
of $7^\circ$, $13'$, and $5'\hspace{-2.5pt}.5$. 
These values correspond to {\sl COBE}, {\sl WMAP}, and 
{\sl Planck} beam sizes, respectively. 
Figure~\ref{fig:skewness} also plots the infinitesimally thin beam case.
We find that {\sl WMAP}, {\sl Planck}, and the ideal experiments 
measure very similar $S_3$ to one another, despite the fact that 
{\sl Planck} and the ideal experiments can use many more modes
than {\sl WMAP}. 
The reason is as follows.
Looking at Eq.~(\ref{eq:skewness}), one finds that $S_3$
is a linear integral of $b_{l_1l_2l_3}$ over $l_i$; thus, integrating
oscillations in $b_{l_1l_2l_3}^{\rm prim}$ around zero 
(see Figure~\ref{fig:bispectrum}) damps the non--Gaussian signal on small 
angular scales, $l\simgt 300$.
Since the Sachs--Wolfe effect, implying no oscillation, 
dominates the {\sl COBE}--scale anisotropy, the cancellation on the 
{\sl COBE} scale affects $S_3$ less significantly than on the {\sl WMAP} 
and {\sl Planck} scales. 
{\sl Planck} suffers from severe cancellation in small angular scales:
{\sl Planck} and the ideal experiments measure only 
the same amount of $S_3$ as {\sl WMAP} does.
As a result, the measured $S_3$ almost saturates at the {\sl WMAP} resolution
scale, $l\sim 500$. 

\begin{figure}
 \begin{center}
  \leavevmode\epsfxsize=9cm \epsfbox{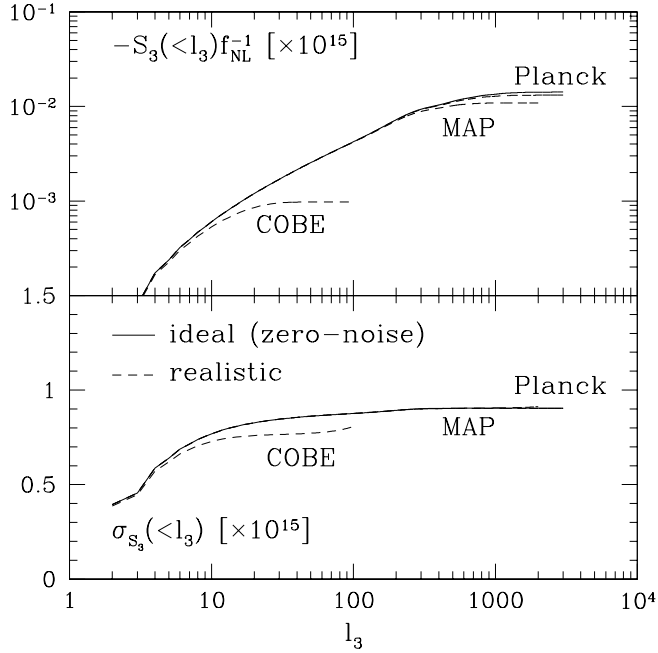}
 \end{center}
 \caption{{\bf Primordial Skewness}~\cite{ks}}
 \mycaption{The top panel shows the primordial CMB skewness 
 (Eq.~(\ref{eq:skewness})) summed up to a certain $l_3$, 
 $-S_3(<l_3)f_{\rm NL}^{-1}\times 10^{15}$.
 The bottom panel shows the error of $S_3$ (Eq.~(\ref{eq:skewvar})) 
 summed up to $l_3$, $\sigma_{S_3}(<l_3)\times 10^{15}$.
 The solid line represents the zero--noise ideal experiment, 
 while the dotted lines show
 {\sl COBE}, {\sl WMAP}, and {\sl Planck} experiments.}
\label{fig:skewness}
\end{figure}

We conclude this section by noting that 
when we can calculate the expected form of the bispectrum, then it becomes
a ``matched filter'' for detecting non--Gaussianity in the data,
and thus much more powerful a tool than the skewness in which the information
is lost through the coarse--graining. 

\subsection{Secondary sources of CMB bispectrum}
\label{sec:secondary}

Even if the CMB bispectrum were significantly detected in the CMB 
map, its origin would not necessarily be primordial, but rather there
would be various secondary sources such as the Sunyaev--Zel'dovich (SZ) 
effect~\cite{zeldovich/snyaev:1969}, the weak lensing effect, and so on, or
foreground sources such as extragalactic radio sources. 
To isolate the primordial origin from the others, we have to know the 
accurate form of bispectra produced by secondary and foreground sources.

\subsubsection{Coupling between the weak lensing and the 
Sunyaev--Zel'dovich effects}

The coupling between the SZ effect and the weak lensing
effect produces an observable effect in the bispectrum 
\cite{goldberg/spergel:1999,cooray/hu:2000}. 
We expand the CMB temperature field including the SZ and the lensing
effect as
\begin{eqnarray}
 \nonumber
  \frac{\Delta T(\hat{\mathbf n})}T
  &=& \frac{\Delta T^{\rm P}\left(
                          \hat{\mathbf n}+\nabla\Theta(\hat{\mathbf n})
                    \right)}T
   +\frac{\Delta T^{\rm SZ}(\hat{\mathbf n})}T\\
 \label{eq:lenscouple}
  &\approx& \frac{\Delta T^{\rm P}(\hat{\mathbf n})}T
  +\nabla\left(\frac{\Delta T^{\rm P}(\hat{\mathbf n})}T\right)\cdot
   \nabla\Theta(\hat{\mathbf n})
  +\frac{\Delta T^{\rm SZ}(\hat{\mathbf n})}T,
\end{eqnarray}
where P denotes the primordial anisotropy, 
$\Theta(\hat{\mathbf n})$ is the lensing potential,
\begin{equation}
 \Theta(\hat{\mathbf n})
  \equiv
  -2\int_0^{r_*} dr \frac{r_*-r}{rr_*}
  \Phi(r,\hat{\mathbf n}r),
\end{equation}
and SZ denotes the SZ effect,
\begin{equation}
  \label{eq:sz}
  \frac{\Delta T^{\rm SZ}(\hat{\mathbf n})}T
  =
  y(\hat{\mathbf n})j_\nu,
\end{equation}
where $j_\nu$ is a spectral function of the SZ effect 
\cite{zeldovich/snyaev:1969}.
$y(\hat{\mathbf n})$ is the Compton $y$-parameter given by
\begin{equation}
  \label{eq:yparam}
  y(\hat{\mathbf n}) \equiv y_0\int \frac{dr}{r_*} 
  \frac{T_\rho(r,\hat{\mathbf n}r)}{\overline{T}_{\rho0}} a^{-2}(r),
\end{equation}
where
\begin{equation}
  \label{eq:y0}
  y_0\equiv \frac{\sigma_T \overline{\rho}_{\rm gas0} k_{\rm B} 
            \overline{T}_{\rho0} r_*}{\mu_e m_p m_e c^2} 
     = 4.3\times 10^{-4}\mu_e^{-1}\left(\Omega_{\rm b} h^2\right) 
       \left(\frac{k_{\rm B} \overline{T}_{\rho0}}{1~{\rm keV}}\right)
       \left(\frac{r_*}{10~{\rm Gpc}}\right).
\end{equation}
$T_\rho\equiv \rho_{\rm gas} T_e/\overline{\rho}_{\rm gas}$ is the 
electron temperature weighted by the gas mass density, the overline denotes
the volume average, and the subscript 0 means the present epoch.
We adopt $\mu_e^{-1}=0.88$, where 
$\mu_e^{-1}\equiv n_e/(\rho_{\rm gas}/m_p)$ is the number of electrons 
per proton mass in the fully ionized medium.
Other quantities have their usual meaning.

Transforming Eq.~(\ref{eq:lenscouple}) into harmonic space, we obtain 
\begin{eqnarray}
  \nonumber
  a_{lm}
  &=&
  a_{lm}^{\rm P}
  +\sum_{l'm'}\sum_{l''m''}(-1)^m
  {\mathcal G}_{l l' l''}^{-m m' m''}\\
 \nonumber
  & &\times
  \frac{l'(l'+1)-l(l+1)+l''(l''+1)}2
  a_{l'm'}^{\rm P}\Theta_{l''m''}
  +a_{lm}^{\rm SZ}\\
 \nonumber
  &=&
  a_{lm}^{\rm P}
  +\sum_{l'm'}\sum_{l''m''}(-1)^{m+m'+m''}
  {\mathcal G}_{l l' l''}^{-m m' m''}\\
 \label{eq:almlens}
  & &\times
  \frac{l'(l'+1)-l(l+1)+l''(l''+1)}2
  a_{l'-m'}^{P*}\Theta^*_{l''-m''}
  +a_{lm}^{\rm SZ},
\end{eqnarray}
where ${\mathcal G}_{l_1l_2l_3}^{m_1m_2m_3}$ is the Gaunt integral
(Eq.~(\ref{eq:gaunt})).
Substituting Eq.~(\ref{eq:almlens}) into Eq.~(\ref{eq:blllmmm}),
and using the identity, ${\mathcal G}_{l_1l_2l_3}^{-m_1-m_2-m_3}
={\mathcal G}_{l_1l_2l_3}^{m_1m_2m_3}$, we obtain the bispectrum,
\begin{eqnarray}
 \nonumber
  B_{l_1l_2l_3}^{m_1m_2m_3}
  &=&{\mathcal G}_{l_1 l_2 l_3}^{m_1 m_2 m_3}
  \left[
  \frac{l_1(l_1+1)-l_2(l_2+1)+l_3(l_3+1)}2
  C_{l_1}^{\rm P} \left<\Theta^*_{l_3m_3} a_{l_3m_3}^{\rm SZ}\right>\right.\\
  \label{eq:szlensbispec}
   & &
   \left. + \mbox{5 permutations}\right].
\end{eqnarray}
The form of Eq.~(\ref{eq:func}) is confirmed;
the reduced bispectrum $b_{l_1l_2l_3}^{\rm sz-lens}$ 
includes the terms in square brackets.

While Eq.~(\ref{eq:szlensbispec}) is complicated, we can understand
the physical effect producing the SZ--lensing bispectrum intuitively.
Figure~\ref{fig:szlens} shows how the SZ--lensing coupling 
produces the three--point correlation. 
Suppose that there are three CMB photons decoupled at the last 
scattering surface (LSS), and
one of these photons penetrates through a SZ cluster between the 
LSS and us; the energy of the photon changes because of the SZ effect.
When the other two photons pass near the SZ cluster, they are
deflected by the gravitational lensing effect, changing their
propagation directions, and coming toward us.
What do we see after all?
We see that the three CMB photons are correlated; we then measure
a non--zero angular bispectrum. 
The cross--correlation strength between the SZ and lensing
effects, $\left<\Theta^*_{l_3m_3} a_{l_3m_3}^{\rm SZ}\right>$,
thus determines the bispectrum amplitude, as indicated by 
Eq.~(\ref{eq:szlensbispec}).

\begin{figure}
\begin{center}
    \leavevmode\epsfxsize=9cm \epsfbox{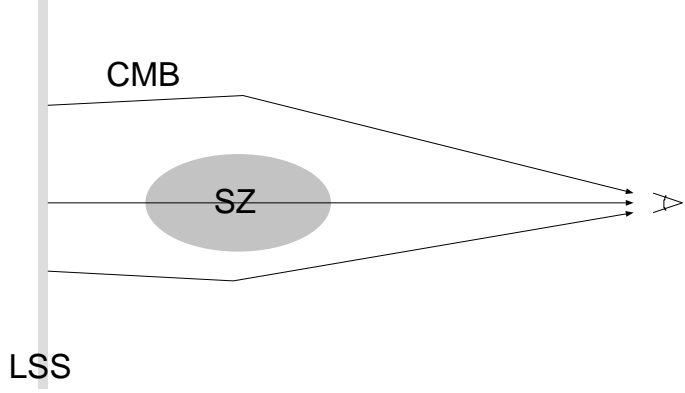}
\end{center}
 \caption{\bf SZ--lensing Coupling}
 \mycaption{A schematic view of the SZ--lensing coupling bispectrum.
 One of the three CMB photons, which are decoupled at the last
 scattering surface (LSS), penetrates through a SZ cluster, 
 changing its temperature, and coming toward us.
 As the other two photons pass near the SZ cluster, 
 they are deflected by the lensing effect, changing their propagation 
 directions, and coming toward us.
 As a result, the three photons are correlated, generating a
 three--point correlation, the bispectrum.} 
\label{fig:szlens}
\end{figure}

In Ref.~\cite{goldberg/spergel:1999} $\left<\Theta^*_{lm} 
a_{lm}^{\rm SZ}\right>$ was derived 
assuming the linear pressure bias model~\cite{persi/etal:1995},
$T_\rho=\overline{T}_\rho b_{\rm gas} \delta$,
and the mean temperature evolution,
$\overline{T}_\rho\simeq \overline{T}_{\rho0}(1+z)^{-1}$, 
for $z<2$, which is roughly suggested by recent hydrodynamic simulations
\cite{cen/ostriker:1999,refregier/etal:2000,springel/white/hernquist:2001}.
They obtained
\begin{equation}
  \label{eq:blsz}
  \left<\Theta^*_{lm} a_{lm}^{\rm SZ}\right>
  \simeq
  -j_\nu\frac{4y_0b_{\rm gas}l^2}{3\Omega_{\rm m}H_0^2}
  \int_0^{z_*} dz \frac{dr}{dz}D^2(z)(1+z)^2
  \frac{r_*-r(z)}{r_*^2r^5(z)}
  P_\Phi\left(k=\frac{l}{r(z)}\right),
\end{equation}
where $D(z)$ is the linear growth factor.
Simulations without non--gravitational heating 
\cite{refregier/etal:2000,springel/white/hernquist:2001} 
suggest that $\overline{T}_{\rho0}\sim 0.2-0.4~{\rm keV}$ and 
$b_{\rm gas}\sim 5-10$; analytic estimations give similar numbers 
\cite{refregier/etal:2000,zhang/pen:2001}.
In the pressure bias model, the free parameters (except cosmological 
ones) are $\overline{T}_{\rho0}$ and $b_{\rm gas}$;  
however, both actually depend upon the cosmological model 
\cite{refregier/etal:2000}.
Since 
$l^3\left<\Theta^*_{lm}a_{lm}^{\rm SZ}\right>\sim 
2\times 10^{-10}j_\nu \overline{T}_{\rho0}b_{\rm gas}$
\cite{goldberg/spergel:1999,cooray/hu:2000} and $l^2C_l^{\rm P}\sim 
6\times 10^{-10}$, we have
\begin{equation}
  \label{eq:orderestszlens}
  b_{lll}^{\rm sz-lens}\sim 
  l^{-3}
  \left[
  \left(l^2C_l^{\rm P}\right)
  \left(l^3\left<\Theta^*_{lm} a_{lm}^{\rm SZ}\right>\right)\times 5/2\right]
  \sim 
  l^{-3}\times
  3\times 10^{-19}j_\nu\overline{T}_{\rho0}b_{\rm gas},
\end{equation}
where $\overline{T}_{\rho0}$ is in units of 1~keV, and
$b_{l_1l_2l_3}=
B_{l_1l_2l_3}^{m_1m_2m_3}\left({\mathcal
G}_{l_1l_2l_3}^{m_1m_2m_3}\right)^{-1}$  
is the reduced bispectrum (Eq.~(\ref{eq:func})).
Comparing this with Eq.~(\ref{eq:orderest}), we obtain
\begin{equation}
  \label{eq:ordercomp}
  \frac{b_{lll}^{\rm prim}}{b_{lll}^{\rm sz-lens}}  
  \sim l^{-1}\times 10
  \left(\frac{f_{\rm NL}}{j_\nu\overline{T}_{\rho0}b_{\rm gas}}\right).
\end{equation}
This estimate suggests that the SZ--lensing bispectrum overwhelms
the primordial bispectrum on small angular scales.
This is why we have to separate the primordial from the SZ--lensing effect.

While the pressure bias model gives a rough estimate of the SZ 
power--spectrum, more accurate predictions exist.
Several authors have predicted the SZ power--spectrum analytically using the 
Press--Schechter approach 
\cite{cole/kaiser:1988,makino/suto:1993,atrio/mucket:1999,komatsu/kitayama:1999,cooray:2000,mosca3}
or the hyper-extended perturbation theory~\cite{zhang/pen:2001}.
The predictions agree with hydrodynamic simulations well
\cite{refregier/etal:2000,seljak/burwell/pen:2001,springel/white/hernquist:2001,refregier/teyssier:2002}. 
While a big uncertainty in the predictions lies in
phenomenological models which describe the SZ surface brightness profile
of halos, the authors of Ref.~\cite{komatsu/seljak:2001} have proposed 
universal gas and temperature profiles and predicted the SZ profile relying 
on a more physical basis; they have then used the universal profiles to 
improve upon the analytic prediction for the SZ power--spectrum.
The universal profiles should describe the SZ profile
in the average sense; on the individual halo--to--halo basis, there could be 
significant deviation from the universal profile, owing to
substructures in halos (see, e.g., Ref.~\cite{komatsu/etal:2001}).

\subsubsection{Extragalactic radio and infrared sources}

The bispectrum from extragalactic radio and infrared sources 
whose fluxes, $F$, are smaller than a certain detection threshold, 
$F_{\rm d}$, is simple to estimate, when we assume the Poisson distribution.
The authors of 
Ref.~\cite{toffolatti/etal:1998,argueso/gonzalez/toffolatti:2003} 
have shown that the Poisson distribution is a good 
approximation at low frequencies ($\nu<100$~GHz).
The Poisson distribution has white--noise power--spectrum; thus, 
the reduced bispectrum (Eq.~(\ref{eq:func})) is constant,
$b_{l_1l_2l_3}^{\rm src}=b^{\rm src}={\rm constant}$, and we obtain
\begin{equation}
  \label{eq:pointsource}
  B_{l_1l_2l_3}^{m_1m_2m_3}
  ={\mathcal G}_{l_l1_2l_3}^{m_1m_2m_3} b^{\rm src},
\end{equation}
where 
\begin{equation}
  \label{eq:Bps}
  b^{\rm src}(<F_{\rm d})
  \equiv g^3(x)
  \int_0^{F_{\rm d}} dF F^3 \frac{dn}{dF}
  =
  g^3(x)\frac{\beta}{3-\beta}n(>F_{\rm d})F_{\rm d}^3.
\end{equation}
Here, $dn/dF$ is the differential source count per unit solid angle, and 
$n(>F_{\rm d})\equiv \int_{F_{\rm d}}^\infty dF (dn/dF)$. 
We have assumed a power--law count, $dn/dF\propto F^{-\beta-1}$, 
for $\beta<2$.
The other symbols mean $x\equiv h\nu/k_{\rm B} T \simeq 
(\nu/56.80~{\rm GHz})(T/2.726~{\rm K})^{-1}$, and
\begin{equation}
  \label{eq:gx}
  g(x)\equiv
  2\frac{(hc)^2}{(k_{\rm B} T)^3}\left(\frac{\sinh x/2}{x^2}\right)^2
  \simeq
  \frac1{67.55~{\rm MJy~sr^{-1}}}\left(\frac{T}{2.726~{\rm K}}\right)^{-3}
  \left(\frac{\sinh x/2}{x^2}\right)^2.
\end{equation}

Using the Poisson angular power--spectrum, $C^{\rm ps}$, given by
\begin{equation}
  \label{eq:Cps}
  C^{\rm ps}(<F_{\rm d})
  \equiv g^2(x)
  \int_0^{F_{\rm d}} dF F^2 \frac{dn}{dF}
  =
  g^2(x)\frac{\beta}{2-\beta}n(>F_{\rm d})F_{\rm d}^2,
\end{equation}
we can rewrite $b^{\rm src}$ into a different form, 
\begin{equation}
  \label{eq:Bps*}
  b^{\rm src}(<F_{\rm d})
  = \frac{(2-\beta)^{3/2}}{\beta^{1/2}(3-\beta)}
  \left[n(>F_{\rm d})\right]^{-1/2}
  \left[C^{\rm ps}(<F_{\rm d})\right]^{3/2}.
\end{equation}

The authors of Ref.~\cite{toffolatti/etal:1998} have 
estimated $n(>F_{\rm d})\sim 
300~{\rm sr^{-1}}$ for $F_{\rm d}\sim 0.2~{\rm Jy}$ at 217~GHz. 
This $F_{\rm d}$ corresponds to $5\sigma$ detection
threshold for the {\sl Planck} experiment at 217~GHz. 
In Ref.~\cite{refregier/spergel/herbig:2000} their estimation 
was extrapolated to 94~GHz, finding 
$n(>F_{\rm d})\sim 7~{\rm sr^{-1}}$ for $F_{\rm d}\sim 2~{\rm Jy}$,
which corresponds to the {\sl WMAP} $5\sigma$ threshold.
These values yield
\begin{eqnarray}
  \label{eq:cl90ghz}
  C^{\rm ps}(90~{\rm GHz},<2~{\rm Jy})&\sim& 2\times 10^{-16},\\
  \label{eq:cl217ghz}
  C^{\rm ps}(217~{\rm GHz},<0.2~{\rm Jy})&\sim& 1\times 10^{-17}.
\end{eqnarray}
Thus, rough estimates for $b^{\rm src}$ are
\begin{eqnarray}
  \label{eq:bl90ghz}
  b^{\rm src}(90~{\rm GHz},<2~{\rm Jy})&\sim& 2\times 10^{-25},\\
  \label{eq:bl217ghz}
  b^{\rm src}(217~{\rm GHz},<0.2~{\rm Jy})&\sim& 5\times 10^{-28}.
\end{eqnarray}
While we have assumed the Euclidean source count ($\beta=3/2$) for 
definiteness, this assumption does not affect order--of--magnitude
estimates here. 

As the primordial reduced bispectrum is $\propto l^{-4}$ 
(Eq.~(\ref{eq:orderest})),
and the SZ--lensing reduced bispectrum is $\propto l^{-3}$
(Eq.~(\ref{eq:orderestszlens})), the point--source bispectrum rapidly
becomes to dominate the total bispectrum on small angular scales:
\begin{eqnarray}
  \label{eq:ordercomp*}
  \frac{b_{lll}^{\rm prim}}{b^{\rm src}}
  &\sim& l^{-4}\times 10^7
  \left(\frac{f_{\rm NL}}{b^{\rm src}/10^{-25}}\right),\\
  \label{eq:ordercomp**}
  \frac{b_{lll}^{\rm sz-lens}}
       {b^{\rm src}}
  &\sim& l^{-3}\times 10^6
  \left(\frac{j_\nu\overline{T}_{\rho0}b_{\rm gas}}
             {b^{\rm src}/10^{-25}}\right).
\end{eqnarray}
For example, the point--sources overwhelm the SZ--lensing bispectrum 
measured by {\sl WMAP} at $l\simgt 100$.

What do the SZ--lensing bispectrum and the point--source bispectrum
look like?
Figure~\ref{fig:reducedb} plots the primordial, the SZ--lensing,
and the point--source reduced bispecta for the equilateral configurations,
$l\equiv l_1=l_2=l_3$.
We have plotted $l^2(l+1)^2b_{lll}/(2\pi)^2$.
We find that these bispecra are very different from each other in shape 
on small angular scales.
It thus suggests that we can separate these three contributions
on the basis of shape difference. 
We study this point in the next section.

\begin{figure}
 \begin{center}
  \leavevmode\epsfxsize=9cm \epsfbox{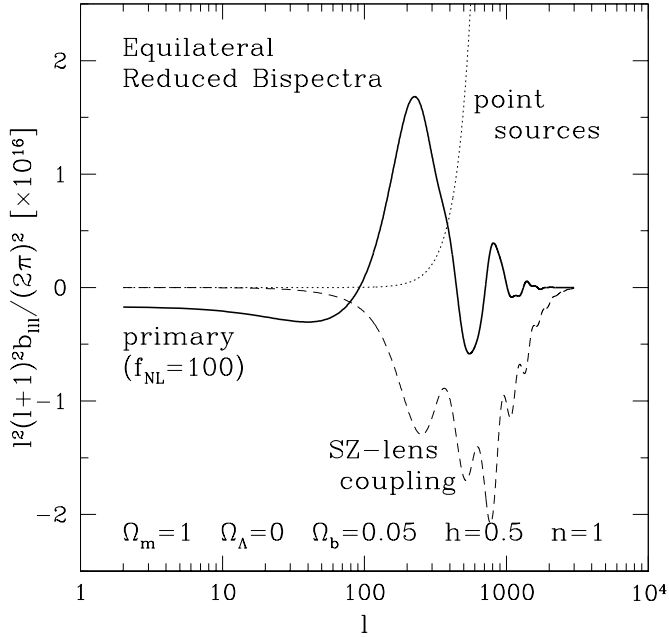}
 \end{center}
 \caption{\bf Equilateral Reduced Bispectra}
 \mycaption{Comparison between the primordial (solid line), 
 the SZ--lensing (dashed line), and the point--source (dotted line) 
 reduced bispectra for the equilateral configurations, $l\equiv l_1=l_2=l_3$.
 We have plotted $\left[l^2(l+1)^2b_{lll}/(2\pi)^2\right]\times 10^{16}$,
 which makes the Sachs--Wolfe plateau of the primordial reduced bispectrum
 on large angular scales, $l\simlt 10$, easily seen.}
\label{fig:reducedb}
\end{figure}

\subsection{Measuring bispectra: signal--to--noise estimation}
\label{sec:measure}

In this section, we study how well we can measure the 
primordial bispectrum, and how well
we can separate it from the secondary bispectra.
Suppose that we fit the observed bispectrum, $B_{l_1l_2l_3}^{\rm obs}$,
by theoretically calculated bispectra, which include both the primordial 
and secondary sources.
We minimize $\chi^2$ defined by
\begin{equation}
  \chi^2
  \equiv 
  \sum_{2\leq l_1\leq l_2\leq l_3}
  \frac{\left(B_{l_1l_2l_3}^{\rm obs}
	     -\sum_i A_i B^{(i)}_{l_1l_2l_3}\right)^2}
  {\sigma^2_{l_1l_2l_3}},
\end{equation}
where $i$ denotes a component such as the primordial, the SZ and 
lensing effects, extragalactic sources, and so on. 
We have removed unobservable modes, $l=0$ and $1$.

As we have shown in Sec.~\ref{sec:npointspectrum}, the variance of the 
bispectrum, $\sigma^2_{l_1l_2l_3}$, is the six--point function of 
$a_{lm}$ \cite{luo:1994,heavens:1998}.
When non--Gaussianity is weak, 
we calculate it as \cite{rs4,gangui/martin:2000}
\begin{equation}
  \sigma^2_{l_1l_2l_3}
  \equiv \left<B_{l_1l_2l_3}^2\right>-\left<B_{l_1l_2l_3}\right>^2
  \approx
  {\mathcal C}_{l_1}{\mathcal C}_{l_2}{\mathcal C}_{l_3}\Delta_{l_1l_2l_3},
\end{equation}
where $\Delta_{l_1l_2l_3}$ takes values 1, 2, or 6 when
all $l$'s are different, two are the same, or all are the same, respectively. 
${\mathcal C}_l\equiv C_l+C_l^{\rm N}$ is the total CMB angular 
power--spectrum,
which includes the power--spectrum of the detector noise, $C_l^{\rm N}$.
We calculate $C_l^{\rm N}$ analytically following 
\cite{knox:1995} with the noise characteristics of relevant experiments.
We do not include $C_l$ from secondary sources, as they are subdominant 
compared with the primordial $C_l$ and $C_l^{\rm N}$ for relevant experiments.
Including $C_l$ from extragalactic sources (Eqs.(\ref{eq:cl90ghz}) 
or (\ref{eq:cl217ghz})) changes our results by less than 10\%.

Taking $\partial\chi^2/\partial A_i=0$, we obtain the equation
\begin{equation}
  \label{eq:fiseq}
  \sum_j
  \left[\sum_{2\leq l_1\leq l_2\leq l_3}
        \frac{B_{l_1l_2l_3}^{(i)}B_{l_1l_2l_3}^{(j)}}{\sigma_{l_1l_2l_3}^2}
  \right]A_j
  =
  \sum_{2\leq l_1\leq l_2\leq l_3}
     \frac{B_{l_1l_2l_3}^{\rm obs}B_{l_1l_2l_3}^{(i)}}{\sigma_{l_1l_2l_3}^2}.
\end{equation}
We then define the Fisher matrix, $F_{ij}$, as
\begin{eqnarray}
 \nonumber
  F_{ij}&\equiv& 
  \sum_{2\leq l_1\leq l_2\leq l_3}
  \frac{B_{l_1l_2l_3}^{(i)}B_{l_1l_2l_3}^{(j)}}{\sigma_{l_1l_2l_3}^2}\\
 \label{eq:fis}
  &=&
  \frac{2}{\pi}\sum_{2\leq l_1\leq l_2\leq l_3}
  \left(l_1+\frac12\right)\left(l_2+\frac12\right)\left(l_3+\frac12\right)
   \left(\begin{array}{ccc}l_1&l_2&l_3\\0&0&0\end{array}\right)^2
  \frac{b_{l_1l_2l_3}^{(i)}b_{l_1l_2l_3}^{(j)}}{\sigma_{l_1l_2l_3}^2},
\end{eqnarray}
where we have used Eq.~(\ref{eq:wigner*}) to replace $B_{l_1l_2l_3}$ by
the reduced bispectrum, $b_{l_1l_2l_3}$ (see Eq.~(\ref{eq:func}) for
definition). 
Since the covariance matrix of $A_i$ is $F_{ij}^{-1}$,
we define the signal--to--noise ratio, $(S/N)_i$, for a component $i$, 
the correlation coefficient, $r_{ij}$, between different components 
$i$ and $j$, and the degradation parameter, $d_i$, of $(S/N)_i$ 
due to $r_{ij}$, as
\begin{eqnarray}
  \label{eq:sn}
  \left(\frac{S}{N}\right)_i &\equiv& \frac1{\sqrt{F_{ii}^{-1}}},\\
  \label{eq:r}
  r_{ij}&\equiv& 
  \frac{F_{ij}^{-1}}{\sqrt{F^{-1}_{ii}F^{-1}_{jj}}},\\
  \label{eq:d}
  d_i&\equiv& F_{ii}F_{ii}^{-1}.
\end{eqnarray}
Note that $r_{ij}$ does not depend upon the amplitude of the bispectra,
but on their shape. 
We have defined $d_i$ so as $d_i=1$ for zero degradation,
while $d_i>1$ for degraded $(S/N)_i$. In Ref.s~\cite{rs4} and 
\cite{cooray/hu:2000} the diagonal component 
of $F_{ij}^{-1}$ has been considered.
We study all the components to look at the separability 
between various bispectra. 

We can give an order--of--magnitude estimate of $S/N$ as a function of 
the angular resolution, $l$, as follows.
Since the number of modes contributing to $S/N$ increases as 
$l^{3/2}$, and 
$l^3\left(\begin{array}{ccc}l&l&l\\0&0&0\end{array}\right)^2
\sim 0.36\times l$, we estimate $(S/N)_i\sim (F_{ii})^{1/2}$ as
\begin{equation}
  \label{eq:snorder}
  \left(\frac{S}{N}\right)_i
  \sim 
  \frac1{3\pi} l^{3/2}
  \times l^{3/2}
  \left|
  \left(\begin{array}{ccc}l&l&l\\0&0&0\end{array}\right)
  \right|\times
   \frac{l^3b_{lll}^{(i)}}
        {(l^2 C_l)^{3/2}}
  \sim l^5b_{lll}^{(i)}\times 4\times 10^{12},
\end{equation}
where we have used $l^2C_l\sim 6\times 10^{-10}$.

Table~\ref{tab:fis} tabulates $F_{ij}$, while Table~\ref{tab:invfis}
tabulates $F_{ij}^{-1}$; Table~\ref{tab:sn} tabulates $(S/N)_i$, while 
table~\ref{tab:corr} tabulates $d_i$ in the diagonal, and $r_{ij}$ 
in the off-diagonal parts.

\begin{table}
 \caption{\bf Fisher Matrix}
 \mycaption{Fisher matrix, $F_{ij}$ (see Eq.~(\ref{eq:fis})): 
 $i$ denotes a component in the first row; 
 $j$ denotes a component in the first column.
 $\overline{T}_{\rho0}$ is in units of 1 keV, 
 $b^{\rm src}_{25}\equiv b^{\rm src}/10^{-25}$, and
 $b^{\rm src}_{27}\equiv b^{\rm src}/10^{-27}$.}
\label{tab:fis}
\begin{center} 
  \begin{tabular}{cccc}\hline\hline
  {\sl COBE} & primordial & SZ--lensing & point--sources \\
  \hline 
  primordial &
  $4.2\times 10^{-6}~f_{\rm NL}^2$ & 
  $-4.0\times 10^{-7}~f_{\rm NL} j_\nu \overline{T}_{\rho0} b_{\rm gas}$ &
  $-1.0\times 10^{-9}~f_{\rm NL} b^{\rm src}_{25}$ \\
  SZ--lensing &
  & 
  $1.3\times 10^{-7}~(j_\nu \overline{T}_{\rho0} b_{\rm gas})^2$ &
  $3.1\times 10^{-10}~j_\nu \overline{T}_{\rho0} b_{\rm gas} 
b^{\rm src}_{25}$ \\
  point--sources&
  & &
  $1.1\times 10^{-12}~(b^{\rm src}_{25})^2$ \\
  \hline
  {\sl WMAP} & & & \\
  \hline 
  primordial &
  $3.4\times 10^{-3}~f_{\rm NL}^2$ & 
  $2.6\times 10^{-3}~f_{\rm NL} j_\nu \overline{T}_{\rho0} b_{\rm gas}$ &
  $2.4\times 10^{-3}~f_{\rm NL} b^{\rm src}_{25}$ \\
  SZ--lensing &
  & 
  $0.14~(j_\nu \overline{T}_{\rho0} b_{\rm gas})^2$ &
  $0.31~j_\nu \overline{T}_{\rho0} b_{\rm gas} b^{\rm src}_{25}$ \\
  point--sources&
  & &
  $5.6~(b^{\rm src}_{25})^2$ \\
  \hline
  {\sl Planck} & & & \\
  \hline 
  primordial &
  $3.8\times 10^{-2}~f_{\rm NL}^2$ & 
  $7.2\times 10^{-2}~f_{\rm NL} j_\nu \overline{T}_{\rho0} b_{\rm gas}$ &
  $1.6\times 10^{-2}~f_{\rm NL} b^{\rm src}_{27}$ \\
  SZ--lensing &
  & 
  $39~(j_\nu \overline{T}_{\rho0} b_{\rm gas})^2$ &
  $5.7~j_\nu \overline{T}_{\rho0} b_{\rm gas} b^{\rm src}_{27}$ \\
  point--sources&
  & &
  $2.7\times 10^3~(b^{\rm src}_{27})^2$\\
   \hline\hline
  \end{tabular}
\end{center}
\end{table}

\begin{table}
 \caption{\bf Inverted Fisher Matrix}
 \mycaption{Inverted Fisher matrix, $F_{ij}^{-1}$.
 The meaning of the symbols is the same as in Table~\ref{tab:fis}.}
\label{tab:invfis}
\begin{center} 
 \begin{tabular}{cccc}\hline\hline
  {\sl COBE} & primordial & SZ--lensing & point--sources \\
  \hline 
  primordial &
  $3.5\times 10^{5}~f_{\rm NL}^{-2}$ & 
  $1.1\times 10^{6}~(f_{\rm NL} j_\nu \overline{T}_{\rho0} b_{\rm gas})^{-1}$ &
  $1.3\times 10^7~(f_{\rm NL} b^{\rm src}_{25})^{-1}$ \\
  SZ--lensing &
  & 
  $3.1\times 10^7~(j_\nu \overline{T}_{\rho0} b_{\rm gas})^{-2}$ &
  $-7.8\times 10^9~(j_\nu \overline{T}_{\rho0} b_{\rm gas} 
  b^{\rm src}_{25})^{-1}$ \\
  point sources&
  & &
  $3.1\times 10^{12}~(b^{\rm src}_{25})^{-2}$ \\
  \hline
  {\sl WMAP} & & & \\
  \hline 
  primordial &
  $3.0\times 10^2~f_{\rm NL}^{-2}$ & 
  $-6.1~(f_{\rm NL} j_\nu \overline{T}_{\rho0} b_{\rm gas})^{-1}$ &
  $0.21~(f_{\rm NL} b^{\rm src}_{25})^{-1}$ \\
  SZ--lensing &
  & 
  $8.4~(j_\nu \overline{T}_{\rho0} b_{\rm gas})^{-2}$ &
  $-0.46~(j_\nu \overline{T}_{\rho0} b_{\rm gas} b^{\rm src}_{25})^{-1}$ \\
  point--sources&
  & &
  $0.21~(b^{\rm src}_{25})^{-2}$ \\
  \hline
  {\sl Planck} & & & \\
  \hline 
  primordial &
  $26~f_{\rm NL}^{-2}$ & 
  $-4.9\times 10^{-2}~(f_{\rm NL} j_\nu 
  \overline{T}_{\rho0} b_{\rm gas})^{-1}$ &
  $-5.7\times 10^{-5}~(f_{\rm NL} b^{\rm src}_{27})^{-1}$ \\
  SZ--lensing &
  & 
  $2.6\times 10^{-2}~(j_\nu \overline{T}_{\rho0} b_{\rm gas})^{-2}$ &
  $-5.4\times 10^{-5}~(j_\nu \overline{T}_{\rho0} b_{\rm gas} 
  b^{\rm src}_{27})^{-1}$ \\
  point--sources&
  & &
  $3.7\times 10^{-4}~(b^{\rm src}_{27})^{-2}$\\
  \hline\hline
 \end{tabular}
\end{center}
\end{table}

\begin{table}
 \caption{\bf Signal--to--noise Ratio}
 \mycaption{Signal--to--noise ratio, $(S/N)_i$ (see Eq.~(\ref{eq:sn})), of 
 detecting the bispectrum.
 $i$ denotes a component in the first row.
 The meaning of the symbols is the same as in Table~\ref{tab:fis}.}
\label{tab:sn}
\begin{center} 
  \begin{tabular}{cccc}\hline\hline
       & primordial & SZ--lensing & point--sources \\
  \hline 
  {\sl COBE} & $1.7\times 10^{-3}~f_{\rm NL}$ 
       & $1.8\times 10^{-4}~\left|j_\nu\right| \overline{T}_{\rho0}b_{\rm gas}$
       & $5.7\times 10^{-7}~b_{25}^{\rm ps}$ \\
  {\sl WMAP}  & $5.8\times 10^{-2}~f_{\rm NL}$ 
       & $0.34~\left|j_\nu\right| \overline{T}_{\rho0}b_{\rm gas}$
       & $2.2~b_{25}^{\rm ps}$ \\
  {\sl Planck} & $0.19~f_{\rm NL}$ 
         & $6.2~\left|j_\nu\right| \overline{T}_{\rho0}b_{\rm gas}$
         & $52~b_{27}^{\rm ps}$\\
   \hline\hline
  \end{tabular}
\end{center}
\end{table}

\begin{table}
 \caption{\bf Signal Degradation and Correlation Matrix}
 \mycaption{Signal degradation parameter, $d_i$ (see Eq.~(\ref{eq:d})), and 
 correlation coefficient, $r_{ij}$ (see Eq.~(\ref{eq:r})), matrix. 
 $i$ denotes a component in the first row; 
 $j$ denotes a component in the first column.
 $d_i$ for $i=j$, while $r_{ij}$ for $i\neq j$.}
\label{tab:corr}
\begin{center} 
  \begin{tabular}{cccc}\hline\hline
  {\sl COBE} & primordial & SZ--lensing & point--sources \\
  \hline 
  primordial &
  $1.46$ & 
  $0.33~{\rm sgn}(j_\nu)$ &
  $1.6\times 10^{-2}$ \\
  SZ--lensing &
  & 
  $3.89$ &
  $-0.79~{\rm sgn}(j_\nu)$ \\
  point--sources&
  & &
  $3.45$ \\
  \hline
  {\sl WMAP} & & & \\
  \hline 
  primordial &
  $1.01$ & 
  $-0.12~{\rm sgn}(j_\nu)$ &
  $2.7\times 10^{-2}$ \\
  SZ--lensing &
  & 
  $1.16$ &
  $-0.35~{\rm sgn}(j_\nu)$ \\
  point--sources&
  & &
  $1.14$ \\
  \hline
  {\sl Planck} & & & \\
  \hline 
  primordial &
  $1.00$ & 
  $-5.9\times 10^{-2}~{\rm sgn}(j_\nu)$ &
  $-5.8\times 10^{-4}$ \\
  SZ--lensing &
  & 
  $1.00$ &
  $-1.8\times 10^{-2}~{\rm sgn}(j_\nu)$ \\
  point--sources&
  & &
  $1.00$\\
   \hline\hline
  \end{tabular}
\end{center}
\end{table}

\subsubsection{Measuring the primordial bispectrum}

Figure~\ref{fig:fis11} shows the signal--to--noise ratio, $S/N$.
The top panel shows the differential $S/N$ for the primordial bispectrum 
at $\ln l_3$ interval, 
$\left[d(S/N)^2/d\ln l_3\right]^{1/2}f_{\rm NL}^{-1}$, and the bottom
panel shows the cumulative $S/N$, $(S/N)(<l_3)f_{\rm NL}^{-1}$, which is 
$S/N$ summed up to a certain $l_3$. 
We have computed the detector noise power--spectrum, $C_l^{\rm N}$, for 
{\sl COBE} four--year map~\cite{bennett/etal:1996}, {\sl WMAP} 90 GHz channel, 
and {\sl Planck} 217 GHz channel, and assumed full sky coverage.
Figure~\ref{fig:fis11} also shows the ideal experiment 
with no noise: $C_l^{\rm N}=0$.
Both $\left[d(S/N)^2/d\ln l_3\right]^{1/2}$ and 
$(S/N)(<l_3)$ increase monotonically with $l_3$, roughly $\propto l_3$, 
up to $l_3\sim 2000$ for the ideal experiment.

\begin{figure}
 \begin{center}
  \leavevmode\epsfxsize=9cm \epsfbox{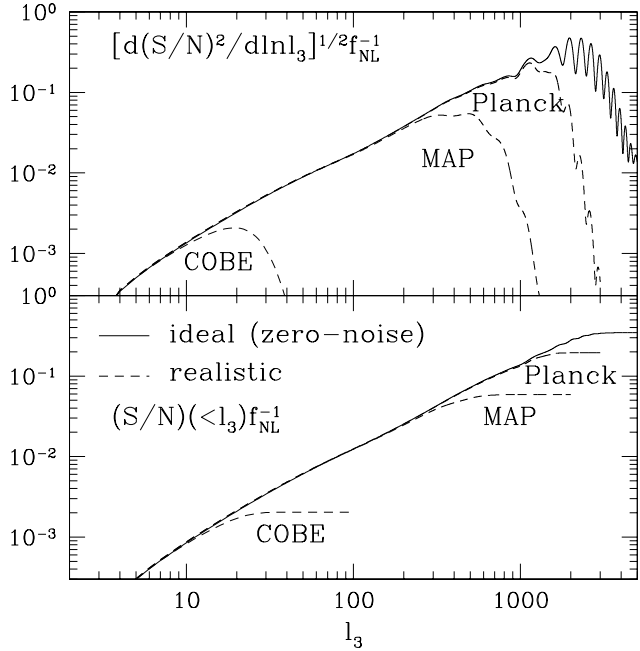}
 \end{center}
 \caption{{\bf Signal--to--noise Ratio}~\cite{ks}}
 \mycaption{The predictions of the signal--to--noise ratio, $S/N$, 
 for {\sl COBE}, {\sl WMAP}, and {\sl Planck} experiments 
 (see Eq.~(\ref{eq:sn})).
 The differential $S/N$ at $\ln l_3$ interval is shown in the upper
 panel, while the cumulative $S/N$ up to a certain $l_3$ is shown
 in the bottom panel. 
 Both are in units of $f_{\rm NL}$.
 Solid line represents the zero-noise ideal experiment, while dotted 
 lines show the realistic experiments mentioned above.
 The total $(S/N)f^{-1}_{\rm NL}$ 
 are $1.7\times 10^{-3}$, $5.8\times 10^{-2}$, and $0.19$
 for {\sl COBE}, {\sl WMAP}, and {\sl Planck} experiments, respectively.}
\label{fig:fis11}
\end{figure}

Beyond $l_3\sim 2000$, an enhancement of the damping tail in $C_l$ 
because of the weak lensing effect~\cite{seljak:1996} stops 
$\left[d(S/N)^2/d\ln l_3\right]^{1/2}$, and hence $(S/N)(<l_3)$,
increasing. 
This leads to an important constraint on observations; 
even for the ideal noise--free, infinitesimally thin beam experiment, 
there is an upper limit on the value of $S/N\simlt 0.3f_{\rm NL}$.
For a given realistic experiment, 
$\left[d(S/N)^2/d\ln l_3\right]^{1/2}$ has a maximum at 
a scale near the beam size.

For {\sl COBE}, {\sl WMAP} and {\sl Planck} experiments,
the total $(S/N)f^{-1}_{\rm NL}$ are $1.7\times 10^{-3}$, 
$5.8\times 10^{-2}$, and $0.19$, respectively (see Table~\ref{tab:sn}).
To obtain $S/N>1$, we need $f_{\rm NL}>600$, 20, and $5$, while the 
ideal experiment requires $f_{\rm NL}>3$ (see Table~\ref{tab:fnl}).
We can also roughly obtain these values by substituting 
Eq.~(\ref{eq:orderest}) into (\ref{eq:snorder}),
\begin{equation}
  \label{eq:snorderprim}
  \left(\frac{S}{N}\right)_{\rm prim}
  \sim 
  l \times 10^{-4}f_{\rm NL}.
\end{equation}
 
The degradation parameters, $d_{\rm prim}$,  
are 1.46, 1.01, and 1.00 for {\sl COBE}, {\sl WMAP}, and {\sl Planck} 
experiments, respectively (see Table~\ref{tab:corr}),
suggesting that {\sl WMAP} and {\sl Planck} experiments will separate the 
primordial bispectrum from the others with 1\% or better accuracy;
however, {\sl COBE} cannot discriminate between them very well, as 
the primordial and the secondary sources change monotonically on 
the {\sl COBE} angular scales.
On the {\sl WMAP} and {\sl Planck} scales, the primordial bispectrum
starts oscillating around zero, being well separated in shape 
from the secondaries that do not oscillate.
This is good news for the forthcoming high angular
resolution CMB experiments.

\subsubsection{Measuring secondary bispectra}

Signal--to--noise ratios for detecting the SZ--lensing bispectrum, 
$(S/N)_{\rm sz-lens}$, in units of 
$\left|j_\nu\right| \overline{T}_{\rho0}b_{\rm gas}$ are 
$1.8\times 10^{-4}$, 0.34, and 6.2 for {\sl COBE}, {\sl WMAP}, and 
{\sl Planck} experiments, respectively (see Table~\ref{tab:sn}), where 
$\overline{T}_{\rho0}$ is in units of 1~keV.
Using Eqs.~(\ref{eq:snorder}) and (\ref{eq:orderestszlens}), 
we can roughly estimate $(S/N)_{\rm sz-lens}$ as
\begin{equation}
  \label{eq:snorderszlens}
  \left(\frac{S}{N}\right)_{\rm sz-lens}
  \sim 
  l^2 \times 10^{-6}\left|j_\nu\right| \overline{T}_{\rho0}b_{\rm gas}.
\end{equation}
Hence, $(S/N)_{\rm sz-lens}$ increases with the angular resolution
more rapidly than the primordial bispectrum (see Eq.~(\ref{eq:snorderprim})).
Since $\left|j_\nu\right| \overline{T}_{\rho0}b_{\rm gas}$ should be
of order unity, {\sl COBE} and {\sl WMAP} cannot detect the 
SZ--lensing bispectrum; however, {\sl Planck} is sensitive enough 
to detect, depending on the frequency, i.e., a value of $j_\nu$.
For example, 217~GHz is insensitive to the SZ effect as
$j_\nu\sim 0$, while $j_\nu=-2$ in the Rayleigh--Jeans regime.

The degradation parameters, $d_{\rm sz-lens}$, are
3.89, 1.16, and 1.00 for {\sl COBE}, {\sl WMAP}, and {\sl Planck} experiments,
respectively (see Table~\ref{tab:corr}); thus, 
{\sl Planck} will separate the SZ--lensing bispectrum from the other effects.
Note that the $(S/N)_{\rm sz-lens}$ values must be understood as 
order--of--magnitude estimates, since our cosmological model is the 
{\sl COBE} normalized SCDM that yields $\sigma_8=1.2$, which is a factor of 
2 greater than the cluster normalization for $\Omega_{\rm m}=1$, and 
$20\%$ greater than the normalization for $\Omega_{\rm m}=0.3$ 
\cite{kitayama/suto:1997}.
Hence, this factor tends to overestimate 
$\left<\Theta^*_{lm} a_{lm}^{\rm SZ}\right>$ (Eq.~(\ref{eq:blsz})) by a 
factor of less than 10; on the other hand, using the linear
$P_\Phi(k)$ power--spectrum rather than the non--linear power--spectrum 
tends to underestimate the effect by a factor of less than 10 at 
$l\sim 3000$ \cite{cooray/hu:2000}.
Yet, our main goal is to discriminate between 
the shapes of various bispectra, not to determine the amplitude, 
so that this factor does not affect our conclusion on the 
degradation parameters, $d_i$.

For the extragalactic radio and infrared sources, one can estimate 
the signal--to--noise ratios as $5.7\times 10^{-7}(b^{\rm src}/10^{-25})$, 
$2.2(b^{\rm src}/10^{-25})$, and $52(b^{\rm src}/10^{-27})$ for 
{\sl COBE}, {\sl WMAP}, and {\sl Planck} experiments, respectively (see
Table~\ref{tab:sn}), and the degradation parameters, $d_{\rm ps}$, as
3.45, 1.14, and 1.00 (see Table~\ref{tab:corr}).
This estimate is consistent with that of 
Ref.~\cite{refregier/spergel/herbig:2000}. 
From Eq.~(\ref{eq:snorder}), we find 
\begin{equation}
  \left(\frac{S}{N}\right)_{\rm ps}
  \sim l^5\times 10^{-13}\left(\frac{b^{\rm src}}{10^{-25}}\right);
\end{equation}
thus, $S/N$ of the point--source bispectrum increases 
very rapidly with the angular resolution.

Although {\sl WMAP} cannot separate the Poisson bispectrum from 
the SZ--lensing bispectrum very well (see 
$r_{ij}$ in Table~\ref{tab:corr}), the SZ--lensing bispectrum is too 
small to be measured by {\sl WMAP} anyway.
{\sl Planck} will do an excellent job on separating all kinds of bispectra,
at least including the primordial signal, SZ--lensing coupling, 
and extragalactic point--sources, on the basis of the shape difference.

\subsubsection{Measuring primordial skewness}

For the skewness, we define $S/N$ as
\begin{equation}
  \label{eq:skew_sn}
  \left(\frac{S}{N}\right)^2\equiv
  \frac{S_3^2}{\sigma^2_{S_3}},
\end{equation}
where the variance is~\cite{srendnicki:1993}
\begin{eqnarray}
  \nonumber
  \sigma_{S_3}^2
  &\equiv& \left<\left(S_3\right)^2\right> =
  6\int_{-1}^{1}\frac{d\cos\theta}2 \left[{\mathcal C}(\theta)\right]^3\\
  \nonumber
  &=&
  6\sum_{l_1l_2l_3}
  \frac{\left(2l_1+1\right)\left(2l_2+1\right)\left(2l_3+1\right)}
        {(4\pi)^3}
  \left(
  \begin{array}{ccc}
  l_1 & l_2 & l_3 \\ 0 & 0 & 0 
  \end{array}
  \right)^2
   {\mathcal C}_{l_1}{\mathcal C}_{l_2}{\mathcal C}_{l_3}
   W^2_{l_1}W^2_{l_2}W^2_{l_3}\\
 \nonumber
  &=&
  \frac{9}{2\pi^3}\sum_{2\leq l_1\leq l_2\leq l_3}
  \left(l_1+\frac12\right)\left(l_2+\frac12\right)\left(l_3+\frac12\right)
  \left(
  \begin{array}{ccc}
  l_1 & l_2 & l_3 \\ 0 & 0 & 0 
  \end{array}
  \right)^2\\
 \label{eq:skewvar}
  & &\times
   {\mathcal C}_{l_1}{\mathcal C}_{l_2}{\mathcal C}_{l_3}
   W^2_{l_1}W^2_{l_2}W^2_{l_3}.
\end{eqnarray}
In the last equality, we have used symmetry of the summed 
quantity with respect to indices (Eq.~(\ref{eq:sumchange})), 
and removed unobservable modes, $l=0$ and $1$.
Typically $\sigma_{S_3}\sim 10^{-15}$, as $\sigma_{S_3}\sim
\left[{\mathcal C}(0)\right]^{3/2}\sim 10^{-15}$, 
where ${\mathcal C}(\theta)$ is the temperature auto--correlation function 
including noise.

The bottom panel of Figure~\ref{fig:skewness} plots $\sigma_{S_3}(<l_3)$,  
which is $\sigma_{S_3}$ summed up to a certain $l_3$, for 
{\sl COBE}, {\sl WMAP}, and {\sl Planck} experiments as well as for the 
ideal experiment.
Since ${\mathcal C}_{l}W^2_l= C_l e^{-l(l+1)\sigma^2_{\rm b}} + w^{-1}$,
where $w^{-1}$ is the white--noise power--spectrum of the detector
noise~\cite{knox:1995}, $w^{-1}$ keeps $\sigma_{S_3}(<l_3)$ 
slightly increasing with $l_3$ beyond the experimental angular resolution 
scale, $l\sim \sigma_{\rm b}^{-1}$.
In contrast, $S_3(<l_3)$ becomes constant beyond $l\sim \sigma_{\rm b}^{-1}$
(see the top panel of Figure~\ref{fig:skewness}). 
As a result, $S/N$ starts slightly decreasing beyond the resolution.
We use the maximum $S/N$ for calculating the minimum value of 
$f_{\rm NL}$ above which the primordial $S_3$ is detectable; 
we find that 
$f_{\rm NL}> 800$, 80, 70, and 60 for {\sl COBE}, {\sl WMAP}, 
{\sl Planck}, and the ideal experiments, respectively, assuming 
full sky coverage.

These $f_{\rm NL}$ values are systematically larger than those for
detecting $B_{l_1l_2l_3}$ by a factor of 1.3, 4, 14, and 20, respectively
(see Table~\ref{tab:fnl}).
The higher the angular resolution is, the less sensitive the primordial
$S_3$ is to non--Gaussianity than $B_{l_1l_2l_3}$.
This is because of the cancellation effect on smaller angular scales
caused by the oscillation of $B_{l_1l_2l_3}$ damps $S_3$.

Figure~\ref{fig:sn} compares the expected signal--to--noise ratio 
of detecting the primordial non--Gaussianity based on the bispectrum
(Eq.~(\ref{eq:sn})) with that based on the skewness (Eq.~(\ref{eq:skew_sn})). 
It shows that the bispectrum is almost an order of magnitude more 
sensitive to the non--Gaussianity than the skewness.
We conclude that when we can compute the predicted form of the
bispectrum, it becomes a ``matched filter'' for detecting the 
non--Gaussianity in data, and thus much more a powerful tool than the 
skewness.
Table~\ref{tab:fnl} summarizes the minimum $f_{\rm NL}$ for 
detecting the primordial non--Gaussianity using the bispectrum or the skewness 
for {\sl COBE}, {\sl WMAP}, {\sl Planck}, and the ideal experiments. 
This shows that even the ideal experiment 
needs $f_{\rm NL}>3$ to detect the primordial bispectrum.

\begin{figure}
 \begin{center}
  \leavevmode\epsfxsize=9cm \epsfbox{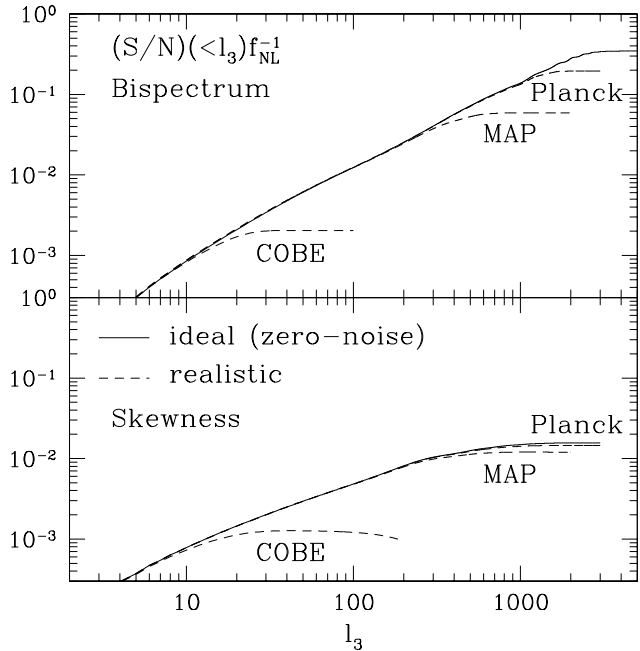}
 \end{center}
 \caption{{\bf Bispectrum vs Skewness}~\cite{ks}}
 \mycaption{Comparison of the signal--to--noise ratio summed up to a certain
 $l_3$, $S/N(<l_3)$, for the bispectrum (top panel; Eq.~(\ref{eq:sn})) 
 and the skewness (bottom panel; Eq.~(\ref{eq:skew_sn})).
 $S/N(<l_3)$ is in units of $f_{\rm NL}$.
 The dotted lines show {\sl COBE}, {\sl WMAP}, and {\sl Planck} 
 experiments (dotted lines), while the solid line shows the ideal experiment.
 See Table~\ref{tab:fnl} for $f_{\rm NL}$ to obtain $S/N>1$.}
\label{fig:sn}
\end{figure}

\begin{table} 
 \caption{\bf Detection Limit for the Non--Linearity Parameter}
 \mycaption{The minimum non--linearity parameter, $f_{\rm NL}$, needed
 for detecting the primordial non--Gaussianity by the bispectrum 
 or the skewness with signal--to--noise ratio greater than 1. 
 These estimates include the effects of cosmic variance, detector noise,
 and foreground sources.}
\label{tab:fnl}
\begin{center} 
  \begin{tabular}{ccc}\hline\hline
  Experiments & $f_{\rm NL}$ (Bispectrum) & $f_{\rm NL}$ (Skewness) \\
  \hline 
  {\sl COBE}   & 600 & 800 \\
  {\sl WMAP}   & 20  & 80 \\
  {\sl Planck} & 5   & 70 \\
  Ideal  & 3   & 60 \\
   \hline\hline
  \end{tabular}
\end{center}
\end{table}

\subsection{Measuring primordial non--Gaussianity in the cosmic 
microwave background}
\label{sec:measure2}

Measuring $f_{\rm NL}$ from nearly full--sky experiments is challenging. 
The bispectrum analysis explained in Sec.~\ref{sec:npointspectrum} 
requires $N^{5/2}$ operations 
($N^{3/2}$ for computing three $l$'s and $N$ for averaging over the sky) 
where $N$ is the number of pixels.
The brute--force analysis is possible for the {\sl COBE} data for which
$N\sim 3000$ \cite{komatsu/etal:2002}, while it is quite challenging for 
mega--pixel experiments 
(e.g., $N\sim 3\times 10^6$ for {\sl WMAP}, $5\times 10^7$ for {\sl Planck}).
In fact, just measuring all configurations of the bispectrum from the 
data is possible. What is challenging is to carry out many Monte
Carlo simulations: in order to quantify the statistical significance of 
the measurements, one needs many simulations . 
It is the simulations that are computationally very expensive.
Since the brute--force trispectrum analysis requires $N^3$, 
it is even more challenging.

Although we measure the individual triangle configurations of the bispectrum
(or quadrilateral configurations of the trispectrum) at first, 
we eventually combine all of them to constrain model parameters
such as $f_{\rm NL}$, as the signal--to--noise per 
configuration is nearly zero.
This may sound inefficient. Measuring all configurations 
is enormously time consuming. 
Is there any statistic which \textit{already} 
combines all the configurations optimally, and fast to compute?
Yes \cite{KSW}.
A physical justification for our methodology is as follows.
A model like Eq.~(\ref{eq:model}) generates non--Gaussianity in real 
space, and the Central--Limit Theorem makes the Fourier modes nearly 
Gaussian; thus, real-space statistics should be more sensitive.
On the other hand, real-space statistics are weighted sum of Fourier-space 
statistics, which are often easier to predict.
Therefore, we need to understand the shape of Fourier-space statistics
to find sensitive real-space statistics, and for this purpose it
is useful to have a specific, physically motivated non--Gaussian 
model, compute Fourier statistics, and find optimal real-space
statistics.

\subsubsection{Reconstructing primordial fluctuations from temperature 
anisotropy}\label{sec:wiener}

We begin with the primordial curvature perturbations
$\Phi\left(\mathbf{x}\right)$ and isocurvature perturbations
$S\left(\mathbf{x}\right)$.
If we can reconstruct these primordial fluctuations from the
observed CMB anisotropy, $\Delta T(\hat\mathbf{n})/T$, then we can 
improve the sensitivity to primordial non--Gaussianity. 
We find that the harmonic coefficients of the CMB anisotropy, 
$a_{lm}=
T^{-1}\int d^2\hat{\mathbf n}\Delta T(\hat\mathbf{n}) 
Y_{lm}^*(\hat\mathbf{n})$,
are related to the primordial fluctuations as
\begin{equation}
 \label{eq:alm}
  a_{lm}= W_l\int r^2 dr 
  \left[ \Phi_{lm}(r)\alpha_l^{adi}(r)
       + S_{lm}(r)\alpha_l^{iso}(r) \right] + n_{lm},
\end{equation}
where $\Phi_{lm}(r)$ and $S_{lm}(r)$ are the harmonic coefficients
of the fluctuations at a given comoving distance, $r=\left|\mathbf{x}\right|$
fom the observer. 
A beam function $W_l$ and the harmonic coefficients of the noise 
$n_{lm}$ represent instrumental effects.
Since noise can be spatially inhomogeneous, the noise covariance matrix
$\left<n_{lm}n_{l'm'}^*\right>$ can be non--diagonal; however, we approximate
it with $\simeq \sigma_0^2\delta_{ll'}\delta_{mm'}$. 
We thus assume the ``mildly inhomogeneous'' noise for which 
this approximation holds.
The function $\alpha_l(r)$ is defined by
\begin{equation}
 \label{eq:alpha_l}
 \alpha_l(r)
  \equiv
  \frac{2}{\pi}\int k^2 dk g_{Tl}(k) j_l(k r),
\end{equation}
where $g_{Tl}(k)$ is the radiation transfer function of either
adiabatic ($adi$) or isocurvature ($iso$) perturbations.
Note that this function is equal to $f_{\rm NL}^{-1}b_l^{\rm NL}(r)$ 
(see Eq.~(\ref{eq:bLr})). 

Next, assuming that $\Phi\left(\mathbf{x}\right)$ dominates,
we try to reconstruct $\Phi\left(\mathbf{x}\right)$ from
the observed $\Delta T(\hat\mathbf{n})$.
A linear filter, ${\mathcal O}_l(r)$, which reconstructs
the underlying field, can be obtained by minimizing the variance of the 
difference between the filtered field ${\mathcal O}_l(r) a_{lm}$ and
the underlying field $\Phi_{lm}(r)$.
By evaluating 
\begin{equation}
 \label{eq:minimization}
  \frac{\partial}{\partial {\mathcal O}_l(r)}
  \langle\left|{\mathcal O}_l(r) a_{lm}-\Phi_{lm}(r)\right|^2\rangle
  =0,
\end{equation}
one obtains a solution for the filter as
\begin{equation}
 \label{eq:Ol}
  {\mathcal O}_l(r)= \frac{\beta_l(r)W_l}{\tilde{C}_l},
\end{equation}
where the function $\beta_l(r)$ is given by
\begin{equation}
 \label{eq:beta_l}
  \beta_l(r) \equiv 
  \frac{2}{\pi}\int k^2 dk P(k) g_{Tl}(k) j_l(k r),
\end{equation}
and $P(k)$ is the power--spectrum of $\Phi$.
Of course, one can replace $\Phi$ with $S$ when $S$ dominates.
This function is equal to $b_l^L(r)$ (see Eq.~\ref{eq:bNLr}).
Here, we put a tilde on a quantity that includes effects
of $W_l$ and noise such that 
$\tilde{C}_l \equiv C_lW_l^2+\sigma_0^2$,
where $C_l$ is the theoretical power--spectrum that uses the same cosmological
model as $g_{Tl}(k)$.

Finally, we transform the filtered field ${\mathcal O}_l(r) a_{lm}$ back to 
pixel space to obtain an Wiener--filtered, reconstructed map of 
$\Phi(r,\hat\mathbf{n})$ or $S(r,\hat\mathbf{n})$.
We have assumed that there is no correlation between $\Phi$ and $S$.
We will return to study the case of non-zero correlation later
(Sec.~\ref{sec:mix}).

Figure~\ref{fig:filters} shows ${\mathcal O}_l(r)$ as a function of $l$ and 
$r$ for 
(a) an adiabatic SCDM ($\Omega_m=1$), 
(b) an adiabatic $\Lambda$CDM ($\Omega_m=0.3$), 
(c) an isocurvature SCDM, and
(d) an isocurvature $\Lambda$CDM.
While we have used $P(k)\propto k^{-3}$ for both adiabatic and
isocurvature modes, the specific choice of $P(k)$ does not affect 
${\mathcal O}_l$ very much as $P(k)$ in $\beta_l$ in the numerator 
approximately cancels out $P(k)$ in $C_l$ in the denominator. 
On large angular scales (smaller $l$) the Sachs--Wolfe (SW) effect
makes ${\mathcal O}_l$ equal to $-3$ for adiabatic modes and $-5/2$ for 
isocurvature modes in SCDM.
For the $\Lambda$CDM models the late--time decay of the gravitational potential
makes this limit different. 
Adiabatic and isocurvature modes are out of phase in $l$.

\begin{figure}
 \begin{center}
 \leavevmode \epsfxsize=9cm \epsfbox{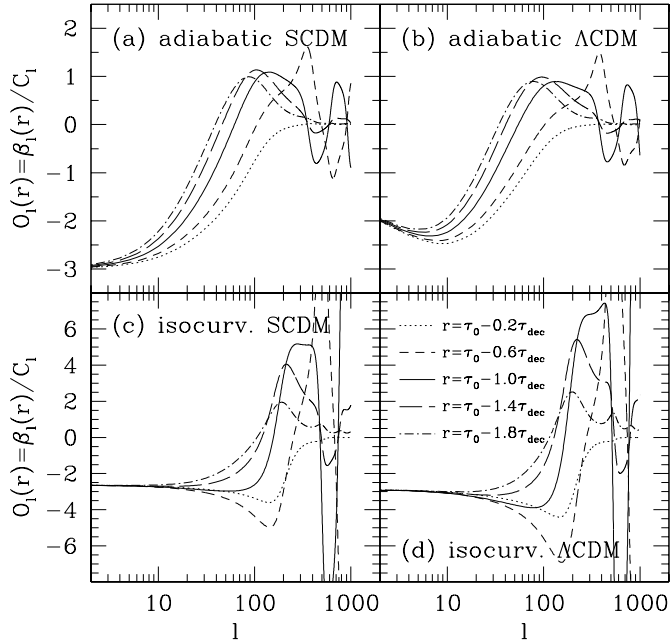}
 \end{center}
\label{fig:filters}
 \caption{{\bf Wiener Filters for the Primordial Fluctuations}~\cite{KSW}}
 \mycaption{
 Wiener filters, ${\mathcal O}_l(r)=\beta_l(r)/C_l$ (Eq.~(\ref{eq:Ol})).
 We plot (a) ${\mathcal O}_l$ for an adiabatic SCDM 
 ($\Omega_m=1$, $\Omega_\Lambda=0$, $\Omega_b=0.05$, $h=0.5$),
 (b) for an adiabatic $\Lambda$CDM ($\Omega_m=0.3$, 
 $\Omega_\Lambda=0.7$, $\Omega_b=0.04$, $h=0.7$), 
 (c) for an isocurvature SCDM, and
 (d) for an isocurvature $\Lambda$CDM.
 The filters are plotted at five conformal distances 
 $r=c(\tau_0-\tau)$ as explained in the bottom-right panel.
 Here $\tau$ is the conformal time ($\tau_0$ at the present).
 The SCDM models have $c\tau_0=11.84$~Gpc and 
 $c\tau_{dec}=0.235$~Gpc, while the $\Lambda$CDM models
 $c\tau_0=13.89$~Gpc and $c\tau_{dec}=0.277$~Gpc,
 where $\tau_{dec}$ is the photon decoupling epoch.}
\end{figure}

The figure shows that ${\mathcal O}_l$ changes 
the sign of the fluctuations as a function of scales.
This indicates that acoustic physics at the last scattering surface
modulates fluctuations so that hot spots in the primordial 
fluctuations can be cold spots in the CMB, for example.
Therefore, the shape of ${\mathcal O}_l$ ``deconvolves'' the sign change,
recovering the phases of fluctuations.
This is an intuitive reason why our cubic statistic derived below
(Eq.~(\ref{eq:skewnessprim})) works, and it proves more advantageous 
to measure primordial non--Gaussianity on a filtered map than on a 
temperature map.

This property should be compared to that of real-space statistics 
measured on a temperature map. 
As we have shown in Sec.~\ref{sec:measure} the skewness of 
a temperature map is much less sensitive to the primordial 
non--Gaussianity than the bispectrum, exactly because of the cancellation 
effect from the acoustic oscillations. 
The skewness of a filtered map, on the other hand,
has a larger signal--to--noise ratio, and more optimal statistics like our 
cubic statistic derived below can be constructed. 
Other real--space statistics such as Minkowski 
functionals~\cite{minkowski:1903,gott/etal:1990,schmalzing/gorski:1998,cab},
peak--peak correlations~\cite{guphea} may also be more sensitive 
to the primordial non--Gaussianity, when measured on the filtered 
maps.

Unfortunately, as $g_{Tl}$ oscillates, our reconstruction of $\Phi$ or 
$S$ from a temperature map alone is not perfect.
While ${\mathcal O}_l$ reconstructs the primordial fluctuations very well 
on large scales via the Sachs--Wolfe effect, 
${\mathcal O}_l\sim 0$ on intermediate
scales ($l\sim 50$ for adiabatic and $l\sim 100$ for isocurvature), 
indicating loss of information on the phases of the underlying fluctuations.
Then, toward smaller scales, we recover information, lose information, 
and so on. 
Exact scales at which ${\mathcal O}_l\sim 0$ depend on $r$ and cosmology.
A good news is that a high signal--to--noise map of the CMB polarization 
anisotropy will enable us to overcome the loss of 
information, as the polarization transfer function is out of phase
in $l$ compared to the temperature transfer function, 
filling up information at which ${\mathcal O}_l\sim 0$.
In other words, the polarization anisotropy has finite information
about the phases of the primordial perturbations, when the temperature 
anisotropy has zero information.

\subsubsection{Measuring primordial non--Gaussianity in adiabatic fluctuations}

Using two functions introduced in the previous section, we construct 
a {\it cubic} statistic which is optimal for the primordial non--Gaussianity.
We apply filters to $a_{lm}$, and then transform the filtered $a_{lm}$'s 
to obtain two maps, $A$ and $B$, given by 
\begin{eqnarray}
 \label{eq:filter1}
 A(r,\hat\mathbf{n}) &\equiv&
  \sum_{lm} \frac{\alpha_l(r)W_l}{\tilde{C}_l}a_{lm}Y_{lm}(\hat\mathbf{n}),\\
 \label{eq:filter2}
 B(r,\hat\mathbf{n}) &\equiv&
  \sum_{lm} \frac{\beta_l(r)W_l}{\tilde{C}_l}a_{lm}Y_{lm}(\hat\mathbf{n}).
\end{eqnarray}
The latter map, $B(r,\hat\mathbf{n})$, is exactly the 
${\mathcal O}_l$-filtered map, a Wiener--filtered map of the underlying 
primordial fluctuations. 
We then form a cubic statistic given by
\begin{equation}
 \label{eq:skewnessprim}
  {\mathcal S}_{\rm prim} \equiv 4\pi \int r^2 dr 
  \int \frac{d^2\hat\mathbf{n}}{4\pi}
  A(r,\hat\mathbf{n}) B^2(r,\hat\mathbf{n}),
\end{equation}
where the angular average is done on the full sky, regardless of 
the sky cut.
We find that ${\mathcal S}_{\rm prim}$ reduces \textit{exactly} to
\begin{equation}
 \label{eq:Sprim}
  {\mathcal S}_{\rm prim}= \sum_{l_1\le l_2\le l_3}
  \frac{\tilde{B}_{l_1l_2l_3}^{obs}\tilde{B}_{l_1l_2l_3}^{\rm prim}}
  {\tilde{C}_{l_1}\tilde{C}_{l_2}\tilde{C}_{l_3}},
\end{equation}
where 
\begin{equation}
\tilde{B}_{l_1l_2l_3} \equiv B_{l_1l_2l_3}W_{l_1}W_{l_2}W_{l_3},
\end{equation}
and $B_{l_1l_2l_3}^{obs}$ is the observed bispectrum with the effect of $W_l$
corrected while $B_{l_1l_2l_3}^{\rm prim}$ is given by
Eqs.~(\ref{eq:blprim}) and (\ref{eq:wigner*}).

The denominator of Eq.~(\ref{eq:Sprim}) is the variance of 
$\tilde{B}_{l_1l_2l_3}^{obs}$ in the limit of weak
non--Gaussianity (say $\left|f_{\rm NL}\right|\lesssim 10^3$)
when all $l$'s are different: 
$\left<\tilde{B}_{l_1l_2l_3}^2\right>= 
\tilde{C}_{l_1}\tilde{C}_{l_2}\tilde{C}_{l_3}\Delta_{l_1l_2l_3}$,
where $\Delta_{l_1l_2l_3}$ is 6 for $l_1=l_2=l_3$, 2 for 
$l_1=l_2\neq l_3$ etc., and 1 otherwise.
The bispectrum configurations are thus summed up nearly optimally with the 
approximate inverse--variance weights, provided that
$\Delta_{l_1l_2l_3}$ is approximated with $\simeq 1$.
The least--square fit of $\tilde{B}_{l_1l_2l_3}^{\rm prim}$ to 
$\tilde{B}_{l_1l_2l_3}^{obs}$ can be performed to yield
\begin{equation}
 \label{eq:Sprim*}
  {\mathcal S}_{\rm prim} \simeq f_{\rm NL}
  \sum_{l_1\le l_2\le l_3}
  \frac{(\tilde{B}_{l_1l_2l_3}^{\rm prim})^2}
  {\tilde{C}_{l_1}\tilde{C}_{l_2}\tilde{C}_{l_3}}.
\end{equation}
This equation gives an estimate of $f_{\rm NL}$ directly 
from ${\mathcal S}_{\rm prim}$.

The most time--consuming part is the back--and--forth harmonic transform
necessary for pre--filtering (see 
Eqs.~(\ref{eq:filter1}) and (\ref{eq:filter2})),
taking $N^{3/2}$ operations times the number of sampling points of $r$,
of order 100, for evaluating the integral (Eq.~(\ref{eq:skewnessprim})).
This is much faster than the full bispectrum analysis which
takes $N^{5/2}$, enabling us to perform a more detailed analysis
of the data in a reasonable amount of computational time.
For example, measurements of all bispectrum configurations up to
$l_{max}=512$ take 8 hours to compute on 16 processors of an SGI Origin 
300; thus, even only 100 Monte Carlo simulations take 1 month to be carried 
out.  
On the other hand, ${\mathcal S}_{\rm prim}$ takes only 30 seconds to compute, 
1000 times faster.
When we measure $f_{\rm NL}$ for $l_{max}=1024$, we speed up by a factor
of 4000: 11 days for the bispectrum vs 4 minutes for ${\mathcal S}_{\rm prim}$.
We can do 1000 simulations for $l_{max}=1024$ in 3 days.

\subsubsection{Mixed fluctuations} \label{sec:mix}

The ${\mathcal O}_l$-filtered map, $B$, is an Wiener--filtered map of
primordial curvature or isocurvature perturbations; however, this is correct 
only when correlations between the two components are negligible.
On the other hand, multi--field inflation models and curvaton models 
naturally predict correlations.
The current CMB data are consistent with, but do not require, a correlated 
mixture of these fluctuation modes  
\cite{amendola/etal:2001,trotta/riazuelo/durrer:2001,ex}.
In this case, the Wiener filter for the primordial fluctuations 
(Eq.~(\ref{eq:Ol})) needs to be modified such that 
${\mathcal O}_l(r)=\beta_l(r)W_l/\tilde{C}_l
\rightarrow \tilde{\beta}_l(r)W_l/\tilde{C}_l$, where
\begin{eqnarray}
 \label{eq:betaadi_l}
 \nonumber
  \tilde{\beta}^{adi}_l(r) &=& 
  \frac{2}{\pi}\int k^2 dk 
  \left[ P_{\Phi}(k) g_{Tl}^{adi}(k) 
       + P_{C}(k)   g^{iso}_{Tl}(k) \right] j_l(k r), \\
 \label{eq:betaiso_l}
 \nonumber
  \tilde{\beta}^{iso}_l(r) &=& 
  \frac{2}{\pi}\int k^2 dk 
  \left[ P_{S}(k)     g_{Tl}^{iso}(k) 
       + P_{C}(k) g^{adi}_{Tl}(k) \right] j_l(k r),
\end{eqnarray}
for curvature ($adi$) and isocurvature ($iso$) perturbations, respectively.
Here $P_{\Phi}$ is the primordial power--spectrum of curvature
perturbations, $P_{S}$ of isocurvature perturbations, and $P_{C}$
of cross correlations. 

For measuring non--Gaussianity from the correlated fluctuations,
we use Eq.~(\ref{eq:model}) as a model for $\Phi$-- and $S$--field
non--Gaussianity to parameterize them with $f_{\rm NL}^{adi}$ and
$f_{\rm NL}^{iso}$, respectively. 
We then form a cubic statistic similar to ${\mathcal S}_{\rm prim}$ 
(Eq.~(\ref{eq:skewnessprim})), using $A(r,\hat\mathbf{n})$ and a new filtered 
map $\tilde{B}(r,\hat\mathbf{n})$ which uses $\tilde{\beta}_l(r)$.
We have two cubic combinations: $A_{adi}\tilde{B}^2_{adi}$ for
measuring $f_{\rm NL}^{adi}$ and $A_{iso}\tilde{B}^2_{iso}$ for
$f_{\rm NL}^{iso}$, each of which comprises four terms including
one $P_{\Phi}^2$ (or $P_{S}^2$), one $P_{C}^2$, 
and two $P_{\Phi}P_{C}$'s (or $P_{S}P_{C}$'s).
In other words, the correlated contribution makes the total number of terms 
contributing to the non--Gaussianity four times more than the 
uncorrelated--fluctuation models (see Ref.~\cite{BMT3} for more 
generic cases).

\subsubsection{Point--source non--Gaussianity}

Next, we show that the filtering method is also useful for 
measuring foreground non--Gaussianity arising from extragalactic 
point--sources.
The residual point--sources left unsubtracted in a map can 
seriously contaminate both the power--spectrum and the bispectrum.
We can, on the other hand, use multi--band observations as well as 
external template maps of dust, free--free, and synchrotron emission, to 
remove diffuse Galactic foreground~\cite{bennett/etal:2003c}.
The radio sources with known positions can be safely masked.

The filtered map for the point--sources is 
\begin{equation}
 D(\hat\mathbf{n}) \equiv  
 \sum_{lm} \frac{W_l}{\tilde{C}_l} a_{lm} Y_{lm}(\hat\mathbf{n}).
\end{equation}
This filtered map was actually used for detecting point--sources
in the {\sl WMAP} maps~\cite{bennett/etal:2003c}.
Using $D(\hat\mathbf{n})$, the cubic statistic is derived as
\begin{equation}
 \label{eq:skewnesssrc}
  {\mathcal S}_{\rm src} \equiv
  \int \frac{d^2\hat\mathbf{n}}{4\pi} D^3(\hat\mathbf{n})
  =
  \frac{3}{2\pi} \sum_{l_1\le l_2\le l_3}
  \frac{\tilde{B}_{l_1l_2l_3}^{obs}\tilde{B}_{l_1l_2l_3}^{\rm src}}
  {\tilde{C}_{l_1}\tilde{C}_{l_2}\tilde{C}_{l_3}}.
\end{equation}
Here, $B_{l_1l_2l_3}^{\rm src}$ is the point--source bispectrum for unit
white--noise bispectrum (i.e., $b^{\rm src}=1$ in Eq.~(\ref{eq:pointsource})).
When the covariance between $B_{l_1l_2l_3}^{\rm prim}$ and
$B_{l_1l_2l_3}^{\rm src}$ is negligible as is the case for {\sl WMAP} and 
{\sl Planck} (see Table~\ref{tab:corr}), we find
\begin{equation}
 \label{eq:Sps}
  {\mathcal S}_{\rm src} \simeq
  \frac{3b^{\rm src}}{2\pi} 
  \sum_{l_1\le l_2\le l_3}
   \frac{(\tilde{B}_{l_1l_2l_3}^{\rm src})^2}
   {\tilde{C}_{l_1}\tilde{C}_{l_2}\tilde{C}_{l_3}}.
\end{equation}
We omit the covariance only for simplicity; however, including
it would be simple~\cite{komatsu/etal:2002}. 

Again, ${\mathcal S}_{\rm src}$ measures $b^{\rm src}$ much faster 
than the full
bispectrum analysis, constraining effects of residual point--sources
on CMB sky maps. Since ${\mathcal S}_{\rm src}$ does not contain the 
extra integral
over $r$, it is even 100 times faster to compute than 
${\mathcal S}_{\rm prim}$.
This statistic is particularly useful because it is sometimes difficult to
tell how much of $C_l$ is due to point--sources.
In Sec.~\ref{sec:obs} we see how ${\mathcal S}_{\rm src}$ 
(i.e., $b^{\rm src}$) is related to $C_l$ due to the 
unsubtracted point--sources.

\subsubsection{Incomplete sky coverage}

Finally, we show how to incorporate incomplete sky coverage and
pixel weights into our statistics. 
Suppose that we weight a sky map by $M(\hat\mathbf{n})$ to measure 
the harmonic coefficients,
\begin{equation}
a_{lm}^{obs}= \frac1{T}\int d^2\hat\mathbf{n} M(\hat\mathbf{n})
\Delta T(\hat\mathbf{n}) Y_{lm}^*(\hat\mathbf{n}).
\end{equation}
A full--sky $a_{lm}$ is related to $a_{lm}^{obs}$ 
through the coupling matrix \linebreak 
$M_{ll'mm'}\equiv \int d^2\hat\mathbf{n} M(\hat\mathbf{n})
Y_{lm}^*(\hat\mathbf{n})Y_{l'm'}(\hat\mathbf{n})$ by
$a_{lm}^{obs}=\sum_{l'm'}a_{l'm'}M_{ll'mm'}$. 
In this case the observed bispectrum is biased by a factor of
$\int d^2\hat\mathbf{n}M^3(\hat\mathbf{n})/(4\pi)$;
thus, we need to divide $S_{\rm prim}$ and $S_{ps}$ by this factor.
If only the sky cut is considered, then this factor is the fraction of 
the sky covered by observations (see Eq.~\ref{eq:biasbl}). 

Monte Carlo simulations of non--Gaussian sky maps 
computed with Eq.~(\ref{eq:alm})
(see Appendix A of Ref.~\cite{k})
show that ${\mathcal S}_{\rm prim}$
reproduces the input $f_{\rm NL}$'s accurately both on full sky and 
incomplete sky with modest Galactic cut and inhomogeneous noise
on the {\sl WMAP} data, i.e., the statistic is unbiased.
The error on $f_{\rm NL}$ from ${\mathcal S}_{\rm prim}$ is 
as small as that from the full bispectrum analysis;
however, one cannot make a sky cut very large, 
e.g., more than $50\%$ of the sky, as for it the covariance matrix of 
$\tilde{B}_{l_1l_2l_3}$ is no longer diagonal.
The cubic statistic does not include the off--diagonal 
terms of the covariance matrix [see Eq.~(\ref{eq:Sprim})];
however, it works fine for {\sl WMAP} sky maps for which one can use
more than $75\%$ of the sky.
Also, Eq.~(\ref{eq:Sps}) correctly
estimates $b^{\rm src}$ using simulated realizations of point--sources
(see Appendix B of Ref.~\cite{k}).

These fast methods allow to carry out extensive Monte Carlo
simulations characterizing the effects of realistic noise properties of 
the experiments, sky cut, foreground sources, and so on.  
A reconstructed map of the primordial fluctuations, which 
plays a key role in the method, potentially gives other real--space 
statistics more sensitivity to primordial non--Gaussianity.
As it has been shown, the method can be applied to the primordial 
non--Gaussianity arising from inflation, gravity, or 
correlated isocurvature fluctuations, as well as the foreground 
non--Gaussianity from radio point--sources, all of which can be important 
sources of non--Gaussian fluctuations on the CMB sky maps.

\subsection{Applications to observational data}
\label{sec:obs}

There are two approaches to testing Gaussianity of the CMB.

\begin{itemize}
\item Blind tests (null tests) which make no assumption about
the form of non--Gaussianity. 
The simplest test would be measurements of deviation of one--point
PDF. from a Gaussian distribution. (Measurements of the skewness,
kurtosis, etc., for example.)
Being model--independent is a merit of this approach,
while the statistical power is weak. If we had no models to test,
this approach would be the only choice.
\item Testing specific models of non--Gaussianity, constraining the 
model parameters. This approach is powerful in putting
{\it quantitative} constraints on non--Gaussianity, at the cost of 
being model--dependent.
If we had a sensible (yet fairly generic) model to test, 
this approach would be more powerful than the blind tests.
\end{itemize}

Both approaches have been applied to the CMB data
on large angular scales ($\sim 7^\circ$) 
\cite{kogut/etal:1996b,heavens:1998,schmalzing/gorski:1998,ferreira/magueijo/gorski:1998,pando/valls/fang:1998,bromley/tegmark:1999,banday/zaroubi/gorski:2000,contaldi/etal:2000,mukherjee/hobson/lasenby:2000,magueijo:2000,novikov/schmalzing/mukhanov:2000,sandvik/magueijo:2001,barreiro/etal:2000,phillips/kogut:2001,komatsu/etal:2002,komatsu:phd,kunz/etal:2001,aghanim/forni/bouchet:2001,cayon/etal:2003}, 
on intermediate scales ($\sim 1^\circ$) 
\cite{park/etal:2001,shandarin/etal:2002}, and on small scales ($\sim 10'$) 
\cite{wu/etal:2001,santos/etal:2003,polenta/etal:2002}.
So far, there is no compelling evidence for the cosmological non--Gaussianity,
and the pre-{\sl WMAP} constraint on $f_{\rm NL}$ was weak, 
$f_{\rm NL}\lesssim (2000-3000)$ at $95\%$ confidence level 
\cite{komatsu/etal:2002,cayon/etal:2003,santos/etal:2003}.

In this section, we briefly review results of Gaussianity tests on the 
{\sl WMAP} data presented in Ref.~\cite{k}.
The {\sl WMAP}, Wilkinson Microwave Anisotropy Probe, has recently produced
clean and precise sky maps of the CMB in 5 microwave bands
\cite{bennett/etal:2003b}, with the angular resolution 30 times better
than that of the Differential Microwave Radiometer (DMR) aboard the 
{\sl COBE} satellite~\cite{bennett/etal:1996}.
Detailed study of these sky maps offers a fundamental test
of cosmology, as various cosmological effects change temperature and 
energy distribution of the CMB at all angular scales (e.g., 
\cite{hu/sugiyama:1995}).
The temperature and polarization power--spectra of the {\sl WMAP} data
\cite{hinshaw/etal:2003a,kogut/etal:2003} have determined the best-fit 
cosmological model with errors in the parameter determinations being 
quite small ($<10\%$) \cite{spergel/etal:2003,page/etal:2003a}.
The systematic errors in the parameter determinations are minimized 
by both the careful instrumental design
\cite{jarosik/etal:2003,page/etal:2003b,barnes/etal:2003,hinshaw/etal:2003b} 
and data analysis techniques~\cite{verde/etal:2003}.

Apart from the CMB, there are a number of non--cosmological,
``foreground'' sources in the microwave sky.
The emission from our Galaxy is the brightest component, which 
must be masked or subtracted out before any cosmological 
analysis of the CMB. Since the {\sl WMAP} observes in 5 frequency bands, 
much of the Galactic emission can be reliably subtracted using the
non--monochromatic nature of the Galaxy~\cite{bennett/etal:2003c}. 
The power--spectra measured in different bands
coincide with each other after the foreground subtraction, which is
reassuring~\cite{hinshaw/etal:2003a}.
Actually, much more problematic a foreground component is the 
extragalactic radio sources. 
Although we can mask those positions of the sky which are known to have
sources brighter than some threshold flux (which is determined by the 
sensitivity of observations), there always remain undetected sources.
The undetected (unmasked) sources potentially contaminate the 
cosmological CMB signals. Since we cannot subtract
them out individually, we must estimate the effect of the sources
in a statistical manner.

The emission from the sources is highly non--Gaussian and only important
on small angular scales; thus, we can use the non--Gaussian signals to 
directly estimate the source contribution.
This example illustrates usefulness of the higher--order statistics 
in a real life.

\subsubsection{Minkowski functionals}

For the first test, one can use (but is not limited to) the Minkowski 
functionals 
\cite{minkowski:1903,gott/etal:1990,schmalzing/gorski:1998,cab},
which measure morphological structures of the CMB, describing 
the properties of regions spatially bounded by a set of contours.
The contours may be specified in terms of fixed temperature thresholds,
$\nu = \Delta T / \sigma$, where $\sigma$ is the standard deviation of 
the map, or in terms of the area.
The three Minkowski functionals are:
(1) the total area above threshold, $A(\nu)$, (2) the total contour 
length, $C(\nu)$, and (3) the genus, $G(\nu)$, which is the number of
hot spots minus the number of cold spots.
Parameterization of contours by threshold is computationally simpler,
while parameterization by area reduces the correlations between the Minkowski 
functionals~\cite{shandarin/etal:2002}; however,
when a joint analysis of the three Minkowski functionals is performed, one 
has to explicitly include their covariance anyway. 
Therefore the simpler threshold parameterization will be used.

In Ref.~\cite{k} the Minkowski functionals at 5 different resolutions
from the pixel size of 3.7 degrees in diameter to 12 arcminutes have been 
measured.  
Figure~\ref{minkowski_res7} shows one example at $28'$ pixel resolution.
The gray band shows the $68\%$ confidence region 
derived from 1000 Gaussian Monte Carlo simulations.
(See section~2.3 of Ref.~\cite{k} for description of 
the simulations.)
The {\sl WMAP} data are in excellent agreement with the Gaussian simulations
at all resolutions.
But, {\it how Gaussian is it?}

 \begin{figure}[t]
 \begin{center}
 \leavevmode \epsfsize=29cm \epsfbox{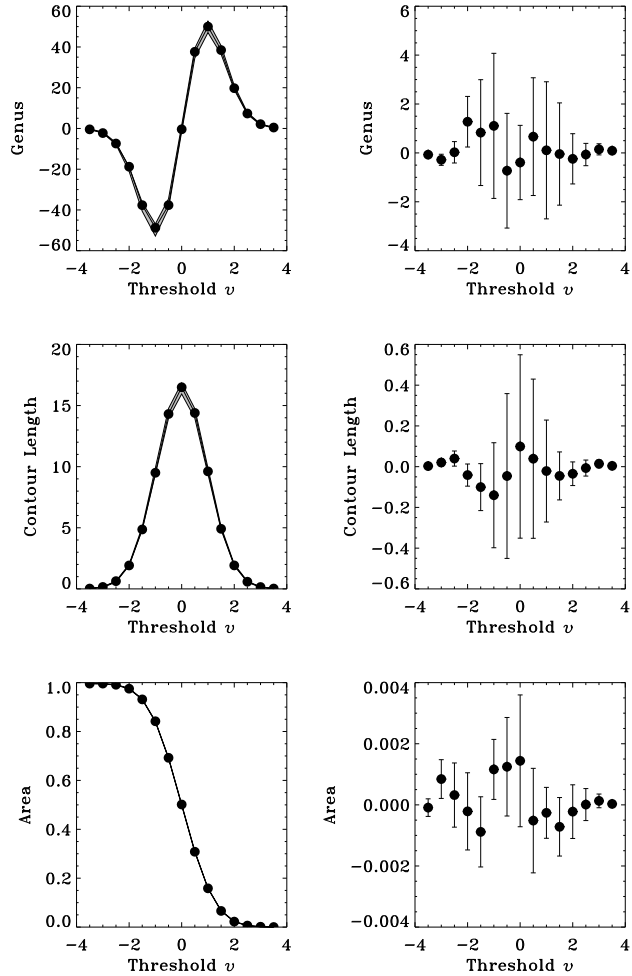} 
 \end{center}
 \caption{{\bf The Minkowski Functionals for the 
Foreground--cleaned {\sl WMAP} Data}~\cite{k}}
 \mycaption{%
 The Minkowski functions at $28'$ pixel resolution (filled circles) 
 and the residuals between the mean of the Gaussian simulations 
 and the {\sl WMAP} data.
 The gray band shows the 68\% confidence interval for the Gaussian 
 Monte Carlo simulations.
 The {\sl WMAP} data are in excellent agreement with the Gaussian simulations.
 \label{minkowski_res7}
 }
 \end{figure}

\subsubsection{Angular bispectrum}

For the second test, we use the fast cubic statistics 
derived in Sec.~\ref{sec:measure2},
which combine three--point (triangle) configurations of the angular bispectrum
that are sensitive to the models under consideration.

Once again, we consider two components. The first one is the primordial 
non--Gaussianity 
from inflation parametrized by $f_{\rm NL}$ (see Sec.~\ref{Phinonline}), 
which determines the amplitude of a 
quadratic term added to Bardeen's curvature perturbations:
$\Phi({\mathbf x})= 
\Phi_{L}({\mathbf x}) + f_{\rm NL}\left[ \Phi_{L}^2({\mathbf x}) 
 	      - \left<\Phi_{L}^2({\mathbf x})\right> \right]$,
It is useful to estimate the {\it r.m.s.} amplitude of $\Phi$ to see
how important the second--order term is. One obtains
$\left<\Phi^2\right>^{1/2}\simeq \left<\Phi^2_{\rm L}\right>^{1/2}
\left(1+f_{\rm NL}^2\left<\Phi^2_{\rm L}\right>\right)$,
where $\left<\Phi^2\right>^{1/2}\simeq 
3.3\times 10^{-5}$ \cite{bennett/etal:1996}; thus, a fractional
contribution from the second term is 
\begin{equation}
f_{\rm NL}^2\left<\Phi^2_{\rm L}\right>\simeq 10^{-5}(f_{\rm NL}/100)^2 \;.
\end{equation}
We are talking about very small effects. 

This parameterization is 
useful to find {\it quantitative} constraints on the amount of 
non--Gaussianity allowed by the CMB data.
Also, the form is general in that
$f_{\rm NL}$ parameterizes the leading--order non--linear 
corrections to $\Phi$. 

Figure~\ref{fig:parameters} shows $f_{\rm NL}$ measured 
from the foreground--cleaned 
Q$+$V$+$W coadded map using the cubic statistic, as a function of the maximum 
multipole $l_{max}$ (for details of measurements, see 
Ref.~\cite{k}).
There is no significant detection of $f_{\rm NL}$ at any angular scale.
There is no significant band--to--band variation, or significant detection
in any band. The best constraint is $-58<f_{\rm NL}<134$ (95\%),
which is equivalent to say that the fractional contribution to the 
{\it r.m.s.} value of 
$\Phi$ from the second--order term is smaller than $2\times 10^{-5}$.
These results support inflationary models, but still do not exclude the
possibility of having a small contribution from non--linearities predicted by
second--order perturbation theory. 

Note that $f_{\rm NL}$ for $l_{max}=265$ has a smaller error than that 
for $l_{max}=512$, because the latter is dominated by the instrumental noise. 
Since all the pixels outside the cut region are uniformly weighted,
the inhomogeneous noise in the map
(pixels on the ecliptic equator are noisier than those on the north and 
south poles) is not accounted for.
This leads to a noisier estimator than a minimum variance estimator. 
The constraint on $f_{\rm NL}$ for $l_{max}=512$ will likely improve with more 
appropriate pixel-weighting schemes~\cite{heavens:1998,santos/etal:2003}.
Apparently, the fact that the constraint actually obtained from the data is 
worse than predicted (c.f., Table~\ref{tab:fnl}) 
should be due to sub-optimalness
of the current estimator.
The simple inverse noise ($N^{-1}$) weighting makes the constraints 
much worse than the uniform weighting, as it increases errors on 
large angular scales where the CMB signal dominates over the 
instrumental noise. (However, it works fine for the point--sources.)
The uniform weighting is thus closer to optimal.

The Minkowski functionals shown in Figure~\ref{minkowski_res7} 
also place constraints on $f_{\rm NL}$, comparing the data to
the predictions derived from Monte Carlo simulations of the non--Gaussian
CMB (for details of the simulations, see Appendix A of 
\cite{k}).
It has been found that $f_{\rm NL}<139$ (95\%), 
remarkably consistent with that from the bispectrum analysis.

\begin{figure}
\begin{center}
\leavevmode \epsfxsize=6.7cm \epsfbox{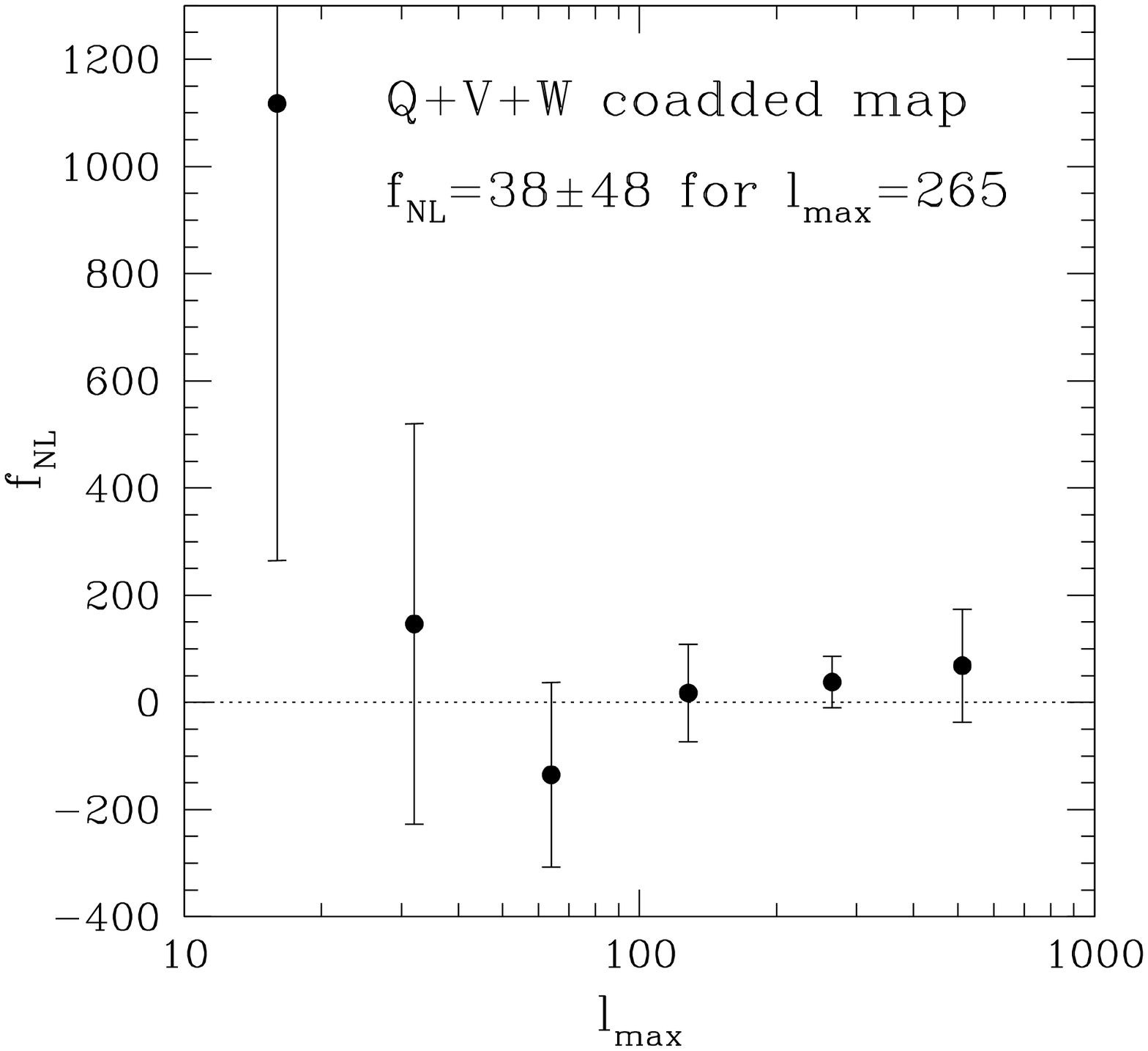} 
 \epsfxsize=6.7cm \epsfbox{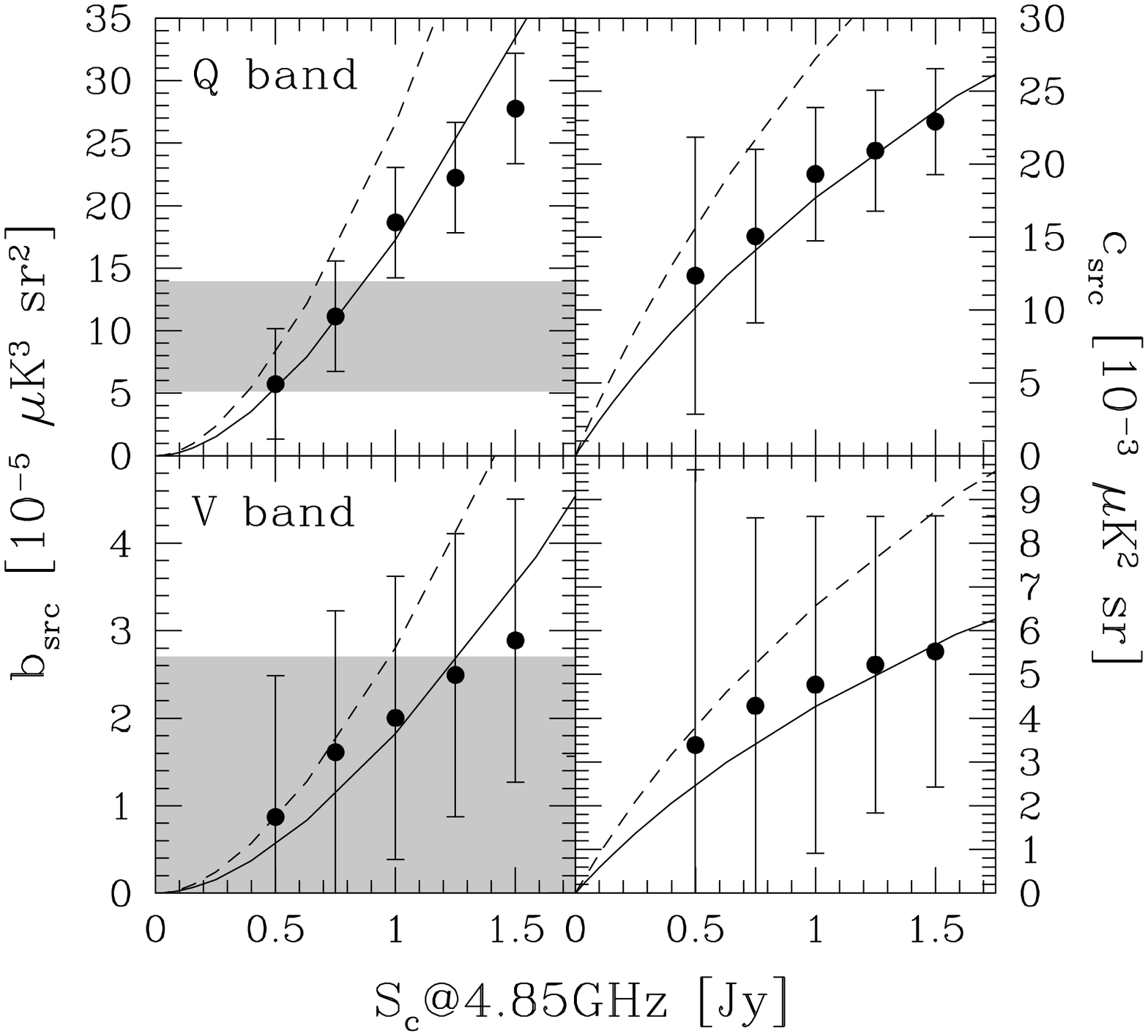}
 \end{center}
 \caption{{\bf Primordial Non--Gaussianity and Point--Source 
 Contribution}~\cite{k}}
 \mycaption{%
 ({\it Left Panel})
 The non--linearity parameter, $f_{\rm NL}$, as a function
 of the maximum multipole $l_{max}$, measured from the Q$+$V$+$W
 coadded map using the bispectrum estimator.
 The error bars at each $l_{max}$ are not independent.
 ({\it Right Panel})
 The point--source angular bispectrum $b^{\rm src}$ and 
 power--spectrum $c^{\rm src}$.
 The left panels show $b^{\rm src}$ in Q band (top panel) and 
 V band (bottom panel).
 The shaded areas show measurements from the {\sl WMAP} sky maps with
 the standard source cut, while the filled circles show those with 
 flux thresholds $S_{c}$ defined at 4.85~GHz. 
 The dashed lines show predictions from the source count model of
 Ref.~\cite{toffolatti/etal:1998}, while the solid lines
 are those multiplied by 0.65 to match the {\sl WMAP} measurements.
 The right panels show $c^{\rm src}$.
 The filled circles are computed from the measured $b^{\rm src}$
 substituted into Eq.~(\ref{eq:bsrc-csrc}).
 The lines are predictions. The error bars are not independent.
 \label{fig:parameters}
 }%
\end{figure}

\subsubsection{Point--source non--Gaussianity}

The second component is the foreground non--Gaussianity from radio 
point--sources, parameterized by the skewness, $b^{\rm src}$.
The filled circles in the right panels of Figure~\ref{fig:parameters} 
show $b^{\rm src}$ measured in Q (top panel) and V (bottom panel) band.
We have used source masks for various flux cuts, $S_{\rm c}$, defined at
4.85~GHz to make these measurements. 
(The masks are made from the GB6$+$PMN~5~GHz source catalogue.)
We find that $b^{\rm src}$ increases as $S_{\rm c}$: the brighter sources
being unmasked, the more non--Gaussianity is detected.
On the other hand one can make predictions for $b^{\rm src}$ using
the source count model.
Comparing the measured values of $b^{\rm src}$ with the predicted 
counts by
\cite{toffolatti/etal:1998} (dashed lines) at 44~GHz, one finds 
that the measured values are smaller than the predicted values by a 
factor of 0.65. The solid lines show the predictions multiplied by 0.65.
Our value for the correction factor matches well the one obtained 
from the {\sl WMAP} source counts for $2-10~{\rm Jy}$ in Q band 
\cite{bennett/etal:2003c}.

The source bispectrum, $b^{\rm src}$, is related to the source 
power--spectrum, $c^{\rm src}$, by an integral relation~\cite{k},
\begin{equation}
 \label{eq:bsrc-csrc}
  c^{\rm src}(S_{c})= b^{\rm src}(S_{c})[g(\nu)S_{c}]^{-1}
  +\int_0^{S_{c}} \frac{dS}S b^{\rm src}(S) [g(\nu)S]^{-1},
\end{equation}
where $g(\nu)$ is a conversion factor from ${\rm Jy~sr^{-1}}$ to $\mu$K  
which depends upon the observing frequency $\nu$ as
$g(\nu) = (24.76~{\rm Jy~\mu K^{-1}~sr^{-1}})^{-1}[(\sinh x/2)/x^2]^2$, 
$x\equiv h\nu/k_{\rm B}T_0\simeq \nu/(56.78~{\rm GHz})$.
One can use this equation combined with the measured $b^{\rm src}$ as a 
function of the flux threshold $S_{c}$ to directly determine 
$c^{\rm src}$ as a function of $S_{c}$, {\it without 
relying on any extrapolations}.
The right panels of Figure~\ref{fig:parameters} also show the estimated
$c^{\rm src}$ as filled circles.
The measurements suggest that $c^{\rm src}$ for the standard
source mask (indicated by the shaded area) is 
$c^{\rm src}=(15\pm 6)\times 10^{-3}~\mu{\rm K^2~sr}$ in Q band.
In V band, $c^{\rm src}=(4.5\pm 4)\times 10^{-3}~\mu{\rm K^2~sr}$.

In addition to the bispectrum, the {\sl WMAP} team has carried out other 
methods to estimate the source
contribution: (1) extrapolation from the number counts 
of detected sources in the {\sl WMAP} data~\cite{bennett/etal:2003c}, 
and (2) the angular power--spectrum on small angular scales
\cite{hinshaw/etal:2003a}. These methods yield consistent results.

In summary, the {\sl WMAP} 1-year data has enormously improved the 
sensitivity for testing the
Gaussianity of the CMB. Yet, we do not have any compelling evidence 
for primordial non--Gaussianity. 
This result is consistent with what is predicted by inflation and 
the second--order perturbation theory. 
There may be some chance to find non--Gaussian signals arising
from second--order perturbations.
Detection can be made possible by the {\sl Planck} experiment combining the 
temperature and polarization anisotropies. 
While we can detect $f_{\rm NL}\sim 5$ by using the temperature alone
(see Table~\ref{tab:fnl}), combining the polarization measurements
increases our sensitivity: we have several observables for the
bispectrum such as $\langle TTT\rangle$,
$\langle TTE\rangle$, $\langle TEE\rangle$, and $\langle EEE\rangle$.
The future polarization-dedicated satellite experiment (e.g., {\sl CMBPol})
in combination with the {\sl Planck} temperature map may enable us to 
detect $f_{\rm NL}\sim 3$.

\section{\bf Conclusions and future prospects}

Testing Gaussianity of the primordial fluctuations is 
and will be one of the most powerful probes of the inflationary paradigm.
Gaussianity tests are complementary to conventional ones using 
the power--spectrum: as we have shown in this review, Gaussianity tests enable 
us to discriminate between different inflationary models which would be 
indistinguishable otherwise. We have examined various examples
including the standard single--field slow--roll inflation,  
the curvaton model, the inhomogeneous reheating scenario, multi--field 
models and some unconvonetional scenarios, which make unique predictions
for the strength of non-Gaussianity and its shape (see Table~1 for a summary).
For the single--field slow--roll inflationary model, on the other hand, 
we have shown that inflation itself produces a negligible amount of 
non--Gaussianity, and the dominant contribution comes from the evolution of 
the ubiquitous second--order perturbations after inflation, which is 
potentially detectable with future observations of temperature and 
polarization anisotropies of the CMB. This effect {\it must exist} regardless
of inflationary models, setting the minimum level of non--Gaussianity in the 
cosmological perturbations. Alternative models for the generation of
perturbations might produce stronger non--Gaussianity
than this minimum amount. Therefore, if we do not find any evidence for this 
ubiquitous non--Gaussianity, then it will challenge our understanding of the 
evolution of cosmological perturbations at a deeper level.
(In other words, no detection of non--Gaussianity at the level of 
$f_{\rm NL}\sim 1$ rules out our standard cosmological model!)   
It is extremely important to keep improving upon our sensitivity to the 
primordial non--Gaussianity until we reach the critical sensitivity, 
$f_{\rm NL} \sim 1$.

We have reviewed in great detail the current constraints on $f_{\rm NL}$ 
from the angular bispectrum of the CMB. Here, let us make a remark on future 
prospects for observational constraints on $f_{\rm NL}$. 
It has been shown that the angular bispectrum of temperature anisotropy 
alone can detect non--Gaussianity, if $|f_{\rm NL}|>5$ \cite{ks}.
This estimate assumes that the {\sl Planck} satellite is the ultimate 
experiment measuring temperature anisotropy 
in terms of primordial non--Gaussianity. Small--scale CMB experiments, such 
as the Atacama Cosmology Telescope \cite{kosowsky:2004}, would detect 
non--Gaussianity from secondary anisotropies (see Sec~10.4 and, e.g.,  
Refs.~\cite{goldberg/spergel:1999,cooray/hu:2000,aghanim/forni:1999}).
If we add polarization information (which is assumed to be measured 
as accurately as temperature up to $l\sim 3000$), then one can improve it to
$f_{\rm NL}>3$. This is still a factor of 3 larger than the critical 
limit; however, fortunately we have many more Gaussianity tests which can, 
in principle, give us independent measurements of $f_{\rm NL}$. 
If fluctuations are Gaussian, then the power--spectrum contains all the 
statistical information, so that one cannot overcome cosmic variance by 
using other statistical tools; however, if fluctuations are non--Gaussian,
then there can be many independent statistical tools measuring different
aspects of the same non--Gaussianity, giving independent constraints on the 
strength and shape of non--Gaussianity. \footnote{
Let us mention here the analysis of the 3--pont function of CMB 
anisotropies in the WMAP data of Ref.~\cite{gazta}. Also interesting 
is a statistical method based on the multivariate empirical distribution 
function of the spherical harmonics, proposed in Refs.~\cite{romani1,romani2}} 
If those statistical tools are orthogonal to the bispectrum, then one can 
improve the limits on $f_{\rm NL}$ by the square--root of the number of 
independent statistical tools (i.e., we need at least 9 completely 
independent methods to measure $f_{\rm NL}=1$.) 

Although numerous statistical estimators have been applied to 
the CMB data for Gaussianity tests, only a few 
of these (Minkowski functionals \cite{k}, Mexican--hat wavelets 
\cite{mukherjee/wang:2004}, 
local curvature \cite{cabella/etal:2004}) have been used to find 
limits on $f_{\rm NL}$. 
Also, the extent to which these statistical tools are independent 
remains unknown (see Ref.~\cite{cab} for the first attempt 
to address this issue); thus, studying statistical power
and complementarity of the statistical tools measuring $f_{\rm NL}$ 
will be one of the most important goals. 
To achieve this goal, it is crucial to have accurate numerical 
simulations of the non--Gaussian 
CMB sky maps (both temperature and polarization), as well as analytical 
calculations of the effects of $f_{\rm NL}$ on the statistics can be very 
complicated. At the time of writing, analytical predictions exist 
only for the bispectrum \cite{ks} and the trispectrum \cite{okamoto/hu:2002}.
Simulations of non-Gaussian temperature fluctuations with the 
$f_{\rm NL}$--model already exist \cite{k,liguori1} 
and can be readily extendend 
to include polarization, as well as arbitrary non--Gaussian initial 
conditions in the primordial curvature and entropy perturbations. 
Moreover, direct simulations of the non--linear dynamics of cosmological 
perturbations, which have been evaluated analytically in this review, 
may be feasible. We clearly need a systematic study of the combined 
statistical power of various methods constraining $f_{\rm NL}$, using 
these simulations.

In addition to the CMB, we have other methods to constrain the primordial 
non--Gaussianity. Galaxy correlations at large distances,
where non--linear clustering is modest, still preserve statistical properties 
of the primordial fluctuations; thus, one can use them to find limits on the 
primordial non-Gaussianity
(see, e.g., Ref.\cite{bernardeau/etal:2002} for a review). 
Using the three--dimensional bispectrum, the authors of 
Ref.~\cite{sefusa} have found a limit $-2000<f_{\rm NL}<1600$ (95\%),  
from the PSCz survey. They conclude that the bispectrum analysis of 
the Sloan Digital Sky Survey can reach $|f_{\rm NL}|\sim (150-200)$, 
and that of an idealistic 
all--sky redshift survey up to $z\sim 1$ can reach $|f_{\rm NL}|\sim 1$. 
In principle, therefore, the LSS data might become
as competetive as the CMB data. One big advantage of the LSS
data is three--dimensional information. 
Since the CMB data give us only two--dimensional information,
the number of modes of the bispectrum that one can measure is fairly 
limited even on the full sky, and this 
is the fundamental limitation of the CMB bispectrum. On the other hand, 
the number of modes available in the three--dimensional bispectrum is 
enormous, and helps to obtain tight limits on $f_{\rm NL}$. 
Of course, there are disadvantages of the LSS data: 
non--linear clustering and bias producing
spurious non--Gaussian signals~\cite{verde/etal:2000}. 
Combination of CMB and LSS data will thus offer 
a systematic error check
and potentially an improved signal--to--noise ratio for detection of 
$f_{\rm NL}\sim 1$.

Yet another tool is the number of massive halos (e.g., clusters of galaxies) 
at high $z$
\cite{lucchin/matarrese:1988,koyama/soda/taruya:1999,majive,robinson/gawiser/silk:2000,willick:2000,chen}.
The number of massive clusters is very sensitive to statistical 
properties of the primordial fluctuations. For example, 
one can calculate the number of clusters corresponding to density 
peaks of, say, 3--$\sigma$, for a Gaussian distribution.
Since these objects are rare, the number is very sensitive to the exact 
shape of the tail of the probability distribution function of density 
fluctuations. 
Even a slight amount of non-Gaussianity can change it rather dramatically. 
This method is powerful when 
density fields are positively skewed, giving more objects for a given mass 
and redshift. (A positive skewness in density fluctuations corresponds 
$f_{\rm NL}<0$.) 
Although the current limits on $f_{\rm NL}$ from the WMAP constrain 
deviation of the number of massive clusters from
the Gaussian prediction to within $\sim$50\% for 
$z<1$ and $M<10^{15}~M_\odot$, the constraints 
rapidly improve as one goes to higher $z$ \cite{k}. Therefore, one needs 
to go to high $z$ to look for
signatures of non--Gaussianity in cluster abundance.
One major problem of this method is, however, as it has been correctly 
pointed out by several authors 
\cite{verde/etal:2001,sefusa}, 
that one needs very accurate (of order a few percent)
determinations of the mass of clusters at $z>1$, in order to find 
$f_{\rm NL}\sim -100$.
It seems rather difficult to achieve this accuracy for many clusters.
Yet, {\it if} one finds {\it one} exceptionally massive cluster 
($\sim 10^{15}~M_\odot$)
 at a very high $z$ ($\sim 3$), then it should tantalizingly indicate 
the presence of non--Gaussianity.
A preliminary, lower $z$ version of this methodology was attempted in 
Ref.~\cite{willick:2000} for a massive cluster, MS~1054--03, at $z=0.83$,  
where evidence for non--Gaussianity was claimed; however, unfortunately 
uncertainties in mass determinations are still too large to claim a robust 
detection of non--Gaussianity (the {\sl WMAP} limit is inconsistent with this 
detection), but in principle one can extend it to higher $z$ with a 
better determination of the mass. 

Non--cosmological non--Gaussianities from Galactic emission and extragalactic 
point--sources are serious contaminants.
Fortunately the shape of the angular bispectrum from point sources is very 
different from that of the primordial bispectrum, and they can be separated 
very well \cite{ks,k}. A problem occurs when we find a non--Gaussian signal,
but we do not know what the origin is. Although many authors claim detection 
of non--Gaussianity in the {\sl WMAP} 1--yr data
\cite{chiang/etal:2003,park:2004,coles/etal:2003,vielva/etal:2003,copi/huterer/starkman:2003,eriksen/etal:2004,hansen/etal:2004,mukherjee/wang:2004,larson/wandelt:2004}, 
none of its can be accounted for by the $f_{\rm NL}$ model, 
and their origin is unclear (also, one should keep in mind that the 
statistical significance of these detections is less than 
around 3--$\sigma$). Clearly, understanding possible foreground contamination 
and other possible systematics in the data are critical issues for 
measurements of primordial non--Gaussianity.

Therefore, testing the Gaussianity of primordial perturbations represents
a challenge for the present and future CMB experiments, as well as for 
LSS observations, which might reveal the ultimate origin of the structures 
we see in the Universe today. 

%
%

\section*{\bf Acknowledgments}

We thank all our colleagues with whom we have discussed many issues 
concerning non--Gaussianity from inflation. In particular,
we are indebted to Viviana Acquaviva, Paolo Creminelli, Domenico Marinucci, 
Andrew Liddle, Michele Liguori, David Lyth, Silvia Mollerach, Jim Peebles, 
Licia  Verde, Nicola Vittorio and David Wands. N.B. acknowledges PPARC  
for financial support.
E.K. acknowledges that Sec.~\ref{sec:obs} is based upon work in collaboration 
with the {\sl WMAP} science team: C.~Barnes, C.~Bennett (PI),
M.~Halpern, R.~Hill, G.~Hinshaw, N.~Jarosik, A.~Kogut, E.~Komatsu,
M.~Limon, S.~Meyer, N.~Odegard, L.~Page, H.~Peiris, D.~Spergel,
G.~Tucker, L.~Verde, J.~Weiland, E.~Wollack, and E.~Wright.
The {\sl WMAP} mission is made possible by the support of the Office of Space
Sciences at NASA Headquarters and by the hard and capable work of scores of
scientists, engineers, technicians, machinists, data analysts, budget analysts,
managers, administrative staff, and reviewers.
S.M. and A.R. thank the INFN Gruppo IV for financial support
within the PD51 Project. S.M. also thanks partial financial support from INAF 
(progetto di ricerca {\it Non--Gaussian primordial perturbations: constraints 
from CMB and redshift surveys}). 

\newpage
\appendix
\section{\bf Second--order gravitational perturbations}
\label{A}

\subsection{Basic notation}
The number of spatial dimensions is $n=3$.
Greek indices ($\alpha, \beta, \dots, \mu, \nu, \dots$)
run from 0 to 3, while latin indices ($a,b,\dots,i,j,k,\dots
m,n,\dots$) run from 1 to 3. 
The total space--time metric $g_{\mu \nu}$ has signature ($-,+,+,+$). 
The connection coefficients are defined as
\be
\label{conness} \Gamma^\alpha_{\beta\gamma}\,=\,
\frac{1}{2}\,g^{\alpha\rho}\left( \frac{\partial
g_{\rho\gamma}}{\partial x^{\beta}} \,+\, \frac{\partial
g_{\beta\rho}}{\partial x^{\gamma}} \,-\, \frac{\partial
g_{\beta\gamma}}{\partial x^{\rho}}\right)\, .
\ee
The Riemann tensor is defined as

\be
R^{\alpha}_{~\beta\mu\nu}=
\Gamma^{\alpha}_{\beta\nu,\mu}-\Gamma^{\alpha}_{\beta\mu,\nu}+
\Gamma^{\alpha}_{\lambda\mu}\Gamma^{\lambda}_{\beta\nu}-
\Gamma^{\alpha}_{\lambda\nu}\Gamma^{\lambda}_{\beta\mu} \,.
\ee

The Ricci tensor is a contraction of the Riemann tensor

\be
R_{\mu\nu}=R^{\alpha}_{~\mu\alpha\nu} \,,
\ee

and in terms of the connection coefficient it reads
\begin{equation}
R_{\mu\nu}\,=\, \partial_\alpha\,\Gamma^\alpha_{\mu\nu} \,-\,
\partial_{\mu}\,\Gamma^\alpha_{\nu\alpha} \,+\,
\Gamma^\alpha_{\sigma\alpha}\,\Gamma^\sigma_{\mu\nu} \,-\,
\Gamma^\alpha_{\sigma\nu} \,\Gamma^\sigma_{\mu\alpha}\,.
\end{equation}
The Ricci scalar is given by contracting the Ricci tensor

\be
R=R^{\mu}_{~\mu} \,.
\ee
The Einstein tensor is defined as

\be
G_{\mu\nu}=R_{\mu\nu}-\frac{1}{2}g_{\mu\nu}R \,.
\ee
The Einstein equations are written as $G_{\mu\nu}=\kappa^2 T_{\mu\nu}$, so 
that $\kappa^2 \equiv 8\pi G_{\rm N}$, 
where $G_{\rm N}$ is Newton's constant.\\ 
In the following expressions we have chosen a specific ordering of the terms. 
In the expressions in which
two spatial indices appear, such as Eq.~(\ref{gammaij}), we have assembled 
together the terms proportional to $\delta_{ij}$. 
The intrinsically second--order terms precede the source terms 
which are quadratic in the first--order perturbations. 
The second--order fluctuations have been listed in the following order as 
$\phi^{(2)}$, $\psi^{(2)}$, $\omega^{(2)}$, $\omega^{(2)}_i$, $\chi^{(2)}$, 
$\chi^{(2)}_i$ and $\chi^{(2)}_{ij}$, respectively. 
This ordering simplifies the analogy between the first--order 
and the second--order equations and allows to obtain immediately the 
expressions in a given gauge. 
\subsection{The connection coefficients}
In a spatially flat FRW background the connection 
coefficients are 
\begin{equation}
\Gamma^0_{00}\,=\, \Aa \,; \qquad
\Gamma^i_{0j}\,=\,\Aa\,\delta^i_{~j}\,; \qquad
\Gamma^0_{ij}\,=\,\Aa\,\delta_{ij}\,;
\end{equation}
\begin{equation}
\Gamma^i_{00}\quad=\quad\Gamma^0_{0i}\quad=\quad\Gamma^i_{jk}\quad=\quad
0 \,.
\end{equation}
The first--order perturbed connection coefficients corresponding to 
first--order metric perturbations in Eq.~(\ref{metric2}) are 
\bea
\deu{\Gamma^0_{00}}&=& {\phi^{(1)}}^{\prime} \, , \\
\deu{\Gamma^0_{0i}}&=& \partial_i\, \phi^{(1)} \,+\,
\frac{a^{\prime}}{a}\partial_i\,\omega^{(1)} \, , \\
\deu{\Gamma^i_{00}} &=& \frac{a^{\prime}}{a}\,\partial^i \omega^{(1)} \,+\,
\partial^i {\omega^{(1)}}^{\prime} \,+\, \partial^i \phi^{(1)} \, , \\
\deu{\Gamma^0_{ij}}&=&
-\,2\,\frac{a^{\prime}}{a}\,\phi^{(1)}\,\delta_{ij} \,-\,
\partial_i \partial_j \omega^{(1)} \,-\,
2\,\frac{a^{\prime}}{a}\,\psi^{(1)}\,\delta_{ij} 
- {\psi^{(1)}}^{\prime}\,\delta_{ij} \\
&-&\frac{a^{\prime}}{a}\,D_{ij} \chi^{(1)} \,+\, 
\frac{1}{2}\,D_{ij}{\chi^{(1)}}^{\prime}
\nonumber \, , \\
\deu{\Gamma^i_{0j}}&=& - \,{\psi^{(1)}}^{\prime}\delta^i_j \,+\,
\frac{1}{2}\,D^i_j{\chi^{(1)}}^{\prime} \, , \\
\deu{\Gamma^i_{jk}} &=& - \partial_j \psi^{(1)} \,\delta_k^i \,-\,
\partial_k \psi^{(1)}\, \delta_j^i \,+\, \partial^i \psi^{(1)} \,\delta_{jk}
\,-\, \frac{a^{\prime}}{a}\,\partial^i \omega^{(1)}
\,\delta_{jk} \\
&+& \frac{1}{2}\,\partial_j D^i_k \chi^{(1)} \,+\, 
\frac{1}{2}\,\partial_k D^i_j \chi^{(1)}  -
\frac{1}{2}\,\partial^i D_{jk} \chi^{(1)} \nonumber \, .
\eea
At second order we get: 
\bea 
\ded{\Gamma^0_{00}}\,&=&\frac{1}{2}\,{\phi^{(2)}}^{\prime}\,-\,
2\,\phi^{(1)}\, {\phi^{(1)}}^{\prime} \,+\, 
\partial^k \phi^{(1)}\,\partial_k \omega^{(1)} 
+\Aa
\partial^k \omega^{(1)}\,\partial_k \omega^{(1)} \\
&+& \partial^k \omega^{(1)}\,\partial_k 
{\omega^{(1)}}^{\prime}\nonumber \,, 
\eea
\bea
\ded{\Gamma^0_{0i}}&=& \frac{1}{2} \partial_i\,\phi^{(2)} \,+\,
\frac{1}{2}\,\Aa (\partial_i \omega^{(2)} + \omega_i^{(2)} ) 
\,-\, 2\,\phi^{(1)}\,\partial_i\,\phi^{(1)}\\
&-& 2\,\Aa \phi^{(1)}\,\partial_i \omega^{(1)}- 
{\psi^{(1)}}^{\prime}\partial_i \omega^{(1)} 
+\frac{1}{2}\,\partial^k \omega^{(1)}\,D_{ik}{\chi^{(1)}}^{\prime} 
\nonumber \,,
\eea
\bea
\ded{\Gamma^i_{00}} \,&=& \,
 \frac{1}{2}\,\partial^i\,\phi^{(2)} \,+\, \frac{1}{2}\,\Aa (\partial^i 
\omega^{(2)} +
 \omega^{i(2)}) \,+\, \frac{1}{2}\,\left( \partial^i {\omega^{(2)}}^{\prime} + 
\left( \omega^{i{(2)}} \right)^{\prime} \right) \\
&+& 2\,\psi^{(1)}\,\partial^i\,\phi^{(1)} \,-\, 
{\phi^{(1)}}^{\prime}\partial^i 
\omega^{(1)} 
+\, 2\,\Aa \psi^{(1)} \,\partial^i \omega^{(1)} 
\,+\, 2\,\psi^{(1)}\,\partial^i
{\omega^{(1)}}^{\prime} \nonumber \\
&-& \partial_k \phi^{(1)} 
\,D^{ik}\chi^{(1)} \,-\, 
\Aa \partial_k \omega^{(1)}
\,D^{ik}\chi^{(1)} 
- \partial_k {\omega^{(1)}}^{\prime}D^{ik}\chi^{(1)} 
\nonumber \,, 
\eea
\bea
\label{gammaij} \ded{\Gamma^0_{ij}} &=& \Big(
-\,\frac{a^{\prime}}{a}\,\phi^{(2)} \,-\,\frac{1}{2}\,{\psi^{(2)}}^{\prime}\,
-\,\frac{a^{\prime}}{a}\,\psi^{(2)} \nonumber \\
&+& 4\,\Aa \,\left( \phi^{(1)} 
\right)^2 \,+\,
2\,\phi^{(1)}\,{\psi^{(1)}}^{\prime} + 4\,\Aa\,\phi^{(1)}\,\psi^{(1)} \,+\,
\partial^k \omega^{(1)}\,\partial_k \psi^{(1)} \\
&-&\, \Aa \partial^k \omega^{(1)}\,\partial_k \omega^{(1)} 
\Big)\,\delta_{ij} \,-\,
\frac{1}{2}\,\partial_i\,\partial_j \omega^{(2)}\,+\, \frac{1}{4}\,\left(
D_{ij} {\chi^{(2)}}^{\prime} + \partial_j\,{\chi_i^{(2)}}^{\prime} +
\partial_i\,{\chi^{(2)}_j}^{\prime} + \left( {\chi^{(2)}_{ij}} 
\right)^{\prime} \right)
\nonumber  \\
&+& \,\frac{1}{2}\,\frac{a^{\prime}}{a}\,\left( D_{ij}\chi^{(2)} +
\partial_i\,\chi^{(2)}_j +
\partial_j\,\chi_i^{(2)} + \chi^{(2)}_{ij}\,\right)
 \,-\, \frac{1}{4}\,\left( \partial_i \omega^{(2)}_j + \partial_j 
\omega_i^{(2)}\right)\nonumber \\
&+&\, 2\,\phi^{(1)}\,\partial_i \partial_j \omega^{(1)}\,-\, 
\partial_i \psi^{(1)}
 \,\partial_j \omega^{(1)} \,-\, \partial_j \psi^{(1)}\,\partial_i 
\omega^{(1)} \,-\,
 \phi^{(1)}\,D_{ij} {\chi^{(1)}}^{\prime} \,+\, \frac{1}{2}\,\partial^k 
\omega^{(1)}\,
\partial_i
 D_{kj}\chi^{(1)} \nonumber   \\
&+&\, \frac{1}{2}\,\partial^k \omega^{(1)}\,\partial_j D_{ik}\chi^{(1)}\,-\,
\frac{1}{2}\,\partial^k \omega^{(1)} \,\partial_k D_{ij}\chi^{(1)} \nonumber 
\, , 
\eea
\bea
\ded{\Gamma^i_{0j}} \,&=& -\,\frac{1}{2}\,{\psi^{(2)}}^{\prime}\delta^i_{~j}
\,+\, \frac{1}{4}\,\left(
 D^i_{~j} {\chi^{(2)}}^{\prime} + \partial_j\left(\chi^{(2)i} 
\right)^{\prime} + \partial^i\left( \chi^{(2)}_j \right)^{\prime} \right.\\ 
&+& \left. \left( {\chi^{(2)i}_{~j}} \right)^{\prime} \right) 
+ \frac{1}{4}\,\left( \partial_j \omega^{i(2)}
- \partial^i \omega^{(2)}_j \right) \nonumber \\
&-& 2\,\psi^{(1)}\,{\psi^{(1)}}^{\prime}\delta^i_{~j}  
\,-\,\partial^i \omega^{(1)}
\,\partial_j \phi^{(1)} \,-\, \Aa \partial^i \omega^{(1)} 
\,\partial_j \omega^{(1)}\nonumber \\ &+&
\psi^{(1)}\,D^i_{~j} {\chi^{(1)}}^{\prime} \,+\, {\psi^{(1)}}^{\prime}D^i_{~j} 
\chi^{(1)} 
-
\frac{1}{2}\,D^{ik}\chi^{(1)}\,D_{kj} {\chi^{(1)}}^{\prime} \nonumber \, ,
\eea
\bea
\ded{\Gamma^i_{jk}} \,&=&\,
\frac{1}{2}\,\left(-\partial_j\,\psi^{(2)}\,\delta^i_{~k}
-\partial_k\,\psi^{(2)}\,\delta^i_{~j} + 
\partial^i\,\psi^{(2)}\,\delta_{jk}\right)
 \,+\, \frac{1}{4} \left( \partial_j D^i_{~k} \chi^{(2)} \right. \\
&+& \left.  \partial_k D^i_{~j} \chi^{(2)}
 - \partial^i D_{jk} \chi^{(2)} \right) \
+ \frac{1}{2}\,\partial_j \partial_k\,\chi^{i(2)}
 + \frac{1}{4}\,\left(\partial_j\,\chi^{i(2)}_{~k} + 
\partial_k\,\chi^{i(2)}_{~j} -
\partial^i\,\chi^{(2)}_{jk}\right) \nonumber \\
&-& \frac{1}{2}\,\Aa \left( \partial^i \omega^{(2)} + 
\omega^{i(2)}\right)\,\delta_{jk} 
+\, 2\,\psi^{(1)}\,\left(-\partial_j\,\psi^{(1)}\,\delta^i_{~k}
-\partial_k\,\psi^{(1)}\,\delta^i_{~j} +
\partial^i\,\psi^{(1)}\,\delta_{jk}\right) \nonumber \\
&+& 2\,\Aa \phi^{(1)}\,\partial^i \omega^{(1)} \,\delta_{jk} \,+\,\partial^i
\omega^{(1)}\,\partial_j \partial_k \omega^{(1)} 
+\,{\psi^{(1)}}^{\prime}\partial^i \omega^{(1)} \,\delta_{jk} \nonumber \\
&+&\,\psi^{(1)}\,\left(
\partial_j D^i_{~k} \chi^{(1)} + \partial_k D^i_{~j} \chi^{(1)} - 
\partial^i D_{jk}\chi^{(1)} \right)
\,+\, \partial_j \psi^{(1)}\,D^i_{~k} \chi^{(1)}  \nonumber \\
&+&\, \partial_k \psi^{(1)}\,D^i_{~j} \chi^{(1)}\,-\, \partial_m
\psi^{(1)}\,D^{im}\chi^{(1)}\,\delta_{jk} \,-\, \Aa 
\partial^i \omega^{(1)}\,D_{jk}\chi^{(1)} \nonumber \\
&+&
\Aa \partial^m \omega^{(1)}\,D^i_{~m} \chi^{(1)}\,\delta_{jk} 
-\, \frac{1}{2}\,\partial^i \omega^{(1)} \,D_{jk}{\chi^{(1)}}^{\prime}\,-\,
\frac{1}{2}\,D^{im}\chi^{(1)}\,\partial_j D_{mk}\chi^{(1)} \nonumber \\
&-&
\frac{1}{2}\,D^{im}\chi^{(1)}\,\partial_k D_{mj}\chi^{(1)} 
+ \frac{1}{2}\,D^{im}\chi^{(1)}\,\partial_m D_{jk}\chi^{(1)} \nonumber \,.
\eea
\subsection{The Ricci tensor components}
In a spatially flat FRW background the components of the 
Ricci tensor $R_{\mu\nu}$ are given by 
\begin{equation}
R_{00}\,=\,-\,3\,\Ac \,+\,3\,\Ab \,; \qquad R_{0i}\,=\,0\,;
\end{equation}
\begin{equation}
R_{ij}\,=\,\left[ \Ac \,+\,\Ab \right]\,\delta_{ij}\,.
\end{equation}
The first--order perturbed Ricci tensor components read
\bea
\deu {R_{00}}&=& \frac{a^{\prime}}{a}\partial_i\partial^i \omega^{(1)} +
\partial_i\partial^i {\omega^{(1)}}^{\prime} + \partial_i\partial^i 
\phi^{(1)} +
3{\psi^{(1)}}^{\prime\prime} + 3\frac{a^{\prime}}{a}{\psi^{(1)}}^{\prime} +
3\frac{a^{\prime}}{a}{\phi^{(1)}}^{\prime} \, ,\\
\deu {R_{0i}}&=& \frac{a^{\prime\prime}}{a}\partial_i \omega^{(1)} +
\left(\frac{a^{\prime}}{a}\right)^2\partial_i \omega^{(1)} +
2\partial_i{\psi^{(1)}}^{\prime} + 2\frac{a^{\prime}}{a}\partial_i \phi^{(1)} +
\frac{1}{2}\partial_k D^k_{~i} {\chi^{(1)}}^{\prime} \, , \\
\deu {R_{ij}}&=& \left[  -\frac{a^{\prime}}{a}{\phi^{(1)}}^{\prime} -
5\frac{a^{\prime}}{a}{\psi^{(1)}}^{\prime} - 
2\frac{a^{\prime\prime}}{a}\phi^{(1)}
-2\left(\frac{a^{\prime}}{a}\right)^2\phi^{(1)} 
-2\frac{a^{\prime\prime}}{a}\psi^{(1)} \right. \\
&-& \left. 2\left(\frac{a^{\prime}}{a}\right)^2\psi^{(1)} - 
{\psi^{(1)}}^{\prime\prime} +
\partial_k\partial^k\psi^{(1)} 
-   \frac{a^{\prime}}{a}\partial_k\partial^k \omega^{(1)} 
\right] \delta_{ij}
-\partial_i\partial_j {\omega^{(1)}}^{\prime} \nonumber \\
&+& \frac{a^{\prime}}{a}D_{ij}{\chi^{(1)}}^{\prime} +
\frac{a^{\prime\prime}}{a}D_{ij}\chi^{(1)} + 
\left(\frac{a^{\prime}}{a}\right)^2 D_{ij}\chi^{(1)} 
+ \frac{1}{2}D_{ij}{\chi^{(1)}}^{\prime\prime} + \partial_i\partial_j 
\psi^{(1)}\nonumber \\
&-& \partial_i\partial_j \phi^{(1)} - 
2\frac{a^{\prime}}{a}\partial_i\partial_j \omega^{(1)} 
+ \frac{1}{2}\partial_k\partial_iD^k_{~j} \chi^{(1)} +
\frac{1}{2}\partial_k\partial_j D^k_{~i} \chi^{(1)} -
\frac{1}{2}\partial_k\partial^k D_{ij} \chi^{(1)} \nonumber \, .
\eea
At second order we obtain
\bea
\ded R_{00}&=& \frac{3}{2} \Aa\ {\phi^{(2)}}^{\prime} \,+\, \frac{1}{2}\,\La\
\phi^{(2)} \,+\, \frac{3}{2} \Aa\ {\psi^{(2)}}^{\prime}\,+\,
\frac{3}{2}\,{\psi^{(2)}}^{\prime\prime}\,+\,\frac{1}{2}\,\Aa\
\partial_k \,\partial^k \omega^{(2)} \\
&+&  \frac{1}{2}\, \partial_k \,\partial^k {\omega^{(2)}}^{\prime} -\, 6
\,\Aa\ \phi^{(1)}\,{\phi^{(1)}}^{\prime} \,-\,
\partial^k \phi^{(1)} \,\partial_k \phi^{(1)} \,-\, 3
{\phi^{(1)}}^{\prime} {\psi^{(1)}}^{\prime} \nonumber \\
&+& \, 2\,\psi^{(1)} 
\,\La\ \phi^{(1)} -\, \partial_k \psi^{(1)} \,
\partial^k \phi^{(1)}\, 
+ 6 \,\Aa\
\psi^{(1)}\,{\psi^{(1)}}^{\prime}\,+\, 
6\psi^{(1)}\,{\psi^{(1)}}^{\prime\prime} \nonumber \\
&+&\, 3 \left( {\psi^{(1)}}^{\prime} \right)^2\,-\,{\phi^{(1)}}^{\prime}\La\ 
\omega^{(1)} + \Aa\ \partial^k 
\omega^{(1)} \,\partial_k \phi^{(1)}\,  
+ \,\Ac\ \partial^k \omega^{(1)}
\partial_k \omega^{(1)} \nonumber \\
&+& \Ab \partial^k \omega^{(1)} \,\partial_k 
\omega^{(1)}\,- \, \Aa \partial_k
\psi^{(1)}\,
\partial^k \omega^{(1)} +\, 2 \,\Aa \psi^{(1)} \,\La\omega^{(1)} \, 
-\, \partial_k \psi^{(1)} \,\partial^k {\omega^{(1)}}^{\prime} \nonumber \\
&+& 2 \,\psi^{(1)}\, \La {\omega^{(1)}}^{\prime} \,+\, 3 \,\Aa\ 
\partial^k \omega^{(1)} \,
\partial_k {\omega^{(1)}}^{\prime} -\partial_k \phi^{(1)} \,\partial_i D^{ik} 
\chi^{(1)}\, -\, \partial_i
\partial_k \phi^{(1)} \, D^{ik} \chi^{(1)} \nonumber \\
&-& \Aa\
\partial_i\partial_k \omega^{(1)}\, D^{ik} \chi^{(1)} \,-\, 
\Aa \partial_k \omega^{(1)}
\,\partial_i D^{ik} \chi^{(1)} -\, \partial_k {\omega^{(1)}}^{\prime} 
\,\partial_i D^{ik} \chi^{(1)} \nonumber \\
&-& \partial_i
\partial_k {\omega^{(1)}}^{\prime} \,D^{ik} \chi^{(1)}   \,+\, 
\frac{1}{2} D^{ik} \chi^{(1)} \,
D_{ki} {\chi^{(1)}}^{\prime\prime}
+\, \frac{1}{4} D^{ik} {\chi^{(1)}}^{\prime}\,D_{ki} {\chi^{(1)}}^{\prime} 
\nonumber \\
&+& \frac{1}{2} \Aa\ D^{ik} \chi^{(1)} \,D_{ki} {\chi^{(1)}}^{\prime} \,. 
\nonumber 
\eea
\bea
\ded R_{0i} &=& \,\Aa \partial_i \phi^{(2)} \,+\, \partial_i
{\psi^{(2)}}^{\prime} \,+\, \frac{1}{4} \partial_k \,D^k_{~i} 
{\chi^{(2)}}^{\prime} \,+\,
\frac{1}{4} 
\LA \,{\chi^{(2)}_i}^{\prime} 
-\frac{1}{4} \LA \,\omega^{(2)}_i \\ 
&+& \frac{1}{2} \Ac (\,\partial_i \omega^{(2)} + \omega^{(2)}_i\,)
+\, \frac{1}{2}
\Ab (\,\partial_i \omega^{(2)} + \omega^{(2)}_i\,) \,-\, 4\,\Aa\, 
\phi^{(1)}\, \partial_i \phi^{(1)} \nonumber \\
&-& 2 {\psi^{(1)}}^{\prime} \partial_i \phi^{(1)} +\,4
\,{\psi^{(1)}}^{\prime}\,\partial_i \psi^{(1)} \,+\, 4 \,\psi^{(1)} 
\,\partial_i
{\psi^{(1)}}^{\prime}\,-\,2\, \Ac \phi^{(1)} \partial_i \omega^{(1)} 
\nonumber \\
&-&2\,\Ab
\phi^{(1)} \,\partial_i \omega^{(1)} 
- \Aa {\phi^{(1)}}^{\prime} \partial_i \omega^{(1)} 
-\, \LA \omega^{(1)}\,\partial_i \phi^{(1)} \nonumber \\
&-&\partial^k \omega^{(1)} \,\partial_i\partial_k \phi^{(1)} \,+\, 
\partial^k \phi^{(1)} \,
\partial_i\partial_k \omega^{(1)}  \,-\, \partial^k \omega^{(1)} 
\,\partial_i\partial_k 
{\omega^{(1)}}^{\prime}- \Aa \LA \omega^{(1)}
\partial_i \omega^{(1)} \nonumber \\
&-& {\psi^{(1)}}^{\prime\prime}\partial_i \omega^{(1)} \,-\,5 \,\Aa
{\psi^{(1)}}^{\prime}\partial_i \omega^{(1)} \,-\, \frac{1}{2} 
\partial^k \phi^{(1)} \,D^{ik}
{\chi^{(1)}}^{\prime}\,+\, \psi^{(1)} \,\partial_k\, D^k_{~i} 
{\chi^{(1)}}^{\prime} \nonumber \\
&+& {\psi^{(1)}}^{\prime} \,\partial_k \,D^k_{~i} \chi^{(1)} 
-\, \frac{1}{2}
\partial_k \psi^{(1)} \,D^k_i {\chi^{(1)}}^{\prime} \,+\,
\partial_k {\psi^{(1)}}^{\prime} \,D^k_{~i} \chi^{(1)}
\,+\, \Aa \partial^k \omega^{(1)} \,D_{ik}{\chi^{(1)}}^{\prime} \nonumber \\  
&+& \frac{1}{2} \partial^k \omega^{(1)} \,D_{ik} {\chi^{(1)}}^{\prime\prime} 
- \frac{1}{2} \partial_k
D^{km} \chi^{(1)} \,D_{mi} {\chi^{(1)}}^{\prime}
 \,-\, \frac{1}{2} D^{km} \chi^{(1)} \,\partial_k D_{mi} 
{\chi^{(1)}}^{\prime} \nonumber \\
&+& \frac{1}{2} D^{km} {\chi^{(1)}}^{\prime} \,\partial_i D_{mk} \chi^{(1)} 
+ \frac{1}{4} D^{km} \chi^{(1)} \,\partial_i D_{mk} {\chi^{(1)}}^{\prime} 
\nonumber \, . 
\eea
The expression for the purely spatial part of $\ded \R_{\mu\nu}$ is very 
long, thus for simplicity we will divide it 
into two parts: the diagonal part $\ded {R^d_{ij}}$, proportional 
to $\delta_{ij}$, and the non--diagonal part ${R^{nd}_{ij}}$. 
\bea
\ded {R^d_{ij}} &=& \Bigg[ -\,
 \Ab \phi^{(2)} \,-\,\frac{1}{2} \Aa {\phi^{(2)}}^{\prime} \,-\, \Ac 
\phi^{(2)}\,-\,
  \,-\, \frac{5}{2}\,\Aa {\psi^{(2)}}^{\prime}
\,-\, \Ab \psi^{(2)} \\
&-&\frac{1}{2} {\psi^{(2)}}^{\prime\prime} 
- \Ac \psi^{(2)} \,+\, \frac{1}{2}\, \LA \psi^{(2)} \,-\,\frac{1}{2} \Aa 
\LA \omega^{(2)} + 4 \left(\Ab + \Ac\right) \left( \phi^{(1)} \right)^2 
\nonumber \\
&+&  4\, \Aa \phi^{(1)}\,{\phi^{(1)}}^{\prime} \nonumber 
+10\,\Aa \phi^{(1)}\,{\psi^{(1)}}^{\prime} \,+\,2\,\Aa 
{\phi^{(1)}}^{\prime}\psi^{(1)} \,+ \,
{\phi^{(1)}}^{\prime}{\psi^{(1)}}^{\prime}
\,+\,2\,\phi^{(1)}\,{\psi^{(1)}}^{\prime\prime} \nonumber \\ 
&+& 4 \left(\Ab + \Ac\right)\phi^{(1)}\,\psi^{(1)} 
+\partial_k \psi^{(1)}\,\partial^k \phi^{(1)} \, +\, 
\left( {\psi^{(1)}}^{\prime} \right)^2 \,+\,
\partial_k \psi^{(1)} \, \partial^k \psi^{(1)} \nonumber \\
&+& 2\,\psi^{(1)} \LA \psi^{(1)} \,+\,\Aa \partial_k \phi^{(1)}\,\partial^k 
\omega^{(1)} + 2\,\Aa \phi^{(1)}\,\LA \omega^{(1)} \, - \Ab
\partial_k \omega^{(1)} \,\partial^k \omega^{(1)} \nonumber \\
&-& \Ac \partial_k 
\omega^{(1)}\,\partial^k \omega^{(1)}
\,-\, \Aa \partial_k \omega^{(1)}\,\partial^k 
{\omega^{(1)}}^{\prime}+ 3\,\Aa
\partial_k \omega^{(1)}\,\partial^k \psi^{(1)} \nonumber \\
&+& 2\,\partial_k
{\psi^{(1)}}^{\prime}\,\partial^k \omega^{(1)} \,+\, 
{\psi^{(1)}}^{\prime}\LA \omega^{(1)} \,+\,
\partial_k \psi^{(1)}\,\partial^k {\omega^{(1)}}^{\prime} 
- \partial_m \psi^{(1)} \,\partial_k D^{km} \chi^{(1)}\nonumber \\
&-& \partial_k
\partial_m \psi^{(1)} \,D^{km} \chi^{(1)}
\,+\, \Aa \partial_m \partial^k \omega^{(1)}\,D^m_k \chi^{(1)} 
+ \Aa \partial^k \omega^{(1)} \,\partial_m D^m_k \chi^{(1)} \nonumber \\
&-& \frac{1}{2} \Aa
D^{mk} \chi^{(1)} \, D_{km}
{\chi^{(1)}}^{\prime} \Bigg]\,\delta_{ij} \,,\nonumber 
\eea
\bea
\ded {R^{nd}_{ij}} &=& 
-\frac{1}{2} \,\partial_i\partial_j \phi^{(2)}
\,+\,\frac{1}{2}\,\partial_i\partial_j \psi^{(2)} \,-\, \Aa
\partial_i\partial_j \omega^{(2)} \,-\, \frac{1}{2}\,\partial_i\partial_j
{\omega^{(2)}}^{\prime} \,-\, \frac{1}{2} \Aa \left( \partial_i 
\,\omega^{(2)}_j +
\partial_j\, \omega^{(2)}_i\right) \\
&-& \frac{1}{4} \left( \partial_i \,{\omega^{(2)}_j}^{\prime} + \partial_j\,
{\omega^{(2)}_i}^{\prime} \right) 
+\,\frac{1}{2} \left(\Ab + \Ac\right)
\left( D_{ij} \chi^{(2)} +
\partial_i \chi^{(2)}_j
+ \partial_j \chi^{(2)}_i + \chi^{(2)}_{ij}\right) \nonumber \\
&+& \frac{1}{2} \Aa \left( D_{ij} {\chi^{(2)}}^{\prime} +
\partial_i {\chi^{(2)}_j}^{\prime}
+ \partial_j {\chi^{(2)}_i}^{\prime} +\left( {\chi^{(2)}_{ij}} 
\right)^{\prime}\right) \,+\,
\frac{1}{2}
\partial_k\partial_i\,D^k_{~j} \chi^{(2)} \,-\,\frac{1}{4} \LA D_{ij} 
\chi^{(2)}\nonumber \\
&-& \frac{1}{4} \LA \chi^{(2)}_{ij} \,+\, \frac{1}{4}\left( D_{ij}
{\chi^{(2)}}^{\prime\prime} +
\partial_i {\chi^{(2)}_j}^{\prime\prime}
+ \partial_j {\chi^{(2)}_i}^{\prime\prime} +
\left({\chi^{(2)}_{ij}}\right)^{\prime\prime}\right) \,
+ \, \partial_i \phi^{(1)}\, \partial_j \phi^{(1)}\nonumber \\
&+& 2\,\phi^{(1)}\,\partial_i\partial_j \phi^{(1)} \,-\,\partial_j \phi^{(1)}\,
\partial_i\psi^{(1)} \,-\, \partial_i \phi^{(1)}\,\partial_j \psi^{(1)} \,+\,
3\,\partial_i \psi^{(1)}\,\partial_j \psi^{(1)} \,+\, 
2 \,\psi^{(1)}\,\partial_i\partial_j \psi^{(1)} 
\nonumber \\
&+& 4 \Aa \phi^{(1)}\,\partial_i\partial_j \omega^{(1)} \,+\,
{\phi^{(1)}}^{\prime}\,\partial_i\partial_j \omega^{(1)} \,+\, 2\,\phi^{(1)}
\,\partial_i\partial_j {\omega^{(1)}}^{\prime} \,+\, \LA \omega^{(1)}
\,\partial_i\partial_j \omega^{(1)} \nonumber \\
&-& \partial_j
\partial^k \omega^{(1)}\,\partial_i \partial_k \omega^{(1)}
\,-\,2 \, \Aa \partial_i \psi^{(1)} \,\partial_j \omega^{(1)} \,-\, 2\,\Aa
\partial_i \omega^{(1)}\,\partial_j \psi^{(1)} \,-\, \partial_i
{\psi^{(1)}}^{\prime}\,\partial_j \omega^{(1)} \nonumber \\
&-& \partial_j {\psi^{(1)}}^{\prime}\,\partial_i \omega^{(1)} \,-\, 
\partial_i \psi^{(1)}
\,\partial_j {\omega^{(1)}}^{\prime} \,-\, \partial_j \psi^{(1)} \,\partial_i
{\omega^{(1)}}^{\prime} \,+\, {\psi^{(1)}}^{\prime}\,\partial_i\partial_j 
\omega^{(1)} \,-\, 2
\,\Ab \phi^{(1)}\,D_{ij} \chi^{(1)} \nonumber \\
&-& 2 \,\Ac \phi^{(1)}\,D_{ij} \chi^{(1)} \,-\, 2\,\Aa \phi^{(1)}\,D_{ij} 
{\chi^{(1)}}^{\prime} \,-\,
\Aa {\phi^{(1)}}^{\prime}\, D_{ij} \chi^{(1)} \, -\,\frac{1}{2} 
{\phi^{(1)}}^{\prime}\,D_{ij}
{\chi^{(1)}}^{\prime} \,-\, \phi^{(1)}\,D_{ij} 
{\chi^{(1)}}^{\prime\prime}\nonumber \\
&+& \frac{1}{2}
\partial_k \phi^{(1)} \, \partial_i D^k_{~j} \chi^{(1)} \,+\, \frac{1}{2} 
\partial_k \phi^{(1)}
\, \partial_j D^k_{~i} \chi^{(1)} \,-\, \frac{1}{2} \partial_k \phi^{(1)} \,
\partial^k D_{ij} \chi^{(1)}  \,-\, 3\,\Aa
{\psi^{(1)}}^{\prime}\,D_{ij} \chi^{(1)} \nonumber \\
&+& \frac{1}{2} {\psi^{(1)}}^{\prime}\, D_{ij} {\chi^{(1)}}^{\prime}\,+\,
\frac{1}{2}
\partial_k \psi^{(1)}\,
\partial_i D^k_{~j} \chi^{(1)} \,+\, \frac{1}{2} \partial_k \psi^{(1)}\,
\partial_j D^k_{~i} \chi^{(1)} \,-\,\frac{3}{2}\, \partial_k \psi^{(1)} 
\,\partial^k D_{ij}
\chi^{(1)}\nonumber \\
&+& \psi^{(1)}\,\partial_k\partial_i D^k_{~j} \chi^{(1)} \,+\,
\psi^{(1)}\,\partial_k\partial_j D^k_{~i} \chi^{(1)} \,-\,
\psi^{(1)}\,\partial_k\partial^k D_{ij} \chi^{(1)} +\,
\partial_i\psi^{(1)}\,\partial_k D^k_{~j} \chi^{(1)}  \nonumber \\
&+&
\partial_j\psi^{(1)}\,\partial_k D^k_{~i} \chi^{(1)}
\, +\, \partial_k\partial_i \psi^{(1)}\,D^k_{~j} \chi^{(1)} \,+\,
\partial_k\partial_j \psi^{(1)}\,D^k_{~i} \chi^{(1)}
\, +\, \frac{1}{2}\,\partial_k\partial_i \omega^{(1)} \, D^k_{~j} 
{\chi^{(1)}}^{\prime}
\nonumber \\
&+& \frac{1}{2}\,\partial_k\partial_j \omega^{(1)} \, D^k_{~i} 
{\chi^{(1)}}^{\prime}
\,-\, \frac{1}{2} \,\partial_k\partial^k \omega^{(1)}\, D_{ij} 
{\chi^{(1)}}^{\prime}
\,+\,\frac{1}{2}\, \partial^k \omega^{(1)}\, \partial_i D_{kj} 
{\chi^{(1)}}^{\prime}
+\,\frac{1}{2}\, \partial^k \omega^{(1)}\, \partial_j D_{ki}
{\chi^{(1)}}^{\prime}\nonumber \\
&-&\partial^k \omega^{(1)} \,\partial_k D_{ij} {\chi^{(1)}}^{\prime}\,+\, 
\frac{1}{2}
\,\partial^k {\omega^{(1)}}^{\prime} \,\partial_i D_{kj} \chi^{(1)} \,+\, 
\frac{1}{2}
\,\partial^k {\omega^{(1)}}^{\prime} \,\partial_j D_{ki} \chi^{(1)} \,-\, 
\frac{1}{2}
\,\partial^k {\omega^{(1)}}^{\prime} \,\partial_k D_{ij} \chi^{(1)}\nonumber \\
&+& \Aa \partial^k \omega^{(1)} \,\partial_i D_{kj} \chi^{(1)} \,+\, \Aa 
\partial^k
\omega^{(1)} \,\partial_j D_{ki} \chi^{(1)} \,-\, \Aa
\partial^k \omega^{(1)} \,\partial_k D_{ij} \chi^{(1)} \,-\, \Aa
\partial_k\partial^k \omega^{(1)} \,D_{ij} \chi^{(1)} \nonumber \\
&-&\frac{1}{2}\, D^k_i {\chi^{(1)}}^{\prime} \, D_{kj} {\chi^{(1)}}^{\prime} 
\,-\,
\frac{1}{2} \partial_i D_{mj} \chi^{(1)} \,
\partial_k D^{km} \chi^{(1)} \,-\, \frac{1}{2} \partial_j D_{mi} \chi^{(1)} \,
\partial_k D^{km} \chi^{(1)}  \nonumber \\
&+& \frac{1}{2} \partial_m D_{ij} \chi^{(1)} \, \partial_k D^{km} \chi^{(1)}
-\,\frac{1}{2} \partial_k\partial_i D_{mj} \chi^{(1)} \, D^{km} \chi^{(1)} 
\,-\,
\frac{1}{2} \partial_k\partial_j D_{mi} \chi^{(1)} \, D^{km} \chi^{(1)} 
\nonumber \\
&+&\frac{1}{2} \partial_k\partial_m D_{ij} \chi^{(1)} \, D^{km} \chi^{(1)} 
\,+\,
\frac{1}{2} D^{km} \chi^{(1)} \,\partial_i\partial_j D_{km} \chi^{(1)} \,+\,
\frac{1}{4} \partial_i D^{mk} \chi^{(1)} \,\partial_j D_{mk} \chi^{(1)} 
\nonumber \,. 
\eea
\subsection{The Ricci scalar} 
At zeroth order the Ricci scalar $R$ reads
\be
R=\frac{6}{a^2} \frac{a^{\prime\prime}}{a} \, .
\ee
The first--order perturbation of $R$ is
\bea
\deu R &=& \frac{1}{a^2} \left( 
-6\frac{a^{\prime}}{a}\partial_i\partial^i \omega^{(1)} -
2\partial_i\partial^i {\omega^{(1)}}^{\prime} - 2\partial_i\partial^i 
\phi^{(1)}
-6{\psi^{(1)}}^{\prime\prime}  \right. \\
&-& \left.  6\frac{a^{\prime}}{a}{\phi^{(1)}}^{\prime} 
-  18\frac{a^{\prime}}{a}{\psi^{(1)}}^{\prime} 
- 12\frac{a^{\prime\prime}}{a}\phi^{(1)} 
+  4\partial_i\partial^i\psi^{(1)} +
\partial_k\partial^i D^k_{~i} \chi^{(1)} \right) \, . \nonumber 
\eea
At second order we find
\bea
\ded R &=& -\, \La \phi^{(2)} \,-\, 3\,\Aa {\phi^{(2)}}^{\prime} \,-\, 
6\,\Ac \phi^{(2)}
\,+\,
2\,\La \psi^{(2)} \,-\, 9\,\Aa {\psi^{(2)}}^{\prime} \,-\, 
3\,{\psi^{(2)}}^{\prime\prime}
\,-\, \La {\omega^{(2)}}^{\prime} \\
&-& 3\,\Aa \La \omega^{(2)}  \,+\, \frac{1}{2}\,\partial_k\partial_i
\,D^{ki}\chi^{(2)} \,+\,24\,\Ac \left( \phi^{(1)} \right)^2 \,+\,
2\,\partial_k \phi^{(1)}\,\partial^k \phi^{(1)} \,+\, 4\,\phi^{(1)}\,\La 
\phi^{(1)}  \nonumber \\
&+& 24\,\Aa \phi^{(1)}\,{\phi^{(1)}}^{\prime}\,+\, 
6\,{\phi^{(1)}}^{\prime}{\psi^{(1)}}^{\prime}
\,+\,36\,\Aa \phi^{(1)}\,{\psi^{(1)}}^{\prime}\,+\, 2\,
\partial_k \psi^{(1)} \,\partial^k \phi^{(1)} \,-\, 4\,\psi^{(1)} 
\,\La \phi^{(1)} \nonumber \\
&+& 12
\,\phi^{(1)}\,{\psi^{(1)}}^{\prime\prime}\, 
-\,12\,\psi^{(1)}\,{\psi^{(1)}}^{\prime\prime}
\,-\,36\,\Aa {\psi^{(1)}}^{\prime}\psi^{(1)} \,+\, 6\,\partial_k
\psi^{(1)}\,\partial^k \psi^{(1)} \,+\, 16\,\psi^{(1)}\,\La 
\psi^{(1)} \nonumber \\
&+& 6\,\Aa
\partial^k \omega^{(1)}\,\partial_k \phi^{(1)} \,+\, 12\,\Aa 
\phi^{(1)}\,\La \omega^{(1)} \,
+\, 4\,\phi^{(1)}\,\La
{\omega^{(1)}}^{\prime} \,+\, 2\,{\phi^{(1)}}^{\prime}\La \omega^{(1)}
 \,\nonumber \\
&-& 5\,\Ac \partial_k \omega^{(1)}\,\partial^k \omega^{(1)} \,-\, 6\,\Aa 
\partial_k
\omega^{(1)}\,\partial^k {\omega^{(1)}}^{\prime} \,+\, \La \omega^{(1)}\,\La 
\omega^{(1)} \,-\,
\partial^i\partial^k \omega^{(1)}\,\partial_i
\partial_k \omega^{(1)} \nonumber \\
&+& 8\,\partial_k \omega^{(1)}\,\partial^k 
{\psi^{(1)}}^{\prime}\,+\,2\,\partial_k
{\omega^{(1)}}^{\prime}\partial^k \psi^{(1)} \,-\,
4\,\psi^{(1)}\,\La {\omega^{(1)}}^{\prime} \,-\, 12\,\Aa \psi^{(1)}\,\La 
\omega^{(1)} \nonumber \\
&+& 4\,{\psi^{(1)}}^{\prime}\La \omega^{(1)} \,+\, 
2\,\partial_k \phi^{(1)}\,\partial_i
D^{ik} \chi^{(1)} \,+\, 2\, \partial_i
\partial_k \phi^{(1)} \,D^{ik} \chi^{(1)} \,
+\, 4\,\psi^{(1)}\,\partial_k\partial_i\,D^{ki} \chi^{(1)} \nonumber \\
&-& 2\,
\partial_k
\partial_i \psi^{(1)}\,D^{ik}\chi^{(1)} \,
+\, 3\, \partial_k \omega^{(1)}\,\partial^i D^k_{~i} {\chi^{(1)}}^{\prime}
 \,+\, 6\,\Aa
\partial^k \omega^{(1)}\,\partial_i D^i_{~k} \chi^{(1)}\,
+\,2\,\partial_i {\omega^{(1)}}^{\prime}\partial_k D^{ik}\chi^{(1)}
 \nonumber \\
&+& 2\, \partial_k\partial_i {\omega^{(1)}}^{\prime}D^{ik}\chi^{(1)}  
\,+\, 6\,\Aa
 \partial_k\partial_i \omega^{(1)}\,D^{ki}\chi^{(1)} \,-\,
 D^{ik}\chi^{(1)}\,D_{ik}{\chi^{(1)}}^{\prime\prime} \,-\,
 \frac{3}{4}\,D^{ik}{\chi^{(1)}}^{\prime}D_{ki}{\chi^{(1)}}^{\prime} 
\nonumber \\
&-& 3\,\Aa
 D^{ik}\chi^{(1)}\,D_{ik}{\chi^{(1)}}^{\prime} \,
-\, 2 \,\partial_k\partial^i\,D_{mi}\chi^{(1)}\,D^{km}\chi^{(1)} \, +\, \LA
 D_{im}\chi^{(1)}\,D^{mi}\chi^{(1)} \nonumber \\
&-&\, \partial_k D^{km}\chi^{(1)}\,\partial^i D_{mi}\chi^{(1)}
 \,+\, \frac{1}{4}\,\partial^i D^{km}\chi^{(1)}\,\partial_i D_{mk}\chi^{(1)} 
\, . \nonumber 
\eea

\subsection{The Einstein tensor components}
The Einstein tensor in a spatially flat FRW background is given by

\bea
G^0_{~0} &=&  -\frac{3}{a^2} \left( \frac{a^{\prime}}{a} \right)^2 \, ,
\eea
\bea
G^i_{~j} &=& -\frac{1}{a^2} \left( 2 \frac{a^{\prime\prime}}{a}
-\Ab \right)~\delta^i_{~j} \, ,
\eea
\bea
G^0_{~i} &=& G^i_{~0} = 0 \, .
\eea
The first--order perturbations of the Einstein tensor components are
\bea
\deu {G_{~0}^{0}}&=& \frac{1}{a^2}\Bigg[ 6\,\Ab \phi^{(1)} 
\,+\,6\,\Aa{\psi^{(1)}}^{\prime}\,+\, 2\,\Aa \La
 \omega^{(1)}\,-\, 2\,\La \psi^{(1)} \\
&-& \frac{1}{2}\,\partial_k 
\partial^i
 \,D^k_i \chi^{(1)} \Bigg] \nonumber \, , 
\eea
\bea
\deu {G^0_{~i}}&=& \frac{1}{a^2}\left( -\,2\,\Aa \partial_i \phi^{(1)} 
\,-\, 2\,\partial_i
{\psi^{(1)}}^{\prime} \,-\, \frac{1}{2}\,\partial_k 
D^k_{~i}{\chi^{(1)}}^{\prime} \right)\, , 
\eea
\bea
\deu {G^i_{~j}} &=& \frac{1}{a^2} \Bigg[ \left( 2\,\Aa 
{\phi^{(1)}}^{\prime} \,+\, 4\,\Ac \phi^{(1)} \,-\,
2\,\Ab \phi^{(1)} \,+\, \LA \phi^{(1)} 
+ 4\,\Aa {\psi^{(1)}}^{\prime} \right. \\
&+& \left.   2\, {\psi^{(1)}}^{\prime\prime} 
- \LA \psi^{(1)} \,+\, 2\,\Aa \LA \omega^{(1)} \,+\, 
\LA {\omega^{(1)}}^{\prime} \,+\,
\frac{1}{2}\partial_k \partial^m D^k_{~m}
\chi^{(1)} \right)\delta_{~j}^{i} \nonumber \\
&-& \partial^i\partial_j \phi^{(1)} 
+  \partial^i\partial_j \psi^{(1)} \,-\,
2\,\Aa \partial^i\partial_j \omega^{(1)} \,-\, \partial^i\partial_j
{\omega^{(1)}}^{\prime}  \nonumber \\
&+& \Aa D^i_{~j} {\chi^{(1)}}^{\prime} 
+ \frac{1}{2}\,D^i_{~j}
{\chi^{(1)}}^{\prime\prime} 
+ \frac{1}{2}\,\partial_k\partial^i \,D^k_{~j} \chi^{(1)} \,+\,
\frac{1}{2}\,\partial_k\partial_j \,D^{ik} \chi^{(1)} \,-\,
\frac{1}{2}\,\partial_k\partial^k \,D^i_{~j} \chi^{(1)}\Bigg] \,. \nonumber 
\eea  
The second--order perturbed Einstein tensor components are given by 
\bea
\ded {G_{~0}^0} &=& \frac{1}{a^2} \Big( 3\Ab \phi^{(2)}\,+\,3\,\Aa 
{\psi^{(2)}}^{\prime} \,
-\, \La \psi^{(2)}
\,+\,\Aa \La \omega^{(2)} \,-\,\frac{1}{4} \partial_k \partial_i\,D^{ki} 
\chi^{(2)}
\\
&-&\, 12 \left( \Aa \right)^2 \left( \phi^{(1)} \right)^2 - 
12\,\Aa\,\phi^{(1)}\,{\psi^{(1)}}^{\prime} 
-\,3\,\partial_i \psi^{(1)}\,\partial^i \psi^{(1)} \,-\, 8\,\psi^{(1)}\ 
\La \psi^{(1)}
\,+\, 12\,\Aa \psi^{(1)}\,{\psi^{(1)}}^{\prime} \nonumber \\
&-& 3\,\left( {\psi^{(1)}}^{\prime} \right)^2 \,+\, 4\,\Aa\,\phi^{(1)}\,\La 
\omega^{(1)}\,-\, 2\,\Aa
\partial_k \omega^{(1)}\,\partial^k \phi^{(1)} \,-\,
\frac{1}{2} \Ac\,\partial_k \omega^{(1)}\,\partial^k \omega^{(1)} \nonumber \\
&+& \frac{1}{2}\,\partial_i\partial_k \omega^{(1)}\,\partial^i\partial^k 
\omega^{(1)}
\,-\, \frac{1}{2}\,\partial_k\partial^k \omega^{(1)}\,\partial_k\partial^k 
\omega^{(1)}
\,-\,2\,\Aa \partial_k \psi^{(1)}\,\partial^k \omega^{(1)} \,+\, 4\,\Aa 
\psi^{(1)}\,\La
\omega^{(1)} \nonumber \\
&-& 2\, \partial_k \omega^{(1)} \,\partial^k {\psi^{(1)}}^{\prime} \,-\,
2{\psi^{(1)}}^{\prime} \La \omega^{(1)} \,-\, 
\phi^{(1)}\,\partial_i\partial^k \,D^i_{~k} \chi^{(1)} \,-\,
2\,\psi^{(1)}
\partial_k\partial^i \,D^k_{~i} \chi^{(1)} \nonumber \\
&+& \partial_k \partial_i \psi^{(1)}\,D^{ki} \chi^{(1)} \,-\,2\,\Aa
\partial_i\partial_k \omega^{(1)}\,D^{ik} \chi^{(1)} \,-\,2\,\Aa \partial_k 
\omega^{(1)} \,\partial_i\,
D^{ik} \chi^{(1)}
\,-\,\partial_k \omega^{(1)}\,\partial^i D^k_{~i} {\chi^{(1)}}^{\prime} 
\nonumber \\
&-& \frac{1}{2}\,\La \,D_{mk} \chi^{(1)} \, D^{km}\chi^{(1)} \,+\,
\partial_m\partial^k \,D_{ik} \chi^{(1)}\,D^{im} \chi^{(1)} \,+\,
\frac{1}{2} \,\partial_k D^{km} \chi^{(1)}\, \partial^i D_{mi} \chi^{(1)} 
\nonumber \\
&-& \frac{1}{8}\,\partial^i D^{km}\chi^{(1)} \, \partial_i D_{km} \chi^{(1)} 
\,+\,
\frac{1}{8}\,D^{ik} {\chi^{(1)}}^{\prime} \,D_{ki} {\chi^{(1)}}^{\prime} 
\,+\, \Aa
D^{ki} \chi^{(1)} \, D_{ik} {\chi^{(1)}}^{\prime} \Big)\, , \nonumber 
\eea
\bea
\ded {G_{~0}^i} &=&\frac{1}{a^2} \Big( \Aa \partial^i \phi^{(2)} 
\,+\,\partial^i {\psi^{(2)}}^{\prime}
\,+\, \frac{1}{4}
\partial_k\,D^{ki} {\chi^{(2)}}^{\prime} \,+\, \frac{1}{4} \La 
{\chi^{i(2)}}^{\prime}
\,-\, \frac{1}{4} \La \omega^{i(2)} \\
&-& \Ac \partial^i \omega^{(2)}  
-\Ac \omega^{i(2)} \,+\,2 \Ab \partial^i \omega^{(2)} \,+\,2 \Ab 
\omega^{i(2)} \,-\, 4\, \Aa
\phi^{(1)}\,\partial^i \phi^{(1)} \,+\, 4\,\Aa\,\psi^{(1)}\,\partial^i 
\phi^{(1)}
\nonumber \\
&-& 2 \,{\psi^{(1)}}^{\prime}\partial^i \phi^{(1)}\,+\,
4\,{\psi^{(1)}}^{\prime}\partial^i \psi^{(1)} \,+\, 8\,\psi^{(1)}\,\partial^i
 {\psi^{(1)}}^{\prime}\,-\, \partial^i \phi^{(1)}\,\La \omega^{(1)} 
\,-\,\partial^k \omega^{(1)} \,
\partial^i\partial_k \phi^{(1)} \nonumber \\
&+& \La \phi^{(1)}
 \,\partial^i \omega^{(1)} \,+\, \partial^i\partial_k \omega^{(1)} 
\,\partial^k \phi^{(1)} \,+\,
 4\,\Ac \phi^{(1)}\,\partial^i \omega^{(1)} \,-\, 8 \Ab 
\phi^{(1)}\,\partial^i \omega^{(1)} \nonumber \\
&+& 2 \,\Aa {\phi^{(1)}}^{\prime} \partial^i \omega^{(1)} \,+\, 
\La {\omega^{(1)}}^{\prime}
 \,\partial^i \omega^{(1)} \,-\,\partial^k \omega^{(1)} \,\partial^i\partial_k
 {\omega^{(1)}}^{\prime}  \,+\,
 2\,{\psi^{(1)}}^{\prime \prime}\partial^i \omega^{(1)} \nonumber \\
&+&\,8\,\Ab \psi^{(1)}\,\partial^i \omega^{(1)}
 \,-\,4\,\Ac\,\psi^{(1)}\,\partial^i \omega^{(1)} \, -\, 2\,\Aa
{\psi^{(1)}}^{\prime}\partial^i \omega^{(1)} \,-\,
 \frac{1}{2}\,\partial^k \phi^{(1)} \,D^i_{~k} {\chi^{(1)}}^{\prime} 
\nonumber \\
&-&2\,\Aa\,\partial_k \phi^{(1)}\,D^{ki}\chi^{(1)}
 \,-\,
 \frac{1}{2}\,\partial_k\psi^{(1)} \,D^{ki} {\chi^{(1)}}^{\prime}\,+\,2\,
 \psi^{(1)}\,\partial_k D^{ki} {\chi^{(1)}}^{\prime}\,+\, 
{\psi^{(1)}}^{\prime} \partial_k D^{ki} \chi^{(1)}
\nonumber \\
&-&\partial_k{\psi^{(1)}}^{\prime} D^{ki} \chi^{(1)}\,+\,\frac{1}{2}
 \,\partial^k \omega^{(1)}\,D^i_{~k} {\chi^{(1)}}^{\prime\prime} 
\,+\, \Aa \partial^k \omega^{(1)} \,
 D^i_{~k} {\chi^{(1)}}^{\prime} \,-\, 4\Ab \partial_k \omega^{(1)} 
\,D^{ik} \chi^{(1)} \nonumber  \\
&+& 2\Ac \,\partial_k \omega^{(1)} \,D^{ik} \chi^{(1)} \,
-\,\frac{1}{2}\,\partial_k D^{km}\chi^{(1)} \,D_{~m}^i 
{\chi^{(1)}}^{\prime} - \frac{1}{2} \, \partial_k D_{~m}^i 
{\chi^{(1)}}^{\prime}\,D^{km} 
\chi^{(1)} \nonumber \\
&+& \frac{1}{4} \partial^i D_{mk} \chi^{(1)} \,D^{km} {\chi^{(1)}}^{\prime}
 \,+\, \frac{1}{2}\, \partial^i D_{mk} {\chi^{(1)}}^{\prime}\,D^{km} 
\chi^{(1)}
 \,-\,\frac{1}{2}\,D^{ik}\chi^{(1)} \,\partial_m D^m_{~k} 
{\chi^{(1)}}^{\prime} \Big) \, , \nonumber 
\eea
\bea
\ded {G^0_{~i}}\, &=& \frac{1}{a^2} \Big( - \,\Aa \partial_i 
\phi^{(2)} \,-\,\partial_i
{\psi^{(2)}}^{\prime} \,-\, \frac{1}{4}
\partial_k\,D^k_{~i} {\chi^{(2)}}^{\prime} \,-\, \frac{1}{4} 
\La {\chi^{(2)}_i}^{\prime}
\,+\, \frac{1}{4} \La \omega^{(2)}_i \\
&+& 8 \Aa \phi^{(1)} 
\partial_i \phi^{(1)} +\,4\,\phi^{(1)}\,\partial_i {\psi^{(1)}}^{\prime} 
+ 2 \,{\psi^{(1)}}^{\prime}\partial_i \phi^{(1)} -\, 
4\,{\psi^{(1)}}^{\prime}\partial_i
\psi^{(1)} \,-\, 4\,\psi^{(1)}\,\partial_i
 {\psi^{(1)}}^{\prime} \nonumber \\
&+& \partial_i \phi^{(1)}\,\La 
\omega^{(1)} - \partial_i\partial_k \omega^{(1)} 
\,\partial^k \phi^{(1)} 
+8 \Ac \phi^{(1)}\,\partial_i \omega^{(1)} \,-\, 4 \Ab 
\phi^{(1)}\,\partial_i \omega^{(1)} \nonumber \\
&-& 2\,\Aa \,\partial^k \omega^{(1)}\,\partial_i\partial_k 
\omega^{(1)}
  \,+\,\La \psi^{(1)}\,\partial_i \omega^{(1)} 
+ \partial^k \omega^{(1)}\,\,\partial_i\partial_k \psi^{(1)}
 \,-\,\partial_k \phi^{(1)}\,D^k_{~i} {\chi^{(1)}}^{\prime} \nonumber \\
&+&  \frac{1}{2}\,\partial^k \phi^{(1)} \,D_{ik} {\chi^{(1)}}^{\prime} 
- \psi^{(1)}\,\partial_k D^k_{~i} {\chi^{(1)}}^{\prime} +\,
 \frac{1}{2}\,\partial_k\psi^{(1)} \,D^k_{~i} {\chi^{(1)}}^{\prime}
 \,-\, {\psi^{(1)}}^{\prime} \partial_k D^k_{~i} \chi^{(1)}  \nonumber \\
&-&\,\partial_k{\psi^{(1)}}^{\prime} D^k_{~i} 
\chi^{(1)}
 \,+\,\partial_i \omega^{(1)}\, \partial_k
 \partial^m\,D^k_{~m} \chi^{(1)} \nonumber \nonumber \\
&-& 2 \,\Ac\, \partial^k \omega^{(1)}\,D_{ik} \chi^{(1)}
 \,+\,\Ab \partial^k \omega^{(1)}\,D_{ik} \chi^{(1)}  \,+\, \partial^k 
\omega^{(1)}\,\partial_m\partial_i 
\,D^m_{~k} \chi^{(1)} \nonumber \\
&-&\, \frac{1}{2}\,\partial^m \omega^{(1)} \,\partial_k \partial^k
 \,D_{im} \chi^{(1)}
  \,+\, \frac{1}{2}\,\partial_k D^{km}\chi^{(1)} \,D_{im} 
{\chi^{(1)}}^{\prime}
 \,+\,\frac{1}{2} \, \partial_k D_{im} {\chi^{(1)}}^{\prime}\,D^{km} 
\chi^{(1)} \nonumber \\
&-&
 \frac{1}{4} \partial_i D_{mk} \chi^{(1)} \,D^{km} {\chi^{(1)}}^{\prime}
 \, -\, \frac{1}{2}\, \partial_i D_{mk} {\chi^{(1)}}^{\prime}\,D^{km} 
\chi^{(1)} \Big)\, , \nonumber 
\eea
\bea
\ded {{G^{d}}^i_j} &=& \frac{1}{a^2} \Big(\frac{1}{2}\La \phi^{(2)} 
\,+\,\Aa
{\phi^{(2)}}^{\prime} \,+\,2\,\Ac\,\phi^{(2)} \,-\,\Ab \phi^{(2)} \, 
-\,\frac{1}{2}\La \psi^{(2)}  \,
+\,{\psi^{(2)}}^{\prime\prime} \\
&+&\, 2\,\Aa\,{\psi^{(2)}}^{\prime}\,+\, \Aa \,\La \omega^{(2)} \,+\, 
\frac{1}{2}
\La {\omega^{(2)}}^{\prime} 
\,-\,\frac{1}{4}\,\partial_k\partial_i\,D^{ki}\chi^{(2)}
 \,+\,4\,\Ab \left( \phi^{(1)} \right)^2\nonumber \\
&-&8\,\Ac \,\left( \phi^{(1)} 
\right)^2\,-\,8\,\Aa\,\phi^{(1)}\,{\phi^{(1)}}^{\prime} \,-\,\partial_k
\phi^{(1)}\,\partial^k \phi^{(1)} 
\,-\,2 \phi^{(1)}\,\La \phi^{(1)} \,-\, 4\,\phi^{(1)}\,
{\psi^{(1)}}^{\prime \prime}
\nonumber \\
&-&2\,{\phi^{(1)}}^{\prime}{\psi^{(1)}}^{\prime} 
\,-\,8\,\Aa\,\phi^{(1)}\,{\psi^{(1)}}^{\prime}
\,-\,2\,\partial_k \psi^{(1)} \,\partial^k \psi^{(1)} -\, 
4\,\psi^{(1)}\,\La \psi^{(1)}
\,+\, \left( {\psi^{(1)}}^{\prime} \right)^2 \nonumber \\
&+& 8\,\Aa \,\psi^{(1)}\,{\psi^{(1)}}^{\prime} 
+4\,\psi^{(1)}\,{\psi^{(1)}}^{\prime \prime}\,+\,
2\,\psi^{(1)}\,\La \phi^{(1)}
\,-\,{\phi^{(1)}}^{\prime}\La \omega^{(1)} \nonumber \\
&-& 2\,\phi^{(1)}\,\La 
{\omega^{(1)}}^{\prime}
\,-\,2\,\Aa\,\partial_k \omega^{(1)}\,\partial^k \phi^{(1)} 
-4\,\Aa \,\phi^{(1)}\,\La \omega^{(1)} \,+\, \frac{3}{2}\Ac \partial_k
\omega^{(1)}\,\partial^k \omega^{(1)} \nonumber \\
&-&  \Ab
\partial_k \omega^{(1)} \,\partial^k \omega^{(1)} \,+\, 2\,\Aa \partial_k 
\omega^{(1)} \,\partial^k 
{\omega^{(1)}}^{\prime}-\frac{1}{2}\,\La \omega^{(1)} \,\La 
\omega^{(1)}\nonumber \\
&+&\frac{1}{2}\,\partial^m\partial^k \omega^{(1)}\,\partial_m\partial_k 
\omega^{(1)}
\,+\, 4\,\Aa\,\psi^{(1)}\,\La \omega^{(1)} \,+\, 2\,\psi^{(1)}\,\La 
{\omega^{(1)}}^{\prime}\nonumber \\
&-& 2\,\partial_k \omega^{(1)}\,\partial^k
{\psi^{(1)}}^{\prime}\,-\,{\psi^{(1)}}^{\prime}\La \omega^{(1)}
 \,-\, \partial_k\partial_m \phi^{(1)}\, D^{km} \chi^{(1)}  \,-\, 
\partial_k \phi^{(1)}\,\partial_m D^{mk} 
\chi^{(1)}
 \nonumber \\
&-&\partial_k \psi^{(1)} \,\partial_m D^{mk} \chi^{(1)} \,-\, \frac{3}{2}
\,\partial_k \omega^{(1)} \,\partial^i D^k_{~i} {\chi^{(1)}}^{\prime}
 \,-\,\partial_k {\omega^{(1)}}^{\prime}\,\partial_m D^{mk} \chi^{(1)} 
\nonumber \\
&-&
\partial_k\partial_m {\omega^{(1)}}^{\prime}\,D^{km} \chi^{(1)} \,-\, 
2\,\Aa \partial^k
\omega^{(1)} \,\partial_m D^m_{~k} \chi^{(1)} \,-\, 2\,\Aa 
\partial_m\partial^k \omega^{(1)}\, D^m_{~k}
\chi^{(1)} \nonumber \\
&+&\frac{3}{4}\,\partial_k\partial^l \,D_{ml}\chi^{(1)}\,D^{km}\chi^{(1)}
\,-\,\frac{1}{2}\,\La\,D_{ml}\chi^{(1)}\,D^{ml}\chi^{(1)}
\,+\,\frac{1}{4}\,\partial_m\partial^k\,D_{lk}\chi^{(1)}\,D^{lm}\chi^{(1)} 
\nonumber \\
&+&\frac{1}{2}\,\partial_k D_{km}\chi^{(1)} \,\partial^l D^{ml}\chi^{(1)} 
- \frac{1}{8}\,\partial^l D_{km}\chi^{(1)} \,\partial_l D^{km}\chi^{(1)} 
\nonumber \\
&+& \,\frac{1}{2}\,D^{mk}\chi^{(1)}\,D_{mk}{\chi^{(1)}}^{\prime\prime} 
+ \frac{3}{8}\, D^{mk}{\chi^{(1)}}^{\prime}\,D_{mk}{\chi^{(1)}}^{\prime} 
\,+\,
\Aa\,D^{mk}\chi^{(1)}\,D_{km}{\chi^{(1)}}^{\prime} \Big) \delta^i_{~j} \, 
.\nonumber 
\eea
\bea
\ded{{G^{nd}}_j^i}&=& \frac{1}{a^2} \Big[ 
-\frac{1}{2} \,\partial^i\partial_j \phi^{(2)}
\,+\,\frac{1}{2}\,\partial^i\partial_j \psi^{(2)} \,-\, \Aa
\partial^i\partial_j \omega^{(2)} \,-\, \frac{1}{2}\,\partial^i\partial_j
{\omega^{(2)}}^{\prime} \\
&-& \frac{1}{2} \Aa \left( \partial^i 
\,\omega^{(2)}_j +
\partial_j\, \omega^{i(2)}\right) 
-\, \frac{1}{4} \left( \partial^i \,{\omega^{(2)}_j}^{\prime} + 
\partial_j\,
{\omega^{i(2)}}^{\prime} \right) \nonumber \\
&+&  \frac{1}{2} \Aa \left( D^i_{~j}
{\chi^{(2)}}^{\prime} +
\partial^i {\chi^{(2)}_j}^{\prime}
+ \partial_j {\chi^{i(2)}}^{\prime} + {\chi^{i(2)}_{~j}}^{\prime}\right)
\,+\, \frac{1}{2} \partial_k\partial^i\,D^k_{~j} \chi^{(2)} \nonumber \\
&-&\,\frac{1}{4} \La D^i_{~j} \chi^{(2)} \,-\, \frac{1}{4} \La 
\chi^{i(2)}_{~j}\,+\,
\frac{1}{4}\left( D^i_{~j} {\chi^{(2)}}^{\prime\prime} +
\partial^i {\chi^{(2)}_j}^{\prime\prime}
+ \partial_j {\chi^{i(2)}}^{\prime\prime} + 
{\chi^{i(2)}_{~j}}^{\prime\prime}\right) \nonumber \\
&+& \, \partial^i \phi^{(1)}\, \partial_j \phi^{(1)} 
+\, 2\,\phi^{(1)}\,\partial^i\partial_j \phi^{(1)} \,-\, 
2\,\psi^{(1)}
\partial^i\partial_j \phi^{(1)}\,-\,
\partial_j \phi^{(1)}\,
\partial^i\psi^{(1)} \,-\, \partial^i \phi^{(1)}\,\partial_j \psi^{(1)} 
\nonumber \\
&+& 3\,\partial^i \psi^{(1)}\,\partial_j \psi^{(1)} \,+\, 4 
\,\psi^{(1)}\,\partial^i\partial_j \psi^{(1)} 
+\,2\,\Aa\,\partial^i \omega^{(1)}\,\partial_j \phi^{(1)} \,+\, 4 \Aa
\phi^{(1)}\,\partial^i\partial_j \omega^{(1)} \nonumber \\
&+& {\phi^{(1)}}^{\prime}\,\partial^i\partial_j \omega^{(1)}
\,+\, 2\,\phi^{(1)} \,\partial^i\partial_j {\omega^{(1)}}^{\prime} 
\,+\, \La \omega^{(1)}
\,\partial^i\partial_j \omega^{(1)}
- \partial_j
\partial^k \omega^{(1)}\,\partial^i \partial_k \omega^{(1)} \nonumber \\
&-& 2 \, \Aa \partial^i \psi^{(1)} \,\partial_j \omega^{(1)} \,-\, 2\,\Aa
\partial^i \omega^{(1)}\,\partial_j \psi^{(1)} \,-\, \partial^i
{\psi^{(1)}}^{\prime}\,\partial_j \omega^{(1)} \,+\, \partial_j
{\psi^{(1)}}^{\prime}\,\partial^i \omega^{(1)} \nonumber \\
&-&\, \partial^i \psi^{(1)} \,\partial_j {\omega^{(1)}}^{\prime} \,-\, 
\partial_j
\psi^{(1)} \,\partial^i {\omega^{(1)}}^{\prime} \,-\, 
2\,\psi^{(1)}\,\partial^i \partial_j
{\omega^{(1)}}^{\prime} \,+\, {\psi^{(1)}}^{\prime}\,\partial^i\partial_j 
\omega^{(1)} \nonumber \\
&-& 4\,\Aa\,\psi^{(1)}\,\partial^i\partial_j \omega^{(1)}
-2\,\Aa \phi^{(1)}\,D^i_{~j} {\chi^{(1)}}^{\prime}\,-\,\frac{1}{2} 
{\phi^{(1)}}^{\prime}\,D^i_{~j}
{\chi^{(1)}}^{\prime} \,-\, \phi^{(1)}\,D^i_{~j} 
{\omega^{(1)}}^{\prime\prime} \nonumber \\
&+&  \frac{1}{2}
\partial_k \phi^{(1)} \, \partial^i D^k_{~j} \chi^{(1)} \,+\, \frac{1}{2} 
\partial_k \phi^{(1)}
\, \partial_j D^{ki} \chi^{(1)} - \frac{1}{2} \partial_k \phi^{(1)} \,
\partial^k D^i_{~j} \chi^{(1)} \,+\,\partial_j \partial_k 
\phi^{(1)}\,D^{ki}\chi^{(1)} \nonumber \\
&+& \frac{1}{2}
{\psi^{(1)}}^{\prime}\, D^i_{~j} {\chi^{(1)}}^{\prime} 
\,+\,{\psi^{(1)}}^{\prime\prime}D^i_{~j} \chi^{(1)}
\,+\, 2\,\Aa {\psi^{(1)}}^{\prime}\,D^i_{~j} \chi^{(1)} 
+ \frac{1}{2} \partial_k \psi^{(1)}\,
\partial^i D^k_{~j} \chi^{(1)} \nonumber \\
&+& 2\,\Aa \,\psi^{(1)}\,D^i_{~j} 
{\chi^{(1)}}^{\prime}
\,+\, \psi^{(1)}\,D^i_{~j} {\omega^{(1)}}^{\prime\prime} 
+ \frac{1}{2} \partial_k
\psi^{(1)}\,
\partial_j D^{ki} \chi^{(1)} \,-\,\frac{3}{2}\, \partial_k \psi^{(1)} 
\,\partial^k D^i_{~j} \chi^{(1)} \nonumber \\
&+& 2\, \psi^{(1)}\,\partial_k\partial^i D^k_{~j} \chi^{(1)} \,+\,2\,
\psi^{(1)}\,\partial_k\partial_j D^{ki} \chi^{(1)}\,-\,2\,
\psi^{(1)}\,\partial_k\partial^k D^i_{~j} \chi^{(1)}\,-\La \psi^{(1)} 
\,D^i_{~j} \chi^{(1)} 
\nonumber \\
&+&\partial^i\psi^{(1)}\,\partial_k D^k_{~j} \chi^{(1)} 
+\,\partial_j\psi^{(1)}\,\partial_k D^{ki} \chi^{(1)} \,+\,
\partial_k\partial^i \psi^{(1)}\,D^k_{~j} \chi^{(1)}
\,+\, \frac{1}{2}\,\partial^i \omega^{(1)}\,\partial_k D^k_{~j} 
{\chi^{(1)}}^{\prime} \nonumber \\
&+& \frac{1}{2}\,\partial_k\partial^i \omega^{(1)} \, D^k_{~j} 
{\chi^{(1)}}^{\prime}
+ \frac{1}{2}\,\partial_k\partial_j \omega^{(1)} \, D^{ki} 
{\chi^{(1)}}^{\prime}
\,-\, \frac{1}{2} \,\partial_k\partial^k \omega^{(1)}\, D^i_j 
{\chi^{(1)}}^{\prime}
\,+\,\frac{1}{2}\, \partial^k \omega^{(1)}\, \partial^i D_{kj} 
{\chi^{(1)}}^{\prime} \nonumber \\
&+&\,\frac{1}{2}\, \partial^k \omega^{(1)}\, \partial_j D^i_{~k}
{\chi^{(1)}}^{\prime} 
-\partial^k \omega^{(1)} \,\partial_k D^i_{~j} {\chi^{(1)}}^{\prime} 
+ \frac{1}{2} \,\partial^k {\omega^{(1)}}^{\prime} \,\partial^i D_{kj} 
\chi^{(1)} 
+ \frac{1}{2} \,\partial^k {\omega^{(1)}}^{\prime} \,\partial_j D_{~k}^i 
\chi^{(1)} \nonumber \\
&-&  \frac{1}{2} \,\partial^k {\omega^{(1)}}^{\prime} \,\partial_k 
D^i_{~j} \chi^{(1)} 
+ \,\partial_k \partial_j {\omega^{(1)}}^{\prime}\,D^{ik}\chi^{(1)} 
\,+\, \Aa\partial^k \omega^{(1)} \,\partial^i
D_{kj} \chi^{(1)} \,+\, \Aa \partial^k \omega^{(1)} \,\partial_j D_{~k}^i 
\chi^{(1)} \nonumber \\
&-& \Aa
\partial^k \omega^{(1)} \,\partial_k D^i_{~j} \chi^{(1)} + 
2\,\Aa\,\partial_k \partial_j
\omega^{(1)}\,D^{ik}\chi^{(1)} \,-\,\frac{1}{2}\, D^{ki} 
{\chi^{(1)}}^{\prime} \, D_{kj}
{\chi^{(1)}}^{\prime}  \nonumber \\
&-& \frac{1}{2} \partial^i D_{mj} \chi^{(1)} \,
\partial_k D^{km} \chi^{(1)} 
- \frac{1}{2} \partial_j D_{~m}^i \chi^{(1)} \,
\partial_k D^{km} \chi^{(1)}  \, +\,
\frac{1}{2} \partial_m D^i_{~j} \chi^{(1)} \, \partial_k D^{km} 
\chi^{(1)} \nonumber \\
&-&\frac{1}{2} \partial_k\partial^i D_{mj} \chi^{(1)} \, D^{km} 
\chi^{(1)}  
- \frac{1}{2} \partial_k\partial_j D^i_{~m} \chi^{(1)} \, D^{km} 
\chi^{(1)} \,+\,
\frac{1}{2} \partial_k\partial_m D^i_{~j} \chi^{(1)} \, D^{km} 
\chi^{(1)} \nonumber \\
&+& \frac{1}{2} D^{km} \chi^{(1)} \,\partial^i\partial_j D_{km} 
\chi^{(1)} 
+\, \frac{1}{4} \partial^i D^{mk} \chi^{(1)} \,\partial_j D_{mk} 
\chi^{(1)}
\,-\,\partial_k\partial^m \,D^k_{~m} \chi^{(1)}\,D^i_{~j} \chi^{(1)} 
\nonumber \\
&-& \Aa
\,D_{kj}{\chi^{(1)}}^{\prime}D^{ik}\chi^{(1)} 
-\frac{1}{2}\,D_{kj}{\omega^{(1)}}^{\prime\prime}D^{ki}\chi^{(1)} \,-\, 
\partial_m
\partial_k \,D^m_{~j} \chi^{(1)}\,D^{ki}\chi^{(1)} \nonumber \\
&+& \frac{1}{2}\,\partial_m\partial^m \,D_{kj}\chi^{(1)}\,D^{ki}\chi^{(1)} 
\Big]\, , \nonumber 
\eea
where $\ded{{G^{d}}^i_{~j}}$ stands for the diagonal part of $\ded{{G}_j^i}$, 
which is proportional to $\delta^i_{~j}$, and $\ded{{G^{nd}}^i_{~j}}$ is 
the non--diagonal contribution.

\section{\bf Perturbing the Klein--Gordon equation}
\label{B}

In the homogeneous background the Klein--Gordon equation for 
the scalar field $\varphi$ is 
\be
\label{KG0}
{\varphi_0}^{\prime\prime}+2 \frac{a^{\prime}}{a}{\varphi_0}^{\prime}=
-\frac{\partial V}{\partial \varphi} a^2
\ee
The perturbed Klein--Gordon equation at first order is 
\bea
\label{KG1App}
&{\deu\varphi}&^{\prime\prime}+2\,\Aa {\deu \varphi}^{\prime}
- \La \deu \varphi- {\phi^{(1)}}^{\prime}{\varphi_0}^{\prime}-
3\,{\psi^{(1)}}^{\prime}{\varphi_0}^{\prime}
-\La {\omega^{(1)}} 
\,{\varphi_0}^{\prime}= \\
&-& \deu \varphi \,\frac{\partial^2 V}{\partial \varphi^2}\,a^2-
2\,\phi^{(1)}\,\frac{\partial V}{\partial \varphi}. \nonumber 
\eea
At second order we get
\bea
\label{KG2App}
&-& \frac{1}{2}{\ded{\varphi}}^{\prime\prime} \,-\,
\,\Aa {\ded{\varphi}}^{\prime} \,+\,\frac{1}{2} \La \ded{\varphi} \,+\,
{\phi^{(2)}}\,{\varphi_0}^{\prime\prime} \,+\, 2\,\Aa 
{\phi^{(2)}}\,{\varphi_0}^{\prime} \,+\,
\frac{1}{2}\,{\phi^{(2)}}^{\prime}{\varphi_0}^{\prime}\\ 
&+& \frac{3}{2}\,{\psi^{(2)}}^{\prime}\,{\varphi_0}^{\prime} \,+\,
\frac{1}{2}\,\La {\omega^{(2)}} \,{\varphi_0}^{\prime} \,-\,
4\,\left( \phi^{(1)} \right)^2\,{\varphi_0}^{\prime\prime} 
\,-\, 8\,\Aa \left( \phi^{(1)} \right)^2\,
{\varphi_0}^{\prime}
\,-\, 4\,\phi^{(1)}\,{\phi^{(1)}}^{\prime}{\varphi_0}^{\prime} 
\nonumber  \nonumber \\
&+& 2\,{\phi^{(1)}}\,{\deu{\varphi}}^{\prime\prime} \, +\,
{\phi^{(1)}}^{\prime}{\deu{\varphi}}^{\prime} \,+\, 4\,\Aa
{\phi^{(1)}}\,{\deu{\varphi}}^{\prime} \,+\,
\partial^k {\phi^{(1)}}\,\partial_k {\deu{\varphi}} \,-\, 6
\,{\phi^{(1)}}\,{\psi^{(1)}}^{\prime}{\varphi_0}^{\prime} \nonumber 
\nonumber \\
&+& 6\,{\psi^{(1)}}\,{\psi^{(1)}}^{\prime}{\varphi_0}^{\prime} \, +\,
3\,{\psi^{(1)}}^{\prime}{\deu{\varphi}}^{\prime}\,-\, \partial^k
{\psi^{(1)}}\,\partial_k {\deu{\varphi}} \,+\, 2\,{\psi^{(1)}}\,\La 
\deu{\varphi} \nonumber  \nonumber \\
&-& 2\,{\phi^{(1)}}\,\La {\omega^{(1)}} \,{\varphi_0}^{\prime} \,-\, 
\partial^k {\omega^{(1)}}\,\partial_k
{\phi^{(1)}} \,{\varphi_0}^{\prime} \,-\,
\partial^k {\omega^{(1)}} \,\partial_k {\psi^{(1)}} \,{\varphi_0}^{\prime}
\,+\, 2\,{\psi^{(1)}}\,\La {\omega^{(1)}} \,{\varphi_0}^{\prime}\nonumber \\
&+& \partial^k {\omega^{(1)}}\,\partial_k {\omega^{(1)}} 
\,{\varphi_0}^{\prime\prime} \,+\,
2\,\Aa \partial^k {\omega^{(1)}} \,\partial_k {\omega^{(1)}}
\,{\varphi_0}^{\prime} \,+\, \partial_k
{\omega^{(1)}} \,\partial^k {\omega^{(1)}}^{\prime}{\varphi_0}^{\prime} \,
+\, 2\,\partial^k
{\omega^{(1)}}\,\partial_k \,{\deu{\varphi}}^{\prime}\nonumber \\
&+& 2\,\Aa
\partial^k {\omega^{(1)}}\,\partial_k \,\deu{\varphi} \,
+\, \La {\omega^{(1)}}\,{\deu{\varphi}}^{\prime}
\,+\, \partial^k {\omega^{(1)}}^{\prime}\partial_k \deu{\varphi} \,
-\, \partial^k
{\omega^{(1)}}\,\partial_i D^i_{~k} {\chi^{(1)}}\,{\varphi_0}^{\prime}
\nonumber \\
&-&
\partial_i\partial_k {\omega^{(1)}}\,D^{ik}{\chi^{(1)}}\,{\varphi_0}^{\prime}
\,-\, \partial_i \partial_k \,\deu{\varphi} \,D^{ik}{\chi^{(1)}} \,-\,
\partial_k\, \deu{\varphi}\, \partial^i D^k_{~i} {\chi^{(1)}}  \nonumber \\
&+&
\frac{1}{2}\,D^{ik}{\chi^{(1)}}\,D_{ki}{\chi^{(1)}}^{\prime}\,
{\varphi_0}^{\prime} = 
\frac{1}{2} \frac{\partial^2 V}{\partial \varphi^2}\,\ded{\varphi}\, a^2
\,+\, \frac{1}{2}\,\frac{\partial^3 V}{\partial \varphi^3} 
(\deu{\varphi})^2 \,a^2 \nonumber \, .
\eea
To obtain the Klein--Gordon equation in the Poisson gauge 
one can simply set $\omega^{(1)}=\omega^{(2)}=0$, $\chi^{(1)}=\chi^{(2)}=0$, 
and $\phi^{(1)}=\psi^{(1)}$.
Thus at first order we find
\be
\label{KG1L}
{\deu\varphi}^{\prime\prime}+2\,\Aa {\deu \varphi}^{\prime}
- \La \deu \varphi- 4 {\phi^{(1)}}^{\prime}{\varphi_0}^{\prime}=
- \deu \varphi \,\frac{\partial^2 V}{\partial \varphi^2}\,a^2-
2\,{\phi^{(1)}}\,\frac{\partial V}{\partial \varphi} \, , 
\ee
while at second order the equation is
\bea
\label{KG2L}
&\frac{1}{2}&{\ded \varphi}^{\prime\prime} \,+\,\Aa {\ded
\varphi}^{\prime} \,-\frac{1}{2} \La \ded\varphi \,
-\, {\phi^{(2)}} \left( {\varphi_0}^{\prime\prime}
+2\,\Aa {\varphi_0}^{\prime} \right) \\
&-& \frac{1}{2}\,{\phi^{(2)}}^{\prime} {\varphi_0}^{\prime} 
-\, \frac{3}{2}\,{\psi^{(2)}}^{\prime}\,{\varphi_0}^{\prime}
-4\,{\phi^{(1)}}\,{\phi^{(1)}}^{\prime}{\varphi_0}^{\prime}
\,-\,4\, {\phi^{(1)}}^{\prime}{\deu \varphi}^{\prime} \nonumber \\
&-&\, 4\,{\phi^{(1)}}\,\La \deu \varphi \,=
-2 {\phi^{(1)}}\,\deu{\varphi}\frac{\partial^2 V}{\partial \varphi^2}\,a^2\,
-\,
\frac{1}{2}\ded \varphi\, \frac{\partial^2 V}{\partial \varphi^2}\,a^2
\,-\, \frac{1}{2}\,(\deu \varphi)^2\,\frac{\partial^3 V}{\partial \varphi^3}  
\,a^2 \, ,\nonumber
\eea
where we have used the background equation (\ref{KG0}) and the first--order 
perturbed equation (\ref{KG1L}) to simplify some terms.

%
%
\section{\bf Wigner 3-$j$ symbol}
\label{app:wigner}

In this appendix, we summarize basic properties of the 
Wigner 3--$j$ symbol, following Ref.~\cite{rotenberg:1959}.
The Wigner 3--$j$ symbol characterizes geometric properties of 
the angular bispectrum.

\subsection{Triangle conditions}

The Wigner 3--$j$ symbol,
\begin{equation}
 \label{eq:3j}
 \left(\begin{array}{ccc}l_1&l_2&l_3\\m_1&m_2&m_3\end{array}\right),
\end{equation}
is related to the Clebsh--Gordan coefficients which describe
coupling of two angular momenta in the quantum mechanics.
In the quantum mechanics, $l$ is the eigenvalue of the angular 
momentum operator, ${\mathbf L}={\mathbf r}\times {\mathbf p}$: 
${\mathbf L}^2 Y_{lm}= l(l+1) Y_{lm}$.
$m$ is the eigenvalue of the $z$-direction component of the 
angular momentum, $L_z Y_{lm}= m Y_{lm}$.

The symbol such as  
\begin{equation}
 (-1)^{m_3}\left(\begin{array}{ccc}l_1&l_2&l_3\\m_1&m_2&-m_3\end{array}\right)
\end{equation}
describes coupling of two angular--momentum states,
${\mathbf L}_1$ and ${\mathbf L}_2$, forming a coupled state,
${\mathbf L}_3={\mathbf L}_1+{\mathbf L}_2$.
It follows from ${\mathbf L}_1+{\mathbf L}_2-{\mathbf L}_3=0$ that
$m_1+m_2-m_3=0$; thus, the Wigner 3--$j$ symbol~(\ref{eq:3j}) 
describes three angular momenta forming a triangle, 
${\mathbf L}_1+{\mathbf L}_2+{\mathbf L}_3=0$, and 
satisfies $m_1+m_2+m_3=0$.

Since ${\mathbf L}_1$, ${\mathbf L}_2$, and ${\mathbf L}_3$ form a 
triangle, they have to satisfy the triangle conditions, 
$\left|L_i-L_j\right|\le L_k\le L_i+L_j$,
where $L_i\equiv \left|{\mathbf L}_i\right|$.
Hence, $l_1$, $l_2$, and $l_3$ also satisfy the triangle conditions,
\begin{equation}
 \left|l_i-l_j\right|\le l_k\le l_i+l_j;
\end{equation}
otherwise, the Wigner 3--$j$ symbol vanishes.
The triangle conditions also include $m_1+m_2+m_3=0$.
These properties may regard ($l$, $m$) as vectors, ${\mathbf l}$,
which satisfy ${\mathbf l}_1+{\mathbf l}_2+{\mathbf l}_3=0$.
Note that, however, ${\mathbf L}\neq {\mathbf l}$.

For $l_1=l_2$ and $l_3=m_3=0$, the Wigner 3--$j$ symbol reduces to
\begin{equation}
 (-1)^m\left(\begin{array}{ccc}l&l&0\\m&-m&0\end{array}\right)
 =
  \frac{(-1)^l}{\sqrt{2l+1}} .
\end{equation}
In Sec.~\ref{sec:npointspectrum}, we have used this relation to reduce 
the covariance matrix of the angular bispectrum and trispectrum.
We have also used this relation to reduce the angular trispectrum
for $L=0$ (see Eq.~(\ref{eq:special})).

\subsection{Symmetry}

The Wigner 3--$j$ symbol is invariant under even permutations,
\begin{equation}
 \left(\begin{array}{ccc}l_1&l_2&l_3\\m_1&m_2&m_3\end{array}\right)
 =\left(\begin{array}{ccc}l_3&l_1&l_2\\m_3&m_1&m_2\end{array}\right)
 =\left(\begin{array}{ccc}l_2&l_3&l_1\\m_2&m_3&m_1\end{array}\right),
\end{equation}
while it changes the phase for odd permutations if
$l_1+l_2+l_3=\mbox{odd}$,
\begin{eqnarray}
 & &
 (-1)^{l_1+l_2+l_3}
	\left(\begin{array}{ccc}l_1&l_2&l_3\\m_1&m_2&m_3\end{array}\right)\\
 &=&\left(\begin{array}{ccc}l_2&l_1&l_3\\m_2&m_1&m_3\end{array}\right)
 =\left(\begin{array}{ccc}l_1&l_3&l_2\\m_1&m_3&m_2\end{array}\right)
 =\left(\begin{array}{ccc}l_3&l_2&l_1\\m_3&m_2&m_1\end{array}\right).
\end{eqnarray}

The phase also changes under the transformation of $m_1+m_2+m_3\rightarrow 
-(m_1+m_2+m_3)$,
if $l_1+l_2+l_3=\mbox{odd}$,
\begin{equation}
 \left(\begin{array}{ccc}l_1&l_2&l_3\\m_1&m_2&m_3\end{array}\right)
 =(-1)^{l_1+l_2+l_3}
  \left(\begin{array}{ccc}l_1&l_2&l_3\\-m_1&-m_2&-m_3\end{array}\right).
\end{equation}
If there is no $z$--direction component of the angular momenta 
in the system, i.e., $m_i=0$, then the Wigner 3--$j$ symbol of the system,
\begin{equation}
 \left(\begin{array}{ccc}l_1&l_2&l_3\\0&0&0\end{array}\right),
\end{equation}
is non--zero only if $l_1+l_2+l_3=\mbox{even}$.
This symbol is invariant under any permutations of $l_i$.

In Sec.~\ref{sec:theory_bl}, we have frequently used the Gaunt 
integral, ${\mathcal G}^{m_1m_2m_3}_{l_1l_2l_3}$, defined by
\begin{eqnarray}
  \nonumber
  {\mathcal G}_{l_1l_2l_3}^{m_1m_2m_3}
  &\equiv&
  \int d^2\hat{\mathbf n}
  Y_{l_1m_1}(\hat{\mathbf n})
  Y_{l_2m_2}(\hat{\mathbf n})
  Y_{l_3m_3}(\hat{\mathbf n})\\
 \nonumber
  &=&\sqrt{
   \frac{\left(2l_1+1\right)\left(2l_2+1\right)\left(2l_3+1\right)}
        {4\pi}
        }
  \left(
  \begin{array}{ccc}
  l_1 & l_2 & l_3 \\ 0 & 0 & 0 
  \end{array}
  \right)\\
 & &\times
  \left(
  \begin{array}{ccc}
  l_1 & l_2 & l_3 \\ m_1 & m_2 & m_3 
  \end{array}
  \right),
\end{eqnarray}
to calculate the angular bispectrum.
By definition, the Gaunt integral is invariant under 
both the odd and the even permutations, and non-zero only if
$l_1+l_2+l_3=\mbox{even}$, $m_1+m_2+m_3=0$, and
$\left|l_i-l_j\right|\le l_k\le l_i+l_j$.
In other words, the Gaunt integral describes fundamental geometric 
properties of the angular bispectrum such as
the triangle conditions.

The Gaunt integral for $m_i=0$ gives the identity for the Legendre polynomials,
\begin{equation}
 \int_{-1}^{1}\frac{dx}2~
  P_{l_1}(x)P_{l_2}(x)P_{l_3}(x)
  =
  \left(\begin{array}{ccc}l_1&l_2&l_3\\0&0&0\end{array}\right)^2.
\end{equation}
In Sec.~\ref{sec:npointspectrum}, we have used this identity to derive 
the bias for the angular bispectrum on the incomplete sky
(Eq.~(\ref{eq:biasbl})).
Here, we have used
\begin{equation}
 Y_{l0}(\hat{\mathbf n})= \sqrt{\frac{4\pi}{2l+1}}P_l(\cos\theta).
\end{equation}

\subsection{Orthogonality}

The Wigner 3--$j$ symbol has the following orthogonality properties:
\begin{equation}
 \sum_{l_3m_3}
  (2l_3+1)
 \left(\begin{array}{ccc}l_1&l_2&l_3\\m_1&m_2&m_3\end{array}\right)
 \left(\begin{array}{ccc}l_1&l_2&l_3\\m'_1&m'_2&m_3\end{array}\right)
 =
  \delta_{m_1m_1'}\delta_{m_2m_2'},
\end{equation}
and
\begin{equation}
 \sum_{m_1m_2}
 \left(\begin{array}{ccc}l_1&l_2&l_3\\m_1&m_2&m_3\end{array}\right)
 \left(\begin{array}{ccc}l_1&l_2&l'_3\\m_1&m_2&m'_3\end{array}\right)
 =
  \frac{\delta_{l_3l_3'}\delta_{m_3m_3'}}{2l_3+1},
\end{equation}
or
\begin{equation}
 \label{eq:normalrelation}
 \sum_{{\rm all}~m}
 \left(\begin{array}{ccc}l_1&l_2&l_3\\m_1&m_2&m_3\end{array}\right)^2
 = 1.
\end{equation}
The orthogonality properties are essential for any basic calculations 
involving the Wigner 3--$j$ symbols.
Note that these orthogonality properties are consistent with 
orthonormality of the angular--momentum eigenstate vectors,
and unitarity of the Clebsh--Gordan coefficients, by definition.

\subsection{Rotation matrix}

A finite rotation operator for the Euler angles
$\alpha$, $\beta$, and $\gamma$, $D(\alpha,\beta,\gamma)$, 
comprises angular momentum operators,
\begin{equation}
 D(\alpha,\beta,\gamma)= e^{-i\alpha L_z}e^{-i\beta L_y}e^{-i\gamma L_z}.
\end{equation}
Since the Wigner 3--$j$ symbol describes coupling of two angular momenta,
it also describes coupling of two rotation operators.
Using the rotation matrix element, 
$D_{m'm}^{(l)}=\left<l,m'\left|D\right|l,m\right>$,
we have
\bea
 \label{eq:rotreduce}
 D_{m'_1m_1}^{(l_1)}D_{m'_2m_2}^{(l_2)}
  =
  \sum_{l_3}(2l_3+1)
  \sum_{m_3m'_3}D_{m'_3m_3}^{(l_3)*}
  \left(\begin{array}{ccc}l_1&l_2&l_3\\m_1&m_2&m_3\end{array}\right)
  \left(\begin{array}{ccc}l_1&l_2&l_3\\m'_1&m'_2&m'_3\end{array}\right)
\, \nonumber \\
\eea
In Sec.~\ref{sec:npointspectrum}, we have used this relation to 
evaluate rotationally invariant harmonic spectra.
Note that the rotation matrix is orthonormal, 
\begin{equation}
 \sum_{m}D_{m'm}^{(l)*}D_{m''m}^{(l)}= \delta_{m'm''}.
\end{equation}

\subsection{Wigner 6-$j$ symbol}

The Wigner 6-$j$ symbol,
\begin{equation}
 \label{eq:6j}
 \left\{\begin{array}{ccc}l_1&l_2&l_3\\l'_1&l'_2&l'_3\end{array}\right\},
\end{equation}
describes coupling of three angular momenta.
We often encounter the Wigner 6-$j$ symbol, when we calculate 
the angular bispectrum which has more complicated geometric structures
(see, e.g., Ref.~\cite{goldberg/spergel:1999}, Appendix C of 
Ref.~\cite{komatsu:phd}).
The angular trispectrum also often includes the Wigner 6-$j$ symbol
\cite{hu:2001}.

The Wigner 6-$j$ symbol is related to the Wigner 3--$j$ symbols through
\begin{eqnarray}
 \nonumber
  & &
  (-1)^{l_1'+l_2'+l_3'}
 \left\{\begin{array}{ccc}l_1&l_2&l_3\\l'_1&l'_2&l'_3\end{array}\right\}
 \left(\begin{array}{ccc}l_1&l_2&l_3\\m_1&m_2&m_3\end{array}\right)\\
 \nonumber
 & &=
 \sum_{{\rm all}~m'}
 (-1)^{m_1'+m_2'+m_3'}\\
 & &\times
 \left(\begin{array}{ccc}l_1&l'_2&l'_3\\m_1&m'_2&-m'_3\end{array}\right)
 \left(\begin{array}{ccc}l'_1&l_2&l'_3\\-m'_1&m_2&m'_3\end{array}\right)
 \left(\begin{array}{ccc}l'_1&l'_2&l_3\\m'_1&-m'_2&m_3\end{array}\right).
\end{eqnarray}
By using Eq.~(\ref{eq:normalrelation}), we also obtain
\begin{eqnarray}
 \nonumber
  (-1)^{l_1'+l_2'+l_3'}
 \left\{\begin{array}{ccc}l_1&l_2&l_3\\l'_1&l'_2&l'_3\end{array}\right\}
 &=&
 \sum_{{\rm all}~mm'}
 (-1)^{m_1'+m_2'+m_3'}\\
 \nonumber
  & &\times
 \left(\begin{array}{ccc}l_1&l_2&l_3\\m_1&m_2&m_3\end{array}\right)
 \left(\begin{array}{ccc}l_1&l'_2&l'_3\\m_1&m'_2&-m'_3\end{array}\right)\\
 & &\times
 \left(\begin{array}{ccc}l'_1&l_2&l'_3\\-m'_1&m_2&m'_3\end{array}\right)
 \left(\begin{array}{ccc}l'_1&l'_2&l_3\\m'_1&-m'_2&m_3\end{array}\right).
 \nonumber \\ 
\end{eqnarray}
           

\end{document}